\RequirePackage[running]{lineno}
\documentclass[aps,prd,nofootinbib,twocolumn,showpacs,superscriptaddress,letterpaper,amsmath,amsfonts,amssymb,preprintnumbers]{revtex4-1}
\pdfoutput=1 

\usepackage{bm,amsmath,amsfonts,amssymb,graphicx,xspace,placeins,multirow,wasysym}
\usepackage[T1]{fontenc}

\usepackage{graphicx}
\usepackage{dcolumn}
\usepackage{bm}
\usepackage{xspace}
\usepackage{siunitx}

\usepackage{adjustbox}
\usepackage{rotating}
\usepackage[autostyle=true]{csquotes}

\usepackage[colorlinks,hyperindex,breaklinks]{hyperref}
\hypersetup{linkcolor=blue,citecolor=blue,filecolor=black,urlcolor=blue}

\def\bxlnu{\ensuremath{B \to X \, \ell^+\, \nu_{\ell}}\xspace}
\def\bulnu{\ensuremath{B \to X_u \, \ell^+\, \nu_{\ell}}\xspace}
\def\bclnu{\ensuremath{B \to X_c \, \ell^+\, \nu_{\ell}}\xspace}

\def\lumi{\ensuremath{711 \, \mathrm{fb}^{-1}}\xspace}
\def\lowqSqCut{\ensuremath{3.0 \, \mathrm{GeV}^2}\xspace}
\def\highqSqCut{\ensuremath{10.0 \, \mathrm{GeV}^2}\xspace}
\def\qSqCut{\ensuremath{q_{\mathrm{sel}}^2 \in [3.0,10.0] \, \mathrm{GeV}^2}\xspace}

\def\bdlnu{\ensuremath{B \to D \, \ell^+\,\nu_{\ell}}\xspace}
\def\bdslnu{\ensuremath{B \to D^* \, \ell^+\,\nu_{\ell}}\xspace}
\def\bddslnu{\ensuremath{B \to D^{**} \, \ell^+\,\nu_{\ell}}\xspace}

\def\bsdlnu{\ensuremath{B_s \to D_s \, \ell^+\,\nu_{\ell}}\xspace}
\def\bsdslnu{\ensuremath{B_s \to D^*_s \, \ell^+\,\nu_{\ell}}\xspace}

\def\Eg{\ensuremath{E_{\gamma}}\xspace}

\begin{document}

\title{Measurements of $q^2$ Moments of Inclusive \bclnu Decays with Hadronic Tagging}

\author{R. van Tonder}
\email{vantonder@physik.uni-bonn.de}
\affiliation{University of Bonn, 53115 Bonn}

\author{L. Cao}
\affiliation{University of Bonn, 53115 Bonn}

\author{W. Sutclif\mbox{}fe}
\affiliation{University of Bonn, 53115 Bonn}

\author{M. Welsch}
\affiliation{University of Bonn, 53115 Bonn}

\author{F. U.\ Bernlochner}
\email{florian.bernlochner@uni-bonn.de}
\affiliation{University of Bonn, 53115 Bonn}

\noaffiliation
\affiliation{Department of Physics, University of the Basque Country UPV/EHU, 48080 Bilbao}
\affiliation{University of Bonn, 53115 Bonn}
\affiliation{Brookhaven National Laboratory, Upton, New York 11973}
\affiliation{Budker Institute of Nuclear Physics SB RAS, Novosibirsk 630090}
\affiliation{Faculty of Mathematics and Physics, Charles University, 121 16 Prague}
\affiliation{Chonnam National University, Gwangju 61186}
\affiliation{University of Cincinnati, Cincinnati, Ohio 45221}
\affiliation{Deutsches Elektronen--Synchrotron, 22607 Hamburg}
\affiliation{Department of Physics, Fu Jen Catholic University, Taipei 24205}
\affiliation{Key Laboratory of Nuclear Physics and Ion-beam Application (MOE) and Institute of Modern Physics, Fudan University, Shanghai 200443}
\affiliation{Justus-Liebig-Universit\"at Gie\ss{}en, 35392 Gie\ss{}en}
\affiliation{Gifu University, Gifu 501-1193}
\affiliation{SOKENDAI (The Graduate University for Advanced Studies), Hayama 240-0193}
\affiliation{Department of Physics and Institute of Natural Sciences, Hanyang University, Seoul 04763}
\affiliation{University of Hawaii, Honolulu, Hawaii 96822}
\affiliation{High Energy Accelerator Research Organization (KEK), Tsukuba 305-0801}
\affiliation{J-PARC Branch, KEK Theory Center, High Energy Accelerator Research Organization (KEK), Tsukuba 305-0801}
\affiliation{National Research University Higher School of Economics, Moscow 101000}
\affiliation{Forschungszentrum J\"{u}lich, 52425 J\"{u}lich}
\affiliation{IKERBASQUE, Basque Foundation for Science, 48013 Bilbao}
\affiliation{Indian Institute of Science Education and Research Mohali, SAS Nagar, 140306}
\affiliation{Indian Institute of Technology Guwahati, Assam 781039}
\affiliation{Indian Institute of Technology Hyderabad, Telangana 502285}
\affiliation{Indian Institute of Technology Madras, Chennai 600036}
\affiliation{Indiana University, Bloomington, Indiana 47408}
\affiliation{Institute of High Energy Physics, Chinese Academy of Sciences, Beijing 100049}
\affiliation{Institute of High Energy Physics, Vienna 1050}
\affiliation{Institute for High Energy Physics, Protvino 142281}
\affiliation{INFN - Sezione di Napoli, I-80126 Napoli}
\affiliation{INFN - Sezione di Roma Tre, I-00146 Roma}
\affiliation{INFN - Sezione di Torino, I-10125 Torino}
\affiliation{Advanced Science Research Center, Japan Atomic Energy Agency, Naka 319-1195}
\affiliation{J. Stefan Institute, 1000 Ljubljana}
\affiliation{Institut f\"ur Experimentelle Teilchenphysik, Karlsruher Institut f\"ur Technologie, 76131 Karlsruhe}
\affiliation{Kitasato University, Sagamihara 252-0373}
\affiliation{Korea Institute of Science and Technology Information, Daejeon 34141}
\affiliation{Korea University, Seoul 02841}
\affiliation{Kyungpook National University, Daegu 41566}
\affiliation{Universit\'{e} Paris-Saclay, CNRS/IN2P3, IJCLab, 91405 Orsay}
\affiliation{P.N. Lebedev Physical Institute of the Russian Academy of Sciences, Moscow 119991}
\affiliation{Liaoning Normal University, Dalian 116029}
\affiliation{Faculty of Mathematics and Physics, University of Ljubljana, 1000 Ljubljana}
\affiliation{Ludwig Maximilians University, 80539 Munich}
\affiliation{Luther College, Decorah, Iowa 52101}
\affiliation{Malaviya National Institute of Technology Jaipur, Jaipur 302017}
\affiliation{Faculty of Chemistry and Chemical Engineering, University of Maribor, 2000 Maribor}
\affiliation{Max-Planck-Institut f\"ur Physik, 80805 M\"unchen}
\affiliation{School of Physics, University of Melbourne, Victoria 3010}
\affiliation{University of Mississippi, University, Mississippi 38677}
\affiliation{University of Miyazaki, Miyazaki 889-2192}
\affiliation{Moscow Physical Engineering Institute, Moscow 115409}
\affiliation{Graduate School of Science, Nagoya University, Nagoya 464-8602}
\affiliation{Kobayashi-Maskawa Institute, Nagoya University, Nagoya 464-8602}
\affiliation{Universit\`{a} di Napoli Federico II, I-80126 Napoli}
\affiliation{Nara Women's University, Nara 630-8506}
\affiliation{National Central University, Chung-li 32054}
\affiliation{National United University, Miao Li 36003}
\affiliation{Department of Physics, National Taiwan University, Taipei 10617}
\affiliation{H. Niewodniczanski Institute of Nuclear Physics, Krakow 31-342}
\affiliation{Nippon Dental University, Niigata 951-8580}
\affiliation{Niigata University, Niigata 950-2181}
\affiliation{Novosibirsk State University, Novosibirsk 630090}
\affiliation{Osaka City University, Osaka 558-8585}
\affiliation{Pacific Northwest National Laboratory, Richland, Washington 99352}
\affiliation{Panjab University, Chandigarh 160014}
\affiliation{Peking University, Beijing 100871}
\affiliation{University of Pittsburgh, Pittsburgh, Pennsylvania 15260}
\affiliation{Punjab Agricultural University, Ludhiana 141004}
\affiliation{Research Center for Nuclear Physics, Osaka University, Osaka 567-0047}
\affiliation{Meson Science Laboratory, Cluster for Pioneering Research, RIKEN, Saitama 351-0198}
\affiliation{Dipartimento di Matematica e Fisica, Universit\`{a} di Roma Tre, I-00146 Roma}
\affiliation{Department of Modern Physics and State Key Laboratory of Particle Detection and Electronics, University of Science and Technology of China, Hefei 230026}
\affiliation{Showa Pharmaceutical University, Tokyo 194-8543}
\affiliation{Soongsil University, Seoul 06978}
\affiliation{Sungkyunkwan University, Suwon 16419}
\affiliation{School of Physics, University of Sydney, New South Wales 2006}
\affiliation{Department of Physics, Faculty of Science, University of Tabuk, Tabuk 71451}
\affiliation{Tata Institute of Fundamental Research, Mumbai 400005}
\affiliation{Department of Physics, Technische Universit\"at M\"unchen, 85748 Garching}
\affiliation{Toho University, Funabashi 274-8510}
\affiliation{Earthquake Research Institute, University of Tokyo, Tokyo 113-0032}
\affiliation{Department of Physics, University of Tokyo, Tokyo 113-0033}
\affiliation{Tokyo Institute of Technology, Tokyo 152-8550}
\affiliation{Tokyo Metropolitan University, Tokyo 192-0397}
\affiliation{Utkal University, Bhubaneswar 751004}
\affiliation{Virginia Polytechnic Institute and State University, Blacksburg, Virginia 24061}
\affiliation{Wayne State University, Detroit, Michigan 48202}
\affiliation{Yamagata University, Yamagata 990-8560}
\affiliation{Yonsei University, Seoul 03722}
  \author{I.~Adachi}\affiliation{High Energy Accelerator Research Organization (KEK), Tsukuba 305-0801}\affiliation{SOKENDAI (The Graduate University for Advanced Studies), Hayama 240-0193} 
  \author{H.~Aihara}\affiliation{Department of Physics, University of Tokyo, Tokyo 113-0033} 
  \author{D.~M.~Asner}\affiliation{Brookhaven National Laboratory, Upton, New York 11973} 
  \author{T.~Aushev}\affiliation{National Research University Higher School of Economics, Moscow 101000} 
 \author{R.~Ayad}\affiliation{Department of Physics, Faculty of Science, University of Tabuk, Tabuk 71451} 
% ...
\collaboration{The Belle Collaboration}

\begin{abstract}
We present the measurement of the first to fourth order moments of the four-momentum transfer squared, $q^2$, of inclusive \bclnu decays using the full Belle data set of \lumi of integrated luminosity at the $\Upsilon(4S)$ resonance where $\ell = e, \mu$. The determination of these moments and their systematic uncertainties open new pathways to determine the absolute value of the CKM matrix element $V_{cb}$ using a reduced set of matrix elements of the heavy quark expansion. In order to identify and reconstruct the $X_c$ system, we reconstruct one of the two $B$-mesons using machine learning techniques in fully hadronic decay modes. The moments are measured with progressively increasing threshold selections on $q^2$ starting with a lower value of \lowqSqCut in steps of 0.5 $\, \mathrm{GeV}^2$ up to a value of \highqSqCut. The measured moments are further unfolded, correcting for reconstruction and selection effects as well as QED final state radiation. We report the moments separately for electron and muon final states and observe no lepton flavor universality violating effects.
\end{abstract}


\preprint{ Belle Preprint 2021-18, KEK Preprint 2021-22}

\date{\today}

\maketitle


\section{introduction}

Precise measurements of the absolute value of the Cabibbo-Kobayashi-Maskawa (CKM) matrix element $V_{cb}$ are important to deepen our understanding of the Standard Model of Particle Physics (SM)~\cite{PhysRevLett.10.531,km_paper}. The CKM matrix is a $3 \times 3$ unitary matrix, whose complex phase is responsible for all known charge-parity (CP) violating effects in the quark sector~\cite{pdg_ckm:2020} through the KM mechanism~\cite{km_paper}. There exists evidence for conceptually similar CP-violating processes to be present in the neutrino sector~\cite{Abe:2019vii}. The quark sector associated CP violation is not sufficient to explain the matter dominance present in the universe today, and it is also unclear if the CP violation in the neutrino sector can produce a baryon asymmetry of the required size. This motivates searches for new CP-violating phenomena e.g. in the form of processes involving heavy exotic particles. If such new exotic states interact with quarks in some notable form, their existence might alter the properties of measurements constraining the unitarity property of the CKM matrix~\cite{Kou:2018nap}.  Precise measurements of $|V_{ub}|$, $|V_{cb}|$, and the CKM angle $\gamma = \phi_3$ are imperative to isolate such effects. These quantities can be measured using tree-level dominated processes, which are expected to remain unaffected by new physics contributions in most models. Thus combining their results one can obtain an unbiased measure for the amount of CP violation in the SM quark sector. 

\begin{figure}[b!]
  \includegraphics[width=0.48\textwidth]{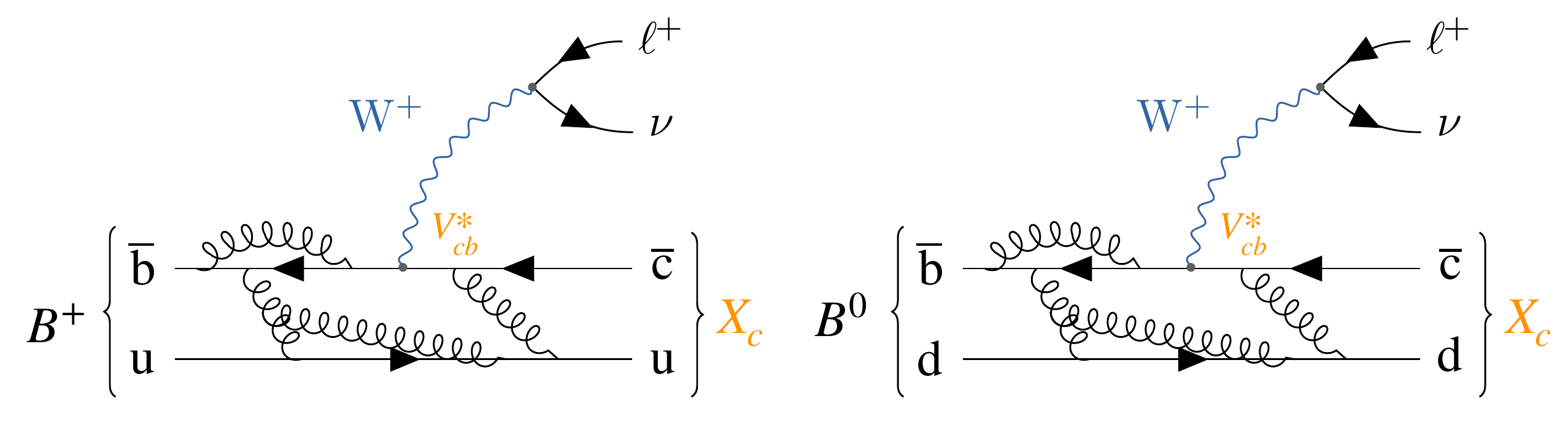} 
\caption{
The inclusive \bclnu semileptonic processes for a $B^+$ (left) or a $B^0$ (right) meson decay.
 }
\label{fig:bclnu}
\end{figure}

Semileptonic \bclnu decays offer a theoretically clean avenue to determine $|V_{cb}|$ and Fig.~\ref{fig:bclnu} depicts the processes involving $B^+$ and $B^0$-meson decays\footnote{Charge conjugation is implied and \bclnu is defined as the average branching fraction of $B^+$ and $B^0$ meson decays and $\ell = e$ or $\mu$.}. Due to the  factorization (up to small known electroweak corrections, cf.~\cite{Sirlin:1981ie}) of the hadronic and lepton-neutrino final states, these processes are theoretically better understood than hadronic transitions involving $V_{cb}$. Furthermore, until future lepton colliders will provide clean samples of $B_c$ decays, measurements involving the theoretically better understood purely leptonic decays are not feasible~\cite{Zheng:2021xuq} even with the large samples of $B_c$ mesons recorded at the LHC. The existing $|V_{cb}|$ determinations either focus on exclusive final states, with \bdslnu and \bdlnu\ transitions providing the most precise values to date~\cite{Amhis:2019ckw}, or on the study of inclusive \bclnu decay. First measurements using \bsdslnu and \bsdlnu decays were recently reported by Ref.~\cite{Aaij:2020hsi}, indicating that in the future such channels will play a prominent role in the precision determination of $|V_{cb}|$ from exclusive final states. To translate measured (differential or partial) branching fractions of exclusive decays into values of $|V_{cb}|$, information regarding the normalization and dynamics of the non-perturbative form factors is needed. Recent studies indicate that the underlying assumptions employed to simplify form factor parametrizations should be re-evaluated, since the use of more generalised and model-independent forms hint at higher values of $|V_{cb}|$, see. e.g. Refs.~\cite{Bigi:2016mdz,Bigi:2017njr,Bernlochner:2017xyx,Ferlewicz:2020lxm}. Inclusive determinations of $|V_{cb}|$ exploit the fact that the total decay rate can be expanded into a manageable number of non-perturbative matrix elements using the heavy quark expansion (HQE). For instance, the total rate of the semileptonic transition can be represented in the HQE in a series of terms proportional to the inverse bottom quark mass times the QCD scale parameter, $\Lambda_{\mathrm{QCD}}/m_b$, of increasing powers and corrections proportional to the strong coupling constant, $\alpha_s(m_b)$, can be systematically included. This approach is independent from the considerations that enter exclusive determinations involving form factors and therefore provide complementary information. Spectral moments such as the moments of the lepton energy, hadronic mass or hadronic energy spectra, can be expressed in the same way as the total rate using the HQE. The expressions describing these observables have been calculated to next-to-next-to-leading order precision in $\alpha_s$ and at leading order in the HQE, at next-to-leading order in $\alpha_s$ and for HQE up to $\mathcal{O}(1/m_{b}^2)$, and up to the HQE of order $\mathcal{O}(1/m_{b}^5)$ at tree-level with respect to the strong interaction~\cite{Jezabek:1988iv,Aquila:2005hq,Pak:2008cp,Melnikov:2008qs,Becher:2007tk,Mannel:2014xza,Alberti:2013kxa,Mannel:2010wj}. With measurements of the total \bclnu branching fraction, the energy and the hadronic mass moments~\cite{Acosta:2005qh,Csorna:2004kp,Abdallah:2005cx,Aubert:2004td,Aubert:2009qda,Urquijo:2006wd,Schwanda:2006nf}, the non-perturbative parameters and $|V_{cb}|$ can be determined using combined fits of all the relevant inputs~\cite{Gambino:2013rza,Alberti:2014yda}. A comprehensive review of this approach can be found in Refs.~\cite{RevModPhys.88.035008,Gambino:2013rza}.

The world averages of $|V_{cb}|$ from exclusive and inclusive determinations are~\cite{Amhis:2019ckw}:
\begin{align}
 |V_{cb}^{\mathrm{excl.}}| & =  \left(39.25 \pm 0.56 \right) \times 10^{-3} \, , \\
 |V_{cb}^{\mathrm{incl.}}| & = \left( 42.19 \pm 0.78 \right) \times 10^{-3} \, .
\end{align}
Here the uncertainties are the sum from experiment and theory. Both world averages exhibit a disagreement of about 3 standard deviations. It is noteworthy that newer measurements of exclusive  $|V_{cb}|$ tend to agree better with the inclusive value, but these also have larger uncertainties. The total uncertainty in the inclusive determination is dominated by theory errors, mainly to cover uncertainties from missing higher-order contributions~\cite{Benson:2003kp,Gambino:2011cq}. This disagreement between the two measurements is limiting the reach of present-day searches for loop-level new physics in the CKM sector of the SM, see e.g. Ref.~\cite{Bona:ICHEP2020} for a recent analysis. 

In Ref.~\cite{Fael:2018vsp} a novel and alternative approach to determine $|V_{cb}|$ from inclusive decays was outlined: by exploiting reparametrization invariance the authors could demonstrate that the full set of 13 non-perturbative matrix elements present in the total rate can be reduced to a set of only 8 parameters at the order of $\mathcal{O}(1/m_{b}^4)$. This reduction eases the proliferation of new parameters when considering higher orders of the HQE. New measurements are needed to determine this reduced set of 8 parameters, as the key prerequisite that gives rise to the reparametrization invariance in the total rate is violated in measurements of moments of lepton momentum, hadronic mass, and hadronic energy spectra. The authors of Ref.~\cite{Fael:2018vsp} thus propose a systematic measurement of moments of the four-momentum transfer squared, $q^2$, of the $B$ meson system to the $X_c$ system with progressively increasing requirements on $q^2$ itself. These measurements either require the identification and explicit reconstruction of the $X_c$ system or the reconstruction of the missing neutrino three momentum. In this paper the former approach is used in combination with the full reconstruction of the second $B$ meson produced in the collision event. This is achieved efficiently with the use of neural networks. We report measurements of the first to fourth moments, $\langle q^2 \rangle$, $\langle q^{4} \rangle$, $\langle q^{6} \rangle$, $\langle q^{8} \rangle$ using a set of threshold selections\footnote{We use natural units: $\hbar = c = 1$.} \qSqCut. The first measurement of the first moment of the $q^2$ spectrum was reported in Ref.~\cite{Csorna:2004kp} with a lepton energy requirement of 1 GeV. However, this requirement reintroduces the full set of non-perturbative matrix elements, since the lepton energy is not a reparametrization invariant quantity. Furthermore, the moments of higher order exhibit greater sensitivity to the higher order terms of the HQE. In this paper a first systematic study is reported with progressively increasing threshold selections on $q^2$. Additionally, the third and fourth order $q^2$ moments are reported for the first time. Due to the relationship of $q^2$ with the lepton momentum, we will restrict our measurement to moments with a minimum threshold selection on $q^2$ of $3 \, \mathrm{GeV}^2$, to ensure that the Belle detector can still reliably reconstruct and identify the lepton from the \bclnu decay. 

The remainder of this paper is organized as follows: Section~\ref{sec:data_set_sim_samples} introduces the data set and simulated event samples, as well as the applied corrections to the simulated samples. In Section~\ref{sec:analysis_strategy} the employed analysis strategy is outlined and the reconstruction of the $X_c$ system and $q^2$, and the background subtraction are discussed. Section~\ref{sec:calibration} discusses the calibration procedure used to reverse detector resolution, selection, and acceptance effects of the measured moments of $q^2$. In Section~\ref{sec:systematics} the systematic uncertainties affecting the moment measurements are discussed, while Section~\ref{sec:results} reports the measured moments and compares them to the expectation from simulated \bclnu decays. Finally, Section~\ref{sec:summary} summarizes our findings and presents our conclusions. 


\section{Data Set and Simulated Samples}\label{sec:data_set_sim_samples}
The results are based on an analysis of the full Belle data set of \mbox{$(772 \pm 10) \times 10^6$} $B$ meson pairs, which were produced at the KEKB accelerator complex~\cite{KEKB} with a center-of-mass energy of $\sqrt{s} = \SI{10.58}{GeV}$ corresponding to the $\Upsilon(4S)$ resonance. An additional $\SI{79}{fb^{-1}}$ of collision events recorded $\SI{60}{MeV}$ below the $\Upsilon(4S)$ resonance peak is used to derive corrections and for cross-checks.

The Belle detector is a large-solid-angle magnetic spectrometer. The detector consists of several sub-detectors: a silicon vertex detector, a 50-layer central drift chamber (CDC), an array of aerogel threshold Cherenkov counters (ACC), a barrel-like arrangement of time-of-flight scintillation counters (TOF), and an electromagnetic calorimeter composed of CsI(Tl) crystals (ECL) located inside a superconducting solenoid coil that provides a \SI{1.5}{T} magnetic field. An iron flux return located outside of the coil is instrumented to detect $K^0_L$ mesons and to identify muons (KLM). More details about the detector layout and performance can be found in Ref.~\citep{Abashian:2000cg} and in references therein.

We identify charged tracks as electron or muon candidates by combining the information of multiple subdetectors into a lepton identification likelihood ratio, $\mathcal{L}_\mathrm{LID}$. The most important identifying features for electrons are the ratio of the energy deposition in the ECL with respect to the reconstructed track momentum, the energy loss in the CDC, the shower shape in the ECL, the quality of the geometrical matching of the track to the shower position in the ECL, and the photon yield in the ACC~\citep{HANAGAKI2002490}. We identify muon candidates from charged track trajectories extrapolated to the outer detector. The most important identifying features are the difference between expected and measured penetration depth as well as the transverse deviation of KLM hits from the extrapolated trajectory~\citep{ABASHIAN200269}. Electron and muon candidates are required to have a minimum transverse momentum of 300 MeV and 600 MeV, respectively, in the laboratory frame of Belle. We further identify charged tracks as pions or kaons using a likelihood ratio $\mathcal{L}_\mathrm{K/\pi \, \mathrm{ID}} = \mathcal{L}_\mathrm{K\, \mathrm{ID}} / \left( \mathcal{L}_\mathrm{K \, \mathrm{ID}} + \mathcal{L}_\mathrm{\pi \, \mathrm{ID}} \right)$. The most important identifying features of the kaon ($\mathcal{L}_\mathrm{K \, \mathrm{ID}} $) and pion ($\mathcal{L}_\mathrm{\pi \, \mathrm{ID}} $) likelihoods for low momentum particles with transverse momentum below 1 GeV in the laboratory frame are the recorded energy loss by ionization, $\mathrm{d}E/\mathrm{d}x$, in the CDC, and the time of flight information from the TOF. Higher-momentum kaon and pion classification relies on the Cherenkov light recorded in the ACC. In order to avoid the difficulties in understanding the efficiencies of reconstructing $K^0_L$ mesons, they are not explicitly reconstructed or used in this measurement.

We identify photons as energy depositions in the ECL, vetoing clusters to which an associated track can be assigned. Only photons with an energy deposition of \mbox{\Eg $ >100 \, \text{MeV}$, $150 \, \text{MeV}$, and $50 \, \text{MeV}$} in the forward endcap, backward endcap and barrel part of the calorimeter, respectively, are selected. We reconstruct $\pi^0$ candidates from photon candidates and require the invariant mass to fall into a window of $m_{\gamma\gamma} \in [0.12, 0.15] \, \text{GeV}$, corresponding to about 2.5 times the $\pi^0$ mass resolution. 

Monte Carlo (MC) samples of $B$ meson decays and continuum processes ($e^+ e^- \to q \bar q$ with $q = u,d,s,c$) are simulated using the \texttt{EvtGen} generator~\citep{EvtGen}. These samples are used to evaluate reconstruction efficiencies and acceptance, and to estimate background contaminations. The sample sizes correspond to approximately three times the Belle collision data for $B$ meson and continuum decays. The interactions of particles traversing the detector are simulated using \texttt{Geant3}~\citep{Geant3}. Electromagnetic final-state radiation is simulated using the \texttt{PHOTOS}~\citep{Photos} software package for all charged final-state particles. To account for, e.g., differences in identification and reconstruction efficiencies in the MC, we employ data-driven methods to derive efficiency corrections. For the particle-identification likelihood ratios these corrections are parametrized as a function of the polar angel and laboratory frame momentum of the particle candidates. 

Inclusive semileptonic  \bclnu\ decays are dominated by \bdlnu\ and \bdslnu\ transitions. We model the \bdlnu\ decays using the BGL parametrization~\cite{Boyd:1994tt} with form factor central values and uncertainties taken from the fit in Ref.~\cite{Glattauer:2015teq}. For \bdslnu\, we use the BGL implementation proposed by Refs.~\cite{Grinstein:2017nlq,Bigi:2017njr} with form factor central values and uncertainties from the fit to the measurement of Ref.~\cite{Waheed:2018djm}. Both processes are normalized to the average branching fraction of Ref.~\cite{Amhis:2019ckw} assuming isospin symmetry. Semileptonic \bddslnu decays with $D^{**} = \{ D_0^*, D_1^*, D_1, D_2^* \}$ denoting the four orbitally excited charmed mesons are modeled using the heavy-quark-symmetry-based form factors proposed in Ref.~\cite{Bernlochner:2016bci}. The $D^{**}$ decays are simulated using masses and widths from Ref.~\cite{pdg:2020}. For the branching fractions the values of Ref.~\cite{Amhis:2019ckw} are adopted and we correct them to account for missing isospin-conjugated and other established decay modes, following the prescription given in Ref.~\cite{Bernlochner:2016bci}. Since all measurements target the $D^{**\,0} \to D^{(*)+} \, \pi^-$ decay modes, where $D^{**\,0}$ refers to all neutral $D^{**}$ states, we correct for the missing isospin modes with a factor of
\begin{align}
 f_\pi = \frac{ \mathcal{B}( \overline{D}{}^{**\,0} \to D^{(*)\, -} \pi^+) }{  \mathcal{B}( \overline{D}{}^{**\,0} \to \overline{D}{}^{(*)} \pi) } =  \frac{2}{3} \, .
\end{align}
The measurement of the $B \to D_2^* \, \ell \bar \nu_\ell$ in Ref.~\cite{Amhis:2019ckw} are converted to be respective to the $\overline{D}{}_2^{*\,0} \to D^{*\, -} \pi^+$ final states only. To also account for $\overline{D}{}_2^{*\,0} \to D^{-} \pi^+$ contributions, a factor of~\cite{pdg:2020} 
\begin{align}
 f_{D_2^*}  = \frac{ \mathcal{B}(\overline{D}{}_2^{*\, 0} \to D^{-} \pi^+) }{ \mathcal{B}(\overline{D}{}_2^{*\, 0} \to D^{*\, -} \pi^+) } = 1.54 \pm 0.15  \, ,
\end{align}
is applied. The world average of $B \to D_1^* \, \ell \bar \nu_\ell$ given in Ref.~\cite{Amhis:2019ckw} combines measurements, which show very poor agreement, and the resulting probability of the combination is below 0.01\%. Notably, the measurement of Ref.~\cite{Liventsev:2007rb} is in conflict with the measured branching fractions of Refs.~\cite{Aubert:2008ea, Abdallah:2005cx} and with the expectation of $\mathcal{B}(B \to D_1^* \, \ell \bar \nu_\ell)$ being of similar size to $\mathcal{B}(B \to D_0^* \, \ell \bar \nu_\ell)$ ~\cite{Bigi:2007qp, Leibovich:1997em}. We perform our own average excluding the conflicting measurement and use
\begin{align}
 \mathcal{B}(B^+ \to \overline{D}{}_1^{*\, 0}(\to D^{*\,-} \pi^+) \, \ell^+ \nu_\ell) & = \left(0.28 \pm 0.06 \right) \times 10^{-2} \, .
\end{align}
This slightly deviates from the treatment in Ref.~\cite{pdg:2020} that omits the measurements of Refs.~\cite{Liventsev:2007rb,Abdallah:2005cx}. The world average of $B \to D_1 \, \ell \bar \nu_\ell$ does not account for contributions from three-body decays of the form $D_1 \to D \pi \pi$. We account for these using a factor~\cite{Aaij:2011rj}
\begin{align}
 f_{D_1} = \frac{ \mathcal{B}(\overline{D}{}_1 \to D^{* \, -} \pi^+ )}{\mathcal{B}( \overline{D}{}_1 \to \overline{D}{}^{0} \pi^+ \pi^- )} = 2.32 \pm 0.54 \, .
\end{align}
To account for missing isospin-conjugated modes of the three-hadron final states we adopt the prescription from Ref.~\cite{Lees:2015eya}, which quotes an average isospin correction factor of 
\begin{align}
 f_{\pi\pi} &= \frac{ \mathcal{B}( \overline{D}{}^{**\,0} \to \overline{D}{}^{(*)\, 0} \pi^+ \pi^-) }{ \mathcal{B}( \overline{D}{}^{**\,0} \to \overline{D}{}^{(*)} \pi \pi) } = \frac{1}{2} \pm \frac{1}{6} \, .
\end{align}
The uncertainty quoted here takes into account the full spread of final states ($f_0(500) \to \pi \pi$ or $\rho \to \pi \pi$ result in $f_{\pi \pi} = 2/3$ and $1/3$, respectively) and the non-resonant three-body decays ($f_{\pi\pi} = 3/7$). We further make the implicit assumption that
\begin{align}
 \mathcal{B}(\overline{D}{}_2^{*} \to \overline D \pi) + \mathcal{B}(\overline{D}{}_2^{*} \to \overline{D}{}^* \pi) = 1 \, , \nonumber \\
 \mathcal{B}(\overline{D}{}_1 \to  \overline{D}{}^{*} \pi) + \mathcal{B}(\overline{D}{}_1 \to \overline D \pi\pi)  = 1 \, , \nonumber \\
  \mathcal{B}(\overline{D}{}_1^{*} \to \overline{D}{}^{*} \pi) = 1 \, , \quad \text{and} \quad
  \mathcal{B}(\overline {D}{}_0^* \to  \overline D \pi)  = 1 \,  \, .
\end{align}
For the remaining $B \to D^{(*)} \, \pi \, \pi \, \ell^+ \, \nu_\ell$ contributions we use the measured value of Ref.~\cite{Lees:2015eya}, while subtracting the contribution of $D_1 \to D \pi \pi$ from the measured non-resonant plus resonant $B \to D \pi \pi \ell \bar \nu_\ell$ branching fraction. The remaining `gap' between the sum of all considered exclusive modes and the inclusive \bclnu branching fraction is filled in equal parts with $B \to D \, \eta \, \ell^+ \, \nu_\ell$ and $B \to D^{*} \, \eta \, \ell^+ \, \nu_\ell$ and for both we assume a 100\% uncertainty. We simulate \mbox{$B \to D^{(*)} \, \pi \, \pi \, \ell^+  \nu_\ell$} and \mbox{$B \to D^{(*)} \, \eta  \, \ell^+  \nu_\ell$} final states assuming that they are produced by the decay of two broad resonant states $D^{**}_{\mathrm{gap}}$ with masses and widths identical to $D_1^{*}$ and $D_0^{*}$. Although there is currently no experimental evidence for decays of charm $1P$ states into these final states or the existence of such an additional broad state (e.g. a $2S$) in semileptonic transitions, this description provides a better kinematic description of the initial three-body decay, $B \to D^{**}_{\mathrm{gap}} \, \ell \bar \nu_\ell$, than e.g. a model based on the equidistribution of all final-state particles in phase space. For the form factors we adapt Ref.~\cite{Bernlochner:2016bci}. Comparisons of kinematic distributions for the different $B \to D^{**}_{\mathrm{gap}} \, \ell \bar \nu_\ell$ models are found in Appendix A. In what follows, we will associate this component with a 100\% uncertainty.

Semileptonic \bulnu decays are modeled as a mixture of specific exclusive modes and non-resonant contributions. They are mixed using a `hybrid' approach proposed by Ref.~\cite{hybrid} and our modeling is identical to the approach detailed in Ref.~\cite{Cao:2021xqf}. 

In Table~\ref{tab:bfs} we summarize the branching fractions and uncertainties for the signal  \bclnu\ processes that we use.

\begin{table}[t!]
\caption{
	Branching fractions for \bclnu signal and \bulnu signal processes that were used are listed. More details on the applied corrections can be found in the text. 
}
\label{tab:bfs}
\vspace{1ex}
\begin{tabular}{lcc}
\hline\hline
 $\mathcal{B}$ & Value $B^+$ & Value $B^0$  \\
 \hline
 \bclnu & \\
 \quad $B \to D \, \ell^+ \, \nu_\ell$ & $\left(2.5 \pm 0.1\right) \times 10^{-2} $ & $\left(2.3 \pm 0.1\right)\times 10^{-2} $ \\
 \quad $B \to D^* \, \ell^+ \, \nu_\ell$ & $\left(5.4 \pm 0.1 \right)\times 10^{-2} $ &$\left( 5.1 \pm 0.1 \right)\times 10^{-2} $ \\
  \quad $B \to D_0^* \, \ell^+ \, \nu_\ell$  & $\left(4.2 \pm 0.8\right) \times 10^{-3}$ & $\left(3.9 \pm 0.7\right) \times 10^{-3}$ \\
  \quad\, $(\hookrightarrow D \pi)$ \\ 
  \quad $B \to D_1^* \, \ell^+ \, \nu_\ell$ & $\left(4.2 \pm 0.8\right) \times 10^{-3}$ & $\left(3.9 \pm 0.8\right) \times 10^{-3}$   \\
    \quad\, $(\hookrightarrow D^* \pi)$ \\
  \quad $B \to D_1 \, \ell^+ \, \nu_\ell$ &  $\left(4.2 \pm 0.3\right) \times 10^{-3}$ & $\left(3.9 \pm 0.3\right) \times 10^{-3}$  \\
  \quad\, $(\hookrightarrow D^* \pi)$ \\
  \quad $B \to D_2^* \, \ell^+ \, \nu_\ell$  & $\left(1.2 \pm 0.1\right) \times 10^{-3}$ & $\left(1.1 \pm 0.1\right) \times 10^{-3}$   \\
  \quad\, $(\hookrightarrow D^* \pi)$ \\ 
    \quad $B \to D_2^* \, \ell^+ \, \nu_\ell$  & $\left(1.8 \pm 0.2\right) \times 10^{-3}$ & $\left(1.7 \pm 0.2\right) \times 10^{-3}$   \\
  \quad\, $(\hookrightarrow D \pi)$ \\ 
    \quad $B \to D_1 \, \ell^+ \, \nu_\ell$ &  $\left(2.4 \pm 1.0\right) \times 10^{-3}$ & $\left(2.3 \pm 0.9\right) \times 10^{-3}$  \\
  \quad\, $(\hookrightarrow D \pi \pi)$ \\
   \quad $B \to D \pi \pi \, \ell^+ \, \nu_\ell$ & $\left(0.6 \pm 0.6 \right) \times 10^{-3}$ & $\left(0.6 \pm 0.6 \right) \times 10^{-3}$  \\
   \quad $B \to D^* \pi \pi \, \ell^+ \, \nu_\ell$ & $\left(2.2 \pm 1.0 \right) \times 10^{-3}$ & $\left(2.0 \pm 1.0 \right) \times 10^{-3}$  \\
   \quad $B \to D \eta \, \ell^+ \, \nu_\ell$ & $\left(4.0 \pm 4.0 \right) \times 10^{-3}$ & $\left(4.0 \pm 4.0 \right) \times 10^{-3}$  \\
   \quad $B \to D^{*} \eta \, \ell^+ \, \nu_\ell$ & $\left(4.0 \pm 4.0 \right) \times 10^{-3}$ & $\left(4.0 \pm 4.0 \right) \times 10^{-3}$  \\
     \hline
   \quad $B \to X_c \, \ell^+ \, \nu_\ell$ & $\left(10.8 \pm 0.4\right) \times 10^{-2} $ & $\left(10.1 \pm 0.4\right) \times 10^{-2} $ \\ \hline
      $B \to X_u \, \ell^+ \, \nu_\ell$ & $\left(2.2 \pm 0.3 \right)  \times 10^{-3}$ & $\left(2.0 \pm 0.3 \right)  \times 10^{-3}$ \\
 \hline\hline
\end{tabular}
\end{table}


\section{Analysis Strategy, $X$ reconstruction, and Background Subtraction}\label{sec:analysis_strategy}

\subsection{Neutral Network Based Tag Side Reconstruction}

The collision events are reconstructed using the hadronic Full Reconstruction (FR) algorithm detailed in Ref.~\cite{Feindt:2011mr}. The algorithm reconstructs one of the two $B$ mesons produced in the collision event fully using hadronic decay channels. This allows for the explicit identification of the constituents of the hadronic $X_{c}$ system of the \bclnu process of interest and we label the $B$ mesons reconstructed in hadronic modes in the following as $B_{\mathrm{tag}}$. Instead of attempting to reconstruct as many $B$ meson decay cascades as possible, the Full Reconstruction algorithm employs a hierarchical approach in four sequential stages: at the first stage, neural networks are trained to identify charged tracks and neutral energy depositions as detector stable particles ($e^+, \mu^+, K^+, \pi^+, \gamma$), neutral $\pi^0$ candidates, or $K_S^0$ candidates. These candidate particles are then combined during a second stage into heavier meson candidates ($J/\psi, D^0, D^{+}, D_s$). For each target final state a neural network is trained to identify probable candidate combinations. The input variables for these neural networks are the output classifiers from the first reconstruction stage, vertex fit probabilities of the candidate combinations, and the four-momenta of all the particles used to reconstruct the composite particle in question. The third stage forms candidates for $D^{*\, 0}, D^{*\, +}$, and $D_s^*$ mesons, and for each a separate neural network is trained to identify viable combinations. The input layer aggregates the output classifiers from all previous reconstruction stages and additional information such as four-momenta of the combined particles. In the final stage the information from all previous stages is used to form $B_{\mathrm{tag}}$ candidates. The viability of such combinations is again assessed by a dedicated neural network that was trained to recognize correctly reconstructed candidates from incorrect combinations and whose output classifier score we denote by $\mathcal{O}_{\mathrm{FR}}$. In total 1104 decay cascades are reconstructed in this way. This results in an efficiency of 0.28\% and 0.18\% for charged and neutral $B$ meson pairs~\cite{Bevan:2014iga}, respectively. As a final step, the output of this classifier is used as an input and combined with a range of event shape variables to train a neural network to distinguish reconstructed $B$ meson candidates from continuum processes. The output classifier score of this neural network is denoted as $\mathcal{O}_{\mathrm{Cont}}$. Both classifier scores are mapped to a range of $[0,1)$ signifying the estimated reconstruction quality from poor to excellent candidates. For the analysis we select $B_{\mathrm{tag}}$ candidates that show at least moderate agreement based on these two outputs, and require that $\mathcal{O}_{\mathrm{FR}} > 10^{-4}$ and  $\mathcal{O}_{\mathrm{Cont}} > 10^{-4}$. 

We use the charges and four momenta of the decay constituents in combination with the known beam-energy to infer the flavor and four-momentum of the $B_{\mathrm{tag}}$ candidate. We require the $B_{\mathrm{tag}}$ candidates to have at least a beam-constrained mass of
\begin{align}
 M_{\mathrm{bc}} = \sqrt{ E_{\mathrm{beam}}^2 - | \bold{p}_{\mathrm{tag}}|^2 }  > 5.27 \, \text{GeV} \, .
\end{align}
Here $ \bold{p}_{\mathrm{tag}}$ denotes the three-momentum of the $B_{\mathrm{tag}}$ candidate in the center-of-mass frame of the colliding $e^+e^-$-pair. Furthermore, $E_{\mathrm{beam}} = \sqrt{s}/2$ denotes half the center-of-mass energy of the colliding $e^+e^-$-pair. The energy difference
\begin{align}\label{eq:deltaE}
  \Delta E = E_{\mathrm{tag}} - E_{\mathrm{beam}} \, ,
\end{align}
is already used in the input layer of the neural network trained in the final stage of the reconstruction, so no further requirements are imposed. Additionally, $E_{\mathrm{tag}}$ denotes the energy of the $B_{\mathrm{tag}}$ candidate in the center-of-mass frame of the colliding $e^+e^-$-pair. In each event a single $B_{\mathrm{tag}}$ candidate is then selected according to the highest $\mathcal{O}_{\mathrm{FR}}$ score value. After the reconstruction of the $B_{\mathrm{tag}}$ candidate, all remaining tracks and clusters are used to define and reconstruct the signal side. 

\subsection{Signal Side Reconstruction}

The signal side of the event is reconstructed by identifying a well-reconstructed lepton using the likelihood described in Section~\ref{sec:data_set_sim_samples}. The signal $B$ rest frame is calculated with the momentum of the $B_{\mathrm{tag}}$ candidate via
\begin{align}
p_{\mathrm{sig}} =  p_{e^+ \, e^-}  - \left( \sqrt{m_B^2 + |\bold{p_{\mathrm{tag}}}|^2 } , \bold{p_{\mathrm{tag}}} \right) \, ,
\end{align}
with $p_{e^+ e^-} = (\sqrt{s}, \bold{0})$ denoting the four-momentum of the colliding electron-positron pair and $m_B = 5.279 \, \mathrm{GeV}$ the nominal $B$ meson mass.

\begin{figure*}
  \includegraphics[width=0.45\textwidth]{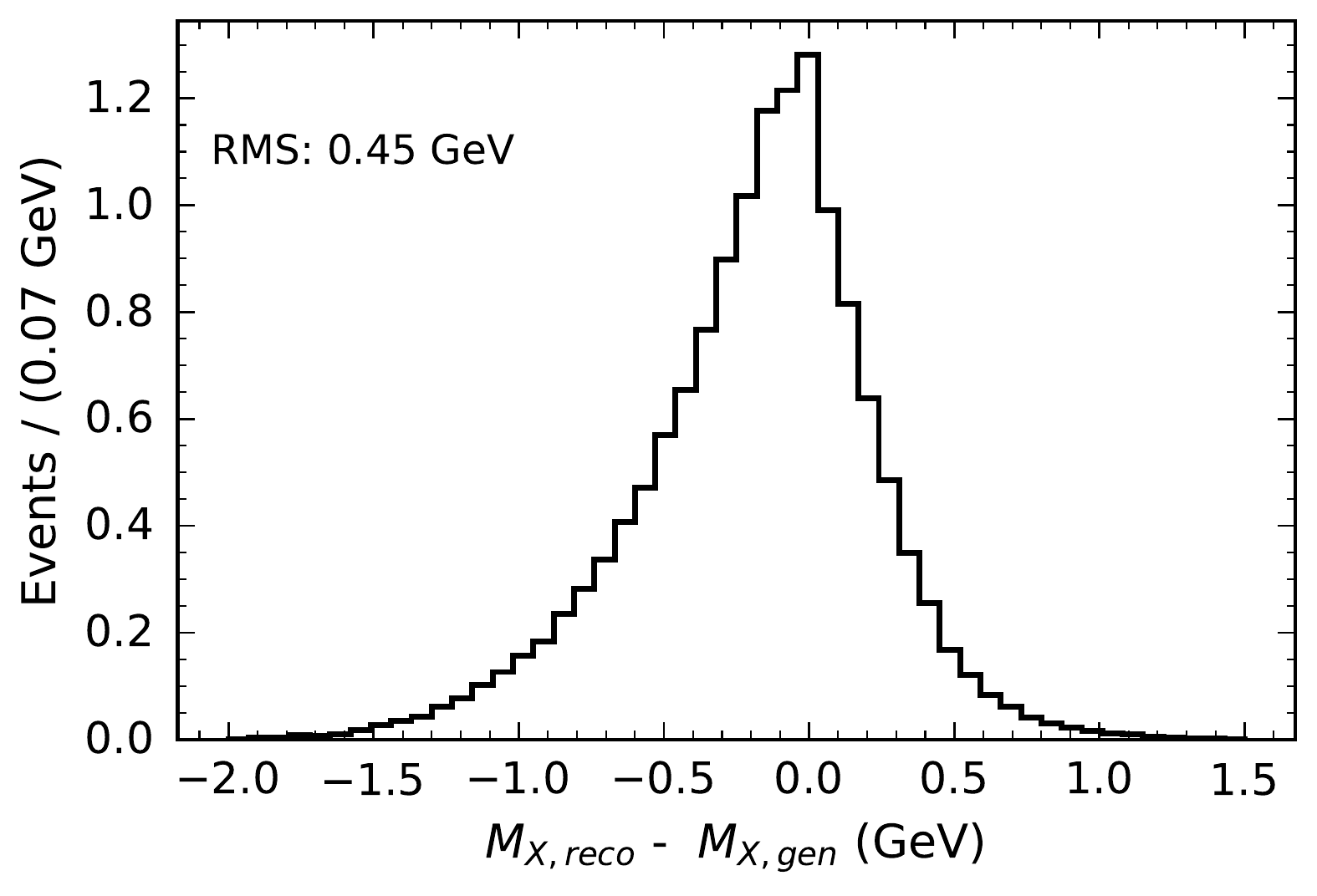}
  \includegraphics[width=0.45\textwidth]{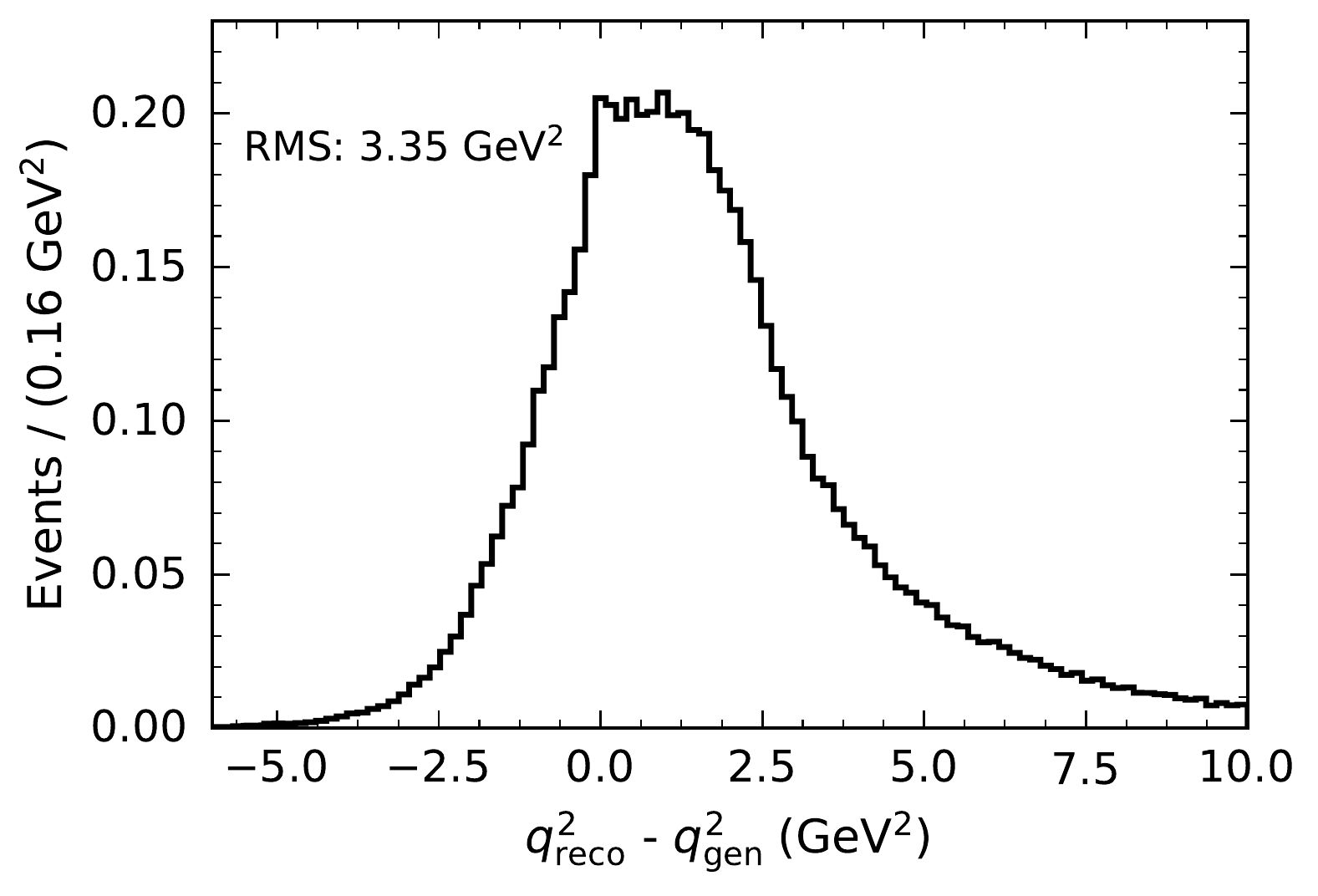}
\caption{
The resolution of the reconstructed $M_X$ and $q^2$ values for \bclnu\ signal is shown as a residual with respect to the generated values. 
 }
\label{fig:mX_q2_resolution}
\end{figure*}

If multiple lepton candidates are present on the signal side, the event is discarded to avoid additional neutrinos on the signal-side, since multiple leptons are likely to originate from a double semileptonic $b \to c \to s$ cascade. For charged  $B_{\mathrm{tag}}$ candidates, we demand that the charge assignment of the signal-side lepton be opposite to that of the $B_{\mathrm{tag}}$ charge. The hadronic $X$ system is reconstructed from the remaining unassigned charged particles and neutral energy depositions. Its four-momentum is calculated as
\begin{align}
  p_X = \sum_i \left( \sqrt{m_\pi^2 + |\bold{p_i}|^2 } , \bold{p_i} \right) + \sum_j \left( E_j , \bold{k_j} \right)  \, ,
\end{align}
with $E_j= |\bold{k_j}|$ the energy of the neutral energy depositions and all charged particles with momentum $\bold{p_i}$ assumed to be pions. With the $X$ system reconstructed, we can also reconstruct the missing four-momentum, 
\begin{align}
 P_{\mathrm{miss}} = (E_{\mathrm{miss}},  \bold{p}_{\mathrm{miss}} ) = p_{\mathrm{sig}} - p_X - p_\ell \, ,
\end{align}
which estimates the four-momentum of the neutrino in the event. Here $E_{\mathrm{miss}}$ and $\bold{p}_{\mathrm{miss}}$ denote the missing energy and momentum, respectively. For correctly reconstructed semileptonic \bclnu\ decays \mbox{$E_{\mathrm{miss}} \approx |\bold{p}_{\mathrm{miss}}|$} and we require 
\begin{align}
 -0.5 \, \mathrm{GeV} < E_{\mathrm{miss}} -  |\bold{p}_{\mathrm{miss}}| < 0.5\, \mathrm{GeV} \, .
\end{align}
The hadronic mass of the $X$ system is later used to discriminate \bclnu decays from backgrounds. It is reconstructed using 
\begin{align}
 M_X = \sqrt{ \left( p_X \right)^\mu \left( p_X \right)_\mu } \, ,
\end{align} 
and exhibits a distinct peak at $\approx 2 \, \mathrm{GeV}$. We require the total observed charge of the event to be \mbox{$|Q_{\mathrm{tot}}| = |Q_{B_{\mathrm{tag}}} + Q_{X} + Q_{\ell}| \leq 1$}, allowing for charge imbalance in events with low-momentum tracks. Finally, we reconstruct the four-momentum transfer squared $q^2$ as
\begin{align}
  q^2 = \left(  p_{\mathrm{sig}} - p_X \right)^2 \, .
\end{align}
The resolution of both variables for \bclnu is shown in Fig.~\ref{fig:mX_q2_resolution} as residuals with respect to the generated values of $q^2$ and $M_X$. The resolution for $M_X$ has a root-mean-square (RMS) deviation of $0.45 \, \text{GeV}$ and exhibits a tail towards negative values of the residuals from not reconstructed constituents of the hadronic $X$ system. The four-momentum transfer squared $q^2$ exhibits a large resolution, which is caused by a combination of the tag-side $B$ and the $X$ reconstruction. The RMS deviation for $q^2$ is $3.35 \, \text{GeV}^2$. The core resolution is dominated by the tagging resolution, whereas the large positive tail is dominated from the resolution of the reconstruction of the $X$ system.

\begin{figure}[ht!]
  \includegraphics[width=0.39\textwidth]{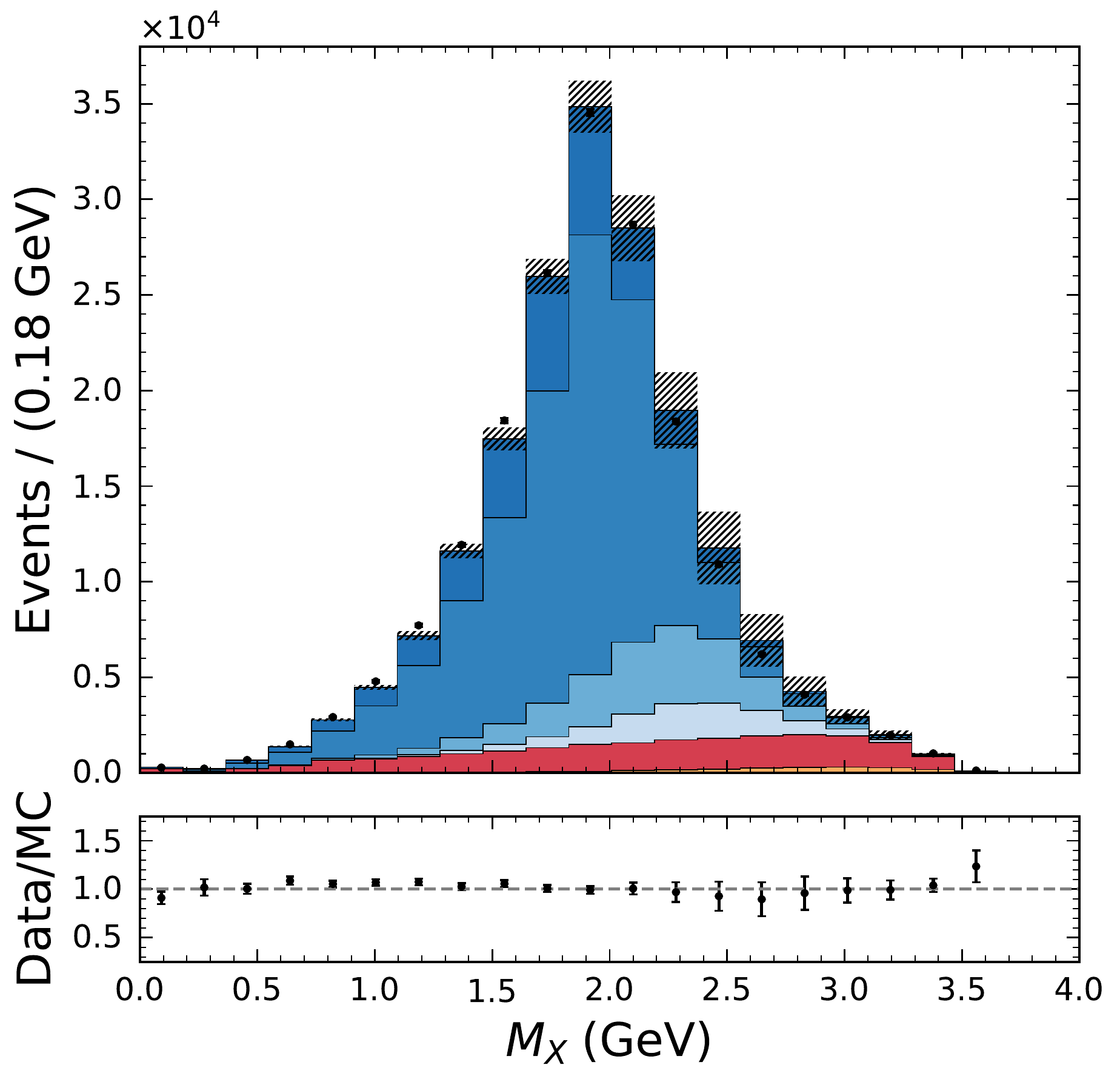}
  \includegraphics[width=0.39\textwidth]{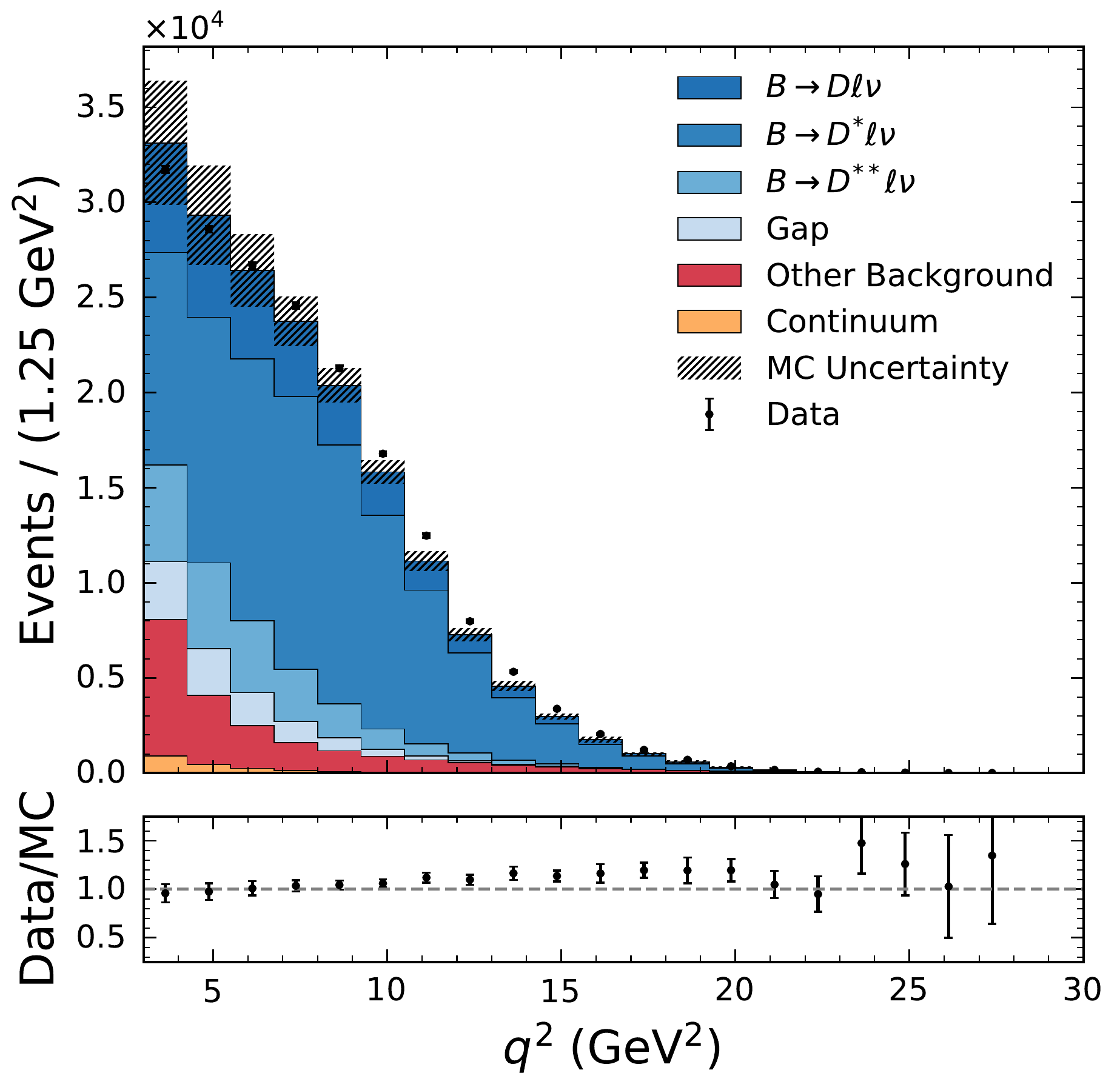}
\caption{
   The reconstructed $M_X$ and $q^2$ distributions are shown. The error band of the simulated samples incorporates the full set of systematic uncertainties discussed in Section~\ref{sec:systematics}.
 }
\label{fig:m_x_q2}
\end{figure}

In the following we analyze only the events with \mbox{$q^2 > 3.0 \, \mathrm{GeV}^2$}, corresponding to a lepton momentum of at least $300 \, \mathrm{MeV}$ in the $B$ rest frame. This corresponds also to the region of phase space in the laboratory frame at which the Belle detector operates efficiently in the identification and reconstruction of leptons. The region of phase space below $3.0 \, \mathrm{GeV}^2$ is dominated by processes other than \bclnu: secondary leptons from cascade decays and fake lepton candidates make up a prominent fraction of the selected event candidates. 

Fig.~\ref{fig:m_x_q2} compares the selected events with the expectation from simulation: the small continuum contribution is normalized using the off-resonance event sample, while the remaining simulated events are normalized to the number of reconstructed events from $\Upsilon(4S) \to B \bar B$. In the following, we separate the electron and muon candidates and analyze them separately. 

\begin{figure*}
  \includegraphics[width=0.4\textwidth]{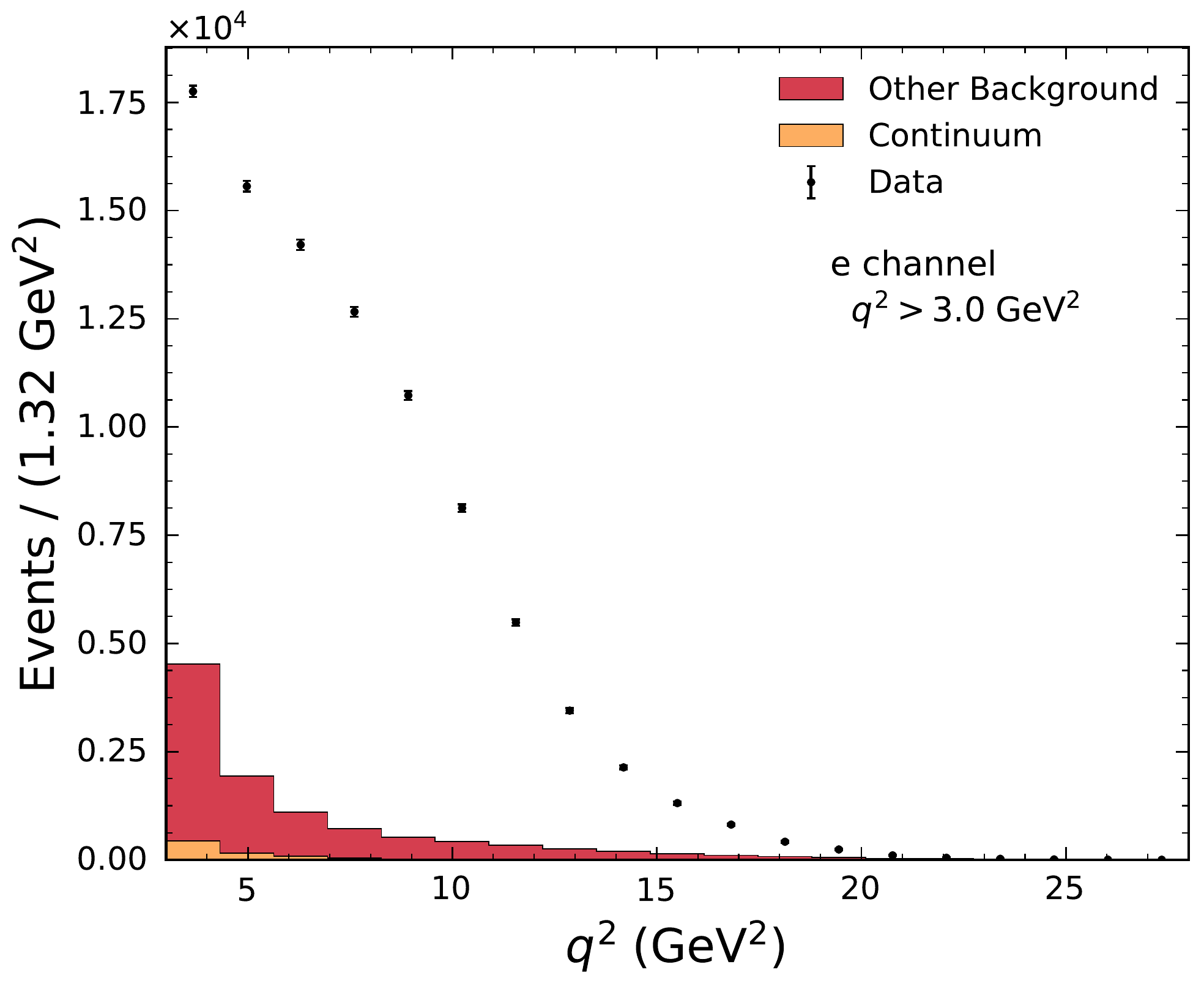}
  \includegraphics[width=0.4\textwidth]{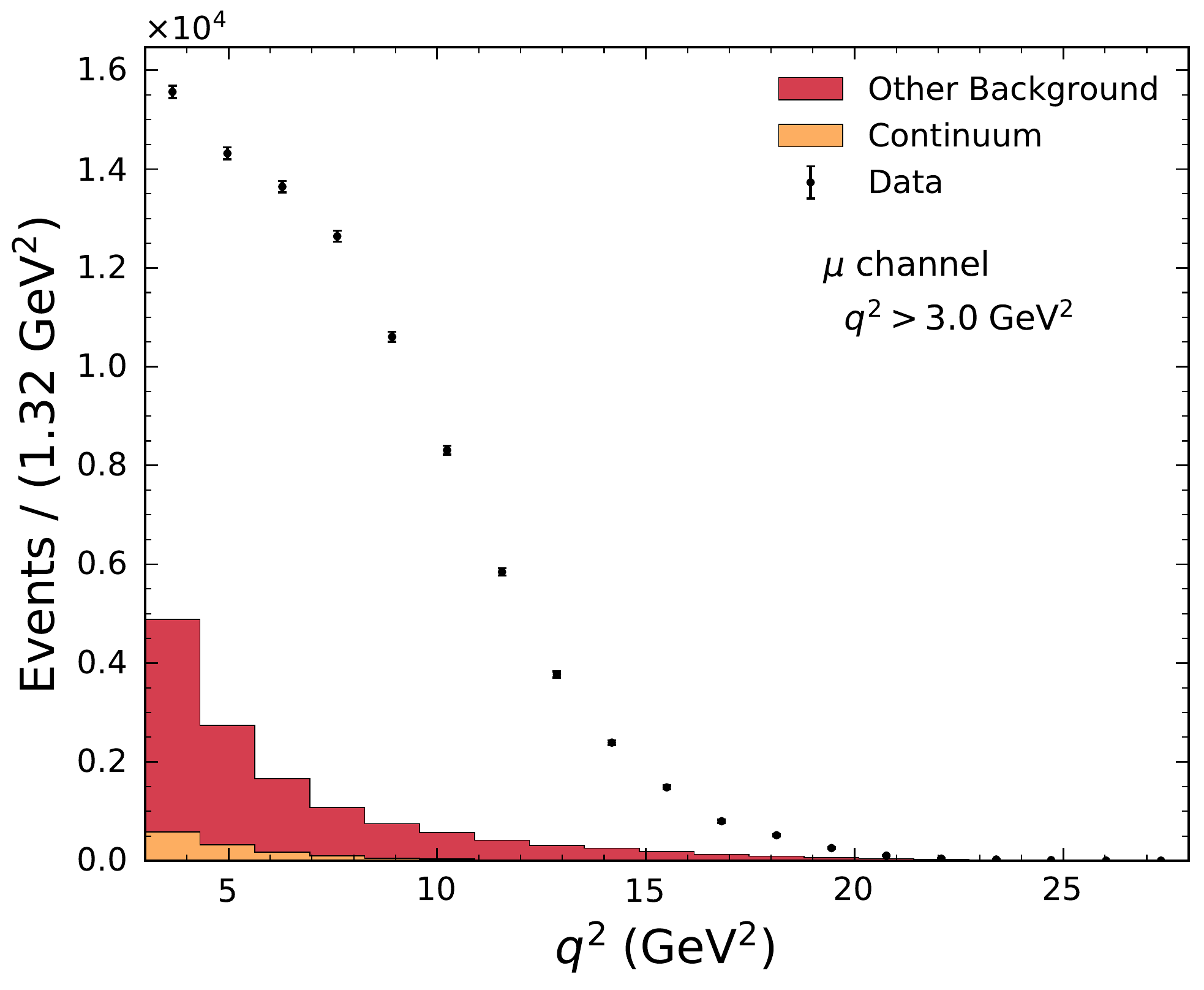}
  \includegraphics[width=0.4\textwidth]{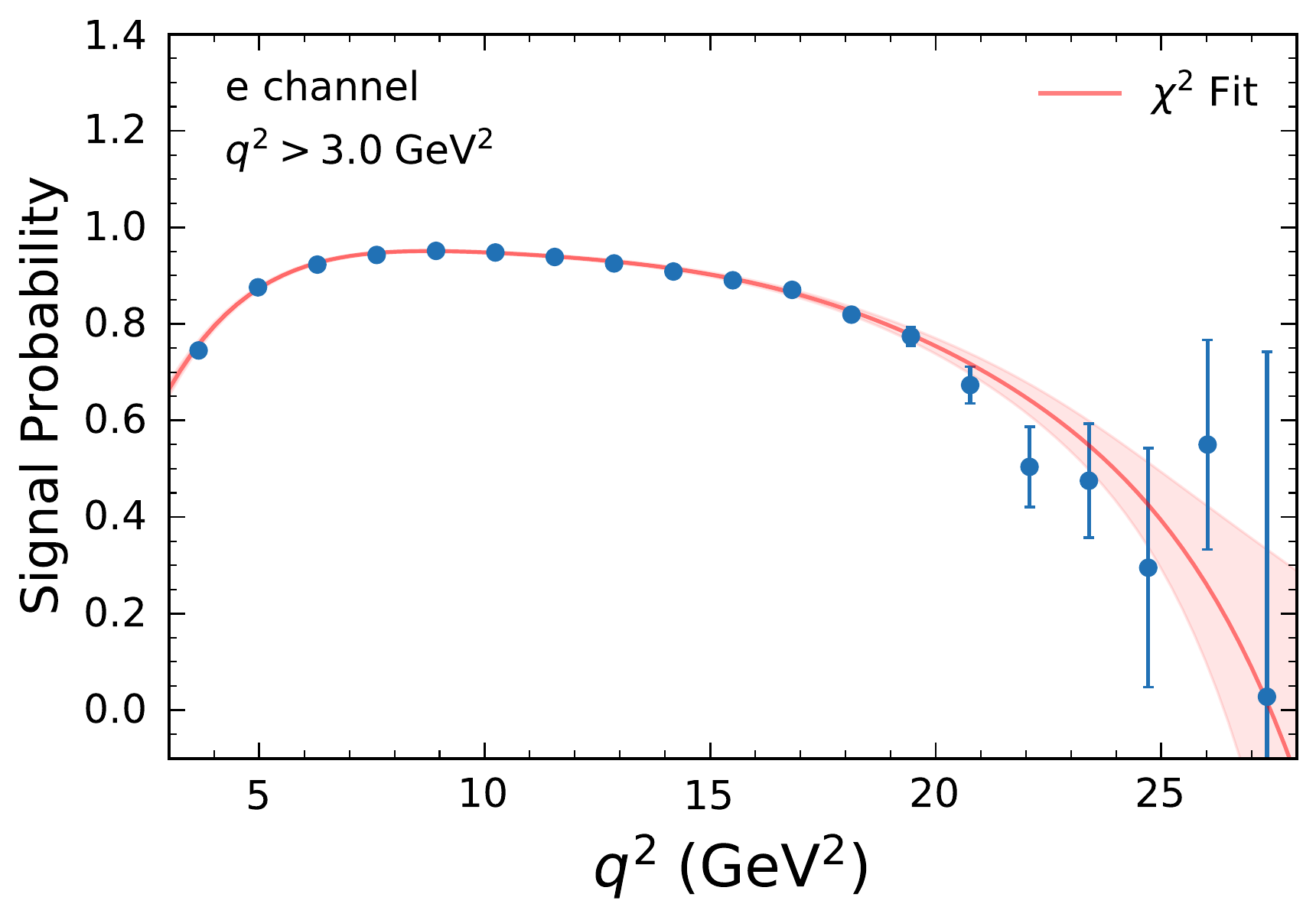}
  \includegraphics[width=0.4\textwidth]{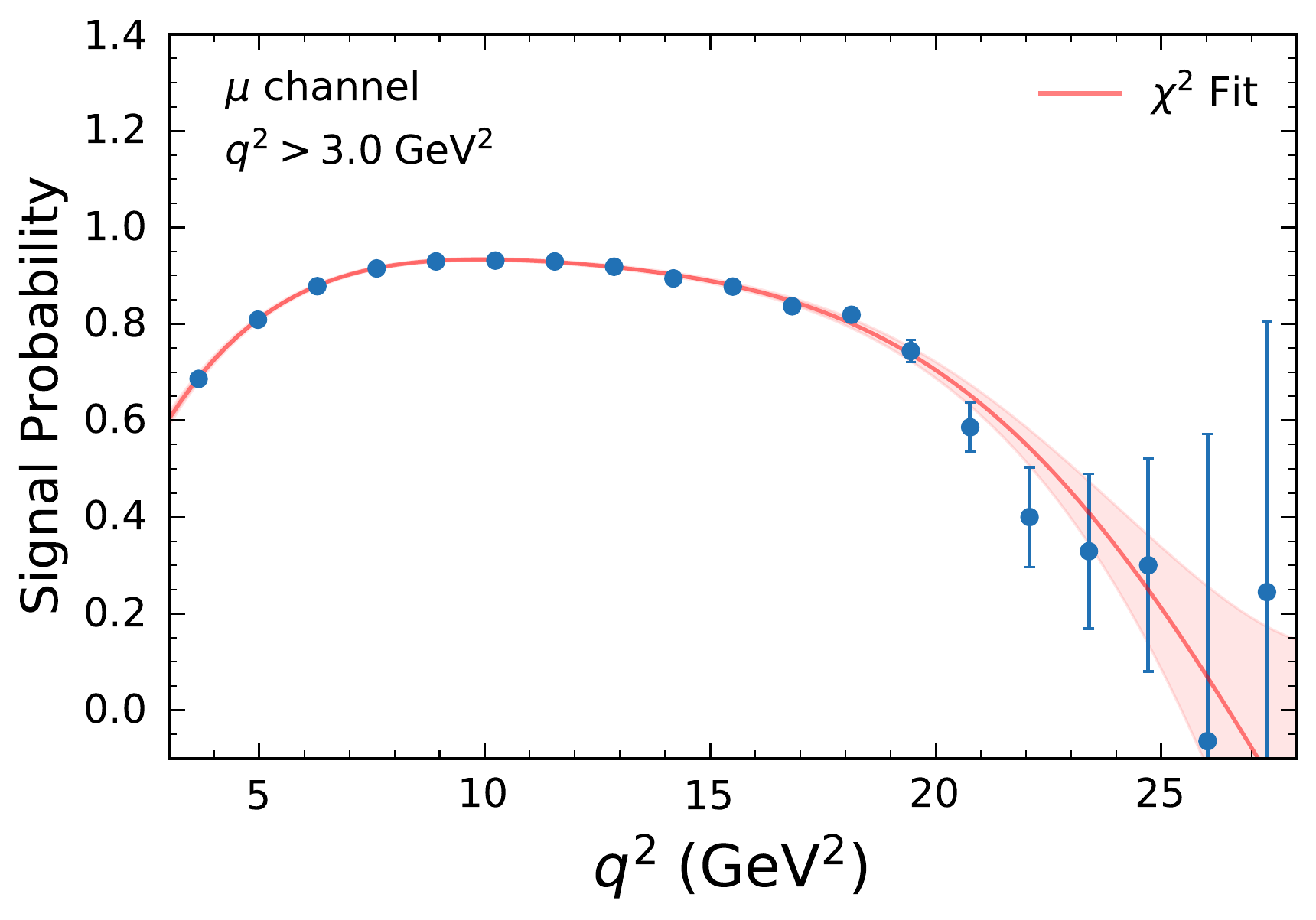}
\caption{
   The reconstructed $q^{2}$ distributions with an example $q^2$ threshold selection of $3.0\, \mathrm{GeV^2}$ (top) for electron (left) and muon (right) candidates and the determined binned signal probabilities (bottom) are shown. The background contributions are scaled to their estimated values using the fit described in the text. The binned signal probabilities are obtained by a fit of a polynomial of a given order $n$ (red curve). 
 }
\label{fig:m_x_w}
\end{figure*}

\subsection{Background Subtraction}\label{sec:bkg_subtraction}

In order to subtract background events, we carry out a two-step procedure. First a binned likelihood fit of the $M_X$ distribution determines the number of expected signal and background events. For this fit we construct a likelihood function $\mathcal{L}$ as the product of individual Poisson distributions $\mathcal{P}$, 
\begin{align} \label{eq:likelihood}
  \mathcal{L} = \prod_i^{\mathrm{bins}} \mathcal{P}(n_i; \nu_i)  \, \times \, \prod_k \mathcal{G}_k \, \times \, \mathcal{P}_{\mathrm{cont}}  \, ,
\end{align}
where $n_i$ denotes the number of observed data events and $\nu_i$ is the total number of expected signal and background events in a given bin $i$. Furthermore, the $\mathcal{G}_k$ denote nuisance-parameter (NP) constraints, whose role is to incorporate systematic uncertainties on e.g. signal and background shapes directly into the fit procedure, with the index $k$ labelling a given uncertainty source. More details of this procedure will be given in Section~\ref{sec:systematics}. The Poisson term $\mathcal{P}_{\mathrm{cont}}$ constrains the normalization of the continuum contribution to its expectation as determined from off-resonance collision events. The number of expected signal and background events in a given bin, $\nu_i$, is estimated using simulated collision events and is given by 
\begin{align}
 \nu_i = \eta^{\mathrm{sig}} f_{i}^{\mathrm{sig}} +  \eta^{B\,\mathrm{bkg}} f_{i}^{B\,\mathrm{bkg}} +  \eta^{\mathrm{cont}} f_{i}^{\mathrm{cont}} \, .
\end{align}
Here, $\eta^{\mathrm{sig}}$ is the total number of \bclnu\ signal events. Furthermore, $\eta^{B \,\mathrm{bkg}}$ denotes the background events stemming from double semileptonic cascades, \bulnu decays, and from hadrons misidentified as leptons originating from $B$ meson decays. The number of continuum events is denoted as $\eta^{\mathrm{cont}}$. Furthermore, $f_i$  denotes the fraction of events being reconstructed in a bin $i$ with shapes as determined by the MC simulation for a given event category. Eq.~\ref{eq:likelihood} is numerically maximized to determine both the total number of \bclnu\ and background events from the observed event yields. This is done using the sequential least squares programming method implementation of Ref.~\cite{James:1975dr,iminuit}. The fit is carried out in 20 equidistant bins of $M_X$ ranging from 0 to 3.5 GeV to determine the number of background events for each studied threshold selection on $q^2$, taking into account systematic uncertainties on the composition of \bclnu and background templates (more details about these will be discussed in Section~\ref{sec:systematics}). The continuum constraint $\mathcal{P}_{\mathrm{cont}}$ is adjusted to reflect the number of continuum events for a given $q^2$ selection value as determined by the off-resonance sample, for which the $M_{\mathrm{bc}}$ selection was adjusted to account for the difference in center-of-mass energies.

\begin{figure*}
  \includegraphics[width=0.39\textwidth]{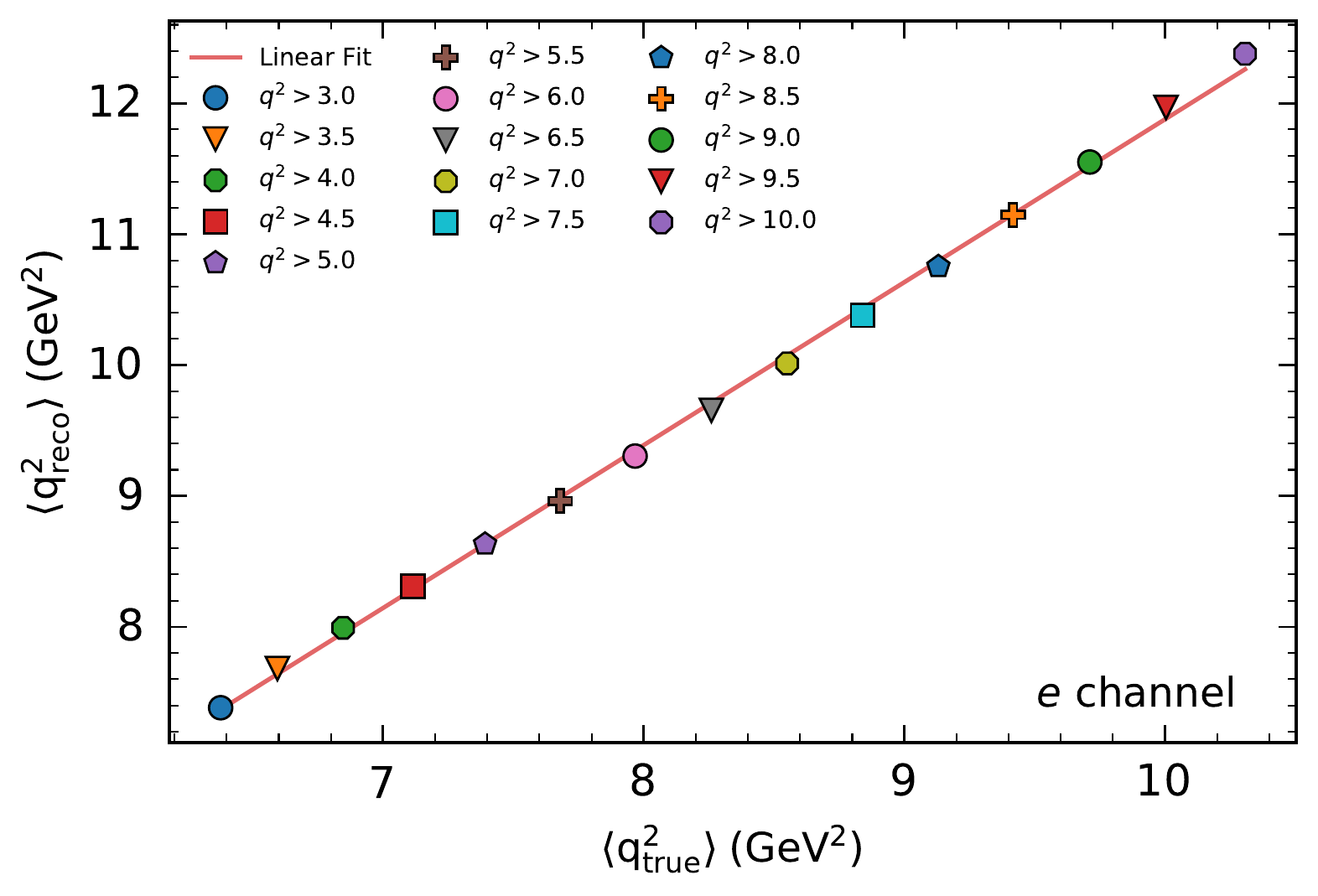}
  \includegraphics[width=0.39\textwidth]{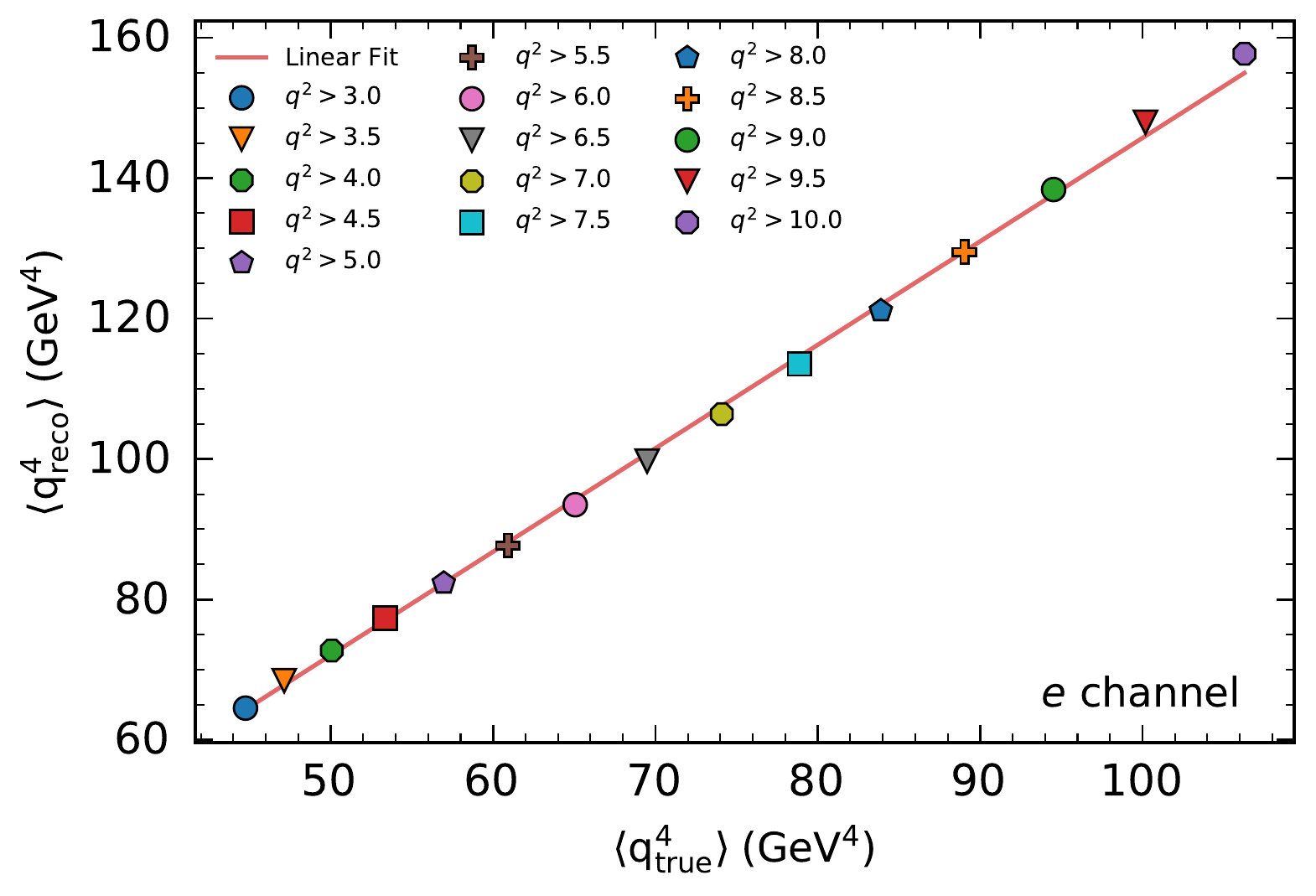}
  \includegraphics[width=0.39\textwidth]{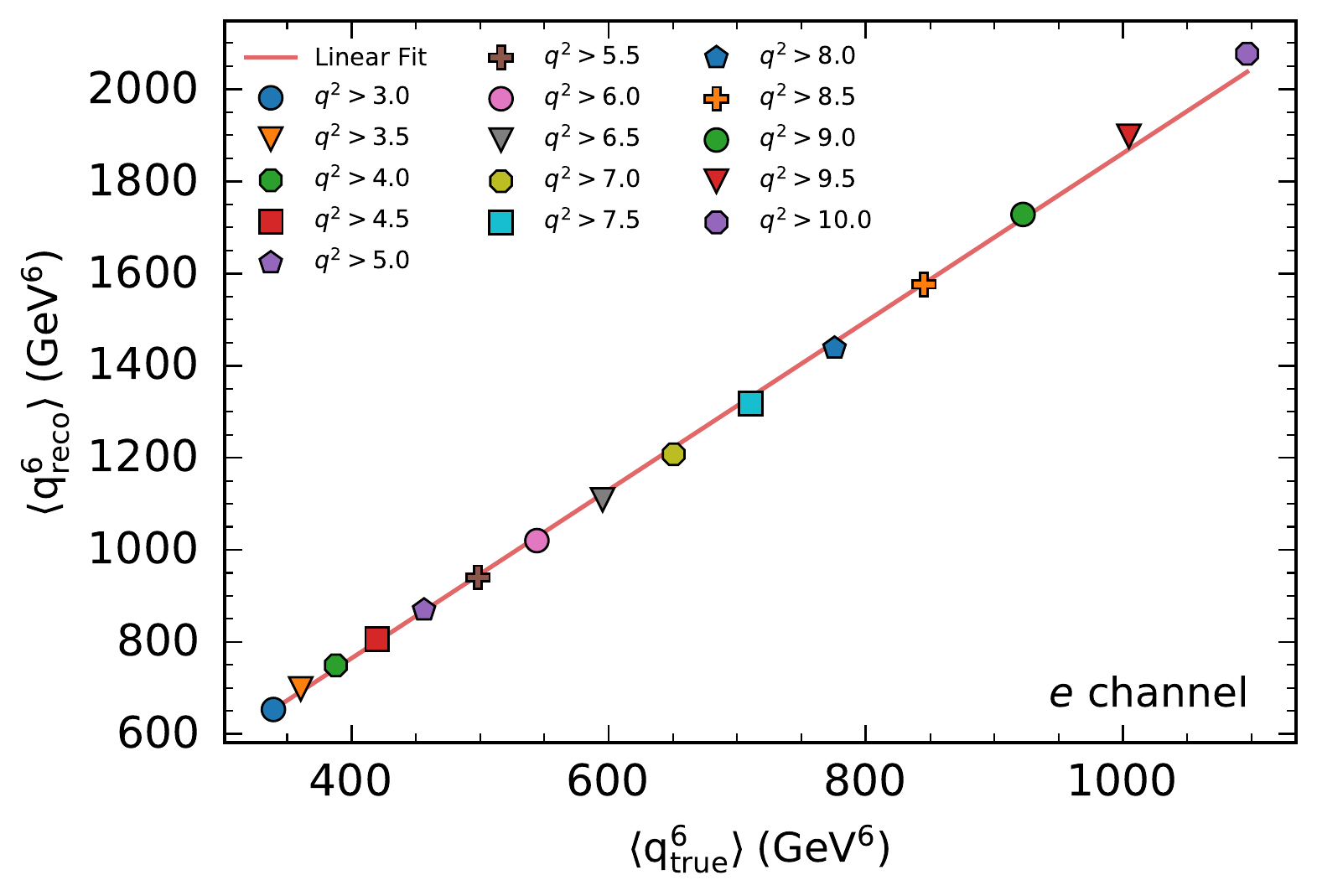}
  \includegraphics[width=0.39\textwidth]{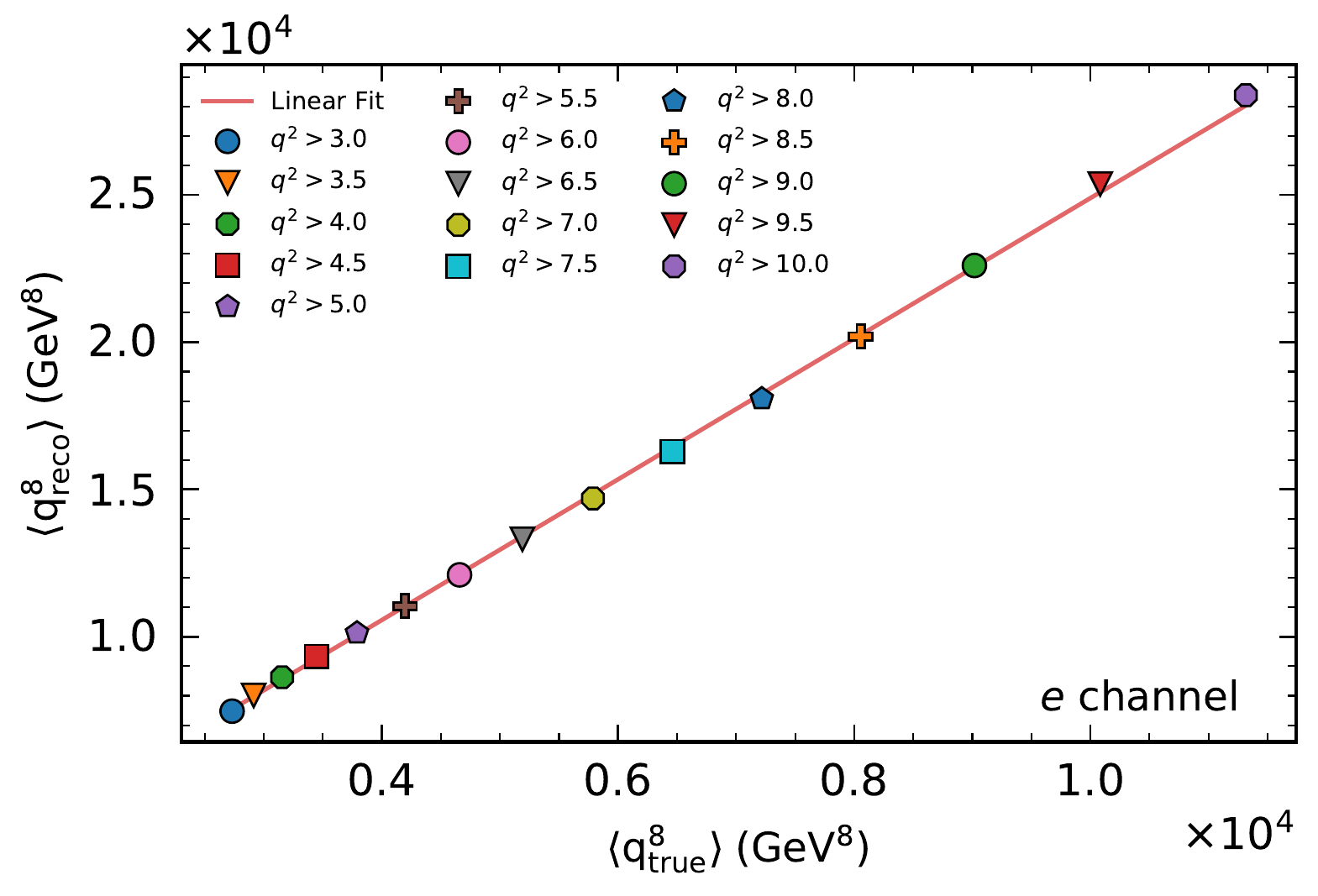}
\caption{
The generator-level, $\langle q^{2m}_{\text{true}} \rangle$, and reconstructed, $\langle q^{2m}_{\text{reco}} \rangle$, \bclnu moments are shown for the first to fourth $q^{2}$ moment for electrons. The different markers represent different threshold selections on the generator-level and reconstructed $q^2$ and the relation between the moments can be described with a linear parametrization of the form \mbox{$a_m +  \langle q^{2m}_{\text{true}} \rangle \cdot b_m =  \langle q^{2m}_{\text{reco}} \rangle $}. The corresponding plots for muons can be found in Appendix B.
}
\label{fig:calib_lin}
\end{figure*}
In a second step, the determined number of background ($\widehat \eta^{\mathrm{bkg}}$) and continuum ($\widehat \eta^{\mathrm{cont}}$) events are used to construct binned signal probabilities as a function of $q^2$, which is defined as 
\begin{align}
 w_i = 1 - \frac{ \widehat \eta^{\mathrm{bkg}} \tilde f_{i}^{\,\mathrm{bkg}} +  \widehat \eta^{\mathrm{cont}} \tilde f_{i}^{\,\mathrm{cont}} }{ n_i } \, .
\end{align}
Here, $ \tilde f_{i}$ denotes the estimated fractions of events being reconstructed in a bin $i$ of $q^2$ for a given background category as determined by the MC simulation.
Fig.~\ref{fig:m_x_w} shows the $q^{2}$ spectrum for electron and muon candidates and the $w_i$ distribution for the threshold selection with $q^2 > 3.0\, \mathrm{GeV}^2$. To avoid dependence on binning effects, we fit the binned signal probabilities for each $q^2$ selection with a polynomial function of a given order $n$ to determine event-by-event weights, $w(q^{2})$, by performing a $\chi^{2}$ minimization. The order of the polynomial is determined using a nested hypothesis test and we accept a polynomial of order $n$ over $n-1$ if the improvement in $\chi^2$ is larger than one. Furthermore, the $\chi^2$ takes into account the full experimental covariance of the background shapes. The resulting polynomial, $w(q^2)$, allows for an unbinned background subtraction and is required to have positive or zero event weights. We set all weights with negative values to zero. The matching figures for other $q^2$ threshold selections can be found in Appendix B.


\section{Moment Calibration Master Formula}\label{sec:calibration}

The reconstructed $q^2$ values are distorted by the reconstruction of the $X_{c}$ system and selection criteria. To measure the first to the fourth moment of the \bclnu $q^2$ spectrum, we need to correct for these effects. 
This is done in three sequential steps:
\begin{itemize}
 \item[-] First, a calibration function is applied event-by-event as a function of the reconstructed $q^2$ value to correct for resolution distortions. This linear correction is determined separately for each order of the moment we wish to measure. For a given event $i$, we denote the calibrated $q^2$ value in the following as $q^{2m}_{\mathrm{cal}\,i}$ with $m$ being the order of the moment. 
 \item[-] In a second step we correct for any residual bias not captured by the linear calibration function for a given moment using a $q^2$ threshold and moment-order dependent correction factor, which we denote as $C_{\mathrm{cal}}$.  
 \item[-] Thirdly, we correct for acceptance and selection effects by applying a correction factor, $C_{\mathrm{acc}}$. This correction is calculated for each $q^{2}$ threshold selection as well as each order of moment.
\end{itemize}
\begin{figure*}
  \includegraphics[width=0.39\textwidth]{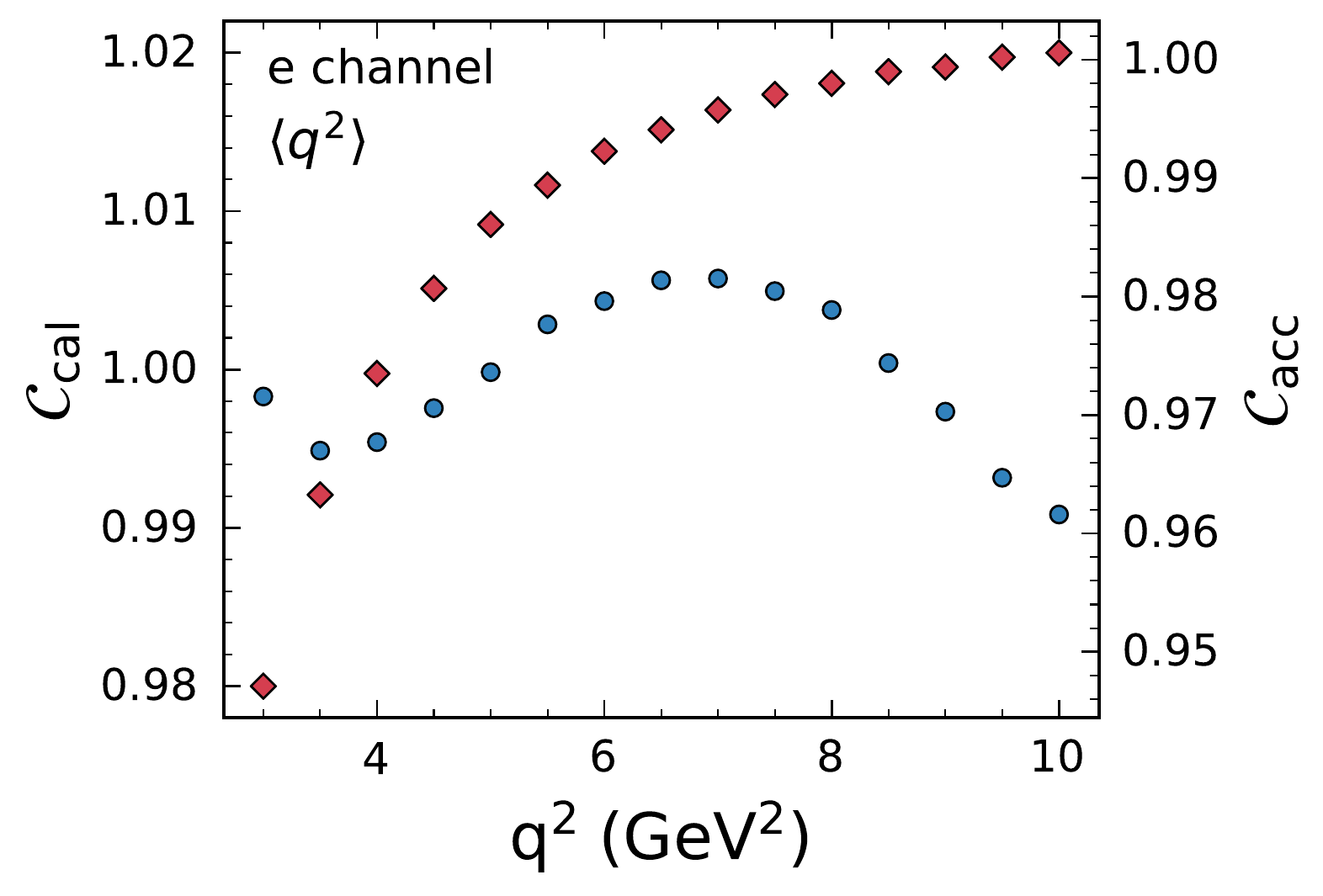}  
  \includegraphics[width=0.39\textwidth]{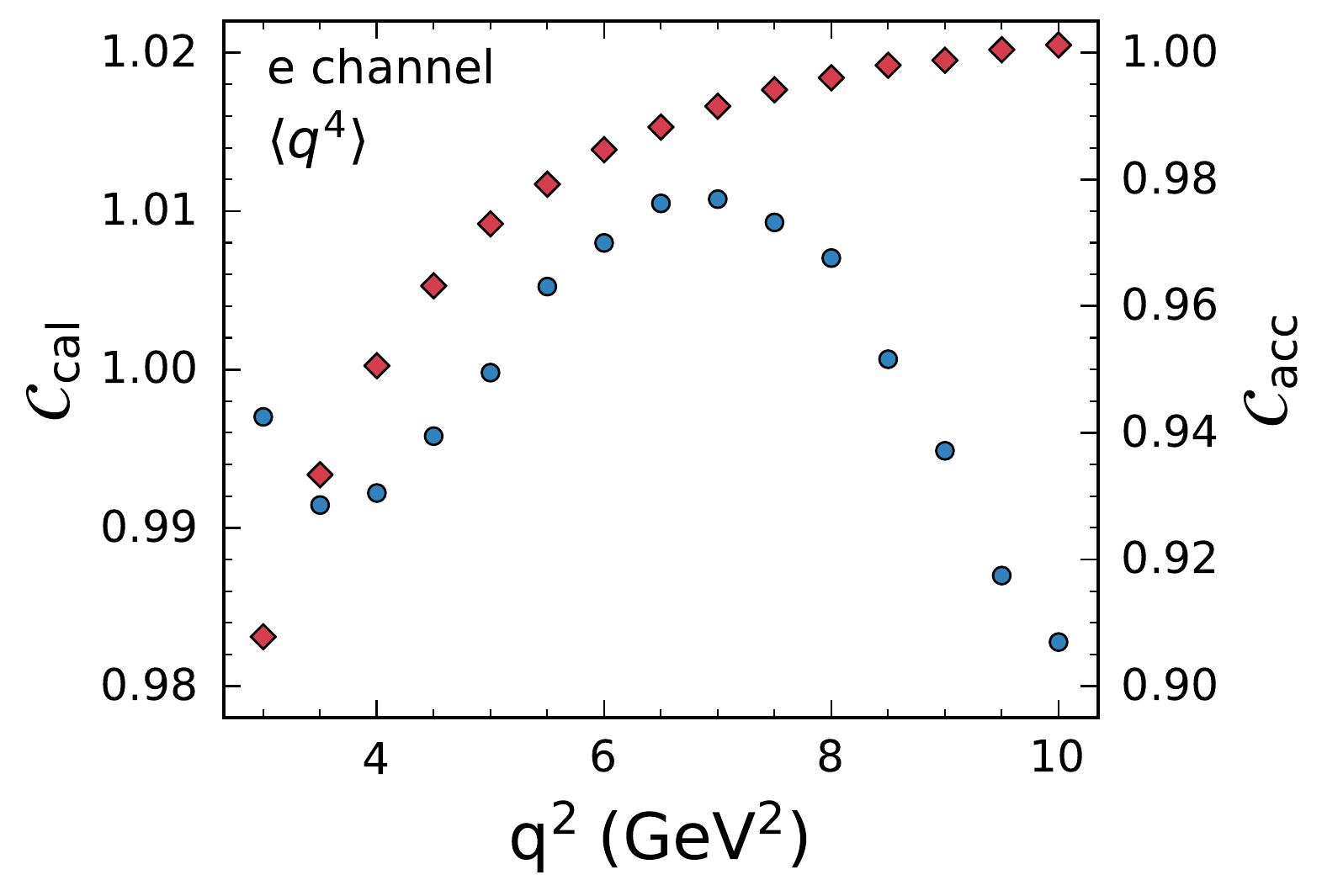}  
  \includegraphics[width=0.39\textwidth]{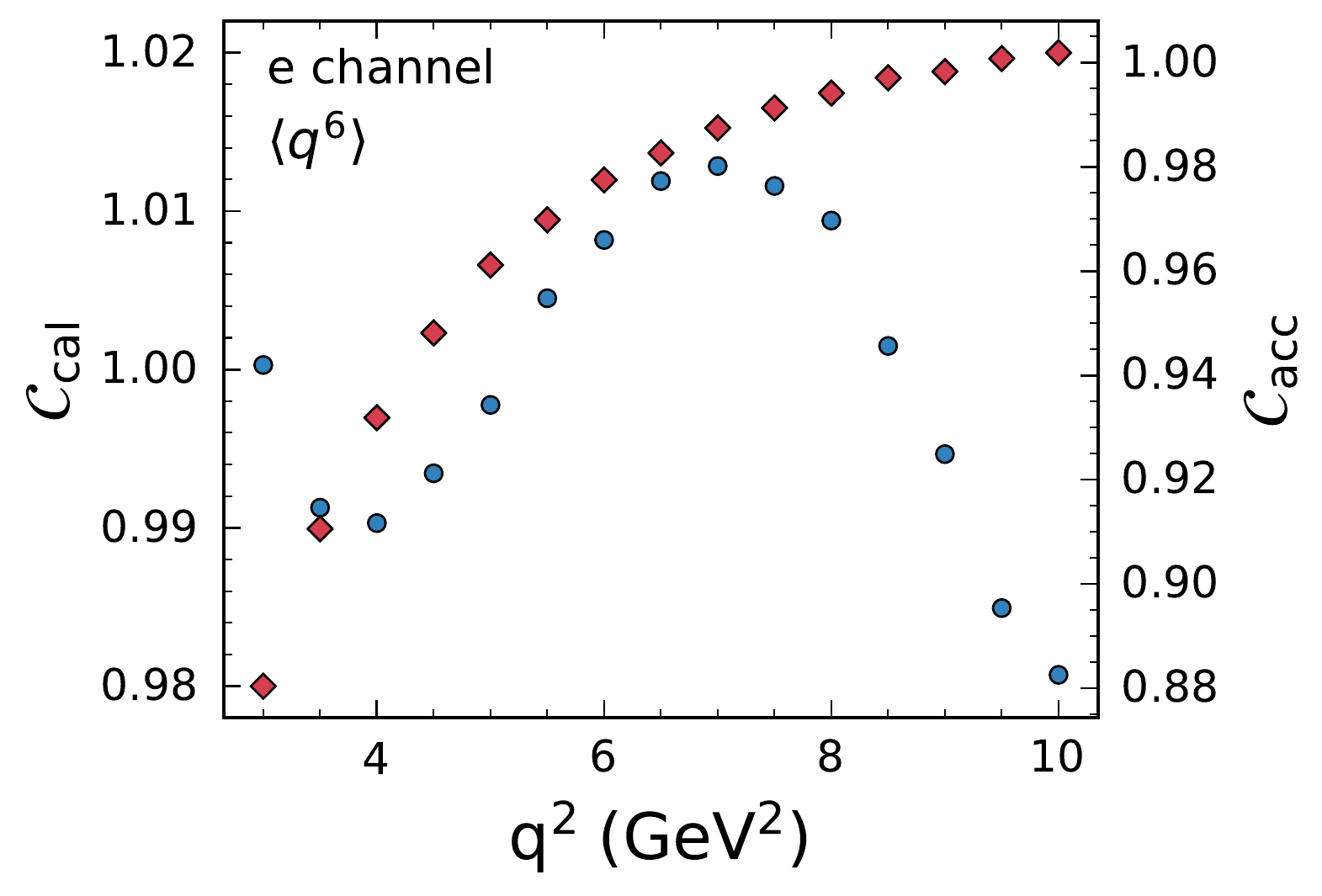}  
  \includegraphics[width=0.39\textwidth]{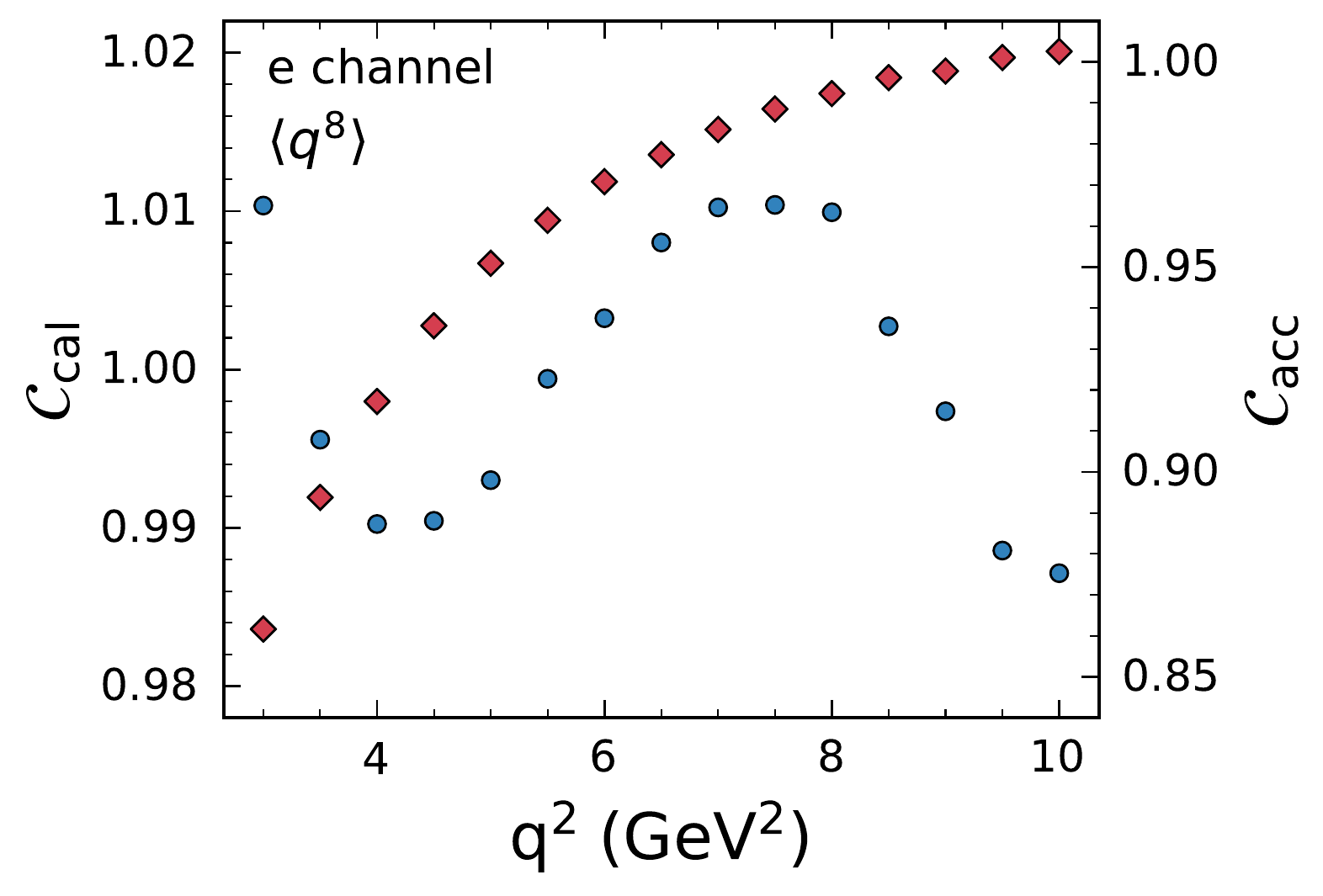}  
\caption{
 The residual bias and acceptance correction factors, denoted as $C_{\mathrm{cal}}$ (circles) and $C_{acc}$ (diamonds), respectively, are shown for the first to fourth $q^{2}$ moment for electrons. The corresponding plots for muons can be found in Appendix B. 
 }
\label{fig:C_cal_C_acc_el}
\end{figure*}
The linear calibration functions and the two correction factors are determined using simulated event samples, which are statistically independent from the simulated samples used in the background subtraction procedure. The calibration function is determined by comparing the reconstructed and generator-level moments. Fig.~\ref{fig:calib_lin} shows the first to fourth generator-level and reconstructed \bclnu moments with different selections on the generator-level and reconstructed $q^2$ value for electrons. There exists a strong linear relationship for the studied order of moments, and we determine a linear calibration function of the form,
\begin{align}
  q^{2m}_{\mathrm{cal}\,i} = \frac{ q^{2m}_{i} - a_m }{ b_m} \, . 
\end{align}
with $a_m$ and $b_m$ denoting the offset and slope for moments of the order $m$. The corresponding figures for muons can be found in Appendix C and show the same linear behavior. In addition, a summary of the fitted parameters of the determined calibration curves is also given in Appendix C. The residual bias correction factor $C_{\mathrm{cal}}$ is determined by comparing simulated samples of the calibrated and generator-level moments for each $q^2$ threshold selection and order of moment under study. Lastly, since different selection efficiencies are observed for different \bclnu\ processes, we determine an additional factor accounting for selection and acceptance effects. The correction factor,  $C_{\mathrm{acc}}$, is calculated by comparing the moments of the generator-level simulated events with a sample without any selection criteria applied. Fig.~\ref{fig:C_cal_C_acc_el} shows the size of both calibration factors for the first to fourth $q^{2}$ moment for electrons. Both factors are close to unity and the corresponding factors for muons, displaying a similar behavior, are found in Appendix B. In addition, selection and acceptance efficiencies for generator-level simulated \bclnu\ samples are shown in Appendix D.

The $q^2$ moments are then given by
\begin{align}
 \langle q^{2m} \rangle = \frac{ C_{\mathrm{cal}} \cdot C_{\mathrm{acc}}}{\sum_i^{\mathrm{events}} w(q^{2}_{i})} \times \sum_i^{\mathrm{events}} w(q^{2}_{i}) \cdot q^{2m}_{\mathrm{cal}\,i} \, .
\end{align}
Here the sums run over all selected events and $q^{2}_{i}$ denotes the measured four-momentum transfer squared of a given event $i$ with a corresponding calibrated four-momentum transfer squared $q^{2m}_{\mathrm{cal}\,i}$ to the power of $m$. The continuous signal probability $w(q^{2}_{i})$ is calculated for each event, while the calculated moments are normalized by the sum of signal probabilities. The full background subtraction and calibration procedure was tested on ensembles of statistically independent simulated samples and no statistical significant biases in the unfolded moments are observed.


\begin{table*}
\caption{
	Summary of statistical and systematic uncertainties for the moments $\langle q^{2,4,6,8} \rangle$ for the electron channel. The values are given as the relative error in permille.
}
\label{tab:q2_q4_q6_q8_el}
\vspace{1ex}
 \footnotesize
 \renewcommand{\arraystretch}{1.05} 
\resizebox{2.\columnwidth}{!}{%
\begin{tabular}{lrrrrrrrrrrrrrrr}
        \hline
        \hline
        $q^{2}$ selection in GeV$^{2}$ &   3.0  &  3.5  &  4.0  &  4.5  &  5.0  &  5.5  &  6.0  &  6.5  &  7.0  &  7.5  &  8.0  &  8.5  &  9.0  &   9.5  &   10.0 \\
        \hline
$\langle q^{2} \rangle$ in GeV$^{2}$          &  6.21 &  6.51 &  6.81 &  7.10 &  7.41 &  7.71 &  8.01 &  8.30 &  8.60 &  8.88 &  9.19 &  9.48 &  9.78 &  10.07 &  10.38 \\
\hline
Stat. error (data)                                 &   1.42 &   1.36 &   1.31 &   1.27 &  1.24 &  1.21 &  1.18 &  1.17 &  1.16 &  1.16 &  1.17 &  1.19 &  1.22 &   1.25 &   1.29 \\
Bkg. subtraction                                   &   1.06 &   0.76 &   0.57 &   0.40 &  0.37 &  0.41 &  0.48 &  0.55 &  0.60 &  0.63 &  0.67 &  0.70 &  0.77 &   0.83 &   0.91 \\
$B \rightarrow X_{u} \ell \nu$ BF                  &   1.80 &   1.66 &   1.52 &   1.10 &  0.78 &  0.57 &  0.42 &  0.29 &  0.23 &  0.19 &  0.16 &  0.13 &  0.12 &   0.15 &   0.10 \\
$B \rightarrow X_{c} \ell \nu$ BF                  &   4.94 &   4.72 &   4.90 &   4.55 &  4.47 &  4.30 &  3.77 &  3.47 &  3.28 &  2.99 &  2.49 &  1.96 &  1.41 &   1.27 &   1.16 \\
Non-resonant model                                 &  13.20 &  11.54 &  10.14 &   8.33 &  6.93 &  5.78 &  4.50 &  3.54 &  2.76 &  2.21 &  1.63 &  1.24 &  0.81 &   0.77 &   0.66 \\
$B \rightarrow X_{c} \ell \nu$ FF                  &   1.63 &   1.47 &   1.27 &   1.10 &  0.94 &  0.84 &  0.79 &  0.75 &  0.69 &  0.63 &  0.56 &  0.49 &  0.42 &   0.40 &   0.37 \\
$N_{\text{tracks}}$ res.                           &   4.67 &   4.40 &   4.17 &   3.91 &  3.72 &  3.50 &  3.31 &  3.13 &  2.95 &  2.83 &  2.66 &  2.50 &  2.40 &   2.24 &   2.13 \\
$N_{\gamma}$ res.                                  &   0.50 &   0.49 &   0.44 &   0.43 &  0.40 &  0.37 &  0.36 &  0.36 &  0.36 &  0.35 &  0.33 &  0.31 &  0.31 &   0.32 &   0.32 \\
$E_{\mathrm{miss}} -  |\bold{p}_{\mathrm{miss}}|$ shape &   0.71 &   0.69 &   0.73 &   0.68 &  0.77 &  0.86 &  0.89 &  0.94 &  0.97 &  1.06 &  1.14 &  1.20 &  1.30 &   1.29 &   1.31 \\
$q^{2}$ scale                                      &   8.98 &   6.77 &   6.12 &   5.77 &   5.70 &  5.50 &  5.12 &  4.86 &  4.72 &  4.62 &  4.25 &  4.33 &  3.91 &   3.90 &   3.94 \\
MC non-closure                                     &   0.06 &   0.09 &   0.03 &   0.01 &  0.01 &  0.03 &  0.02 &  0.01 &  0.01 &  0.01 &  0.00 &  0.03 &  0.00 &   0.01 &   0.00 \\
Cal. function                                      &   0.13 &   0.08 &   0.03 &   0.02 &  0.08 &  0.12 &  0.17 &  0.22 &  0.26 &  0.30 &  0.34 &  0.38 &  0.42 &   0.45 &   0.49 \\
Stat. bias corr.                                   &   1.22 &   1.18 &   1.14 &   1.11 &  1.08 &  1.06 &  1.04 &  1.03 &  1.02 &  1.02 &  1.02 &  1.03 &  1.05 &   1.07 &   1.10 \\
PID eff.                                           &   0.17 &   0.16 &   0.14 &   0.13 &  0.13 &  0.11 &  0.10 &  0.09 &  0.09 &  0.08 &  0.08 &  0.08 &  0.07 &   0.06 &   0.05 \\
Track eff.                                         &   0.42 &   0.39 &   0.36 &   0.33 &  0.31 &  0.29 &  0.27 &  0.25 &  0.23 &  0.22 &  0.20 &  0.19 &  0.18 &   0.16 &   0.15 \\
$B^{0}/B^{\pm}$ tag eff.                            &   0.21 &   0.25 &   0.35 &   0.41 &  0.52 &  0.44 &  0.45 &  0.37 &  0.55 &  0.74 &  0.72 &  0.73 &  0.60 &   0.47 &   0.45 \\
\hline
Sys. error (total)                                 &  17.62 &  15.11 &  13.72 &  11.98 &  10.88 &  9.89 &  8.66 &  7.83 &  7.27 &  6.85 &  6.16 &  5.87 &  5.29 &   5.17 &   5.13 \\
\hline
Total rel. error in \permil                        &  17.68 &  15.17 &  13.78 &  12.04 &  10.95 &  9.96 &  8.74 &  7.92 &  7.36 &  6.95 &  6.27 &  5.99 &  5.43 &   5.32 &   5.29 \\
    \hline
	\hline
	
$\langle q^{4} \rangle $ in GeV$^{4}$          &  42.99 &  46.27 &  49.74 &  53.41 &  57.42 &  61.64 &  65.96 &  70.46 &  75.15 &  79.92 &  85.20 &  90.50 &  96.07 &  101.73 &  108.10 \\
\hline
Stat. error (data)                                 &   3.23 &   3.13 &   3.04 &   2.96 &   2.88 &   2.82 &   2.77 &   2.74 &   2.72 &   2.72 &   2.74 &   2.78 &   2.84 &    2.92 &    3.01 \\
Bkg. subtraction                                   &   1.78 &   1.35 &   1.14 &   0.99 &   1.06 &   1.17 &   1.33 &   1.49 &   1.58 &   1.65 &   1.71 &   1.78 &   1.93 &    2.05 &    2.23 \\
$B \rightarrow X_{u} \ell \nu$ BF                  &   3.84 &   3.52 &   3.16 &   2.29 &   1.64 &   1.21 &   0.90 &   0.63 &   0.52 &   0.43 &   0.37 &   0.31 &   0.29 &    0.34 &    0.24 \\
$B \rightarrow X_{c} \ell \nu$ BF                  &   9.45 &   9.45 &   9.93 &   9.48 &   9.41 &   9.10 &   8.13 &   7.54 &   7.11 &   6.45 &   5.40 &   4.31 &   3.18 &    2.88 &    2.62 \\
Non-resonant model                                 &  25.19 &  22.36 &  19.84 &  16.57 &  13.95 &  11.72 &   9.25 &   7.36 &   5.81 &   4.68 &   3.49 &   2.69 &   1.82 &    1.73 &    1.49 \\
$B \rightarrow X_{c} \ell \nu$ FF                  &   3.13 &   2.84 &   2.51 &   2.22 &   1.95 &   1.81 &   1.72 &   1.64 &   1.52 &   1.41 &   1.26 &   1.12 &   0.98 &    0.92 &    0.85 \\
$N_{\text{tracks}}$ res.                           &  10.44 &   9.95 &   9.49 &   8.98 &   8.56 &   8.09 &   7.67 &   7.26 &   6.84 &   6.52 &   6.13 &   5.75 &   5.48 &    5.11 &    4.84 \\
$N_{\gamma}$ res.                                  &   1.18 &   1.16 &   1.07 &   1.05 &   1.00 &   0.93 &   0.91 &   0.90 &   0.90 &   0.86 &   0.82 &   0.79 &   0.78 &    0.81 &    0.80 \\
$E_{\mathrm{miss}} -  |\bold{p}_{\mathrm{miss}}|$ shape &   2.17 &   2.11 &   2.15 &   2.05 &   2.20 &   2.32 &   2.36 &   2.43 &   2.47 &   2.62 &   2.75 &   2.83 &   3.00 &    2.95 &    2.93 \\
$q^{2}$ scale                                      &  18.61 &  15.14 &  14.05 &  13.28 &  13.10 &  12.57 &  11.79 &  11.15 &  10.79 &  10.59 &   9.81 &   9.95 &   9.04 &    9.03 &    9.15 \\
MC non-closure                                     &   0.03 &   0.01 &   0.01 &   0.01 &   0.01 &   0.01 &   0.01 &   0.01 &   0.01 &   0.01 &   0.00 &   0.01 &   0.00 &    0.00 &    0.00 \\
Cal. function                                      &   0.27 &   0.15 &   0.03 &   0.08 &   0.20 &   0.31 &   0.42 &   0.52 &   0.61 &   0.70 &   0.79 &   0.88 &   0.95 &    1.03 &    1.10 \\
Stat. bias corr.                                   &   2.68 &   2.63 &   2.57 &   2.52 &   2.47 &   2.43 &   2.40 &   2.37 &   2.35 &   2.35 &   2.35 &   2.37 &   2.40 &    2.45 &    2.50 \\
PID eff.                                           &   0.36 &   0.33 &   0.31 &   0.29 &   0.28 &   0.24 &   0.22 &   0.20 &   0.20 &   0.19 &   0.18 &   0.17 &   0.14 &    0.12 &    0.10 \\
Track eff.                                         &   0.90 &   0.85 &   0.80 &   0.75 &   0.71 &   0.66 &   0.62 &   0.58 &   0.53 &   0.50 &   0.46 &   0.43 &   0.40 &    0.36 &    0.34 \\
$B^{0}/B^{\pm}$ tag eff.                            &   0.81 &   0.89 &   1.03 &   1.12 &   1.30 &   1.14 &   1.15 &   1.00 &   1.32 &   1.65 &   1.60 &   1.58 &   1.30 &    1.03 &    0.97 \\
\hline
Sys. error (total)                                 &  34.95 &  30.89 &  28.49 &  25.42 &  23.45 &  21.54 &  19.21 &  17.53 &  16.35 &  15.48 &  14.04 &  13.42 &  12.18 &   11.90 &   11.82 \\
\hline
Total rel. error in \permil                          &  35.10 &  31.04 &  28.65 &  25.59 &  23.63 &  21.72 &  19.41 &  17.74 &  16.58 &  15.72 &  14.30 &  13.71 &  12.50 &   12.25 &   12.20 \\
    \hline
	\hline
	
$\langle q^{6} \rangle $ in GeV$^{6}$          &  326.23 &  355.51 &  387.59 &  423.92 &  465.17 &  510.57 &  558.65 &  610.98 &  667.93 &  728.29 &  797.82 &  870.48 &  949.14 &  1031.95 &  1128.69 \\
\hline
Stat. error (data)                                 &    5.84 &    5.70 &    5.56 &    5.40 &    5.26 &    5.13 &    5.03 &    4.96 &    4.91 &    4.89 &    4.89 &    4.94 &    5.03 &     5.14 &     5.27 \\
Bkg. subtraction                                   &    2.55 &    2.18 &    2.20 &    2.14 &    2.38 &    2.54 &    2.81 &    3.05 &    3.14 &    3.22 &    3.30 &    3.37 &    3.62 &     3.81 &     4.11 \\
$B \rightarrow X_{u} \ell \nu$ BF                  &    6.43 &    5.82 &    5.07 &    3.66 &    2.62 &    1.97 &    1.45 &    1.04 &    0.88 &    0.74 &    0.65 &    0.55 &    0.51 &     0.59 &     0.42 \\
$B \rightarrow X_{c} \ell \nu$ BF                  &   13.83 &   14.41 &   15.25 &   14.81 &   14.75 &   14.32 &   12.97 &   12.09 &   11.37 &   10.31 &    8.70 &    7.01 &    5.31 &     4.81 &     4.38 \\
Non-resonant model                                 &   35.93 &   32.23 &   28.84 &   24.43 &   20.79 &   17.60 &   14.08 &   11.34 &    9.04 &    7.33 &    5.55 &    4.33 &    3.03 &     2.85 &     2.48 \\
$B \rightarrow X_{c} \ell \nu$ FF                  &    4.42 &    4.10 &    3.73 &    3.41 &    3.12 &    2.96 &    2.84 &    2.72 &    2.54 &    2.36 &    2.12 &    1.90 &    1.68 &     1.58 &     1.44 \\
$N_{\text{tracks}}$ res.                           &   17.45 &   16.76 &   16.09 &   15.31 &   14.63 &   13.87 &   13.17 &   12.46 &   11.75 &   11.16 &   10.46 &    9.79 &    9.28 &     8.63 &     8.12 \\
$N_{\gamma}$ res.                                  &    2.15 &    2.11 &    1.99 &    1.95 &    1.87 &    1.76 &    1.72 &    1.69 &    1.68 &    1.61 &    1.54 &    1.48 &    1.47 &     1.49 &     1.47 \\
$E_{\mathrm{miss}} -  |\bold{p}_{\mathrm{miss}}|$ shape &    4.56 &    4.44 &    4.43 &    4.27 &    4.41 &    4.53 &    4.53 &    4.58 &    4.59 &    4.75 &    4.88 &    4.93 &    5.10 &     4.96 &     4.86 \\
$q^{2}$ scale                                      &   29.62 &   25.28 &   23.94 &   22.61 &   22.35 &   21.35 &   20.15 &   19.01 &   18.35 &   18.07 &   16.85 &   17.06 &   15.55 &    15.56 &    15.83 \\
MC non-closure                                     &    0.01 &    0.00 &    0.00 &    0.00 &    0.00 &    0.00 &    0.00 &    0.00 &    0.00 &    0.00 &    0.00 &    0.00 &    0.00 &     0.00 &     0.00 \\
Cal. function                                      &    0.42 &    0.22 &    0.02 &    0.18 &    0.38 &    0.57 &    0.75 &    0.92 &    1.09 &    1.24 &    1.39 &    1.53 &    1.65 &     1.77 &     1.87 \\
Stat. bias corr.                                   &    4.68 &    4.60 &    4.52 &    4.44 &    4.36 &    4.29 &    4.23 &    4.18 &    4.14 &    4.11 &    4.10 &    4.12 &    4.15 &     4.20 &     4.27 \\
PID eff.                                           &    0.57 &    0.53 &    0.50 &    0.47 &    0.45 &    0.40 &    0.37 &    0.33 &    0.33 &    0.30 &    0.29 &    0.27 &    0.23 &     0.19 &     0.16 \\
Track eff.                                         &    1.47 &    1.40 &    1.33 &    1.25 &    1.18 &    1.10 &    1.04 &    0.97 &    0.90 &    0.84 &    0.78 &    0.72 &    0.67 &     0.61 &     0.57 \\
$B^{0}/B^{\pm}$ tag eff.                            &    1.79 &    1.87 &    2.03 &    2.13 &    2.34 &    2.11 &    2.09 &    1.86 &    2.29 &    2.71 &    2.58 &    2.50 &    2.06 &     1.64 &     1.52 \\
\hline
Sys. error (total)                                 &   52.77 &   47.68 &   44.62 &   40.49 &   37.88 &   35.07 &   31.80 &   29.25 &   27.40 &   26.06 &   23.83 &   22.86 &   20.84 &    20.39 &    20.28 \\
\hline
Total rel. error in \permil                         &   53.09 &   48.01 &   44.96 &   40.85 &   38.24 &   35.44 &   32.20 &   29.67 &   27.84 &   26.52 &   24.33 &   23.39 &   21.44 &    21.03 &    20.95 \\
    \hline
	\hline
	
$\langle q^{8} \rangle $ in GeV$^{8}$          &  2663.52 &  2914.20 &  3192.80 &  3531.01 &  3925.20 &  4375.31 &  4862.26 &  5414.29 &  6036.81 &  6723.88 &  7545.39 &  8439.78 &  9431.26 &  10512.44 &  11818.99 \\
\hline
Stat. error (data)                                 &     9.77 &     9.58 &     9.38 &     9.11 &     8.84 &     8.58 &     8.40 &     8.23 &     8.09 &     8.00 &     7.93 &     7.93 &     8.01 &      8.13 &      8.25 \\
Bkg. subtraction                                   &     4.02 &     3.88 &     4.31 &     4.24 &     4.67 &     4.85 &     5.20 &     5.52 &     5.51 &     5.58 &     5.65 &     5.71 &     6.05 &      6.31 &      6.77 \\
$B \rightarrow X_{u} \ell \nu$ BF                  &     9.95 &     8.85 &     7.47 &     5.32 &     3.79 &     2.87 &     2.11 &     1.52 &     1.32 &     1.13 &     1.01 &     0.87 &     0.80 &      0.91 &      0.64 \\
$B \rightarrow X_{c} \ell \nu$ BF                  &    19.15 &    20.11 &    21.08 &    20.54 &    20.40 &    19.78 &    18.08 &    16.93 &    15.92 &    14.46 &    12.29 &    10.03 &     7.78 &      7.06 &      6.42 \\
Non-resonant model                                 &    45.70 &    41.23 &    37.11 &    31.73 &    27.22 &    23.21 &    18.80 &    15.31 &    12.33 &    10.08 &     7.73 &     6.11 &     4.40 &      4.13 &      3.61 \\
$B \rightarrow X_{c} \ell \nu$ FF                  &     5.57 &     5.32 &     5.03 &     4.75 &     4.49 &     4.33 &     4.19 &     4.01 &     3.76 &     3.50 &     3.16 &     2.85 &     2.54 &      2.37 &      2.16 \\
$N_{\text{tracks}}$ res.                           &    25.81 &    24.89 &    24.00 &    22.90 &    21.91 &    20.80 &    19.77 &    18.71 &    17.64 &    16.71 &    15.64 &    14.60 &    13.78 &     12.78 &     11.97 \\
$N_{\gamma}$ res.                                  &     3.51 &     3.45 &     3.30 &     3.22 &     3.10 &     2.94 &     2.87 &     2.81 &     2.77 &     2.66 &     2.55 &     2.44 &     2.41 &      2.42 &      2.37 \\
$E_{\mathrm{miss}} -  |\bold{p}_{\mathrm{miss}}|$ shape &     7.95 &     7.73 &     7.63 &     7.37 &     7.46 &     7.52 &     7.44 &     7.42 &     7.34 &     7.44 &     7.51 &     7.47 &     7.59 &      7.30 &      7.07 \\
$q^{2}$ scale                                      &    42.51 &    37.31 &    35.84 &    33.74 &    33.46 &    31.82 &    30.22 &    28.43 &    27.42 &    27.12 &    25.46 &    25.75 &    23.55 &     23.65 &     24.18 \\
MC non-closure                                     &     0.00 &     0.00 &     0.00 &     0.00 &     0.00 &     0.00 &     0.00 &     0.00 &     0.00 &     0.00 &     0.00 &     0.00 &     0.00 &      0.00 &      0.00 \\
Cal. function                                      &     0.62 &     0.31 &     0.01 &     0.31 &     0.62 &     0.93 &     1.20 &     1.47 &     1.72 &     1.95 &     2.18 &     2.39 &     2.57 &      2.74 &      2.89 \\
Stat. bias corr.                                   &     7.59 &     7.45 &     7.31 &     7.16 &     7.02 &     6.89 &     6.76 &     6.65 &     6.56 &     6.48 &     6.43 &     6.40 &     6.40 &      6.44 &      6.51 \\
PID eff.                                           &     0.79 &     0.74 &     0.70 &     0.66 &     0.63 &     0.56 &     0.53 &     0.48 &     0.47 &     0.43 &     0.41 &     0.37 &     0.32 &      0.27 &      0.22 \\
Track eff.                                         &     2.11 &     2.02 &     1.93 &     1.83 &     1.73 &     1.62 &     1.53 &     1.43 &     1.33 &     1.24 &     1.15 &     1.06 &     0.98 &      0.89 &      0.83 \\
$B^{0}/B^{\pm}$ tag eff.                            &     3.03 &     3.10 &     3.26 &     3.34 &     3.56 &     3.24 &     3.18 &     2.86 &     3.35 &     3.81 &     3.59 &     3.43 &     2.81 &      2.24 &      2.05 \\
\hline
Sys. error (total)                &    72.27 &    66.18 &    62.60 &    57.37 &    54.26 &    50.52 &    46.44 &    42.99 &    40.44 &    38.66 &    35.64 &    34.31 &    31.44 &     30.82 &     30.73 \\
\hline
Total rel. error in \permil                      &    72.93 &    66.86 &    63.30 &    58.09 &    54.97 &    51.25 &    47.19 &    43.77 &    41.24 &    39.48 &    36.51 &    35.21 &    32.44 &     31.87 &     31.81 \\
    \hline
	\hline
\end{tabular}
}
\end{table*}

\begin{table*}
\caption{
	Summary of statistical and systematic uncertainties for the moments $\langle q^{2,4,6,8} \rangle$ for the muon channel. The values are given as the relative error in permille.
}
\label{tab:q2_q4_q6_q8_mu}
\vspace{1ex}
 \renewcommand{\arraystretch}{1.05}
 \footnotesize 
\resizebox{2.\columnwidth}{!}{%
\begin{tabular}{lrrrrrrrrrrrrrrr}
        \hline
        \hline
        $q^{2}$ selection in GeV$^{2}$ &   3.0  &  3.5  &  4.0  &  4.5  &  5.0  &  5.5  &  6.0  &  6.5  &  7.0  &  7.5  &  8.0  &  8.5  &  9.0  &   9.5  &   10.0 \\
        \hline
$\langle q^{2} \rangle $ in GeV$^{2}$          &   6.25 &   6.54 &   6.83 &   7.13 &   7.42 &  7.72 &  8.02 &  8.32 &  8.61 &  8.91 &  9.21 &  9.51 &  9.80 &  10.09 &  10.40 \\
\hline
Stat. error (data)                                 &   1.51 &   1.45 &   1.39 &   1.34 &   1.30 &  1.27 &  1.24 &  1.22 &  1.21 &  1.20 &  1.20 &  1.22 &  1.23 &   1.26 &   1.29 \\
Bkg. subtraction                                   &   1.34 &   1.12 &   0.90 &   0.71 &   0.59 &  0.53 &  0.49 &  0.49 &  0.57 &  0.63 &  0.65 &  0.70 &  0.76 &   0.77 &   0.82 \\
$B \rightarrow X_{u} \ell \nu$ BF                  &   2.18 &   2.04 &   1.75 &   1.48 &   1.35 &  1.04 &  0.76 &  0.54 &  0.38 &  0.29 &  0.19 &  0.16 &  0.12 &   0.05 &   0.05 \\
$B \rightarrow X_{c} \ell \nu$ BF                  &   4.82 &   5.02 &   5.14 &   5.14 &   5.05 &  5.00 &  4.67 &  4.05 &  3.51 &  3.11 &  2.66 &  2.21 &  1.75 &   1.36 &   1.16 \\
Non-resonant model                                 &  14.25 &  12.72 &  11.04 &   9.28 &   7.83 &  6.62 &  5.42 &  4.00 &  3.02 &  2.28 &  1.65 &  1.43 &  1.04 &   0.86 &   0.78 \\
$B \rightarrow X_{c} \ell \nu$ FF                  &   1.43 &   1.30 &   1.16 &   1.03 &   0.91 &  0.85 &  0.82 &  0.74 &  0.69 &  0.62 &  0.54 &  0.48 &  0.42 &   0.39 &   0.35 \\
$N_{\text{tracks}}$ res.                           &   5.66 &   5.31 &   4.96 &   4.65 &   4.36 &  4.06 &  3.78 &  3.52 &  3.29 &  3.06 &  2.85 &  2.66 &  2.51 &   2.38 &   2.20 \\
$N_{\gamma}$ res.                                  &   0.39 &   0.38 &   0.34 &   0.31 &   0.30 &  0.28 &  0.30 &  0.31 &  0.32 &  0.28 &  0.27 &  0.26 &  0.25 &   0.27 &   0.29 \\
$E_{\mathrm{miss}} -  |\bold{p}_{\mathrm{miss}}|$ shape &   1.29 &   1.26 &   1.21 &   1.17 &   1.15 &  1.11 &  1.04 &  1.05 &  1.06 &  1.09 &  1.16 &  1.20 &  1.30 &   1.33 &   1.29 \\
$q^{2}$ scale                                      &   9.48 &   7.15 &   6.65 &   6.65 &   6.12 &   5.91 &   5.83 &  5.48 &  5.26 &  4.69 &  4.27 &  4.42 &  3.91 &   3.94 &   4.38 \\
MC non-closure                                     &   0.19 &   0.11 &   0.12 &   0.11 &   0.11 &  0.05 &  0.05 &  0.06 &  0.08 &  0.07 &  0.11 &  0.04 &  0.04 &   0.06 &   0.02 \\
Cal. function                                      &   0.13 &   0.08 &   0.03 &   0.02 &   0.07 &  0.12 &  0.17 &  0.22 &  0.26 &  0.31 &  0.35 &  0.39 &  0.43 &   0.47 &   0.51 \\
Stat. bias corr.                                   &   1.32 &   1.27 &   1.23 &   1.19 &   1.16 &  1.13 &  1.10 &  1.08 &  1.07 &  1.06 &  1.06 &  1.06 &  1.07 &   1.09 &   1.11 \\
PID eff.                                           &   0.16 &   0.14 &   0.14 &   0.13 &   0.13 &  0.12 &  0.11 &  0.10 &  0.10 &  0.10 &  0.09 &  0.08 &  0.08 &   0.07 &   0.06 \\
Track eff.                                         &   0.44 &   0.42 &   0.39 &   0.36 &   0.34 &  0.31 &  0.29 &  0.27 &  0.25 &  0.23 &  0.21 &  0.20 &  0.18 &   0.17 &   0.15 \\
$B^{0}/B^{\pm}$ tag eff.                            &   0.46 &   0.58 &   0.50 &   0.44 &   0.51 &  0.40 &  0.28 &  0.34 &  0.36 &  0.38 &  0.29 &  0.23 &  0.20 &   0.12 &   0.47 \\
\hline
Sys. error (total)                                 &  18.99 &  16.65 &  15.03 &  13.62 &  12.22 &  11.19 &  10.17 &  8.86 &  7.97 &  7.06 &  6.30 &  6.09 &  5.44 &   5.27 &   5.50 \\
\hline
Total rel. error in \permil                           &  19.05 &  16.71 &  15.09 &  13.68 &  12.29 &  11.26 &  10.25 &  8.94 &  8.06 &  7.16 &  6.41 &  6.21 &  5.58 &   5.42 &   5.65 \\
            \hline
	\hline
$\langle q^{4} \rangle $ in GeV$^{4}$          &  43.52 &  46.73 &  50.16 &  53.81 &  57.65 &  61.75 &  66.09 &  70.71 &  75.36 &  80.34 &  85.63 &  90.96 &  96.39 &  102.18 &  108.50 \\
\hline
Stat. error (data)                                 &   3.45 &   3.33 &   3.22 &   3.12 &   3.03 &   2.95 &   2.88 &   2.83 &   2.80 &   2.79 &   2.79 &   2.81 &   2.86 &    2.91 &    2.97 \\
Bkg. subtraction                                   &   2.41 &   2.10 &   1.76 &   1.48 &   1.36 &   1.34 &   1.30 &   1.34 &   1.53 &   1.65 &   1.67 &   1.80 &   1.90 &    1.90 &    2.00 \\
$B \rightarrow X_{u} \ell \nu$ BF                  &   4.66 &   4.39 &   3.80 &   3.20 &   2.90 &   2.23 &   1.61 &   1.14 &   0.81 &   0.61 &   0.39 &   0.34 &   0.26 &    0.11 &    0.10 \\
$B \rightarrow X_{c} \ell \nu$ BF                  &   9.55 &  10.30 &  10.68 &  10.78 &  10.67 &  10.50 &   9.84 &   8.63 &   7.52 &   6.67 &   5.69 &   4.73 &   3.78 &    2.97 &    2.52 \\
Non-resonant model                                 &  27.34 &  24.67 &  21.67 &  18.48 &  15.72 &  13.36 &  11.00 &   8.24 &   6.30 &   4.82 &   3.53 &   3.05 &   2.27 &    1.88 &    1.69 \\
$B \rightarrow X_{c} \ell \nu$ FF                  &   2.73 &   2.51 &   2.29 &   2.08 &   1.90 &   1.80 &   1.75 &   1.61 &   1.49 &   1.36 &   1.20 &   1.07 &   0.95 &    0.87 &    0.79 \\
$N_{\text{tracks}}$ res.                           &  12.45 &  11.81 &  11.13 &  10.51 &   9.89 &   9.25 &   8.63 &   8.04 &   7.51 &   6.98 &   6.49 &   6.04 &   5.66 &    5.34 &    4.90 \\
$N_{\gamma}$ res.                                  &   0.95 &   0.92 &   0.85 &   0.79 &   0.78 &   0.74 &   0.77 &   0.78 &   0.78 &   0.71 &   0.69 &   0.66 &   0.64 &    0.67 &    0.71 \\
$E_{\mathrm{miss}} -  |\bold{p}_{\mathrm{miss}}|$ shape &   3.35 &   3.26 &   3.12 &   3.01 &   2.92 &   2.81 &   2.66 &   2.65 &   2.64 &   2.67 &   2.76 &   2.81 &   2.95 &    2.98 &    2.85 \\
$q^{2}$ scale                                      &  19.97 &  16.30 &  15.38 &  15.26 &  14.19 &  13.68 &  13.33 &  12.59 &  11.97 &  10.81 &   9.86 &  10.03 &   8.97 &    9.17 &   10.05 \\
MC non-closure                                     &   0.06 &   0.04 &   0.04 &   0.03 &   0.03 &   0.02 &   0.02 &   0.02 &   0.03 &   0.02 &   0.03 &   0.01 &   0.01 &    0.01 &    0.01 \\
Cal. function                                      &   0.27 &   0.16 &   0.04 &   0.07 &   0.19 &   0.30 &   0.41 &   0.52 &   0.62 &   0.71 &   0.81 &   0.90 &   0.98 &    1.06 &    1.14 \\
Stat. bias corr.                                   &   2.92 &   2.85 &   2.77 &   2.70 &   2.64 &   2.58 &   2.53 &   2.48 &   2.45 &   2.43 &   2.42 &   2.42 &   2.43 &    2.46 &    2.49 \\
PID eff.                                           &   0.35 &   0.32 &   0.31 &   0.29 &   0.29 &   0.26 &   0.25 &   0.23 &   0.22 &   0.22 &   0.20 &   0.18 &   0.18 &    0.16 &    0.13 \\
Track eff.                                         &   0.97 &   0.91 &   0.86 &   0.81 &   0.76 &   0.70 &   0.65 &   0.61 &   0.56 &   0.52 &   0.48 &   0.44 &   0.41 &    0.38 &    0.34 \\
$B^{0}/B^{\pm}$ tag eff.                            &   0.92 &   1.09 &   0.95 &   0.84 &   0.94 &   0.74 &   0.51 &   0.59 &   0.62 &   0.63 &   0.43 &   0.29 &   0.22 &    0.42 &    1.12 \\
\hline
Sys. error (total)                &  38.08 &  34.23 &  31.41 &  28.91 &  26.30 &  24.24 &  22.18 &  19.65 &  17.77 &  15.90 &  14.27 &  13.73 &  12.36 &   12.08 &   12.51 \\
\hline
Total rel. error in \permil                           &  38.23 &  34.39 &  31.58 &  29.08 &  26.48 &  24.42 &  22.37 &  19.85 &  17.99 &  16.14 &  14.54 &  14.02 &  12.69 &   12.43 &   12.86 \\
            \hline
	\hline
$\langle q^{6} \rangle $ in GeV$^{6}$          &  331.70 &  360.44 &  392.58 &  428.49 &  467.70 &  511.77 &  560.28 &  614.62 &  670.90 &  733.57 &  803.64 &  876.13 &  952.91 &  1038.72 &  1135.12 \\
\hline
Stat. error (data)                                 &    6.22 &    6.03 &    5.84 &    5.66 &    5.49 &    5.34 &    5.20 &    5.08 &    5.01 &    4.96 &    4.94 &    4.96 &    5.02 &     5.07 &     5.16 \\
Bkg. subtraction                                   &    3.62 &    3.28 &    2.94 &    2.67 &    2.67 &    2.77 &    2.71 &    2.77 &    3.08 &    3.26 &    3.24 &    3.48 &    3.57 &     3.52 &     3.68 \\
$B \rightarrow X_{u} \ell \nu$ BF                  &    7.78 &    7.30 &    6.34 &    5.26 &    4.74 &    3.62 &    2.57 &    1.81 &    1.27 &    0.95 &    0.59 &    0.52 &    0.41 &     0.15 &     0.14 \\
$B \rightarrow X_{c} \ell \nu$ BF                  &   14.82 &   16.14 &   16.68 &   16.81 &   16.67 &   16.30 &   15.29 &   13.54 &   11.86 &   10.53 &    9.00 &    7.50 &    6.04 &     4.78 &     4.06 \\
Non-resonant model                                 &   39.15 &   35.58 &   31.54 &   27.20 &   23.35 &   19.95 &   16.53 &   12.55 &    9.72 &    7.52 &    5.59 &    4.82 &    3.64 &     3.02 &     2.71 \\
$B \rightarrow X_{c} \ell \nu$ FF                  &    3.86 &    3.65 &    3.42 &    3.19 &    3.00 &    2.88 &    2.82 &    2.61 &    2.44 &    2.23 &    1.98 &    1.78 &    1.58 &     1.44 &     1.30 \\
$N_{\text{tracks}}$ res.                           &   20.45 &   19.54 &   18.55 &   17.59 &   16.63 &   15.59 &   14.58 &   13.60 &   12.70 &   11.79 &   10.93 &   10.14 &    9.46 &     8.85 &     8.10 \\
$N_{\gamma}$ res.                                  &    1.77 &    1.73 &    1.63 &    1.54 &    1.52 &    1.45 &    1.46 &    1.47 &    1.46 &    1.34 &    1.30 &    1.25 &    1.21 &     1.24 &     1.29 \\
$E_{\mathrm{miss}} -  |\bold{p}_{\mathrm{miss}}|$ shape &    6.30 &    6.10 &    5.85 &    5.63 &    5.43 &    5.22 &    4.94 &    4.87 &    4.79 &    4.76 &    4.83 &    4.82 &    4.95 &     4.92 &     4.65 \\
$q^{2}$ shift                                      &   32.17 &   27.63 &   26.32 &   26.02 &   24.37 &   23.45 &   22.66 &   21.49 &   20.24 &   18.50 &   16.91 &   16.97 &   15.32 &    15.86 &    17.17 \\
MC non-closure                                     &    0.01 &    0.01 &    0.01 &    0.01 &    0.01 &    0.00 &    0.00 &    0.00 &    0.01 &    0.00 &    0.01 &    0.00 &    0.00 &     0.00 &     0.00 \\
Cal. function                                      &    0.42 &    0.24 &    0.04 &    0.16 &    0.35 &    0.54 &    0.73 &    0.91 &    1.08 &    1.24 &    1.40 &    1.55 &    1.68 &     1.80 &     1.92 \\
Stat. bias corr.                                   &    5.10 &    4.98 &    4.86 &    4.75 &    4.64 &    4.53 &    4.43 &    4.34 &    4.27 &    4.21 &    4.17 &    4.15 &    4.14 &     4.16 &     4.19 \\
PID eff.                                           &    0.57 &    0.53 &    0.51 &    0.48 &    0.48 &    0.44 &    0.41 &    0.39 &    0.36 &    0.36 &    0.33 &    0.29 &    0.28 &     0.25 &     0.21 \\
Track eff.                                         &    1.57 &    1.50 &    1.42 &    1.34 &    1.26 &    1.18 &    1.09 &    1.02 &    0.94 &    0.87 &    0.80 &    0.73 &    0.67 &     0.62 &     0.56 \\
$B^{0}/B^{\pm}$ tag eff.                            &    1.16 &    1.36 &    1.18 &    1.03 &    1.15 &    0.85 &    0.53 &    0.63 &    0.65 &    0.63 &    0.32 &    0.10 &    0.02 &     0.97 &     1.98 \\
\hline
Sys. error (total)                                 &   58.03 &   53.09 &   49.34 &   45.97 &   42.29 &   39.22 &   36.10 &   32.45 &   29.49 &   26.65 &   24.04 &   23.03 &   20.88 &    20.59 &    21.17 \\
\hline
Total rel. error in \permil                           &   58.36 &   53.43 &   49.68 &   46.32 &   42.65 &   39.58 &   36.47 &   32.84 &   29.91 &   27.11 &   24.54 &   23.56 &   21.47 &    21.21 &    21.79 \\
            \hline
	\hline
$\langle q^{8} \rangle $ in GeV$^{8}$          &  2717.22 &  2963.88 &  3248.31 &  3578.45 &  3947.44 &  4384.73 &  4878.23 &  5458.95 &  6072.92 &  6780.95 &  7616.67 &  8497.60 &  9466.03 &  10603.31 &  11917.23 \\
\hline
Stat. error (data)                                 &    10.35 &    10.07 &     9.78 &     9.47 &     9.19 &     8.89 &     8.63 &     8.36 &     8.19 &     8.05 &     7.94 &     7.91 &     7.94 &      7.95 &      7.99 \\
Bkg. subtraction                                   &     5.57 &     5.25 &     4.98 &     4.80 &     4.99 &     5.23 &     5.02 &     5.06 &     5.51 &     5.71 &     5.58 &     6.00 &     5.96 &      5.81 &      6.02 \\
$B \rightarrow X_{u} \ell \nu$ BF                  &    11.94 &    11.10 &     9.61 &     7.82 &     7.00 &     5.31 &     3.66 &     2.53 &     1.76 &     1.30 &     0.79 &     0.69 &     0.59 &      0.20 &      0.16 \\
$B \rightarrow X_{c} \ell \nu$ BF                  &    21.51 &    22.91 &    23.24 &    23.14 &    22.84 &    22.14 &    20.76 &    18.50 &    16.31 &    14.53 &    12.43 &    10.40 &     8.44 &      6.74 &      5.74 \\
Non-resonant model                                 &    49.93 &    45.52 &    40.56 &    35.22 &    30.45 &    26.13 &    21.80 &    16.75 &    13.12 &    10.26 &     7.73 &     6.66 &     5.10 &      4.25 &      3.79 \\
$B \rightarrow X_{c} \ell \nu$ FF                  &     4.91 &     4.76 &     4.60 &     4.40 &     4.23 &     4.12 &     4.03 &     3.75 &     3.52 &     3.23 &     2.88 &     2.59 &     2.31 &      2.09 &      1.89 \\
$N_{\text{tracks}}$ res.                           &    29.72 &    28.51 &    27.15 &    25.82 &    24.47 &    22.99 &    21.54 &    20.09 &    18.76 &    17.40 &    16.09 &    14.89 &    13.83 &     12.86 &     11.73 \\
$N_{\gamma}$ res.                                  &     2.95 &     2.89 &     2.75 &     2.62 &     2.58 &     2.46 &     2.46 &     2.44 &     2.39 &     2.22 &     2.16 &     2.07 &     2.00 &      2.01 &      2.06 \\
$E_{\mathrm{miss}} -  |\bold{p}_{\mathrm{miss}}|$ shape &    10.18 &     9.83 &     9.42 &     9.05 &     8.69 &     8.33 &     7.89 &     7.70 &     7.50 &     7.35 &     7.33 &     7.21 &     7.26 &      7.11 &      6.66 \\
$q^{2}$ scale                                      &    46.61 &    41.26 &    39.53 &    39.00 &    36.70 &    35.23 &    33.82 &    32.22 &    30.11 &    27.83 &    25.47 &    25.28 &    23.04 &     24.16 &     25.90 \\
MC non-closure                                     &     0.00 &     0.00 &     0.00 &     0.00 &     0.00 &     0.00 &     0.00 &     0.00 &     0.00 &     0.00 &     0.00 &     0.00 &     0.00 &      0.00 &      0.00 \\
Cal. function                                      &     0.63 &     0.34 &     0.04 &     0.26 &     0.55 &     0.86 &     1.14 &     1.42 &     1.68 &     1.92 &     2.17 &     2.37 &     2.56 &      2.74 &      2.90 \\
Stat. bias corr.                                   &     8.21 &     8.00 &     7.79 &     7.59 &     7.38 &     7.18 &     6.99 &     6.83 &     6.67 &     6.54 &     6.43 &     6.35 &     6.29 &      6.27 &      6.27 \\
PID eff.                                           &     0.81 &     0.77 &     0.74 &     0.70 &     0.69 &     0.64 &     0.60 &     0.56 &     0.52 &     0.51 &     0.47 &     0.42 &     0.40 &      0.35 &      0.30 \\
Track eff.                                         &     2.25 &     2.16 &     2.05 &     1.95 &     1.84 &     1.72 &     1.60 &     1.49 &     1.38 &     1.27 &     1.16 &     1.06 &     0.98 &      0.89 &      0.81 \\
$B^{0}/B^{\pm}$ tag eff.                            &     0.97 &     1.19 &     0.98 &     0.80 &     0.94 &     0.58 &     0.18 &     0.29 &     0.30 &     0.26 &     0.14 &     0.42 &     0.59 &      1.79 &      3.06 \\
\hline
Sys. error (total)                &    79.98 &    73.90 &    69.18 &    64.95 &    60.23 &    56.09 &    51.86 &    47.19 &    43.06 &    39.32 &    35.61 &    34.06 &    31.06 &     30.93 &     31.65 \\
\hline
Total rel. error in \permil                           &    80.64 &    74.58 &    69.87 &    65.64 &    60.93 &    56.79 &    52.58 &    47.93 &    43.83 &    40.13 &    36.48 &    34.97 &    32.06 &     31.93 &     32.64 \\
            \hline
	\hline
\end{tabular}
}
\end{table*}


\section{Systematic Uncertainties}\label{sec:systematics}

\begin{figure*}
  \includegraphics[width=0.35\textwidth]{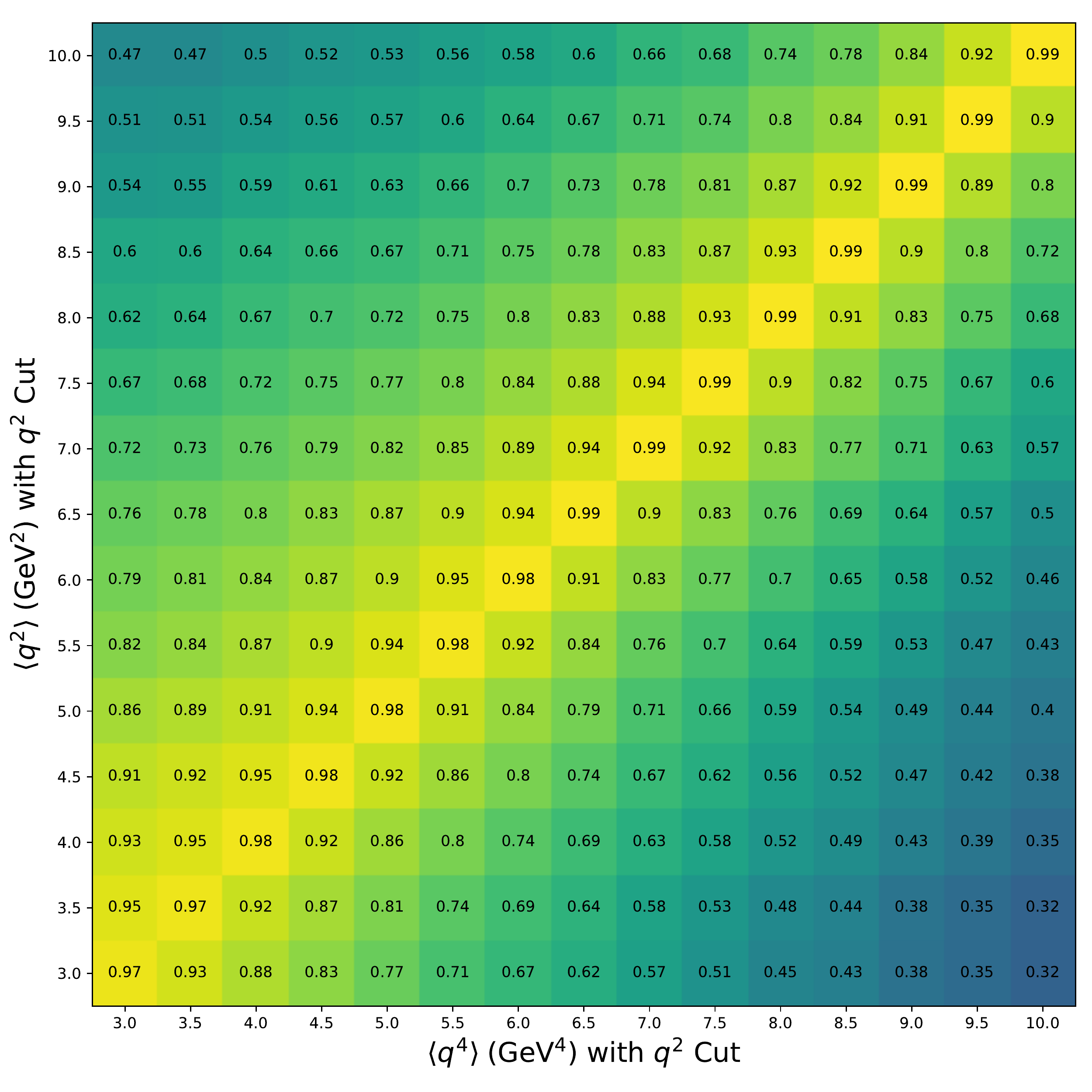}
  \includegraphics[width=0.35\textwidth]{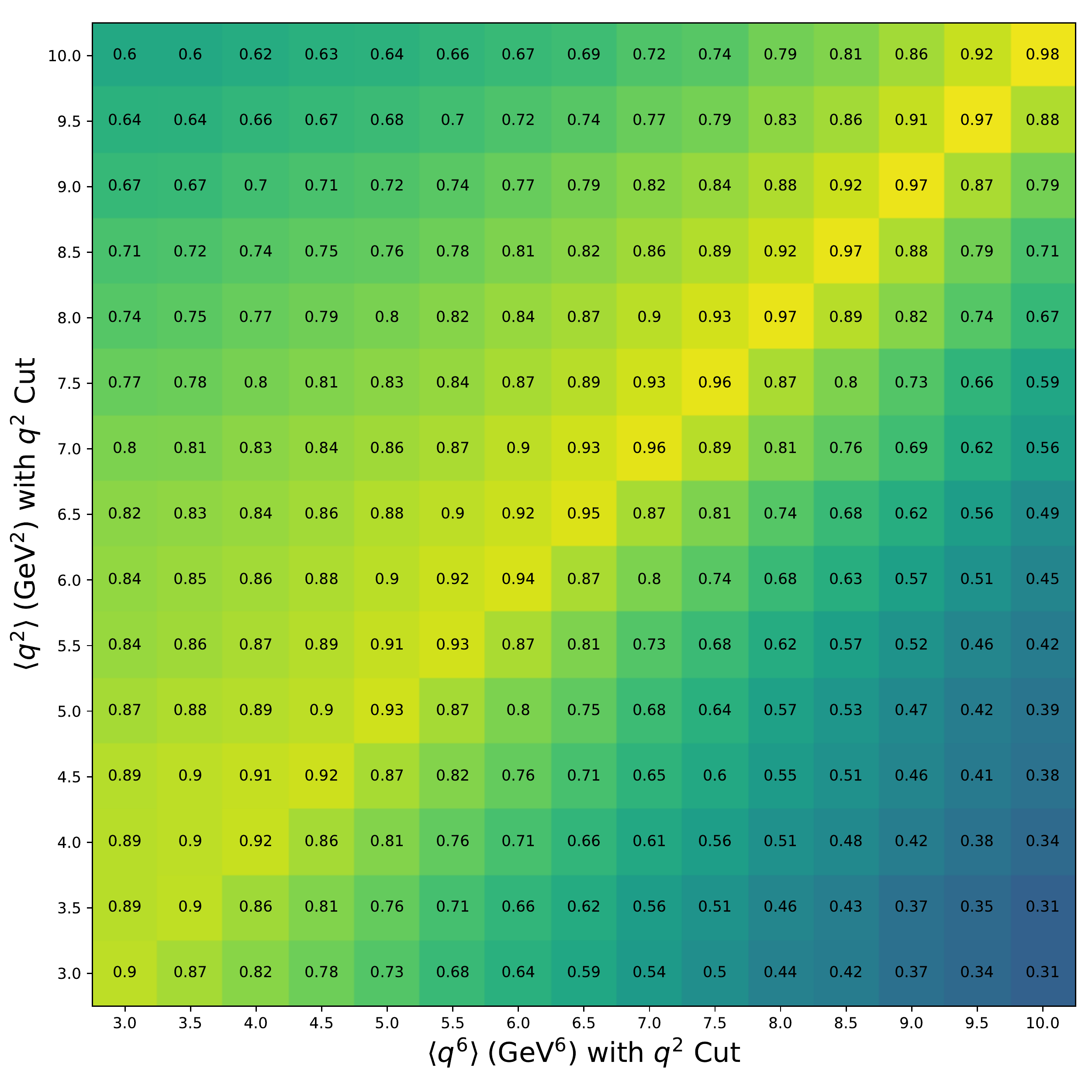}
  \includegraphics[width=0.35\textwidth]{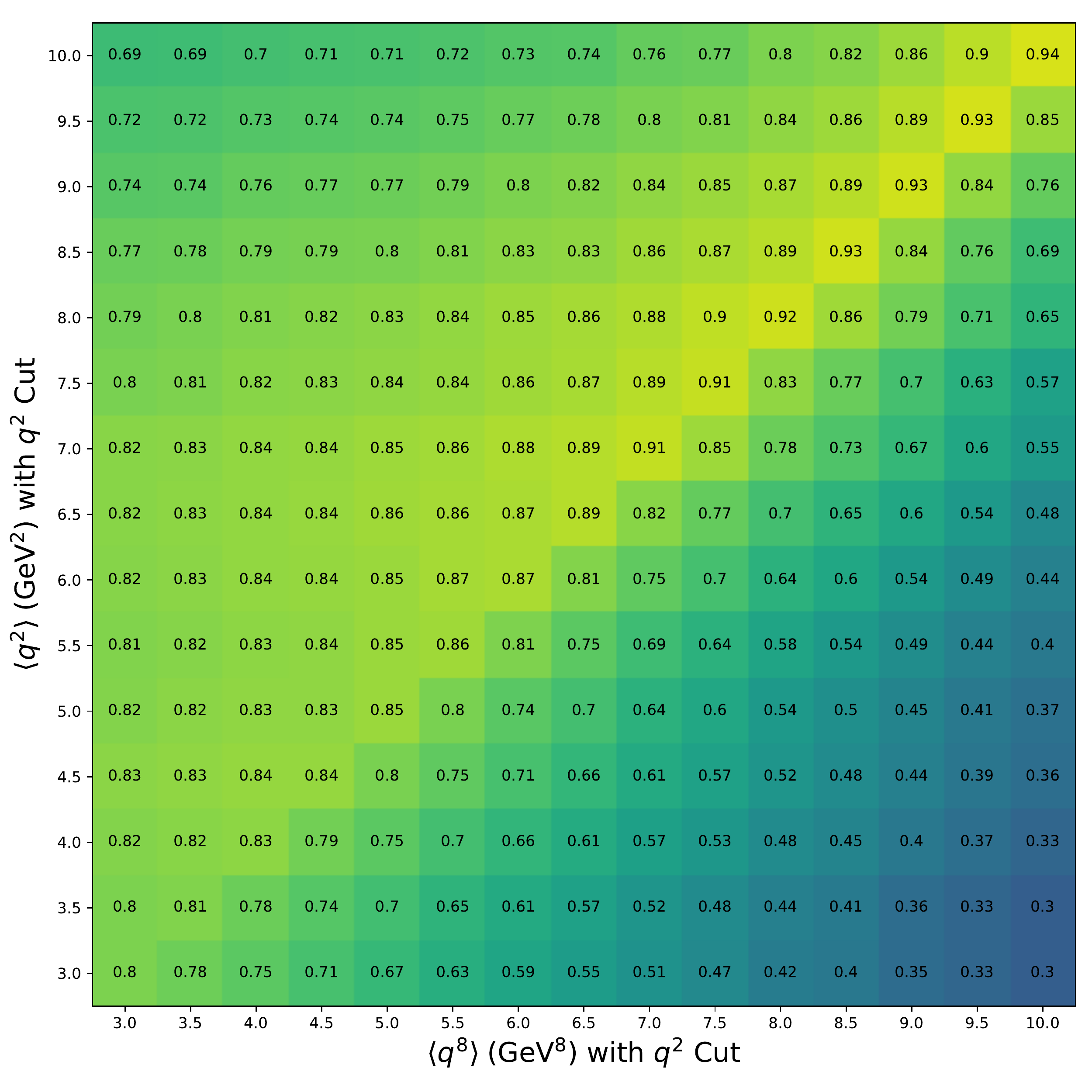}
\caption{
 Statistical correlations between the first and the second, third, and fourth moment for the electron final state.
}
\label{fig:stat_correlation}
\end{figure*}

Several systematic uncertainties affect the measured $q^2$ moments and their impact on the background subtraction and calibration procedure are discussed in the following sections. The most important sources of systematic uncertainty are associated with the assumed composition of the \bclnu process: the decays involving the higher mass states beyond the $1S$ $D$ and $D^*$ meson are poorly known and the composition affects the background subtraction as well as the calibration of the measured moments. In addition, we observe systematic shifts in the energy and momentum of the $X$ system, whose size we use to estimate a $q^2$ scale uncertainty. Tables~\ref{tab:q2_q4_q6_q8_el} and ~\ref{tab:q2_q4_q6_q8_mu} summarize the relative statistical and systematic uncertainties on the measured $q^2$ moments for electron and muon final states, given in permille.

\subsection{Background Subtraction}

We evaluate the uncertainties on the background subtraction by considering various sources of systematic uncertainty included as NP constraints in the binned likelihood fits. By taking into account the full experimental covariance of the background shapes when performing the $\chi^2$ minimization, we directly propagate these sources of uncertainty into the signal probability functions, $w(q^2)$. Subsequently, we determine orthogonal variations for the parameters of the fitted polynomial curves and extract varied sets of moments to evaluate the total uncertainty (labeled \enquote{Bkg. subtraction} in Tables~\ref{tab:q2_q4_q6_q8_el} and ~\ref{tab:q2_q4_q6_q8_mu} ). Specifically, we consider signal modeling uncertainties by variations of the BGL parameters and heavy quark form factors of \bdlnu, \bdslnu, and \bddslnu\  within their uncertainties. In addition, we propagate the branching fraction uncertainties, cf. Table~\ref{tab:bfs}. The uncertainties on the \bclnu gap branching fractions are taken to be large enough to account for the difference between the sum of all exclusive branching fractions measured and the inclusive branching fraction measured. We also evaluate the impact on the efficiency of the particle-identification uncertainties on the background shapes, and the overall tracking efficiency uncertainty. The statistical uncertainty on all generated MC samples is also evaluated and propagated into the systematic errors. The \bulnu background component is varied within its branching fraction uncertainty (\enquote{$B \rightarrow X_{u} \ell \nu$ BF}).

\subsection{Calibration}

For the calibration we change the composition of the assumed \bclnu process and redetermine the calibration functions as well as the bias and acceptance correction factors (\enquote{$B \rightarrow X_{c} \ell \nu$ BF}). Specifically, for the \bddslnu and gap contributions, we redetermine these factors by completely removing their respective contributions, to account for the poor knowledge of their precise decay processes. We vary the \bdlnu\, and \bdslnu\ within their respective branching fractions. In addition, we evaluate the impact of the modeling of non-resonant decays by replacing the model described in Section~\ref{sec:data_set_sim_samples} with a model based on the equidistribution of all final-state particles in phase space (\enquote{Non-resonant model}). The signal modeling is evaluated in a similar manner as described above by using the variations of the BGL parameters and heavy quark form factors (\enquote{$B \rightarrow X_{c} \ell \nu$ FF}). To estimate an uncertainty associated with the modeling of the resolution, we calculate bin-wise weight functions in a signal enriched region by comparing distributions of the number of charged particles and neutral clusters in the $X$ system to the number observed in the recorded collision events. The weight functions are subsequently applied to the simulated MC samples and we redetermine the calibration functions and correction factors (\enquote{$N_{\text{tracks}}$ res.} \& \enquote{$N_{\gamma}$ res.}). Additionally, we employ a similar strategy to take into account a potential mismodeling of the shape of the missing energy and momentum distribution (\enquote{$E_{\mathrm{miss}} -  |\bold{p}_{\mathrm{miss}}|$ shape}). To estimate an uncertainty on the modeling of the scale of $q^2$ (\enquote{$q^2$ scale}), we shift the reconstructed value in the simulation by 1\%, and redetermine the background subtraction weights and the calibration functions. The shift in $q^2$ corresponds to the observed differences in the mean values of the energy and momentum of the reconstructed $X$ system, and we take the full difference to the moments determined with no correction as an uncertainty. The additional remaining bias due to the moment extraction method is estimated by making use of the ensemble tests described in Section~\ref{sec:calibration} (\enquote{MC non-closure}). We propagate the statistical uncertainty due to the determination of the linear calibration functions by determining orthogonal variations for each of the parameters of the fitted curves and extracting new, varied sets of moments (\enquote{Cal. function}). Additionally, we estimate the statistical uncertainty due to the bias and acceptance correction factors by varying the correction factors within one standard deviation and repeating the moment calculation (\enquote{Stat. bias corr.}). The impact of the efficiency of the particle-identification uncertainties (\enquote{PID eff.}), and the overall track finding efficiency uncertainty (\enquote{Track eff.}), on the calibration curves are estimated by deriving varied calibrations for each efficiency uncertainty under consideration. Lastly, we account for the different reconstruction efficiencies of neutral and charged $B$ mesons in data and simulated samples. We correct for the difference in MC and repeat the calibration procedure to extract varied sets of $q^{2}$ moments (\enquote{$B^{0}/B^{\pm}$ tag eff.}) and take the full difference as the systematic uncertainty. 

\subsection{Statistical Correlations}

The statistical correlation of the measured moments is determined using a bootstrapping approach~\cite{booststrap,PhysRevD.39.274}: replicas of the measured collision data sets are created using sampling by replacement and the entire analysis procedure is repeated to estimate the statistical correlations between the different moments. An example for the electron final state is shown in Fig.~\ref{fig:stat_correlation}. Typically moments of the same order with similar $q^2$ threshold selections are highly correlated. The moments of higher order with identical $q^2$ threshold selections contain more independent information the higher the difference in the order.  

\begin{figure*}
  \includegraphics[width=0.45\textwidth]{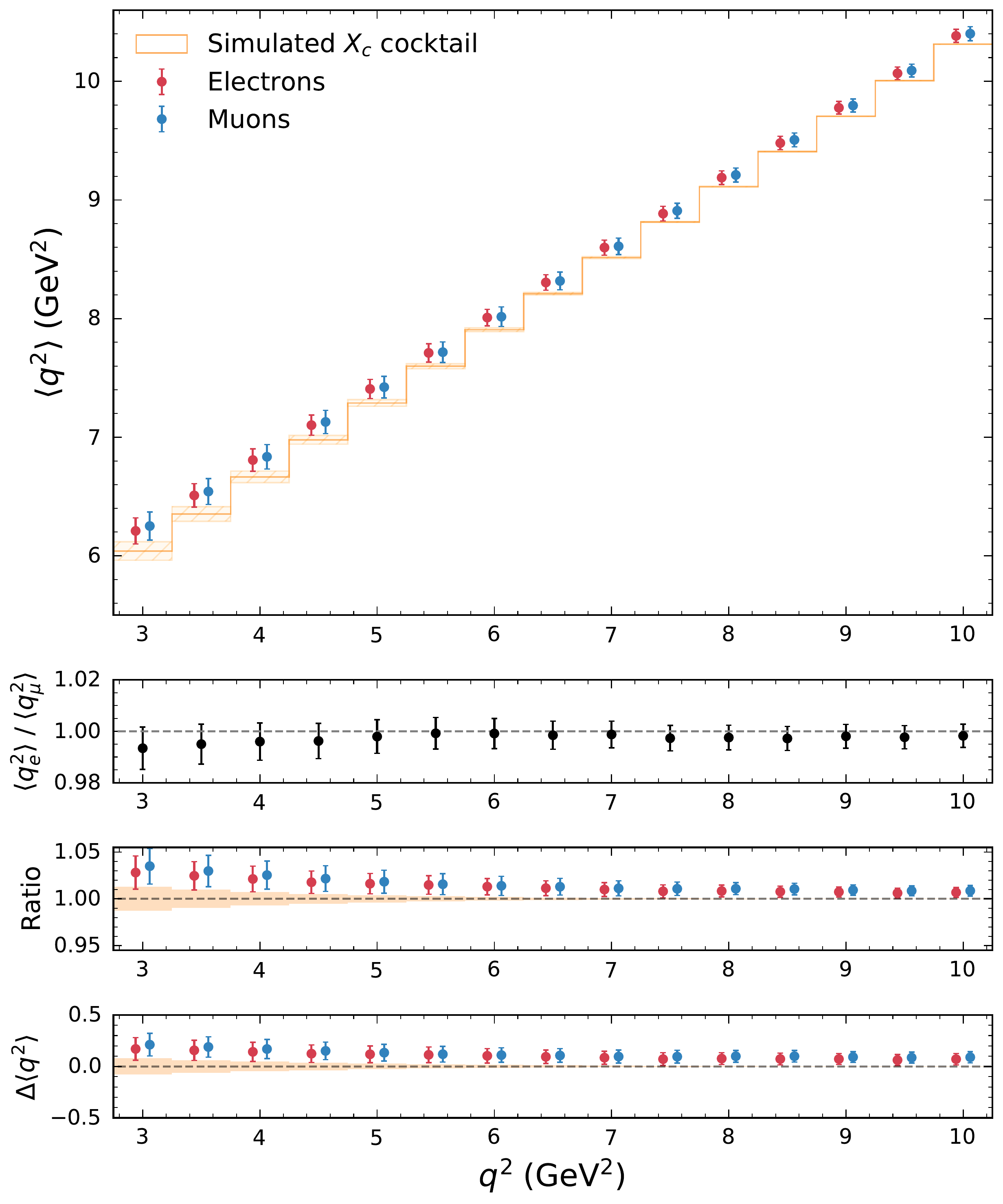}
 \includegraphics[width=0.45\textwidth]{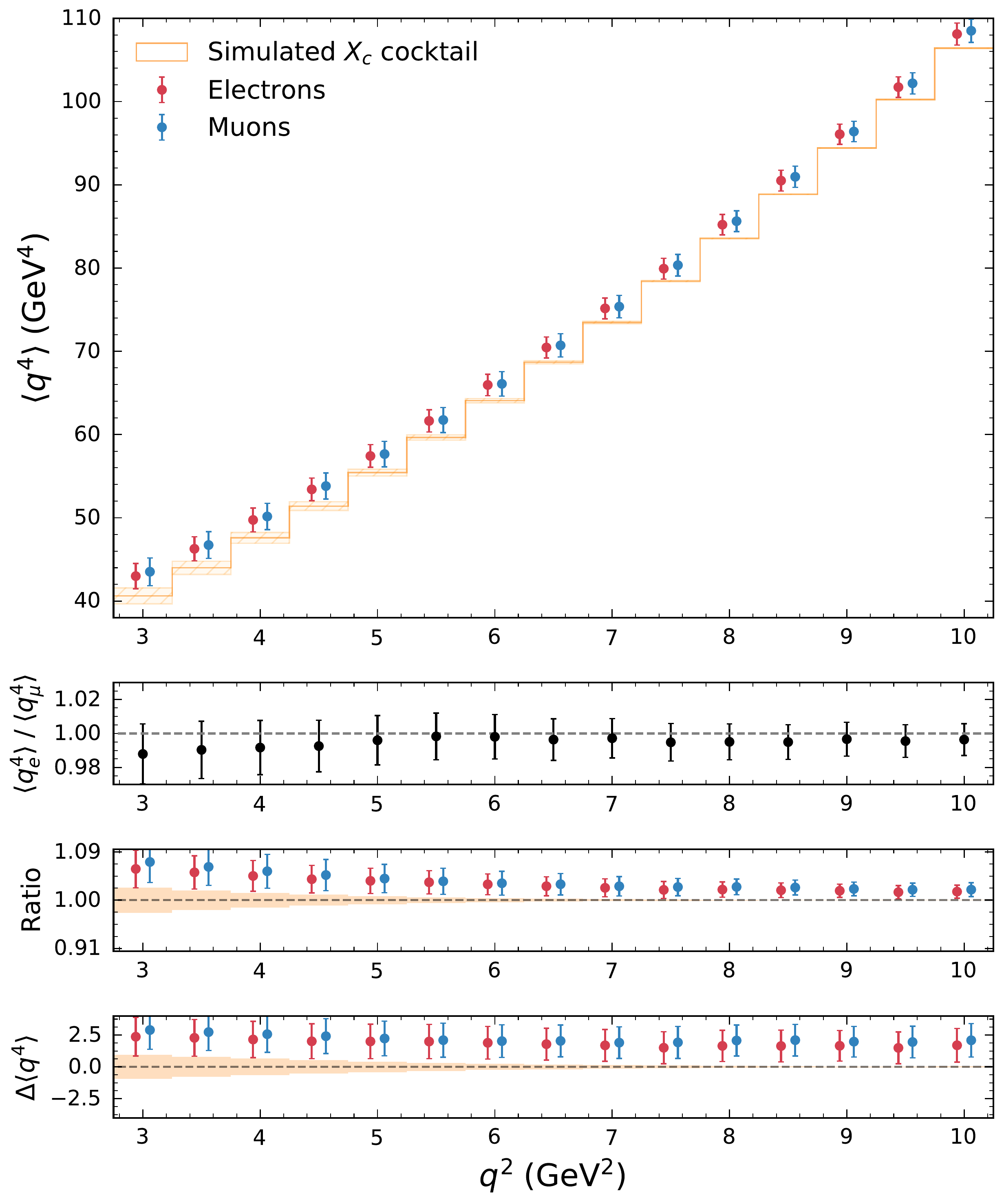}
 \includegraphics[width=0.45\textwidth]{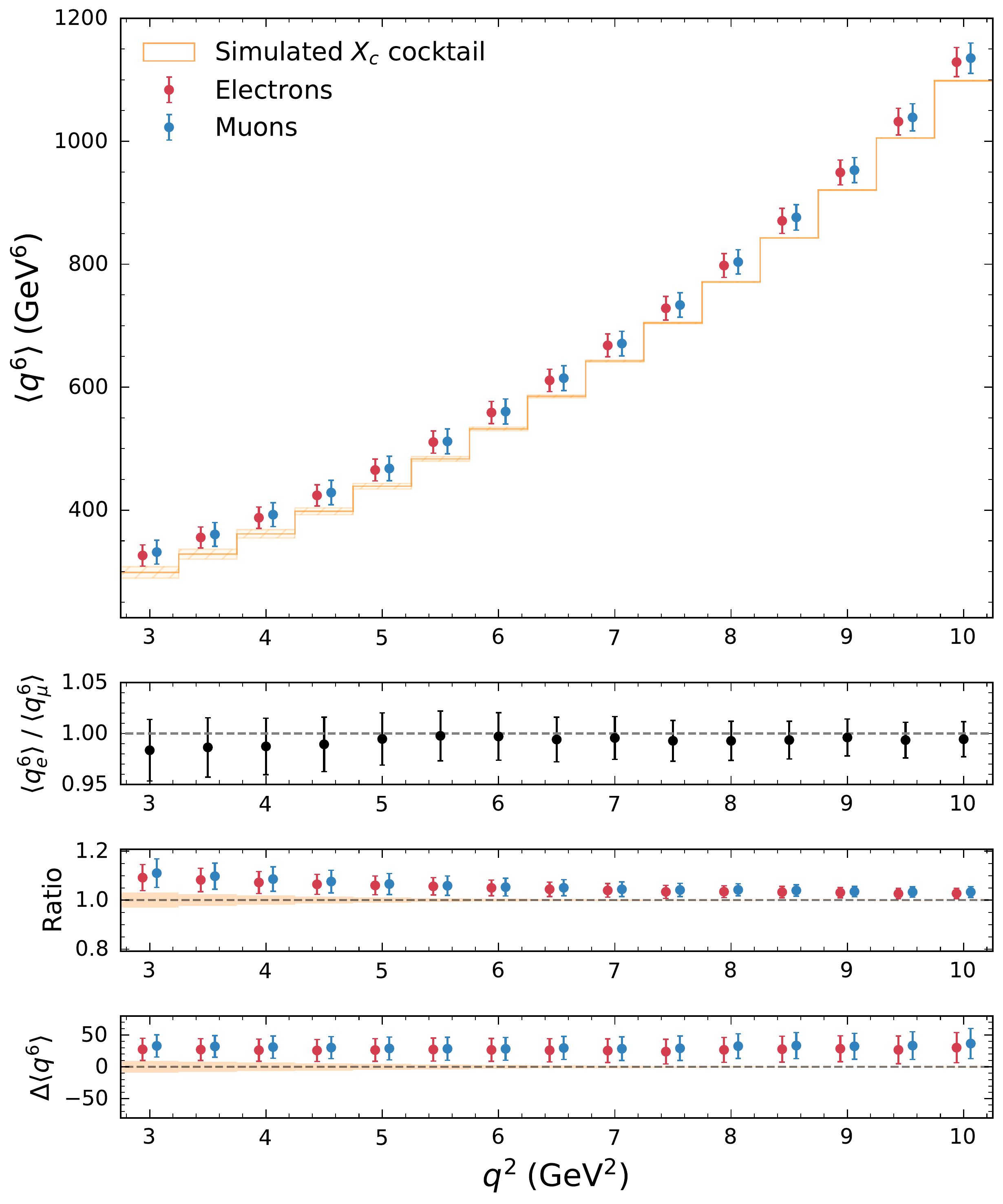}
 \includegraphics[width=0.45\textwidth]{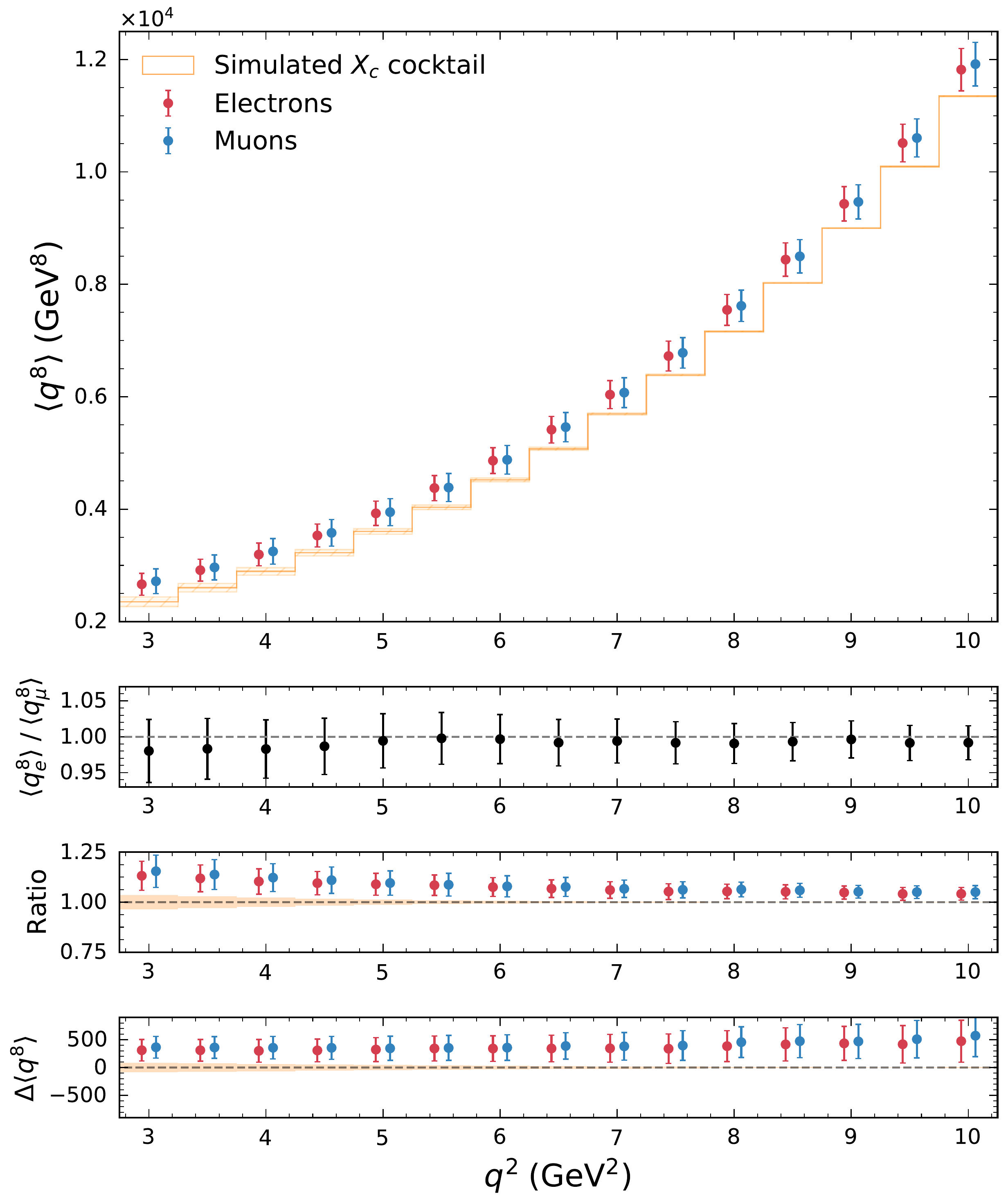}
\caption{
 The expectation of lepton flavor universality of the moments are tested for the first to fourth $q^{2}$ moments: in the ratio of electron to muon moments many of the associated systematic uncertainties cancel and all reported moments are compatible with the expectation of lepton flavor universality (bottom top). Note that the individual electron and muon moments are highly correlated. Furthermore, the measured and generated-level moments for all the 
threshold selections on $q^2$ are compared as a ratio (bottom middle) and difference (bottom lower) for both electrons and muons.
}
\label{fig:lfu_test}
\end{figure*}

\subsection{Systematic Correlations}

We estimate the systematic covariance from the efficiency of particle identification by creating replicas of the data-driven efficiency corrections, while taking into account correlated and uncorrelated systematic uncertainty sources. We fully correlate or anti-correlate all other identical sources of systematic uncertainty across all $q^2$ threshold selections and order of moments. For the comparison between the measured electron and muon $q^2$ moments, we assume that all uncertainties are fully correlated between both sets of measurements with the exception of the contributions that are derived independently for the two different measurements. These sources of systematic uncertainty include: the uncertainty on the background subtraction, the MC non-closure error, the uncertainty on the calibration function, the statistical errors on the bias and acceptance correction factors, the particle-identification efficiency uncertainty, the modeling uncertainty of the resolution, and the modeling of the missing energy and momentum distribution. The full systematic covariance matrix is then determined by combining the statistical and all the individual systematic covariance matrices. These correlation matrices for the moments for both the electron and muon final states are given in Appendix E.

\subsection{Summary}

The dominant systematic uncertainties stem from the uncertainty associated with the modeling of the \bclnu\ composition, especially the non-resonant contributions. The estimated uncertainties associated with these sources for the first moment, with a minimum threshold selection of $q^2 > 3.0 \, \mathrm{GeV}^2$ for the electron final state, are found to be 0.49\% and 1.32\%, respectively. This is followed by the uncertainty associated with the modeling of the number of charged particles in the $X$ system, which remains a leading systematic uncertainty across all $q^2$ selections for both the muon and electron final states. On the other hand, the uncertainty due to the modeling of the \bulnu\ background component is found to be a leading systematic uncertainty mainly for low $q^2$ selections. While the statistical uncertainties due to the additional correction factors and the determination of the linear calibration functions are small contributions to the overall systematic error for low $q^2$ selections, these sources increase as the selection criteria progress to higher values of $q^2$ and the number of events gradually decrease. A similar trend is observed for the uncertainty due to the modeling of the missing energy and momentum, as well as the uncertainty due to the different reconstruction efficiencies of neutral and charged $B$ mesons. Conversely, the uncertainties involving the efficiency of the particle-identification, the overall tracking efficiency and the modeling of the number of neutral clusters in the $X$ system are relatively small for the first selections and gradually decrease as tighter $q^2$ threshold selections are imposed. The smallest source of systematic error across all $q^2$ selections is the estimated residual bias due to the extraction method.

\section{Results}\label{sec:results}
The measured $q^2$ moments for electrons and muons along with a detailed breakdown of the systematic uncertainties are given in Tables~II and III. A determination of $\left| V_{cb} \right|$ from the measured moments is beyond the scope of this paper and left for future work, as the method proposed in Ref.~\cite{Fael:2018vsp} has not yet been
fully implemented and no public code exists. Furthermore, the treatment of theoretical uncertainties on the extraction of $\left| V_{cb} \right|$ is central due to the high precision of the measured $q^2$ moments.

The first moment with the loosest threshold selection of $q^2 > 3.0 \, \mathrm{GeV}^2$ is about one standard deviation higher in data than in our simulated samples. Fig.~\ref{fig:lfu_test} shows the comparison for both the electron and muon measurement for this and higher selection criteria on $q^2$ and also for higher moments. The generator-level moments include uncertainties on the \bclnu\ composition and form factor variations. The higher the threshold selection on $q^2$, the better the agreement becomes. This picture is consistent between the electron and muon channels and also for higher moments. This indicates that the \bclnu\ composition used tends to overestimate the number of signal events at low $q^2$ values, which also has been reported by Ref.~\cite{TheBABAR:2016lja}: there the inclusive lepton spectrum is analyzed using a \bclnu\ model similar to the one used in this measurement. The data prefer to increase the yield of the \bclnu\ constituents that produce a more energetic lepton momentum spectrum, resulting in a higher value of the first moment of $q^2$. Similar observations have been reported by Ref.~\cite{Bernlochner:2014dca}, in which the moments of the lepton energy, hadronic mass, and hadronic energy spectra were analyzed to determine the exclusive composition of \bclnu. The direct measurement of the $q^2$ moments reported here support this picture.

We test the expectation of lepton flavor universality of identical moments. Figure~\ref{fig:lfu_test} compares the measured moments for electrons and muons. In the shown ratio many of the systematic uncertainties cancel and we observe no deviation from the expectation of unity. In addition, we estimate a qualitative measure to test the agreement between the electron and muon moment measurement by calculating the $\chi^{2}$ for each order of measured $q^{2}$ moments. The calculated values range between 2.21 and 0.96 with a corresponding p-value of 0.99 for all order of $q^{2}$ moments.

We also extract the central, or normalized moments. The central moments have the advantage of becoming slightly less correlated with respect to the correlations of the nominal moments, especially for the higher moments. To calculate the central moments directly from the nominal measured moments, the following non-linear transformations are applied:
\begin{equation}
\label{central_eqn}
    \begin{pmatrix}
        \langle q^{2} \rangle \\
        \langle q^{4} \rangle \\
        \langle q^{6} \rangle \\
        \langle q^{8} \rangle
    \end{pmatrix}
    \qquad
    \rightarrow 
    \qquad
    \begin{pmatrix}
        \langle q^{2} \rangle \\
        \langle ( q^{2}- \langle q^{2} \rangle)^{2}\rangle\\
        \langle ( q^{2}- \langle q^{2} \rangle)^{3}\rangle\\
        \langle ( q^{2}- \langle q^{2} \rangle)^{4}\rangle\\
    \end{pmatrix} \, .
\end{equation}
The new systematic covariance matrix $\mathcal{C}^{\prime}$ for the vector of central moments is determined by making use of the Jacobian matrix $\mathcal{J}$ for the transformation, together with the initial systematic covariance matrix $\mathcal{C}$ describing the nominal moments, such that
\begin{equation}
\label{central_err}
\mathcal{C}^{\prime} = \mathcal{J} \mathcal{C} \mathcal{J}^{T} \, .
\end{equation}
This approximation for the uncertainties of the central moments yields the same results as Gaussian error propagation. Not only are slightly lower correlations between threshold selections observed for central moments of the same order, but also negative correlations between different orders of moments. Appendix F compares the second, third and fourth measured central moments for electrons and muons. 


\section{Summary and Conclusions}\label{sec:summary}

In this paper we report the first systematic study of the first to the fourth moment of the \bclnu\ $q^2$ spectra with progressively increasing threshold selections on $q^2$. The measured moments are crucial experimental inputs for a novel and alternative approach determining $|V_{cb}|$ from inclusive decays, which was outlined in Ref.~\cite{Fael:2018vsp}: using reparametrization invariance the set of non-perturbative matrix elements can be strongly reduced. As no public code is available to determine $|V_{cb}|$, this is left for future work. The reported measurement uses the full Belle data set and employs a neural network assisted hadronic tag reconstruction. Explicit reconstruction of the tag $B$ meson allows for the explicit reconstruction of the hadronic $X$ system of the semileptonic decay, and thus the reconstruction of the four-momentum transfer squared $q^2$. Background from other processes is subtracted using the reconstructed hadronic mass spectrum $M_X$ in an unbinned approach using event-weights. The reconstructed moments are then calibrated in a three step procedure to account for the finite detector resolution and acceptance effects. The background subtraction and calibration procedure introduces systematic uncertainties, most dominantly from the assumed \bclnu\ composition. The moments are measured separately for electron and muon \bclnu\ final states. This allows for testing of lepton flavor universality in inclusive processes involving electrons and muons, and no deviation in the measured moments from the expectation of unity are observed. In addition to measuring the nominal $q^2$ moments, a non-linear transformation is applied to extract the central moments, which are slightly less correlated compared to the statistical and systematic correlations of the nominal moments. The measured moments are also compared to the expectation from the exclusive make-up of \bclnu. Here, contributions from heavier charmed final states and high multiplicity decays are poorly constrained by current measurements. The measured $q^2$ spectrum has higher moments than the generator-level \bclnu\ model for low $q^2$ threshold selections. As the \bclnu\ decay gets increasingly dominated at high $q^2$ by the well measured $D$ and $D^*$ final states, this points towards a problem in the modeling of the other components. Similar observations have been reported by Refs.~\cite{TheBABAR:2016lja,Bernlochner:2014dca}. One of the leading sources of systematics for low $q^{2}$ threshold selections is observed to be the modeling of the \bulnu\ decays. Events originating from this background component are first rejected by imposing selection criteria on decay kinematics, after which the remaining component is further suppressed by making use of the background subtraction procedure. A possible future improvement of the analysis strategy as well as the overall precision of the measurement is outlined in Ref.~\cite{Mannel:2021mwe}. Rather than subtracting the \bulnu\ component, the authors suggest measuring the full \bxlnu\ spectrum and obtaining the \bulnu\ contribution precisely from within the HQE. This strategy has the potential to reduce the uncertainty on the measured value of $|V_{cb}|$ determined from the $q^2$ moments even further. The numerical values and full covariance matrices of the measured moments and central moments will be made available on HEPData (https://www.hepdata.net).


\acknowledgments

We thank Keri Vos, Kevin Olschewsky, and Matteo Fael for useful discussions about the subject matter of this manuscript. RvT, LC, WS, and FB were supported by the DFG Emmy-Noether Grant No.\ BE~6075/1-1. We thank the KEKB group for the excellent operation of the accelerator; the KEK cryogenics group for the efficient operation of the solenoid; and the KEK computer group, and the Pacific Northwest National Laboratory (PNNL) Environmental Molecular Sciences Laboratory (EMSL) computing group for strong computing support; and the National Institute of Informatics, and Science Information NETwork 5 (SINET5) for valuable network support.  We acknowledge support from
the Ministry of Education, Culture, Sports, Science, and Technology (MEXT) of Japan, the Japan Society for the  Promotion of Science (JSPS), and the Tau-Lepton Physics Research Center of Nagoya University; the Australian Research Council including grants
DP180102629, 
DP170102389, 
DP170102204, 
DP150103061, 
FT130100303; 
Austrian Federal Ministry of Education, Science and Research (FWF) and
FWF Austrian Science Fund No.~P~31361-N36;
the National Natural Science Foundation of China under Contracts
No.~11435013,  
No.~11475187,  
No.~11521505,  
No.~11575017,  
No.~11675166,  
No.~11705209;  
Key Research Program of Frontier Sciences, Chinese Academy of Sciences (CAS), Grant No.~QYZDJ-SSW-SLH011; 
the  CAS Center for Excellence in Particle Physics (CCEPP); 
the Shanghai Pujiang Program under Grant No.~18PJ1401000;  
the Shanghai Science and Technology Committee (STCSM) under Grant No.~19ZR1403000; 
the Ministry of Education, Youth and Sports of the Czech
Republic under Contract No.~LTT17020;
Horizon 2020 ERC Advanced Grant No.~884719 and ERC Starting Grant No.~947006 ``InterLeptons'' (European Union);
the Carl Zeiss Foundation, the Deutsche Forschungsgemeinschaft, the
Excellence Cluster Universe, and the Volkswagen Stiftung;
the Department of Atomic Energy (Project Identification No. RTI 4002) and the Department of Science and Technology of India; 
the Istituto Nazionale di Fisica Nucleare of Italy; 
National Research Foundation (NRF) of Korea Grant
Nos.~2016R1\-D1A1B\-01010135, 2016R1\-D1A1B\-02012900, 2018R1\-A2B\-3003643,
2018R1\-A6A1A\-06024970, 2018R1\-D1A1B\-07047294, 2019K1\-A3A7A\-09033840,
2019R1\-I1A3A\-01058933;
Radiation Science Research Institute, Foreign Large-size Research Facility Application Supporting project, the Global Science Experimental Data Hub Center of the Korea Institute of Science and Technology Information and KREONET/GLORIAD;
the Polish Ministry of Science and Higher Education and 
the National Science Center;
the Ministry of Science and Higher Education of the Russian Federation, Agreement 14.W03.31.0026, 
and the HSE University Basic Research Program, Moscow; 
University of Tabuk research grants
S-1440-0321, S-0256-1438, and S-0280-1439 (Saudi Arabia);
the Slovenian Research Agency Grant Nos. J1-9124 and P1-0135;
Ikerbasque, Basque Foundation for Science, Spain;
the Swiss National Science Foundation; 
the Ministry of Education and the Ministry of Science and Technology of Taiwan;
and the United States Department of Energy and the National Science Foundation.


\bibliographystyle{apsrev4-1}
%


\begin{appendix}

\clearpage
\onecolumngrid

\section*{Appendix A. Comparison of $B \to D^{**}_{\mathrm{gap}} \, \ell \bar \nu_\ell$ decay models}

\begin{figure}[h!]
  \includegraphics[width=0.45\textwidth]{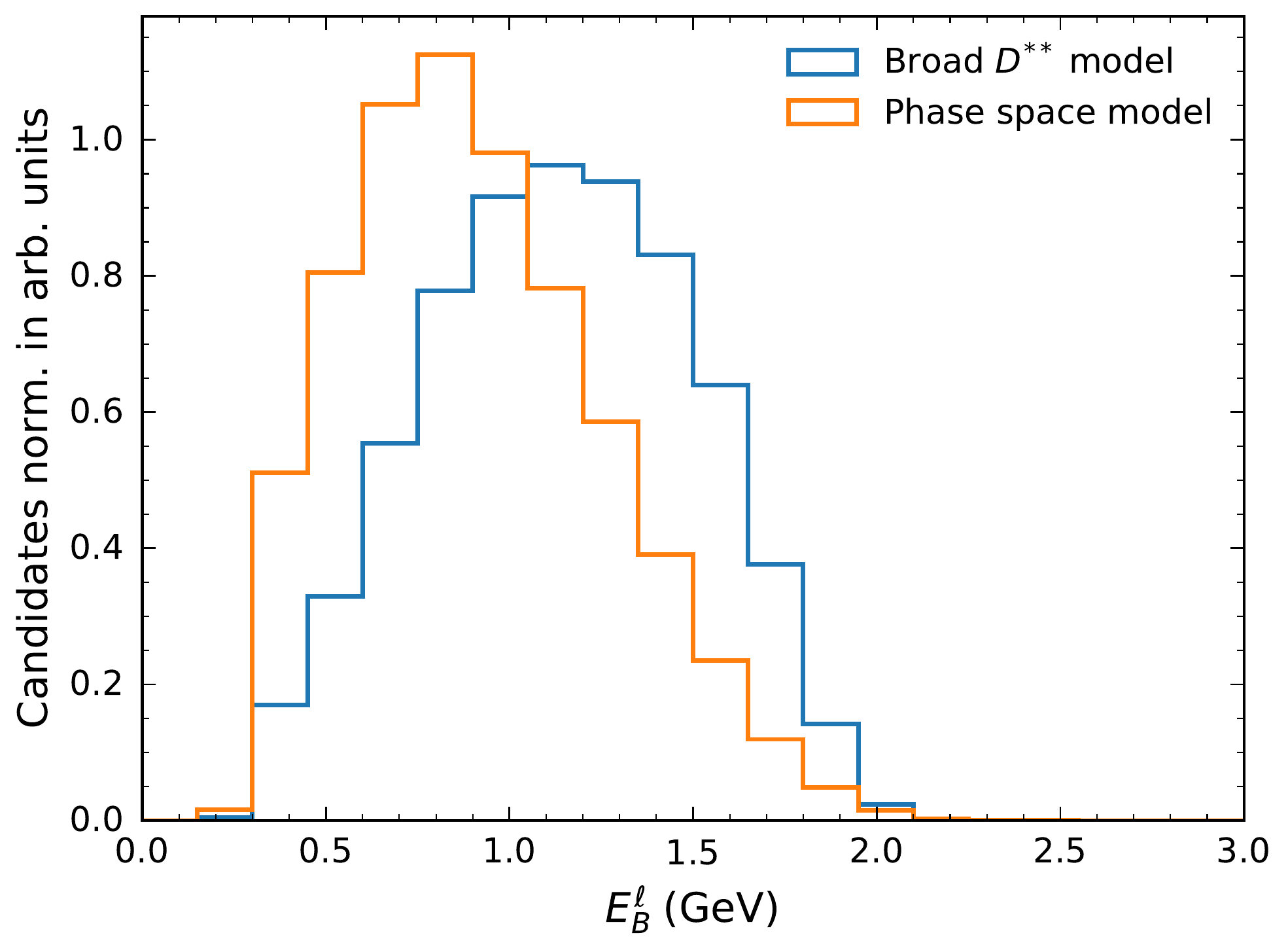}
 \includegraphics[width=0.45\textwidth]{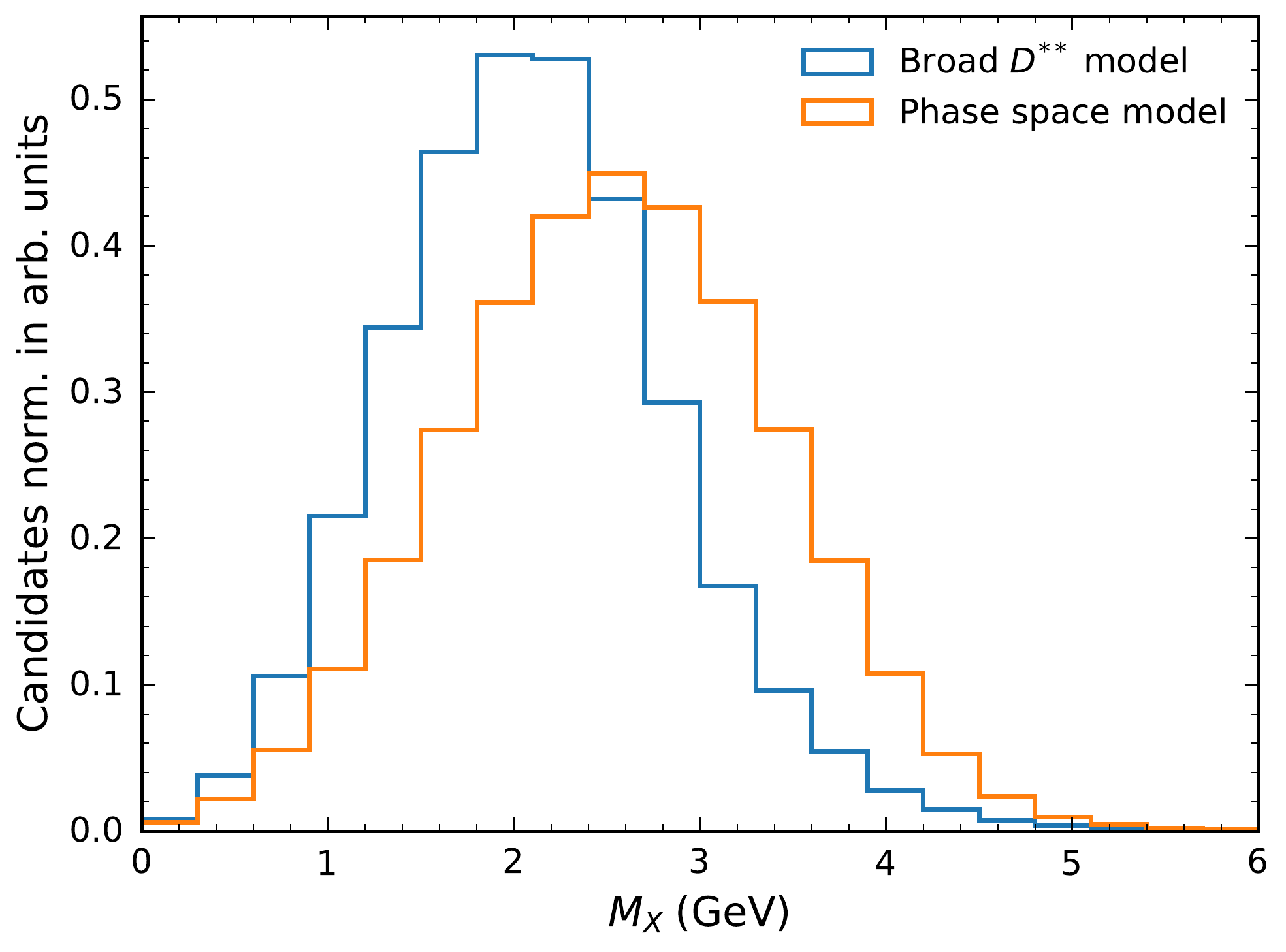}
 \includegraphics[width=0.45\textwidth]{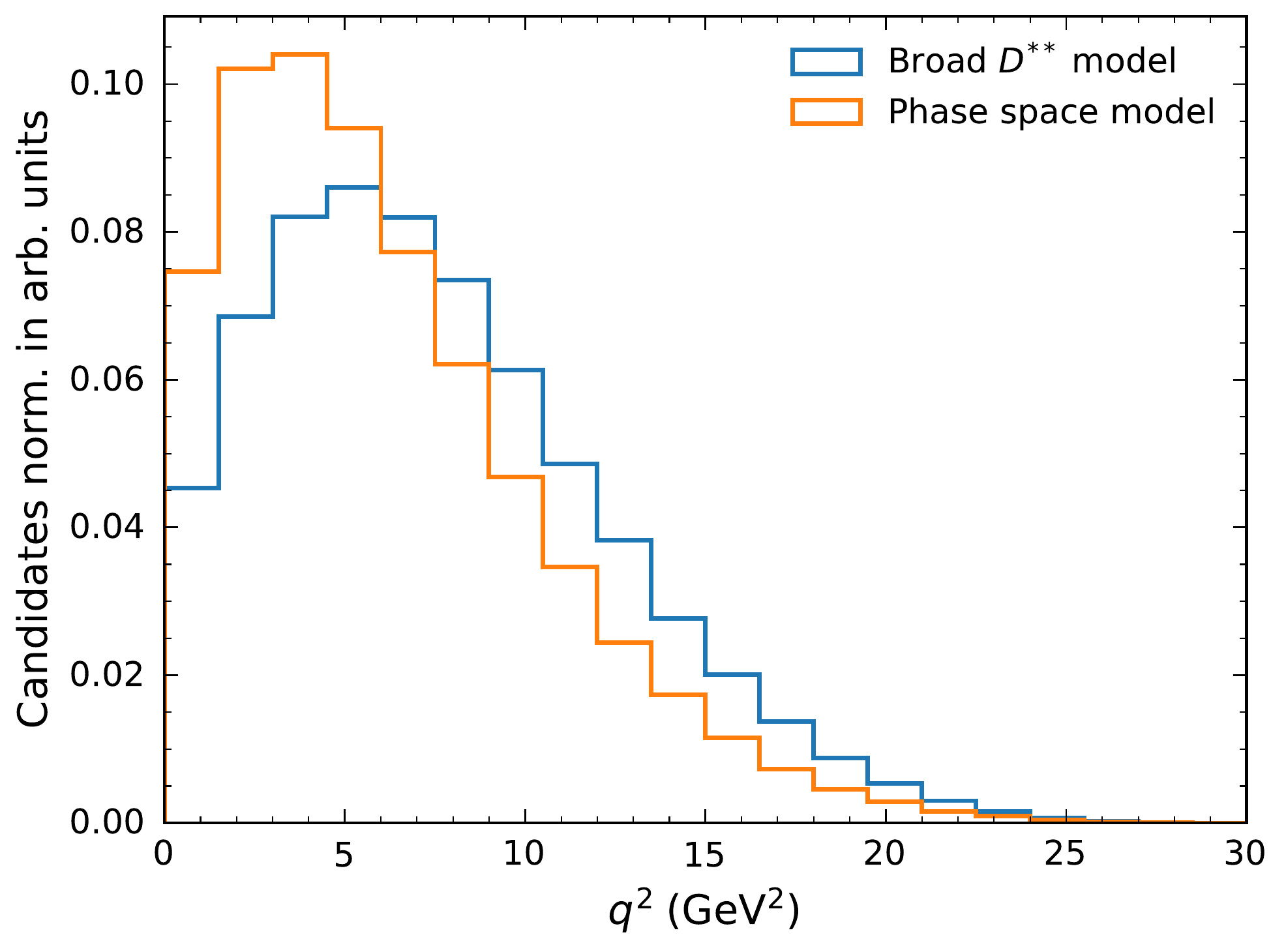}
\caption{Comparisons between the different $B \to D^{**}_{\mathrm{gap}} \, \ell \bar \nu_\ell$ decay models for the reconstructed lepton energy in the signal $B$ rest frame $E_{\ell}^{B}$ (upper left), the $M_X$ (upper right), and the $q^{2}$ (bottom) distributions.
}
\label{fig:gap_compare}
\end{figure}

As discussed in Section~\ref{sec:data_set_sim_samples}, the remaining `gap' between the sum of all considered exclusive modes and the inclusive \bclnu branching fraction is filled in equal parts with $B \to D \, \eta \, \ell^+ \, \nu_\ell$ and $B \to D^{*} \, \eta \, \ell^+ \, \nu_\ell$ final states assuming that they are produced by the decay of two broad resonant states $D^{**}_{\mathrm{gap}}$ with masses and widths identical to $D_1^{*}$ and $D_0^{*}$. This model provides a better kinematic description of the initial three-body decay $B \to D^{**}_{\mathrm{gap}} \, \ell \bar \nu_\ell$ than a model based on the equidistribution of all final-state particles in phase space. Comparisons of kinematic distributions for the different $B \to D^{**}_{\mathrm{gap}} \, \ell \bar \nu_\ell$ decay models are shown in Figure~\ref{fig:gap_compare}.

\clearpage

\section*{Appendix B. Background Subtraction Fits}

\begin{figure}[b!]
  \includegraphics[width=0.25\textwidth]{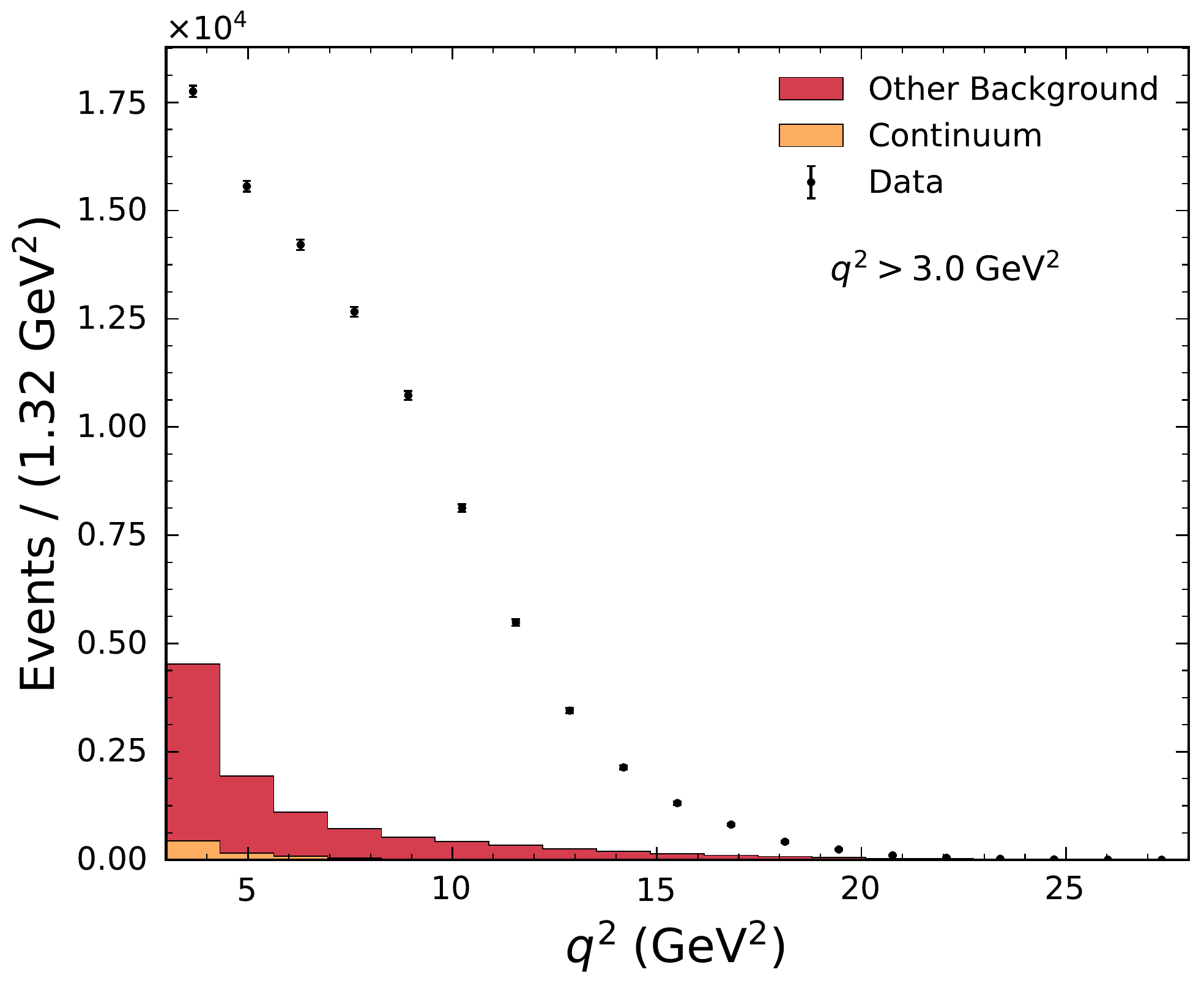}  
  \includegraphics[width=0.25\textwidth]{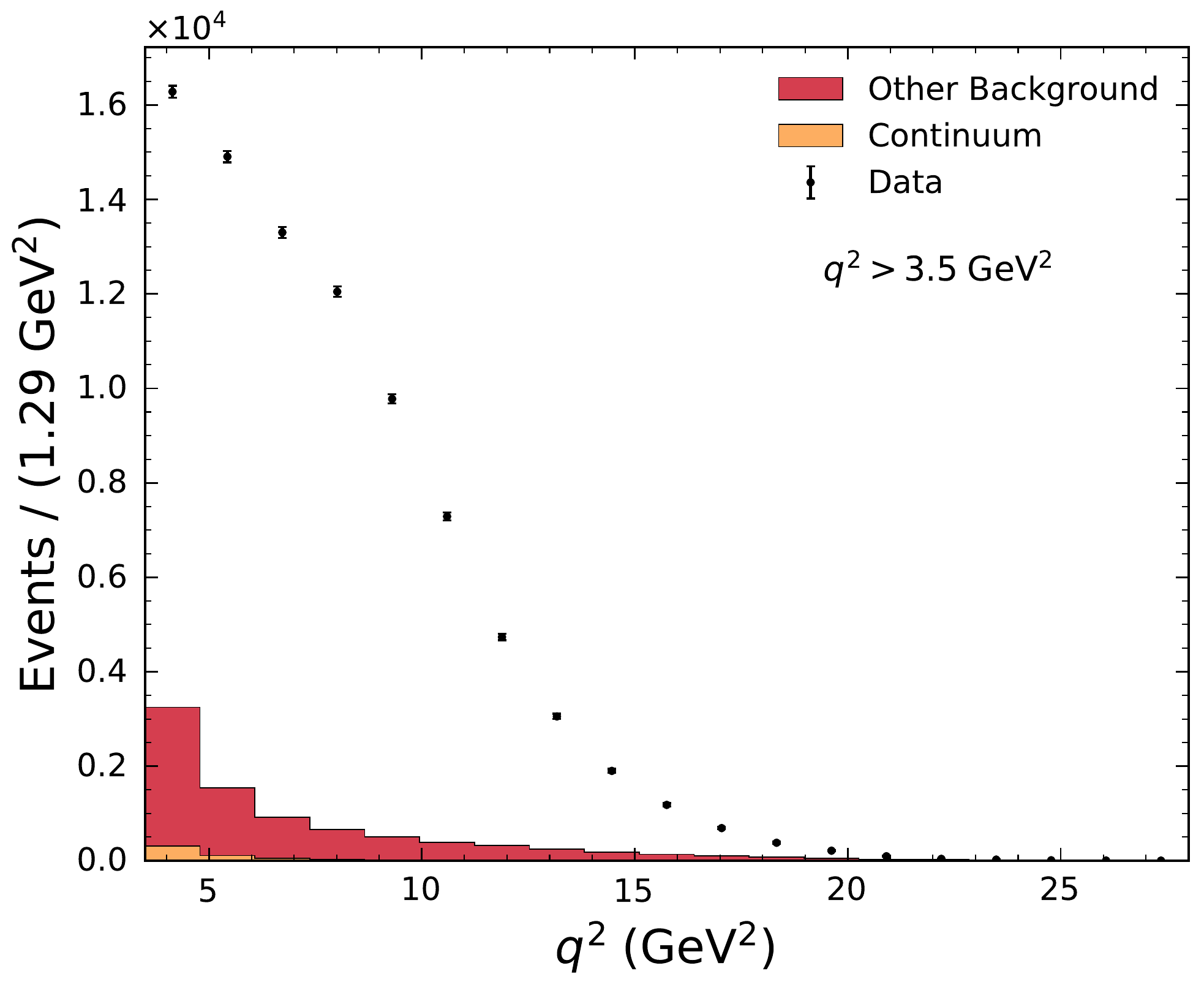}  
  \includegraphics[width=0.25\textwidth]{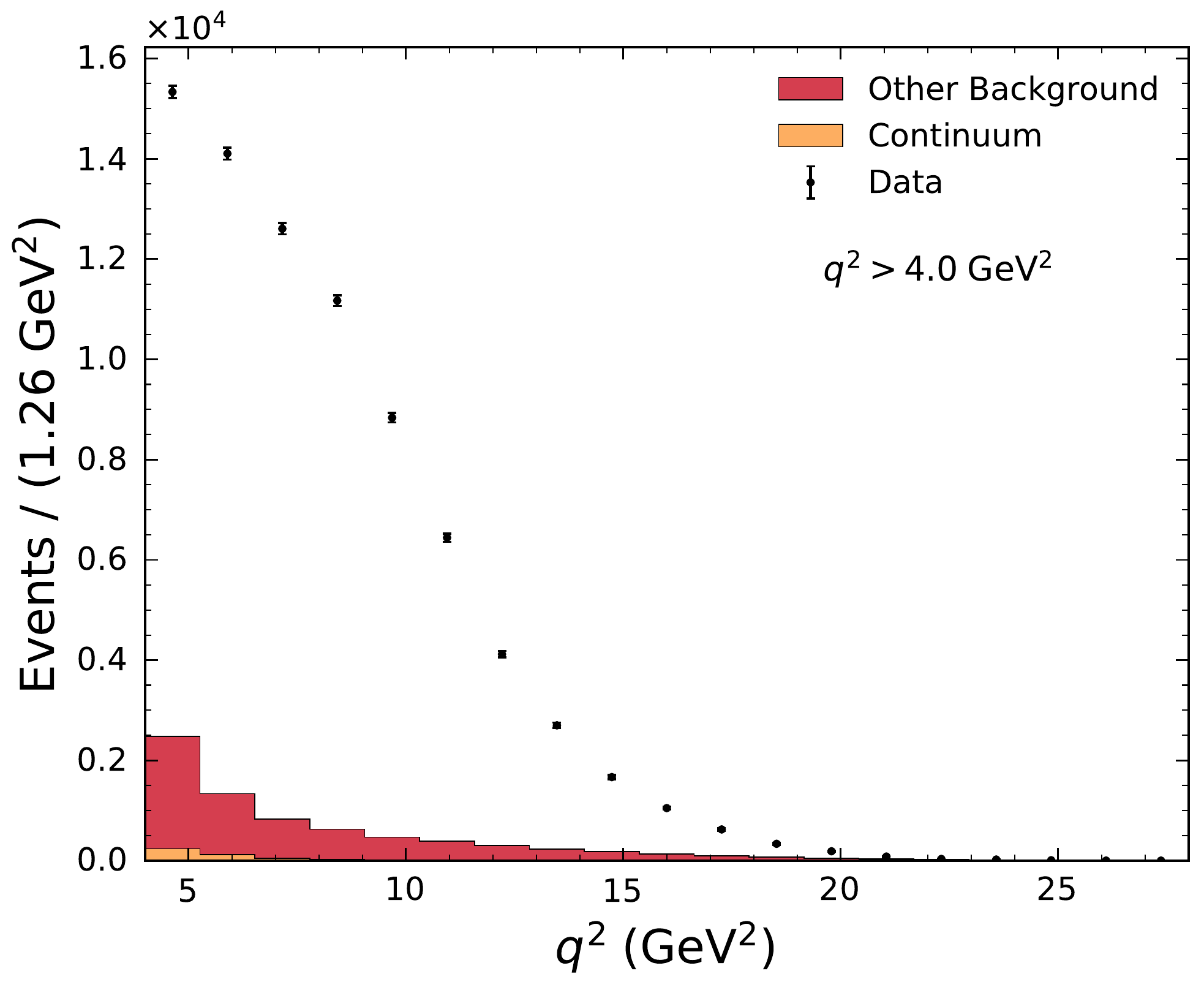}  
  \includegraphics[width=0.25\textwidth]{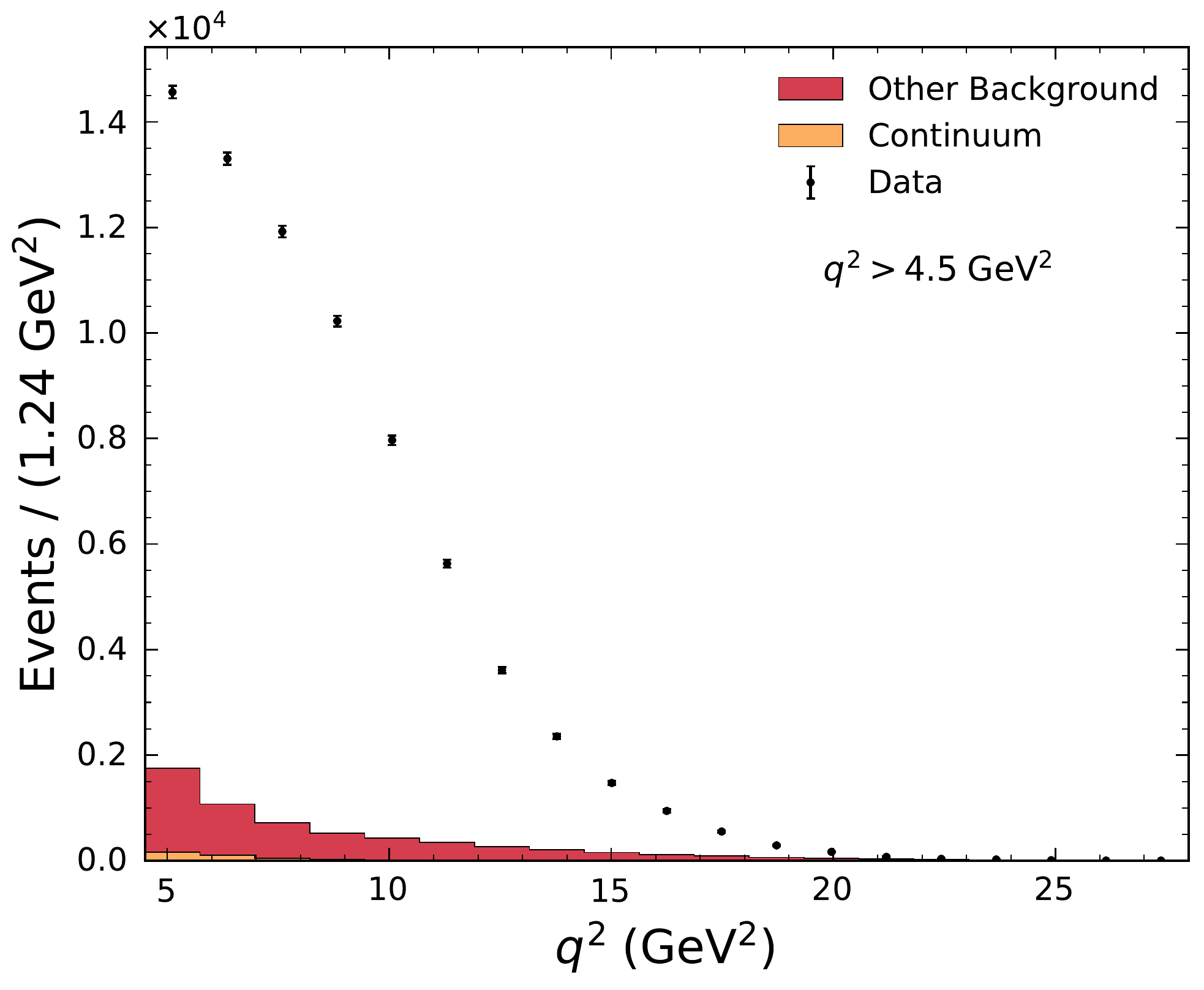}  
  \includegraphics[width=0.25\textwidth]{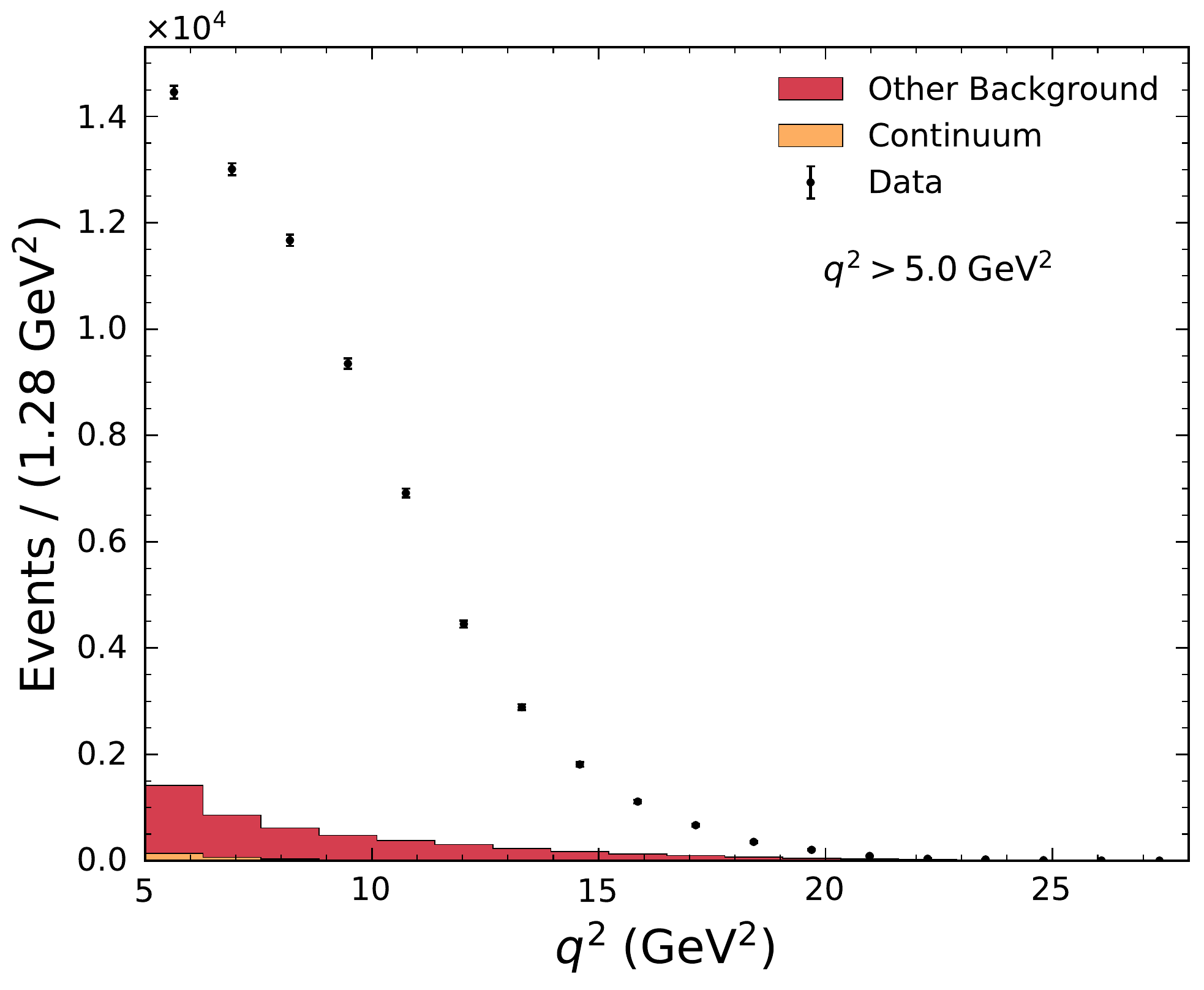}  
  \includegraphics[width=0.25\textwidth]{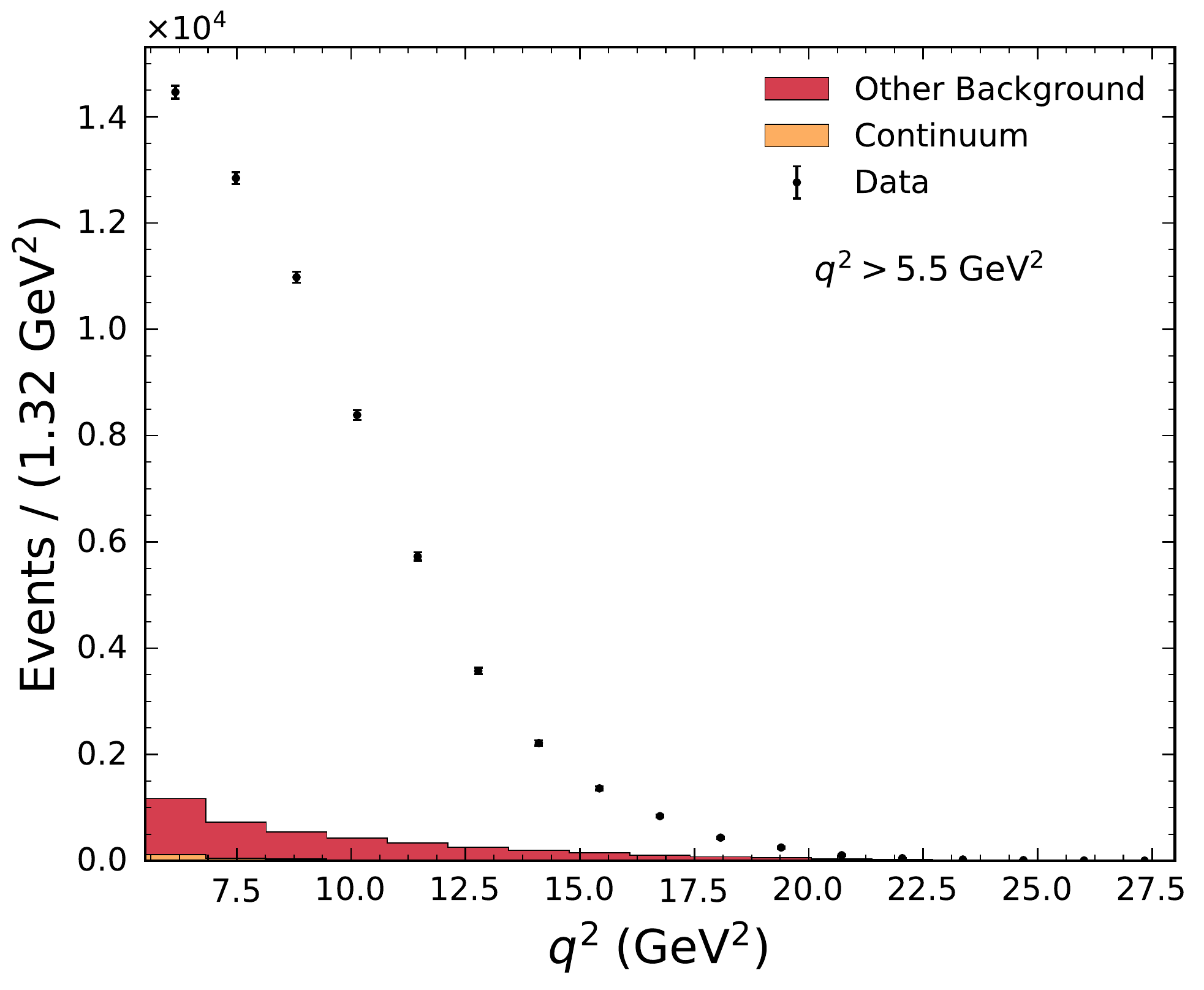}  
  \includegraphics[width=0.25\textwidth]{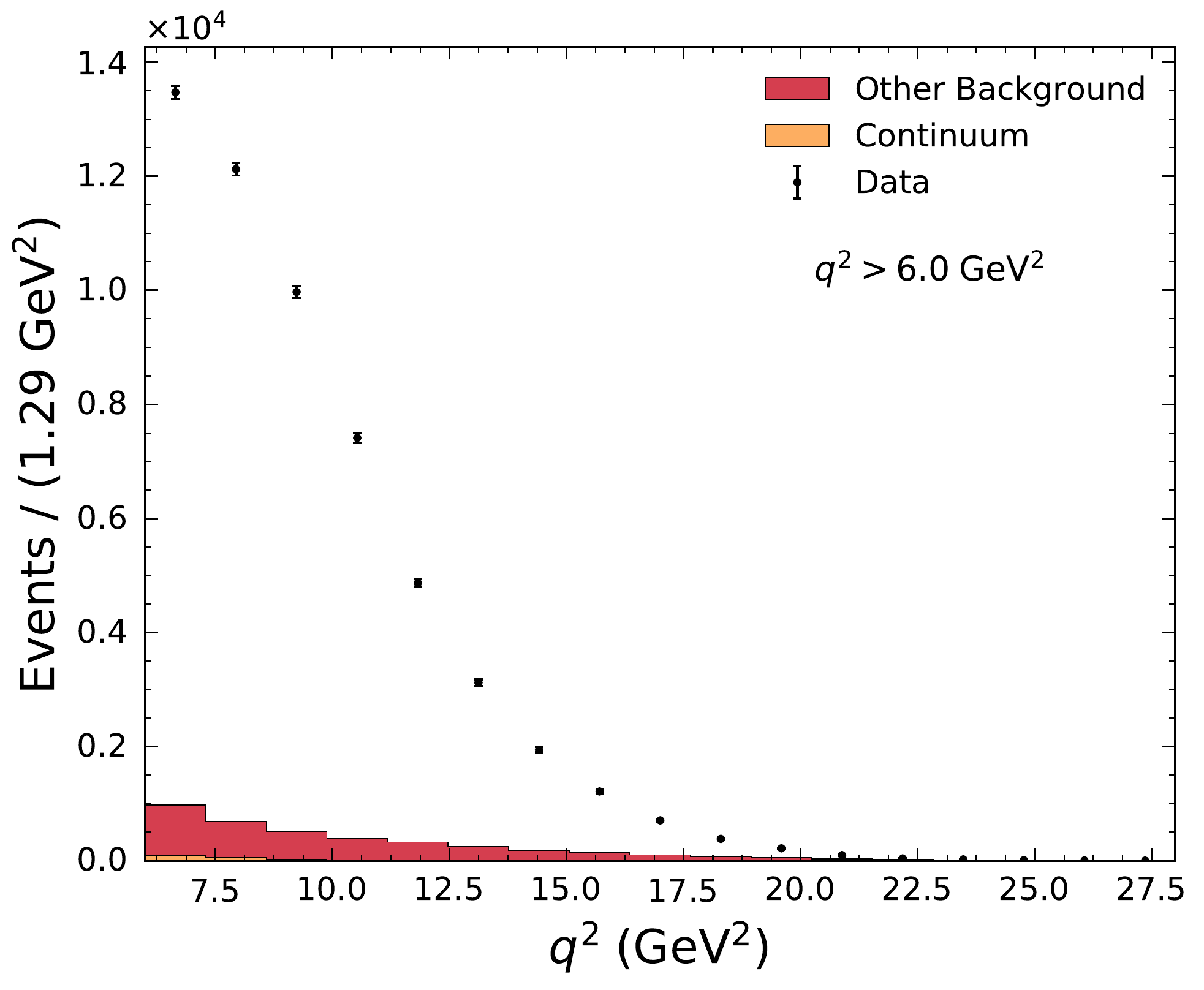}  
  \includegraphics[width=0.25\textwidth]{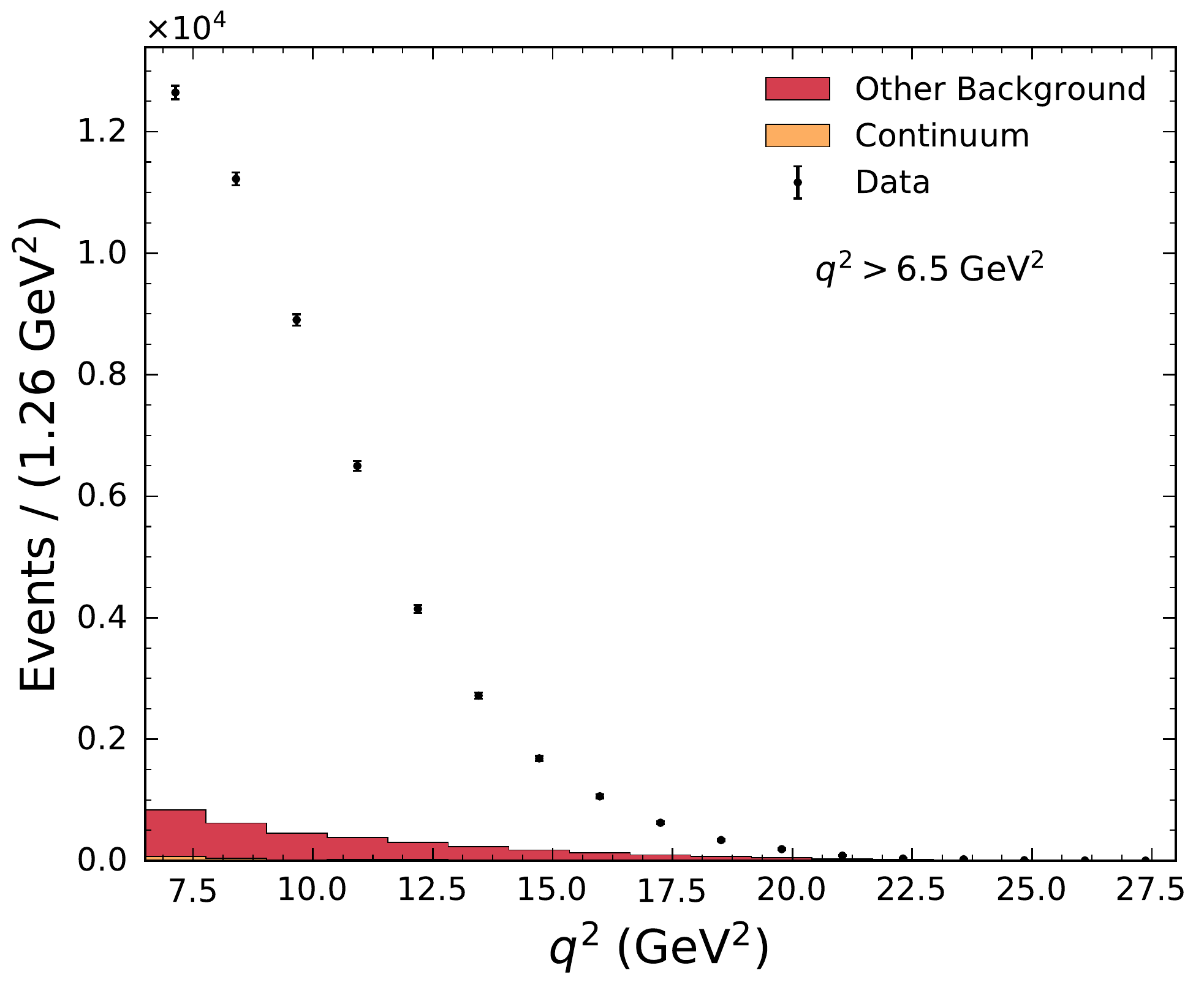}  
  \includegraphics[width=0.25\textwidth]{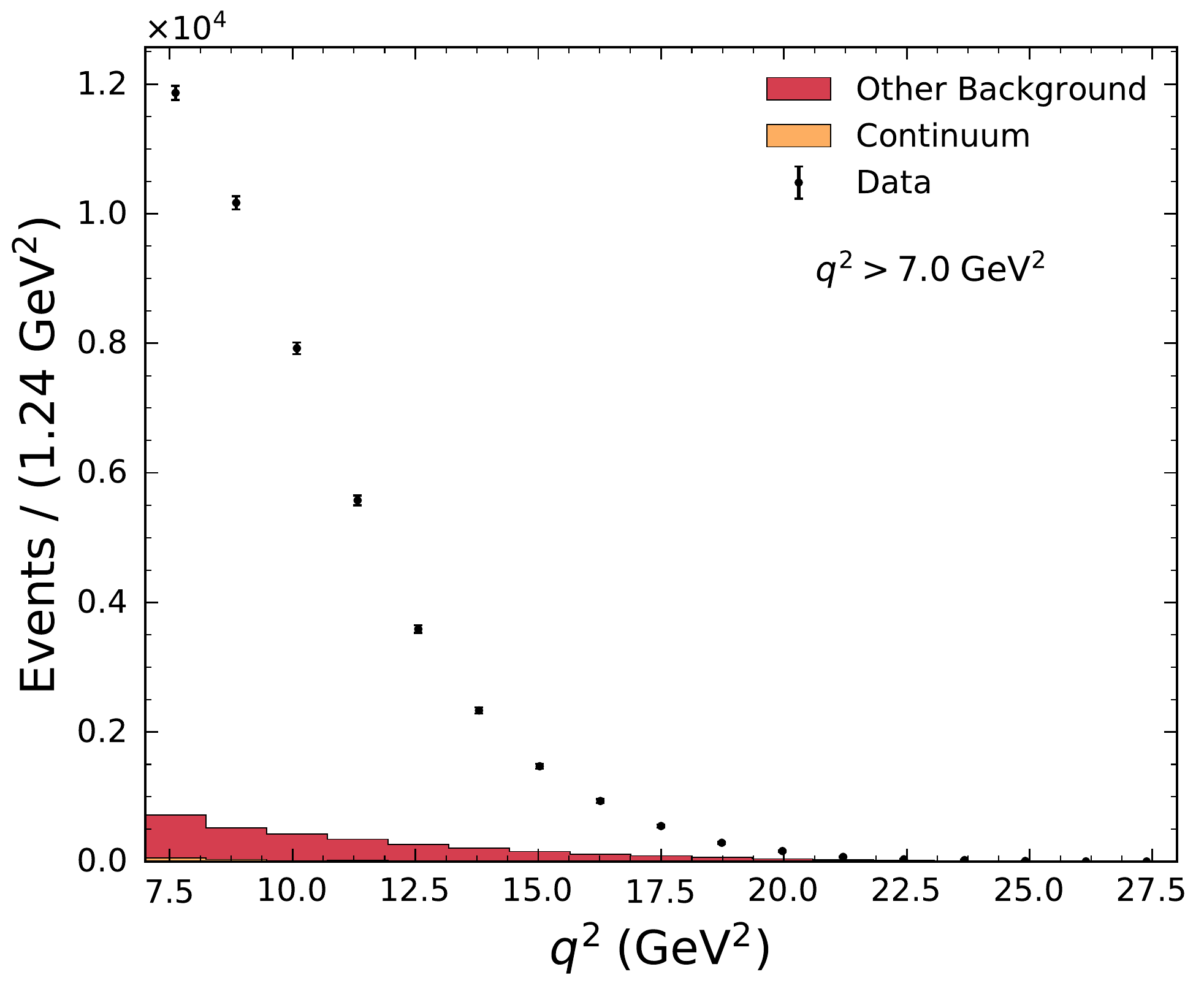}  
  \includegraphics[width=0.25\textwidth]{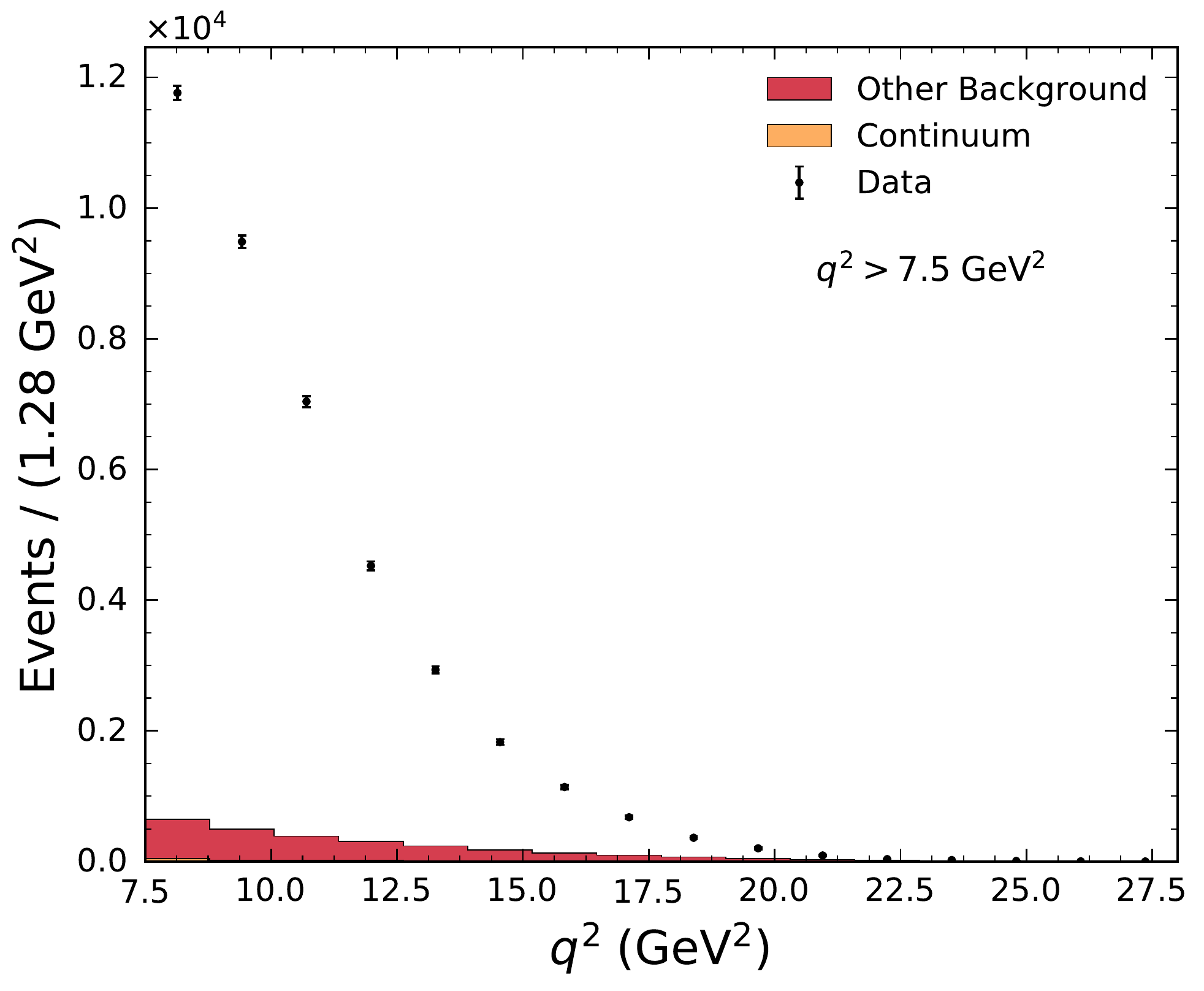}  
  \includegraphics[width=0.25\textwidth]{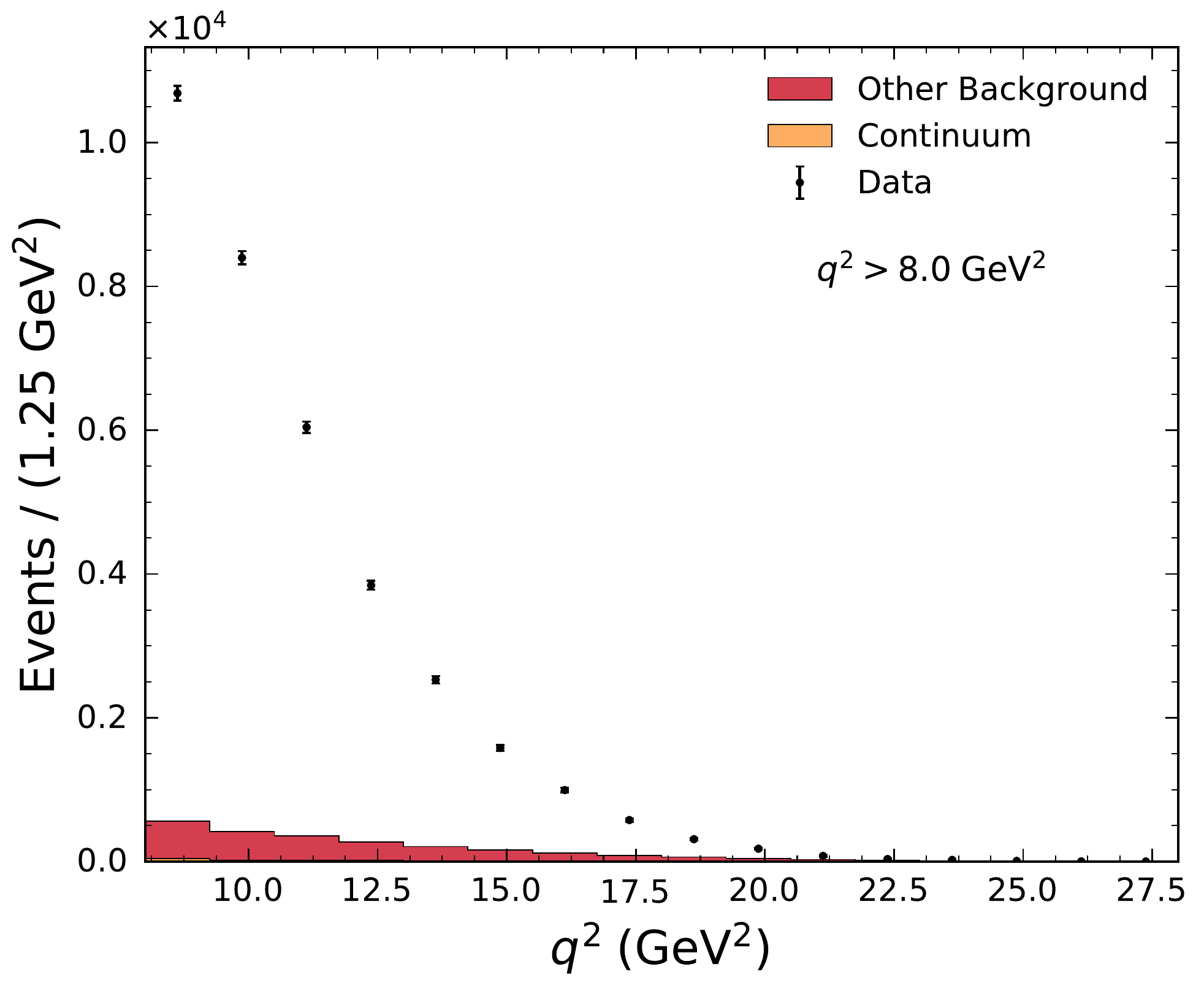}  
  \includegraphics[width=0.25\textwidth]{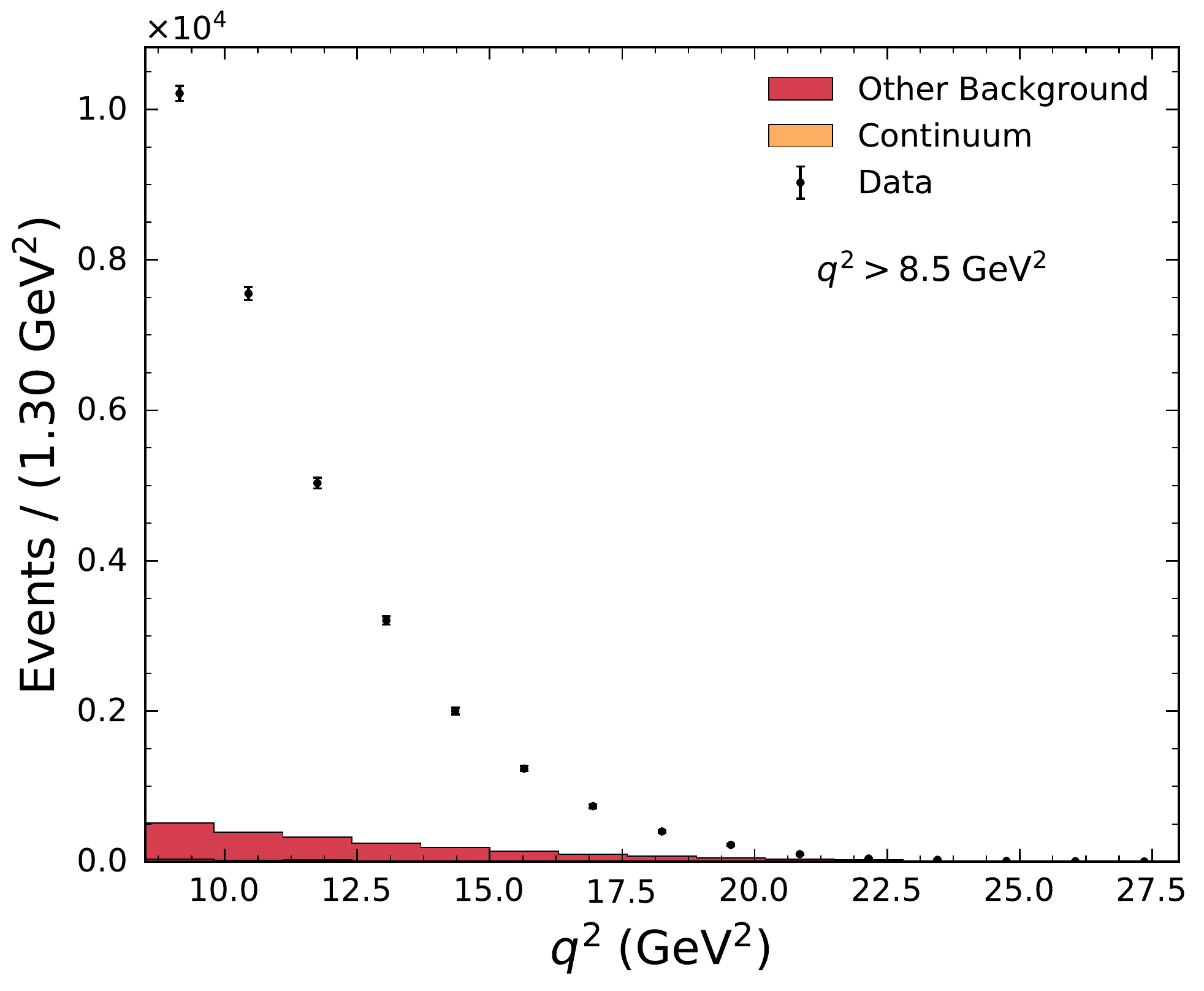}  
  \includegraphics[width=0.25\textwidth]{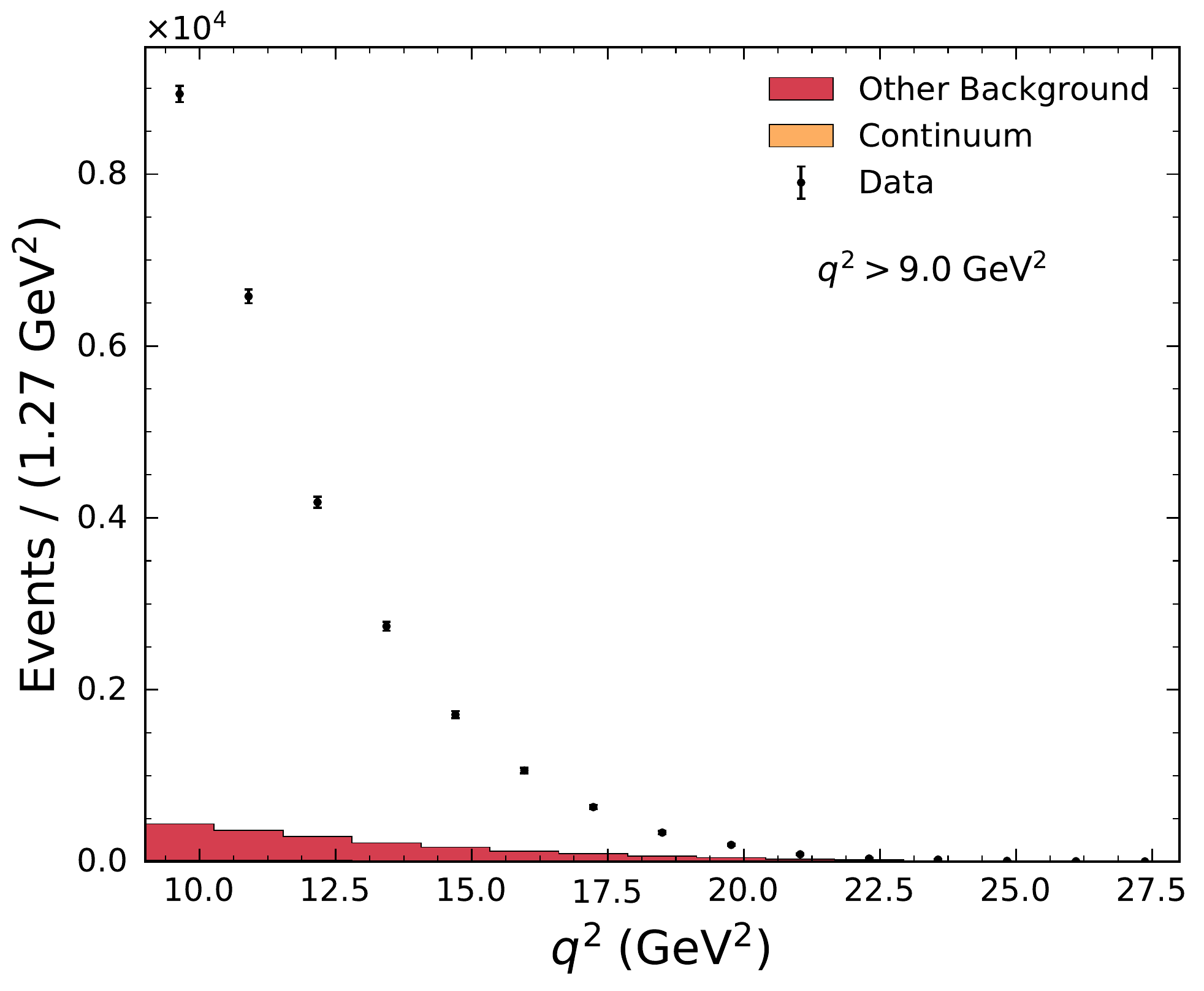}  
  \includegraphics[width=0.25\textwidth]{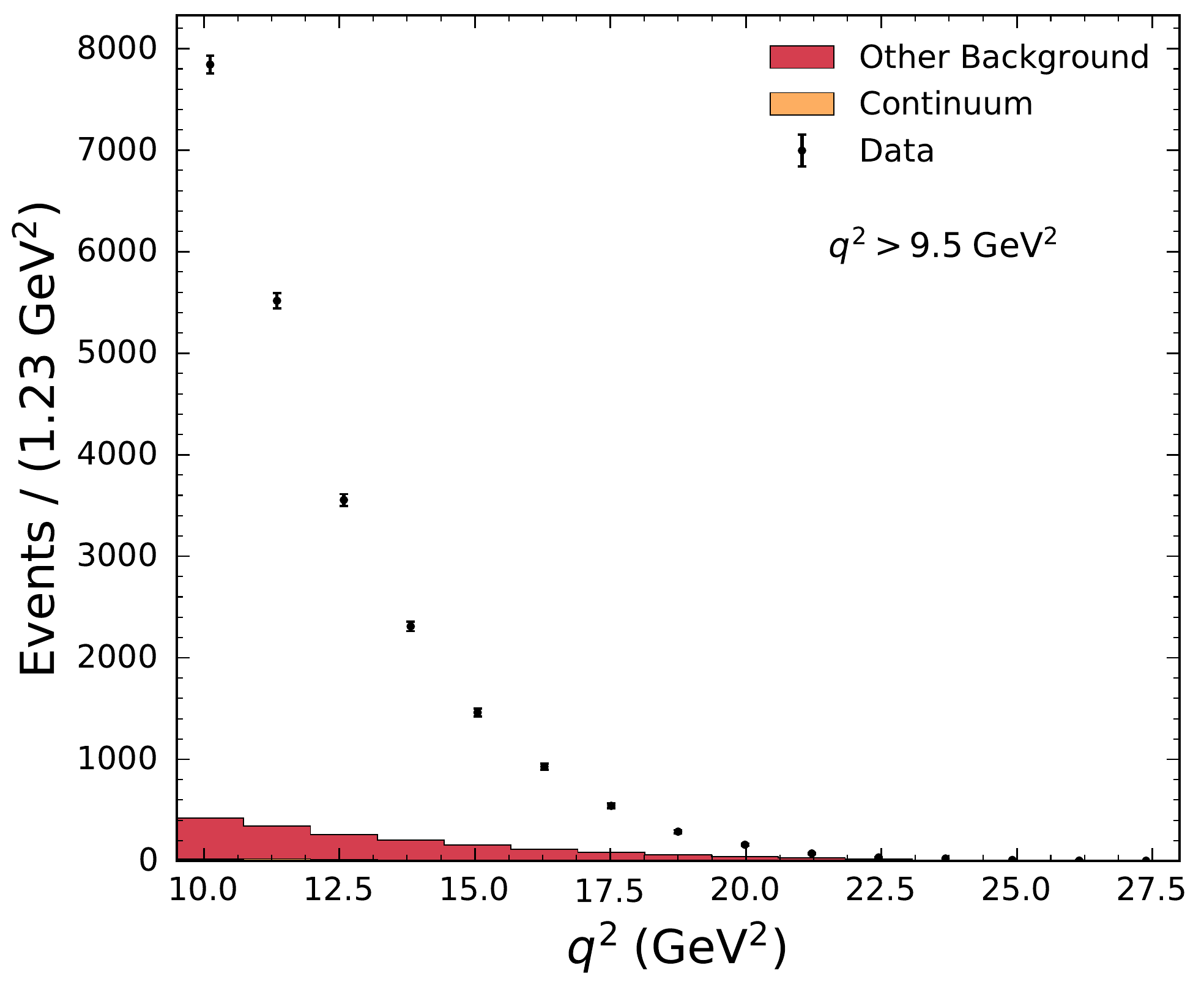}  
  \includegraphics[width=0.25\textwidth]{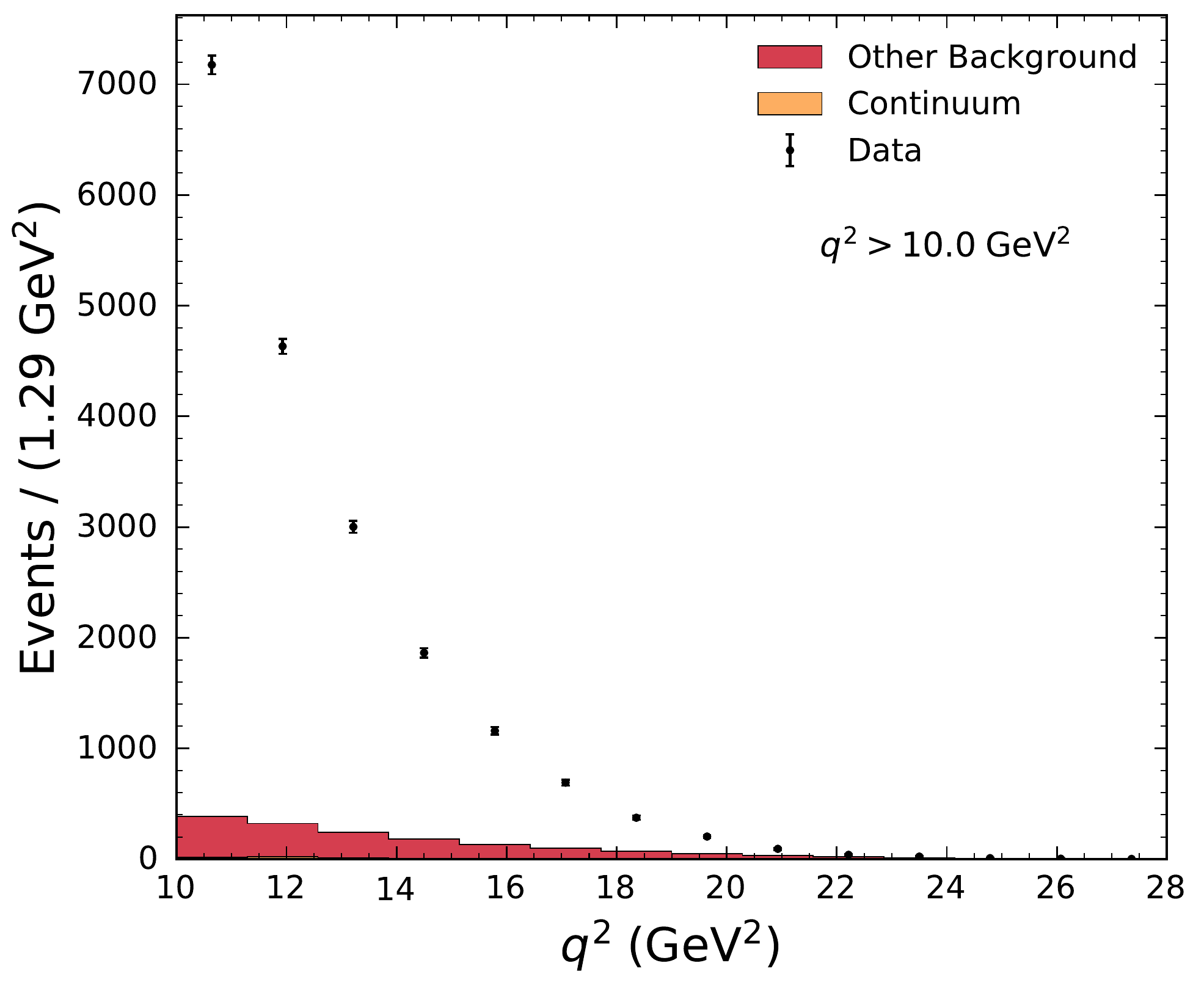}  
\caption{
 The determined background in the $q^{2}$ spectrum, after obtaining the expected background yields from the fit to the $M_X$ spectrum, for different $q^2$ threshold selections for electrons are shown.
 }
\label{fig:bkg_sub_el}
\end{figure}

\begin{figure}[b!]
  \includegraphics[width=0.25\textwidth]{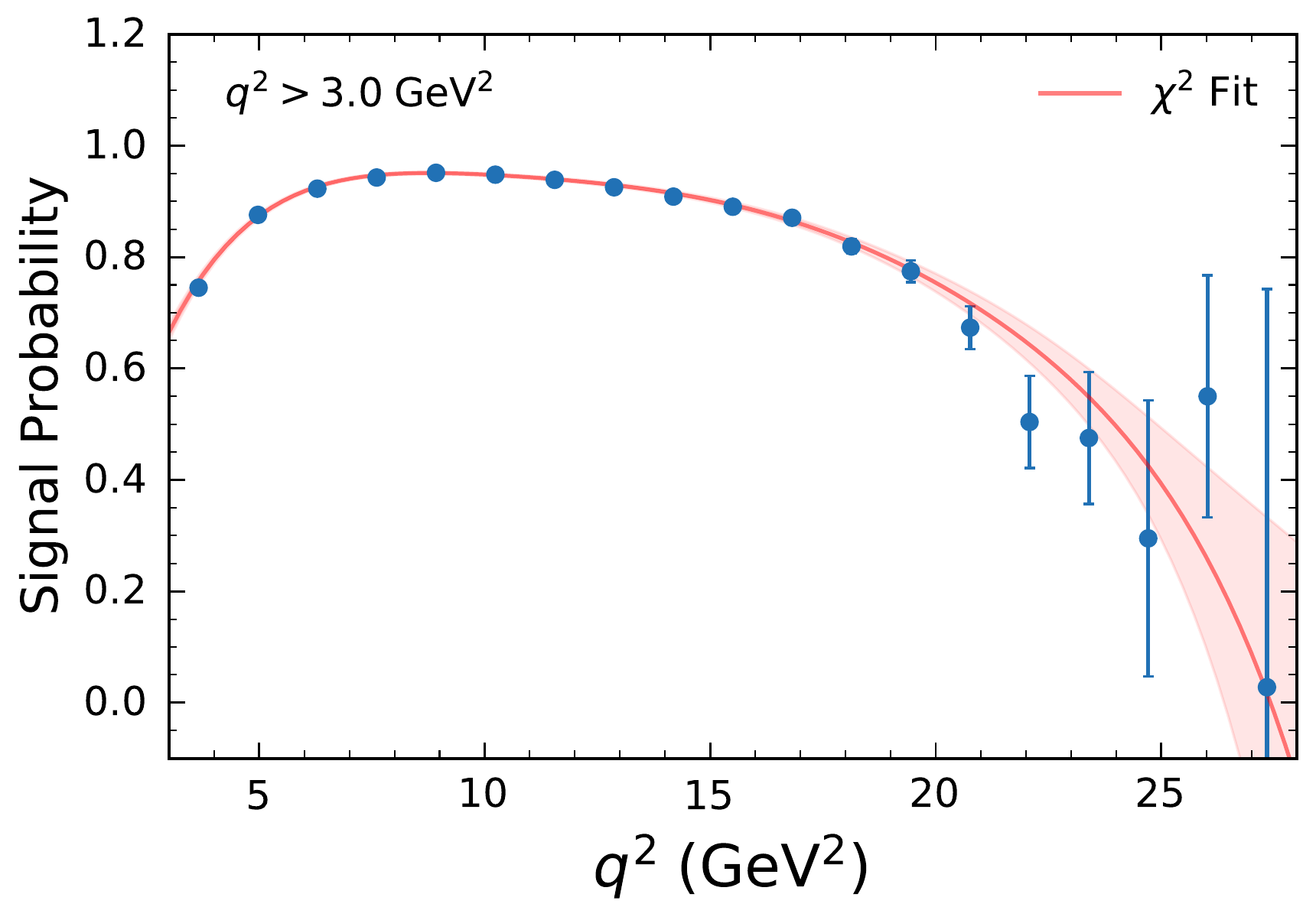}  
  \includegraphics[width=0.25\textwidth]{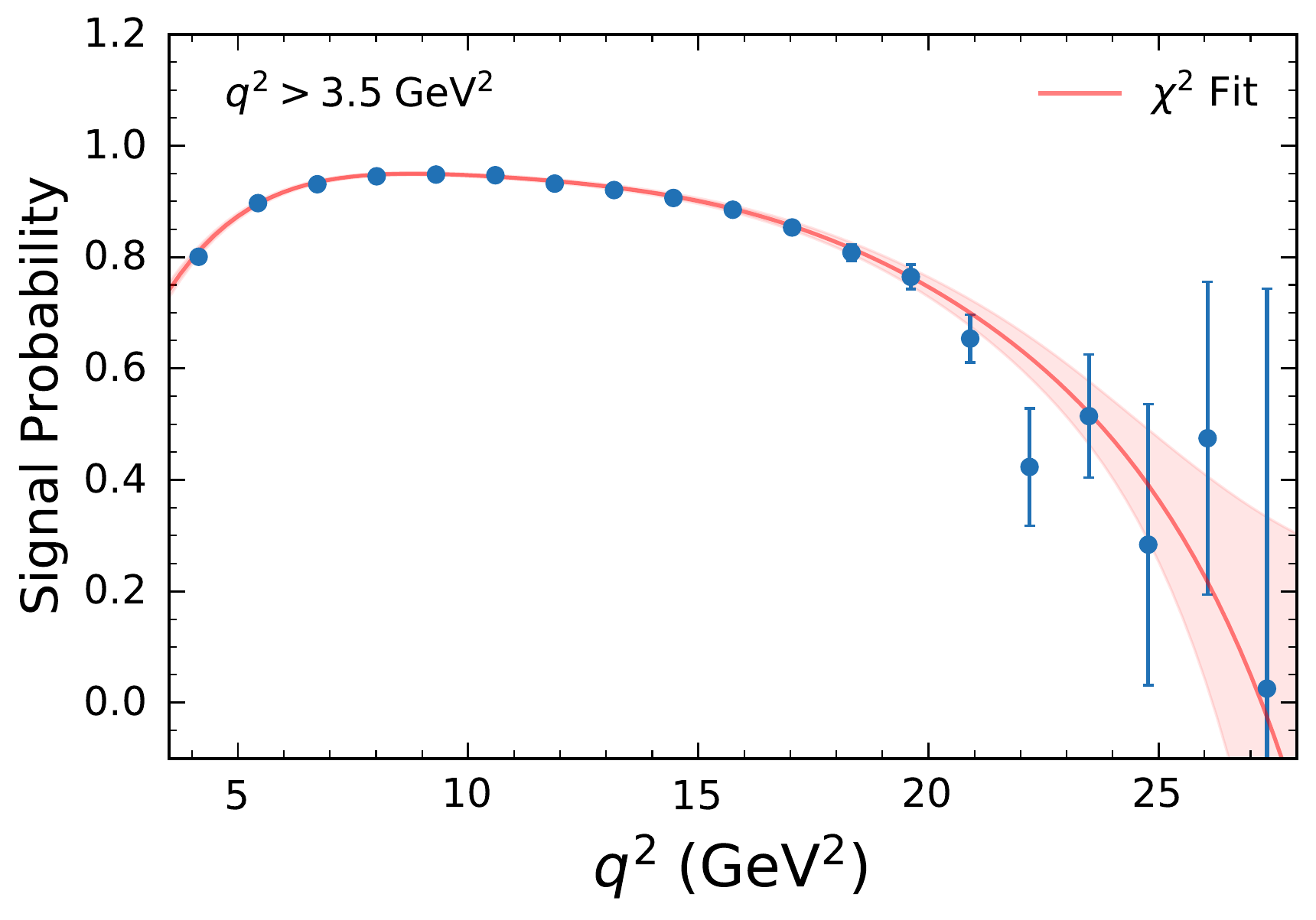}  
  \includegraphics[width=0.25\textwidth]{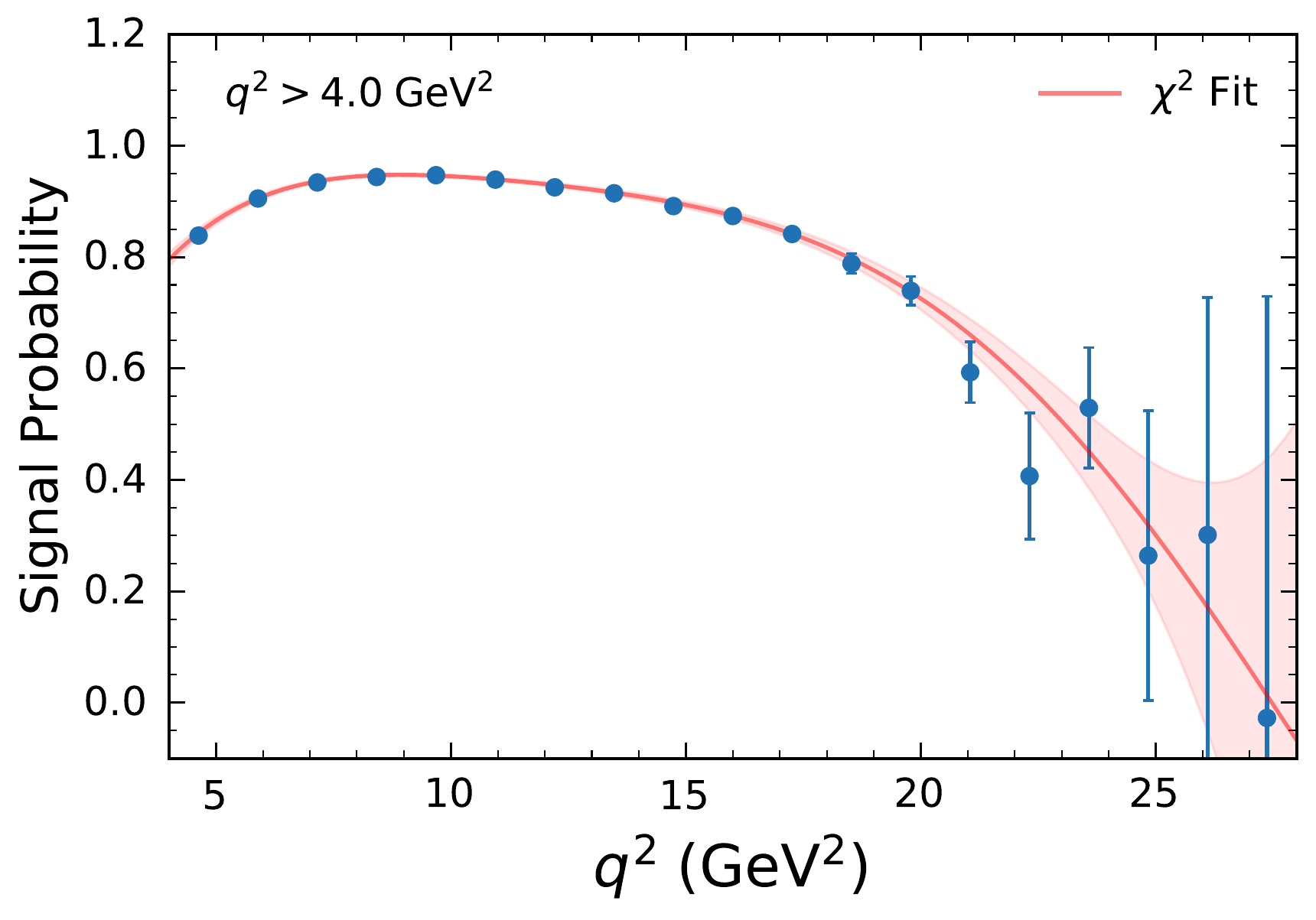}  
  \includegraphics[width=0.25\textwidth]{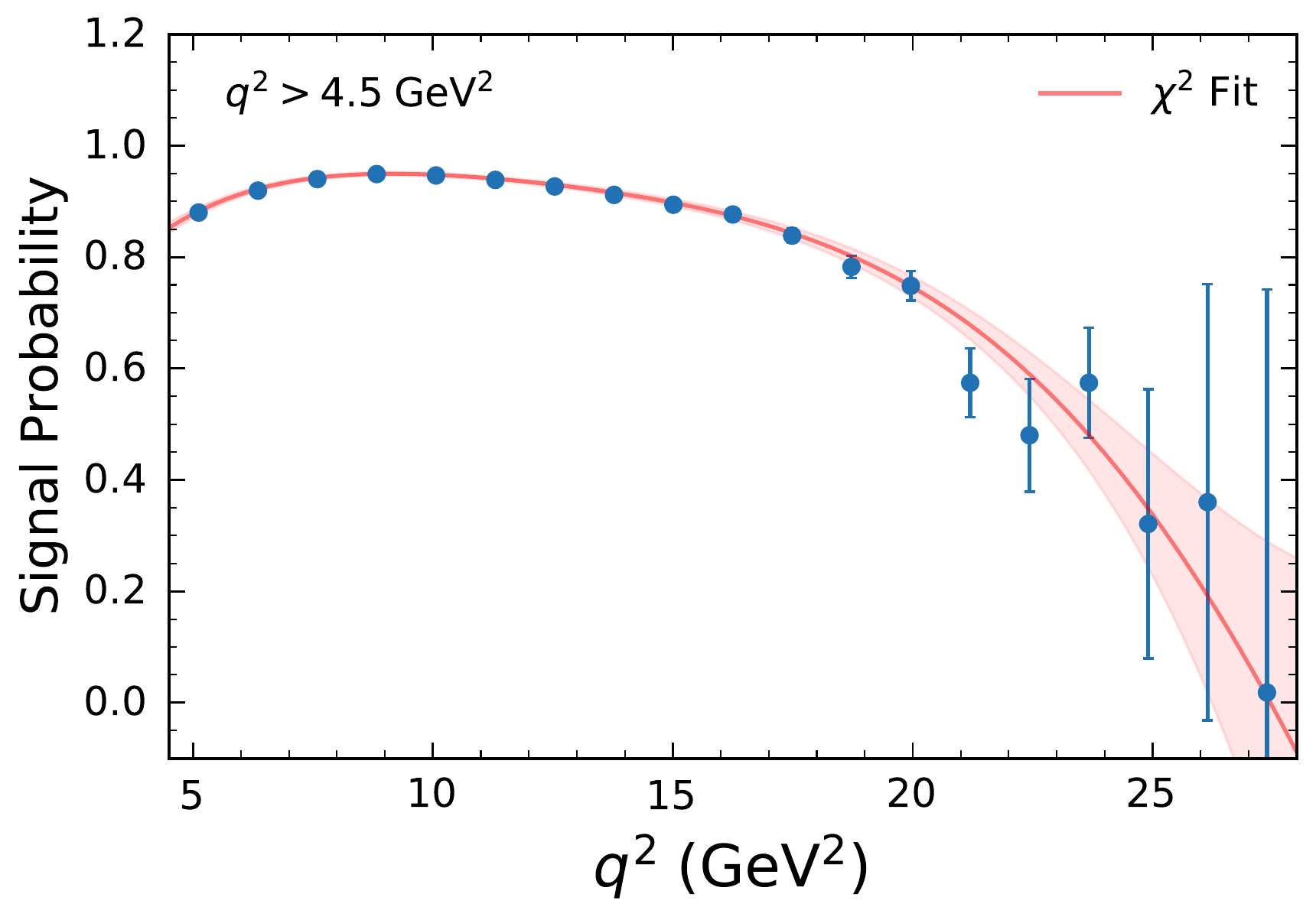}  
  \includegraphics[width=0.25\textwidth]{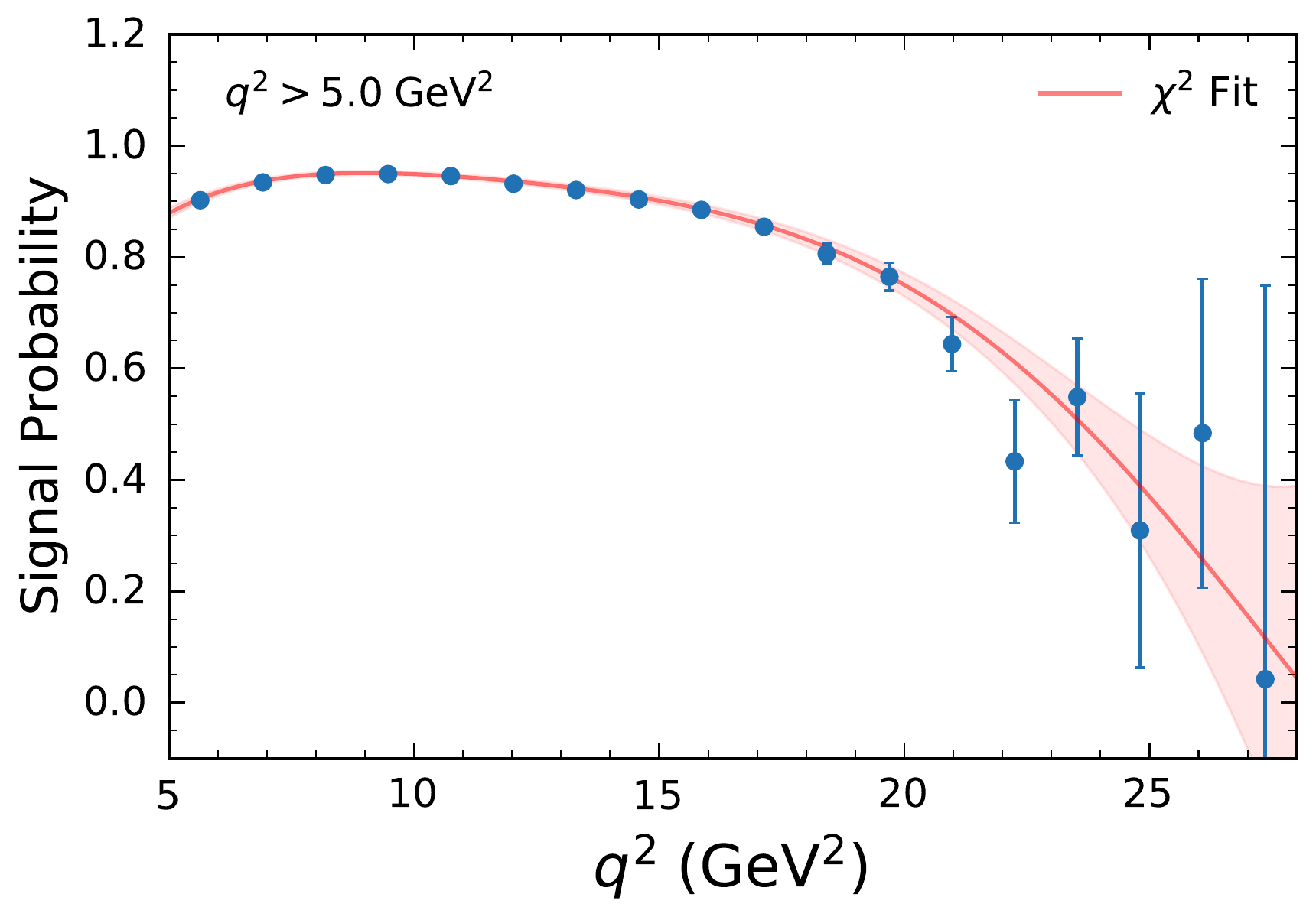}  
  \includegraphics[width=0.25\textwidth]{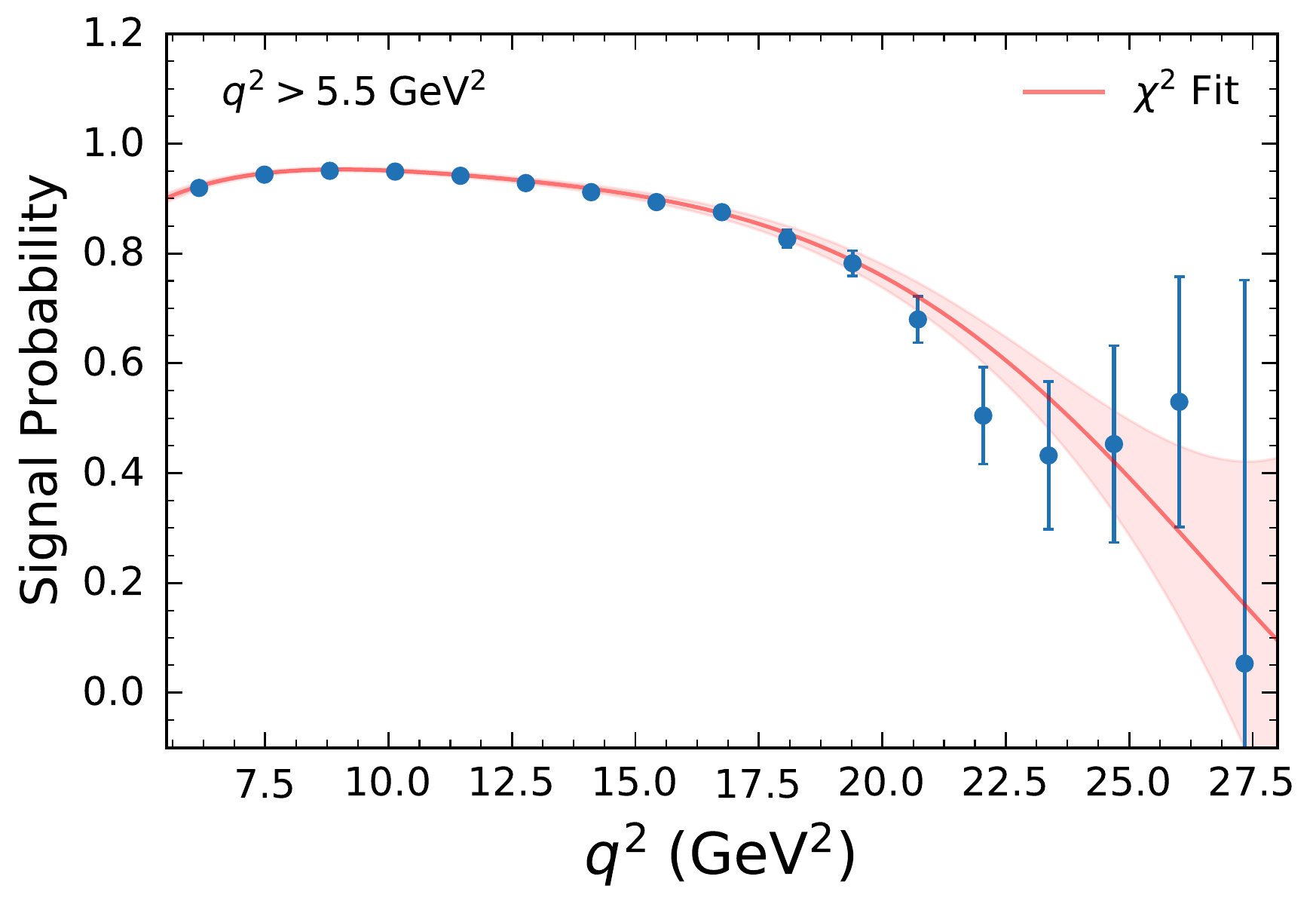}  
  \includegraphics[width=0.25\textwidth]{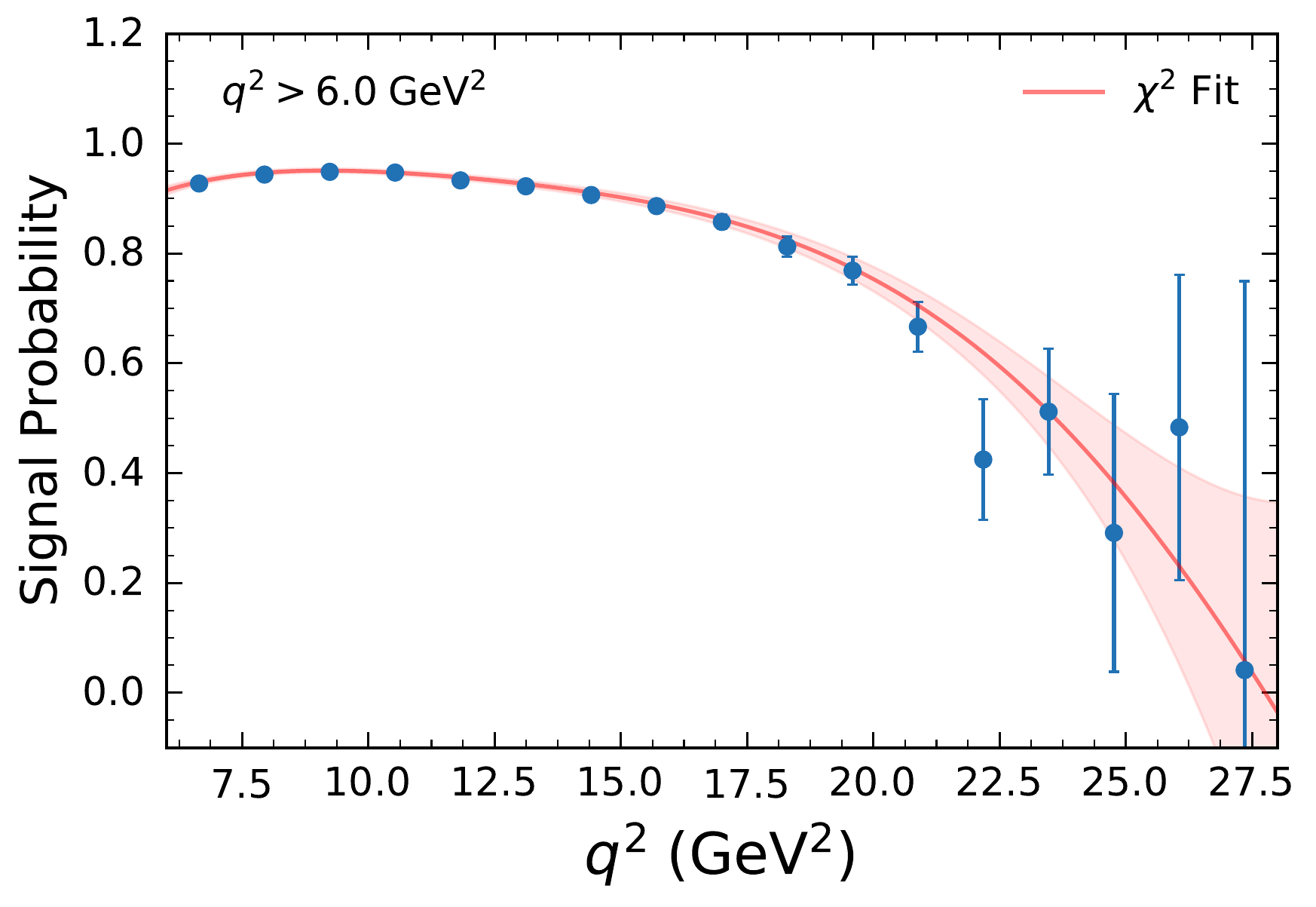}  
  \includegraphics[width=0.25\textwidth]{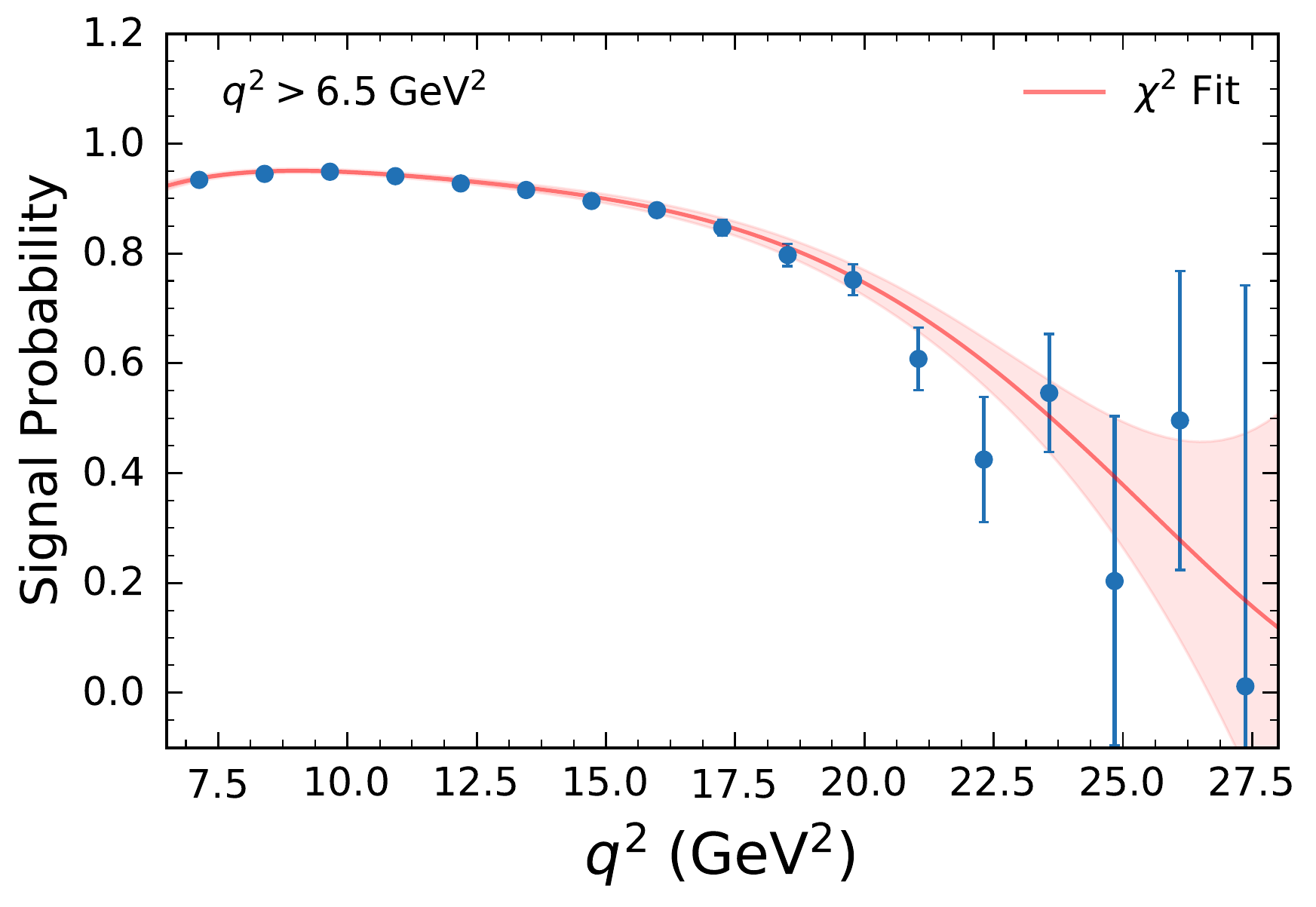}  
  \includegraphics[width=0.25\textwidth]{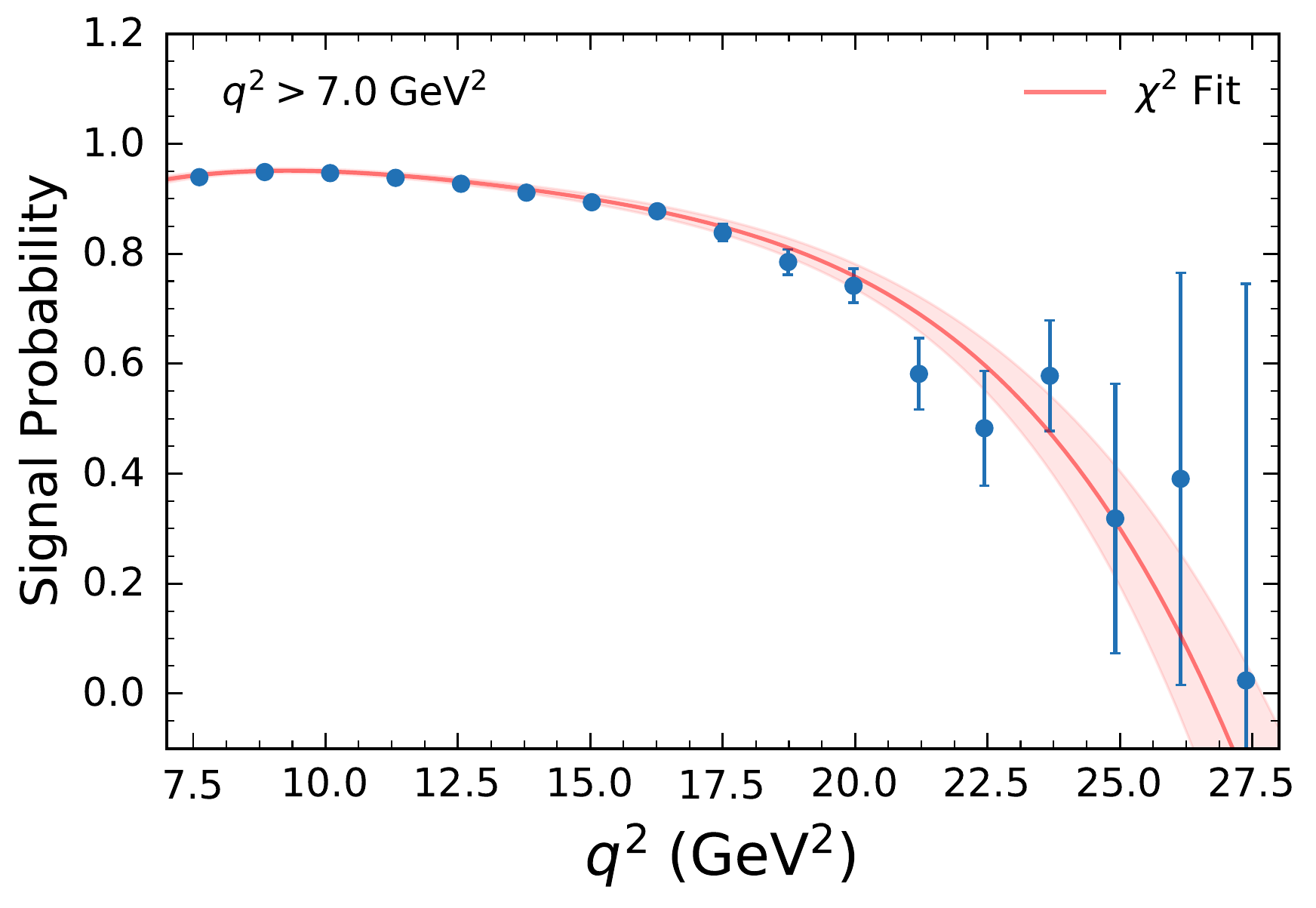}  
  \includegraphics[width=0.25\textwidth]{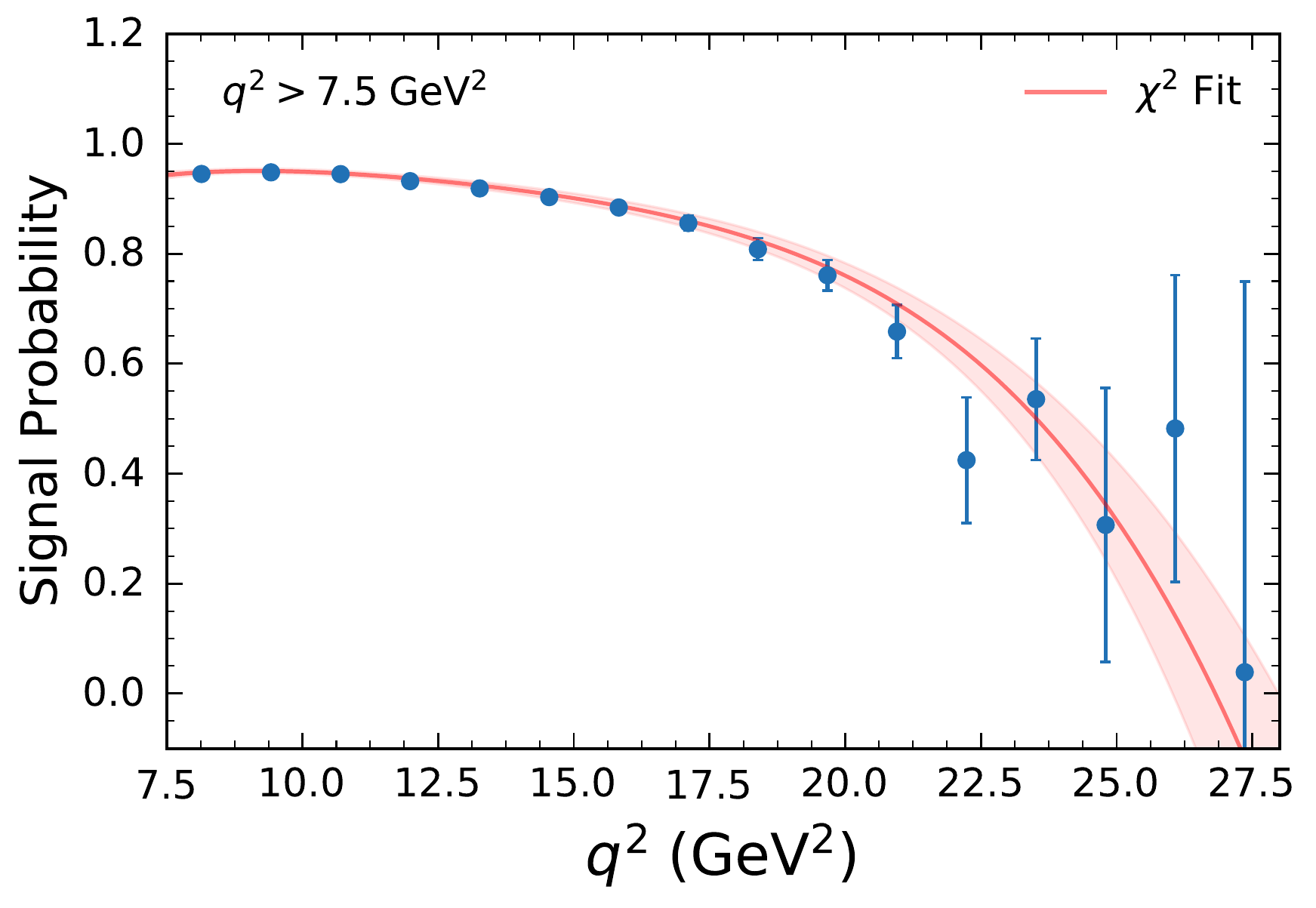}  
  \includegraphics[width=0.25\textwidth]{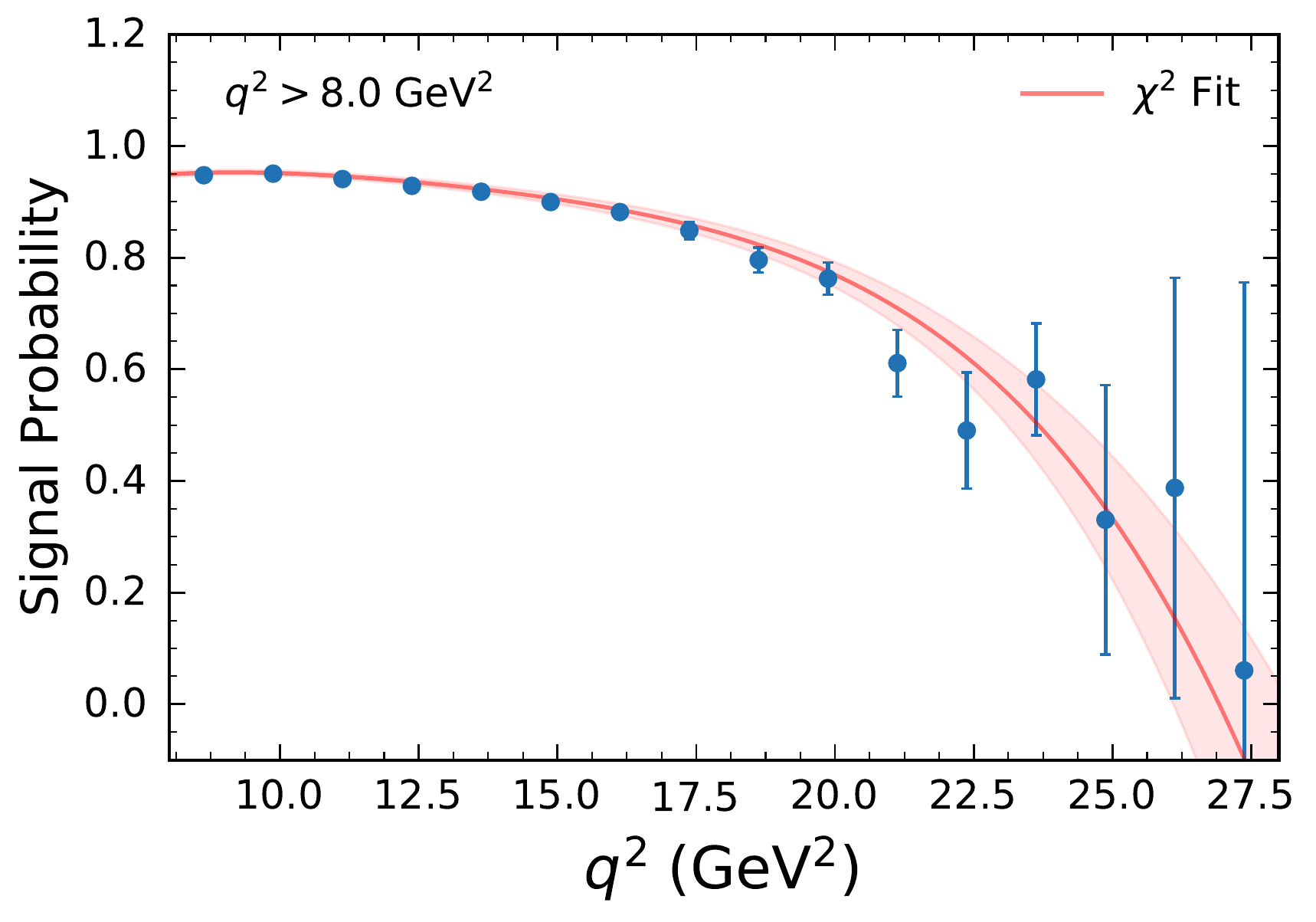}  
  \includegraphics[width=0.25\textwidth]{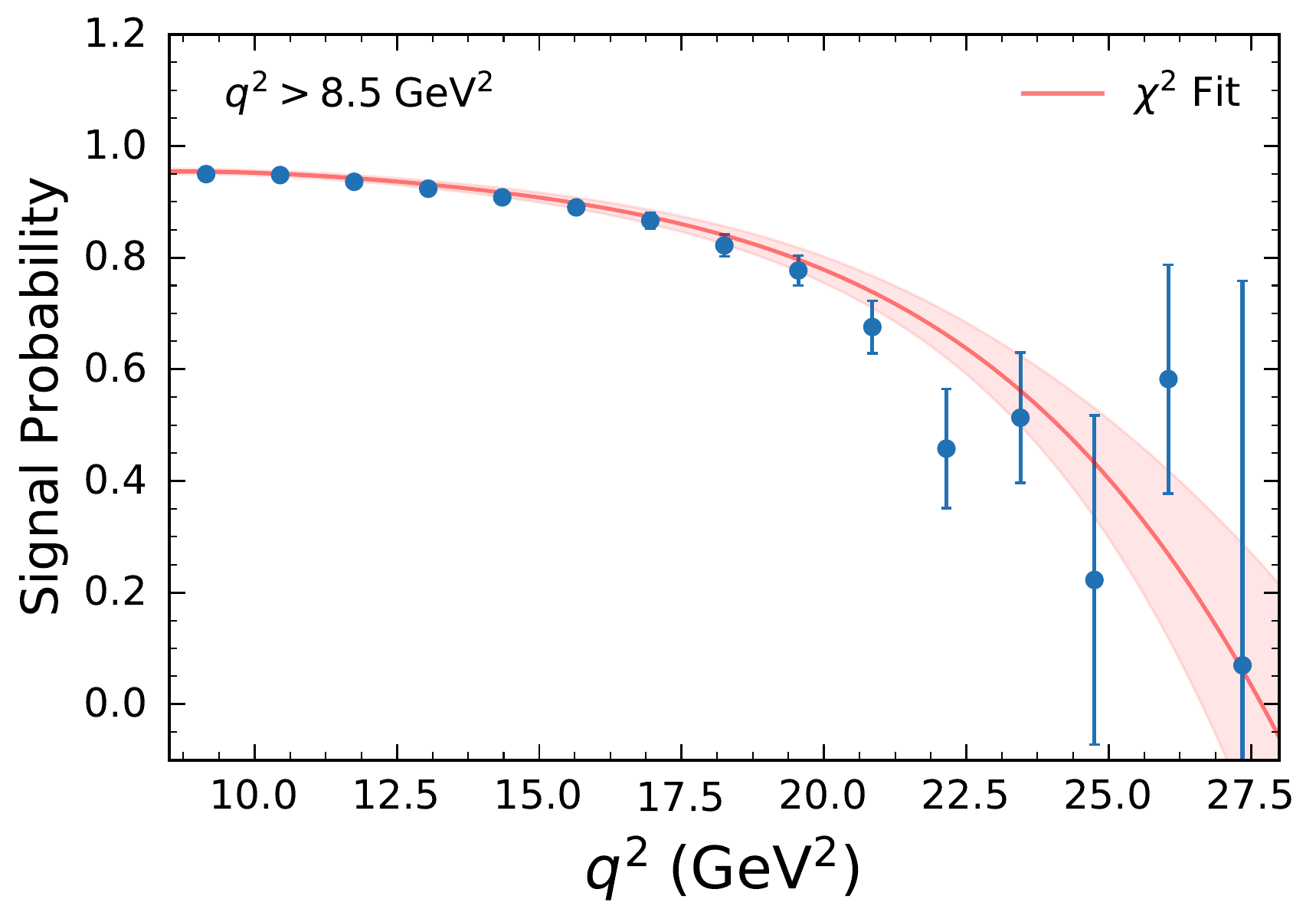}  
  \includegraphics[width=0.25\textwidth]{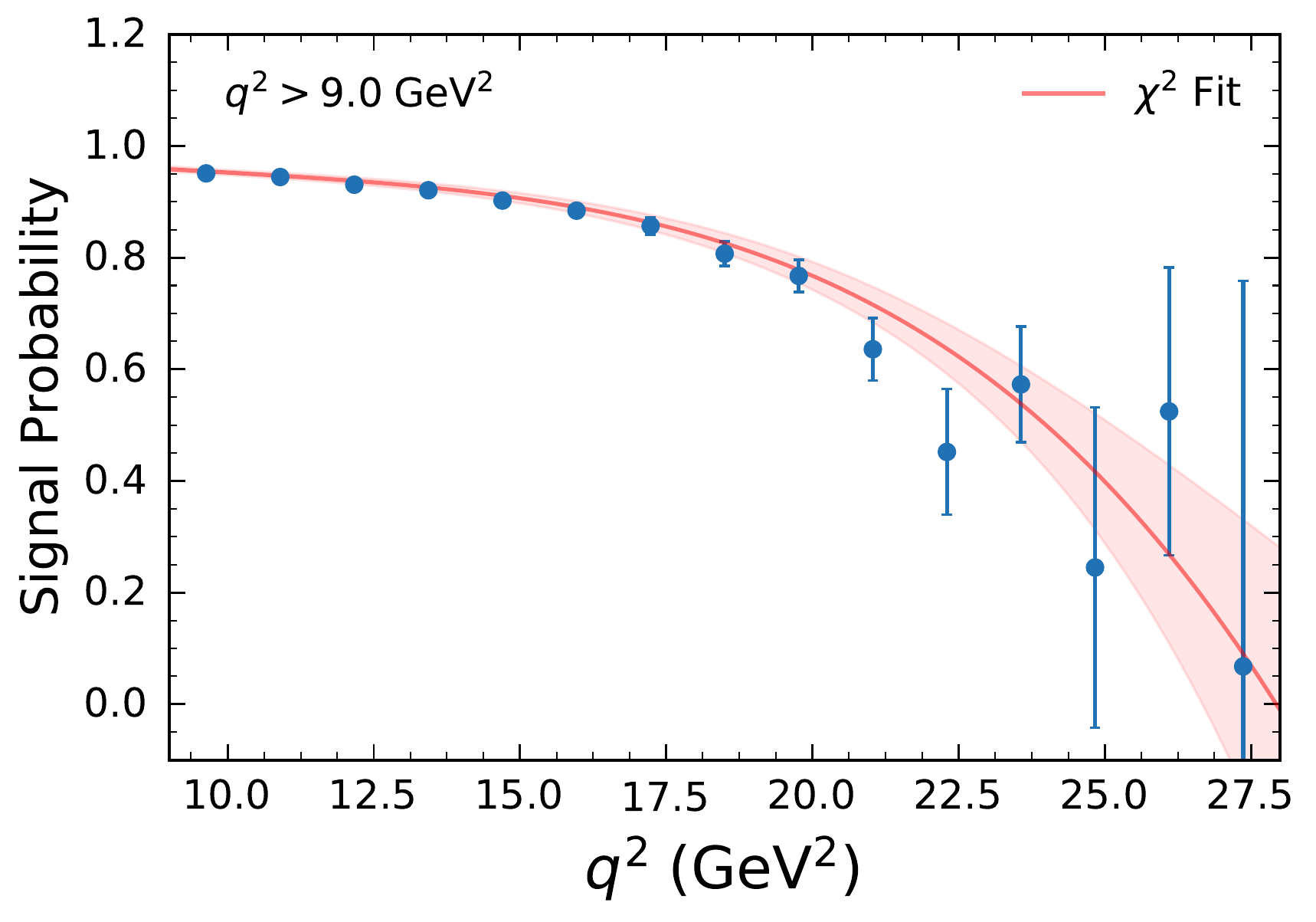}  
  \includegraphics[width=0.25\textwidth]{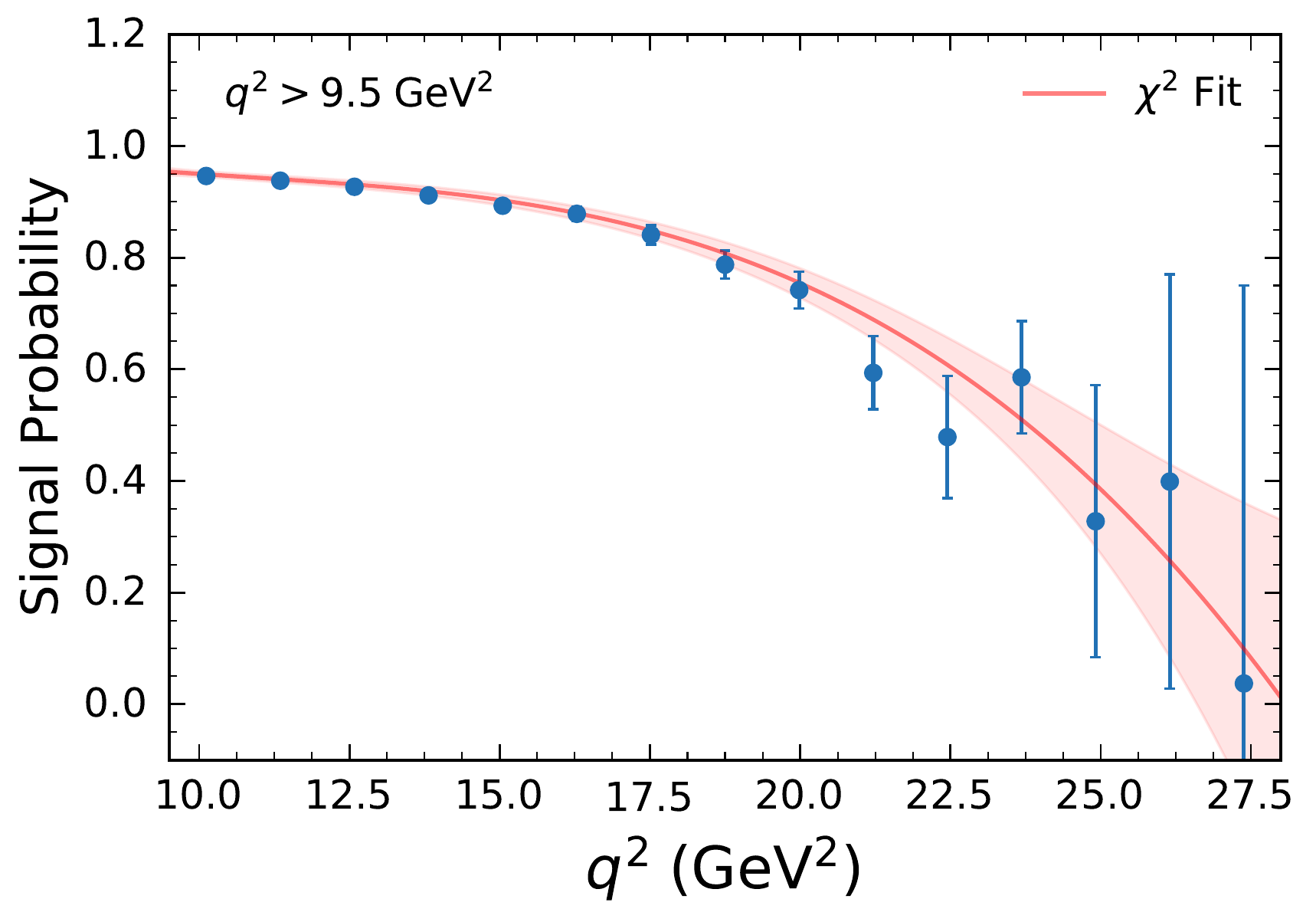}
  \includegraphics[width=0.25\textwidth]{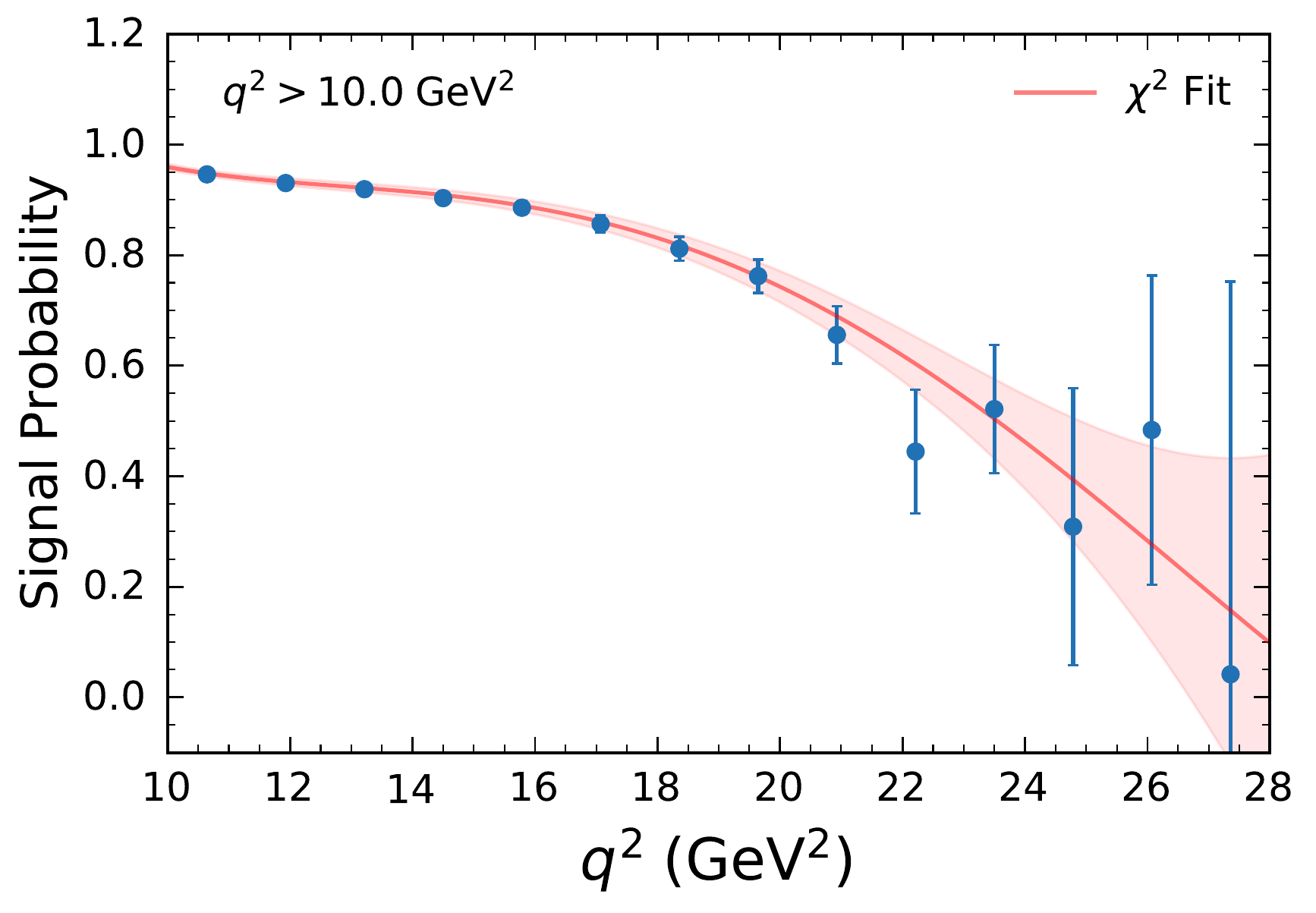}    
\caption{
 The determined signal probabilities for different $q^2$ threshold selections for electrons are shown.
 }
\label{fig:w_el}
\end{figure}

\begin{figure}[b!]
  \includegraphics[width=0.25\textwidth]{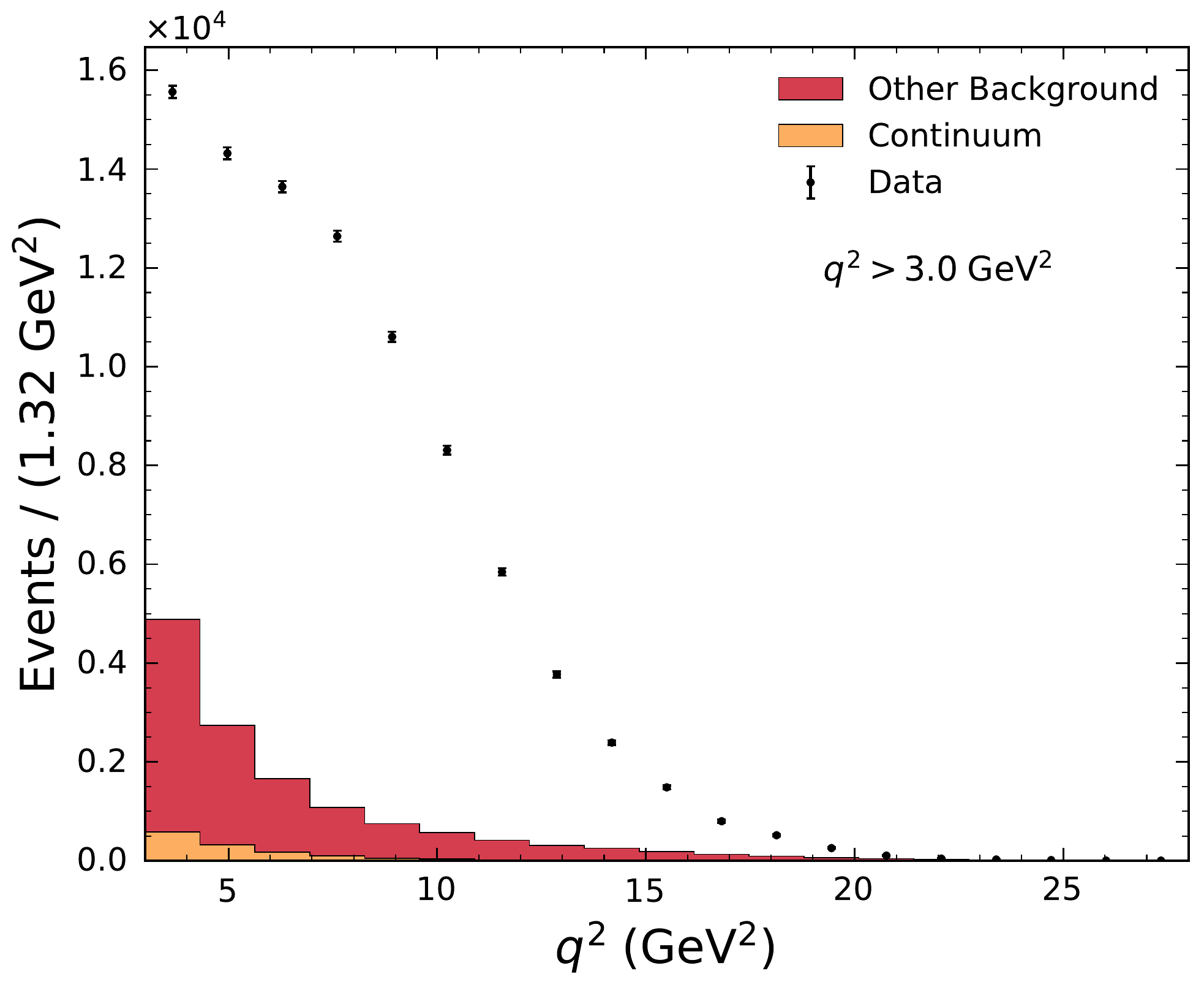}  
  \includegraphics[width=0.25\textwidth]{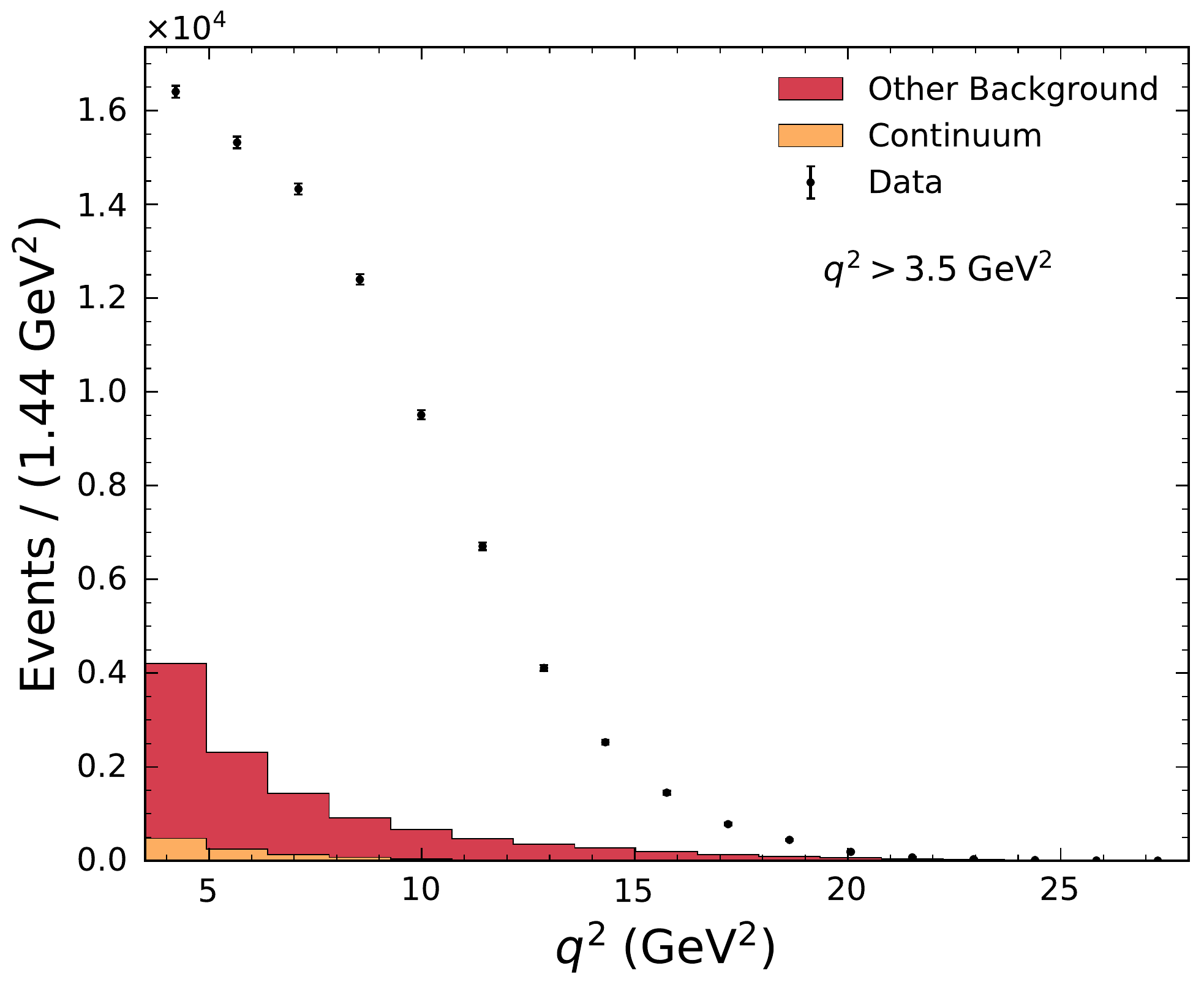}  
  \includegraphics[width=0.25\textwidth]{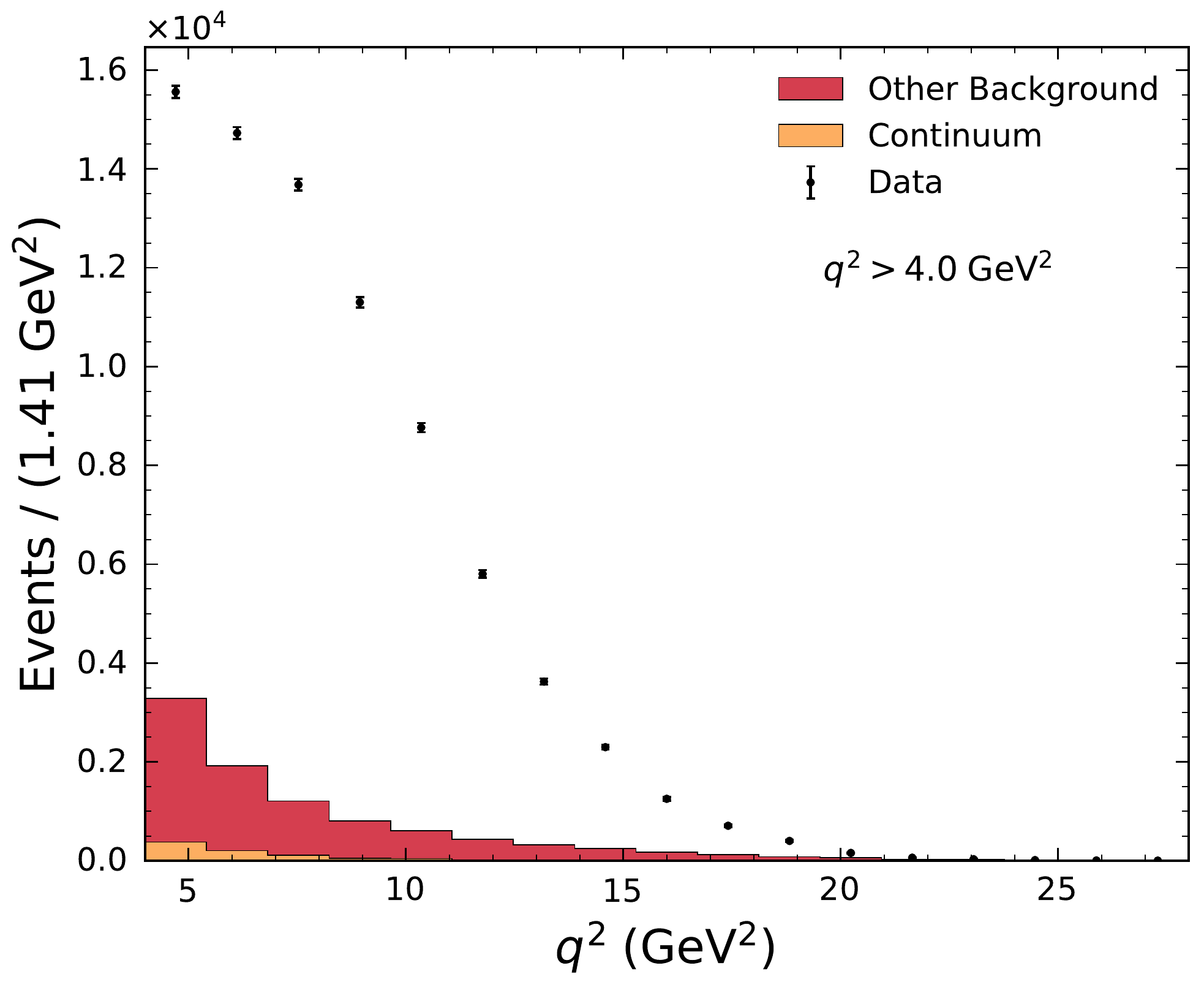}  
  \includegraphics[width=0.25\textwidth]{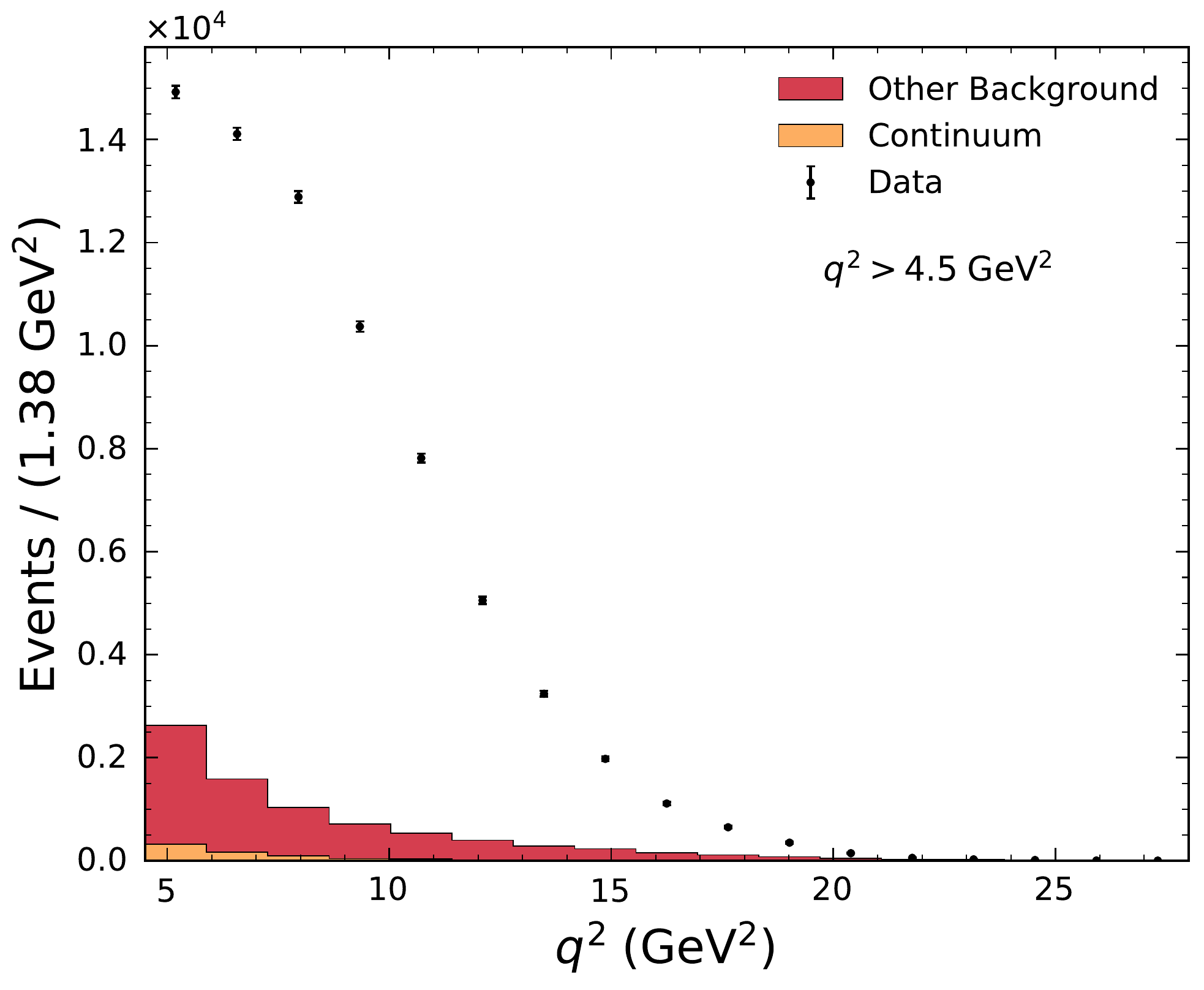}  
  \includegraphics[width=0.25\textwidth]{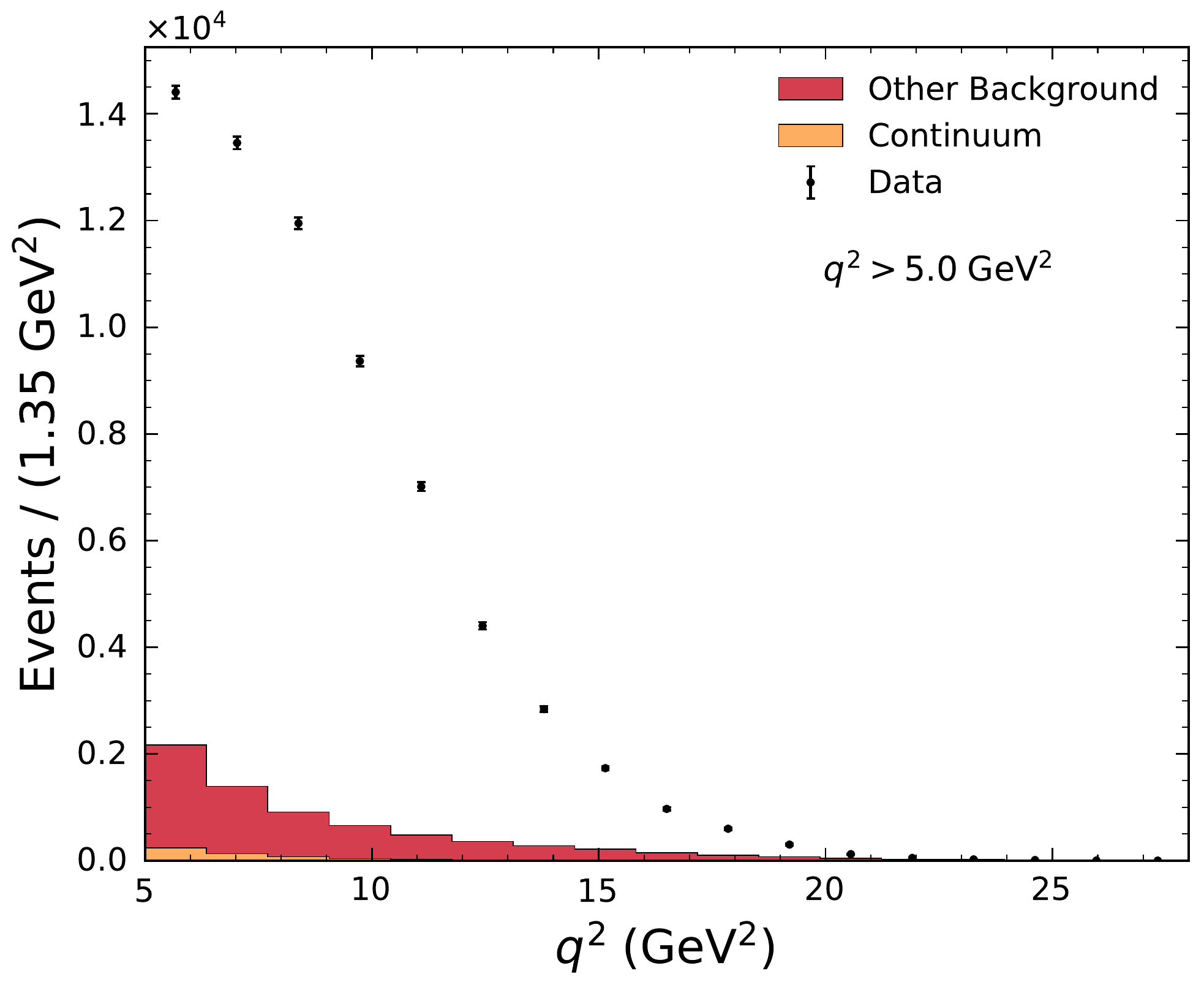}  
  \includegraphics[width=0.25\textwidth]{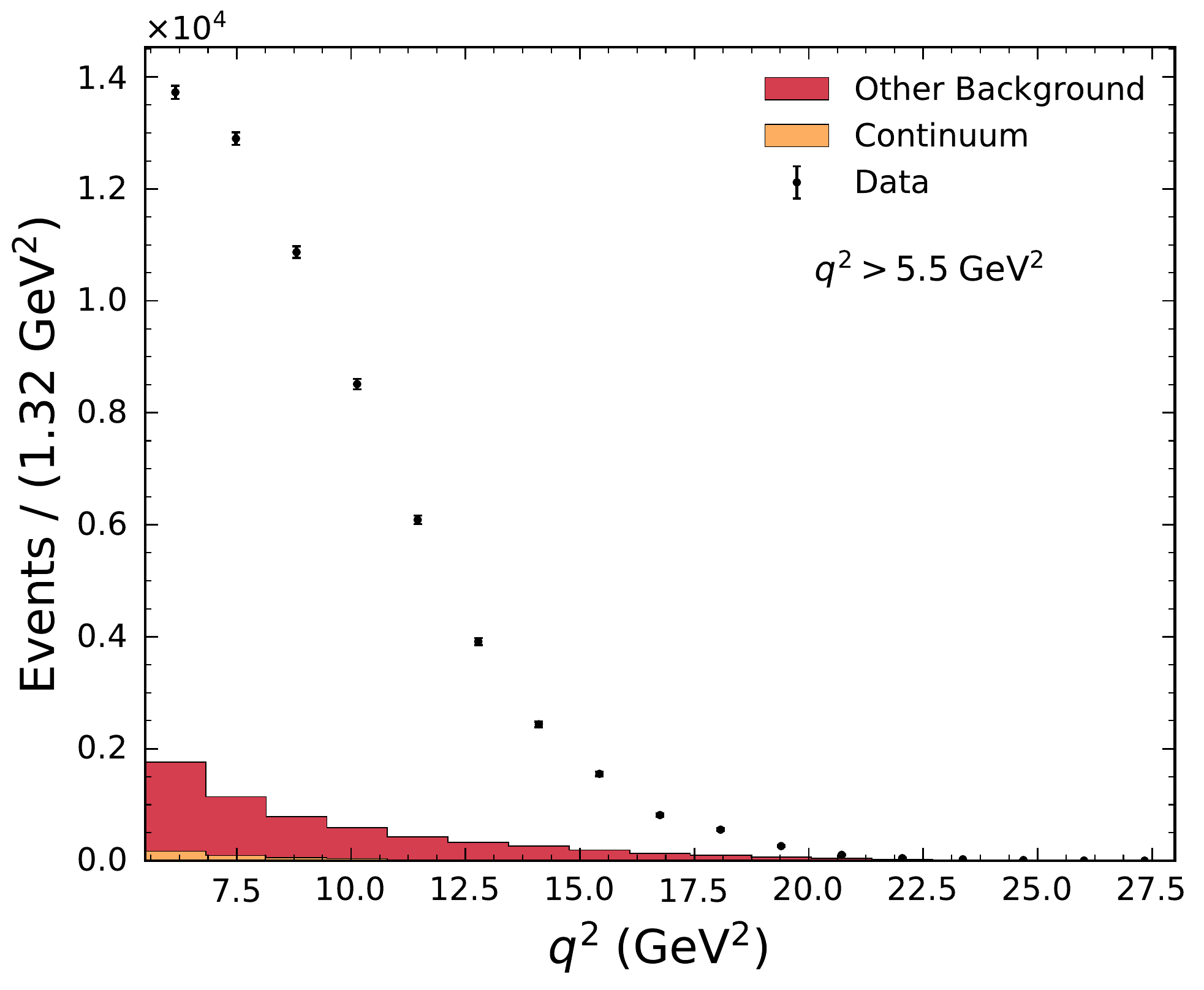}  
  \includegraphics[width=0.25\textwidth]{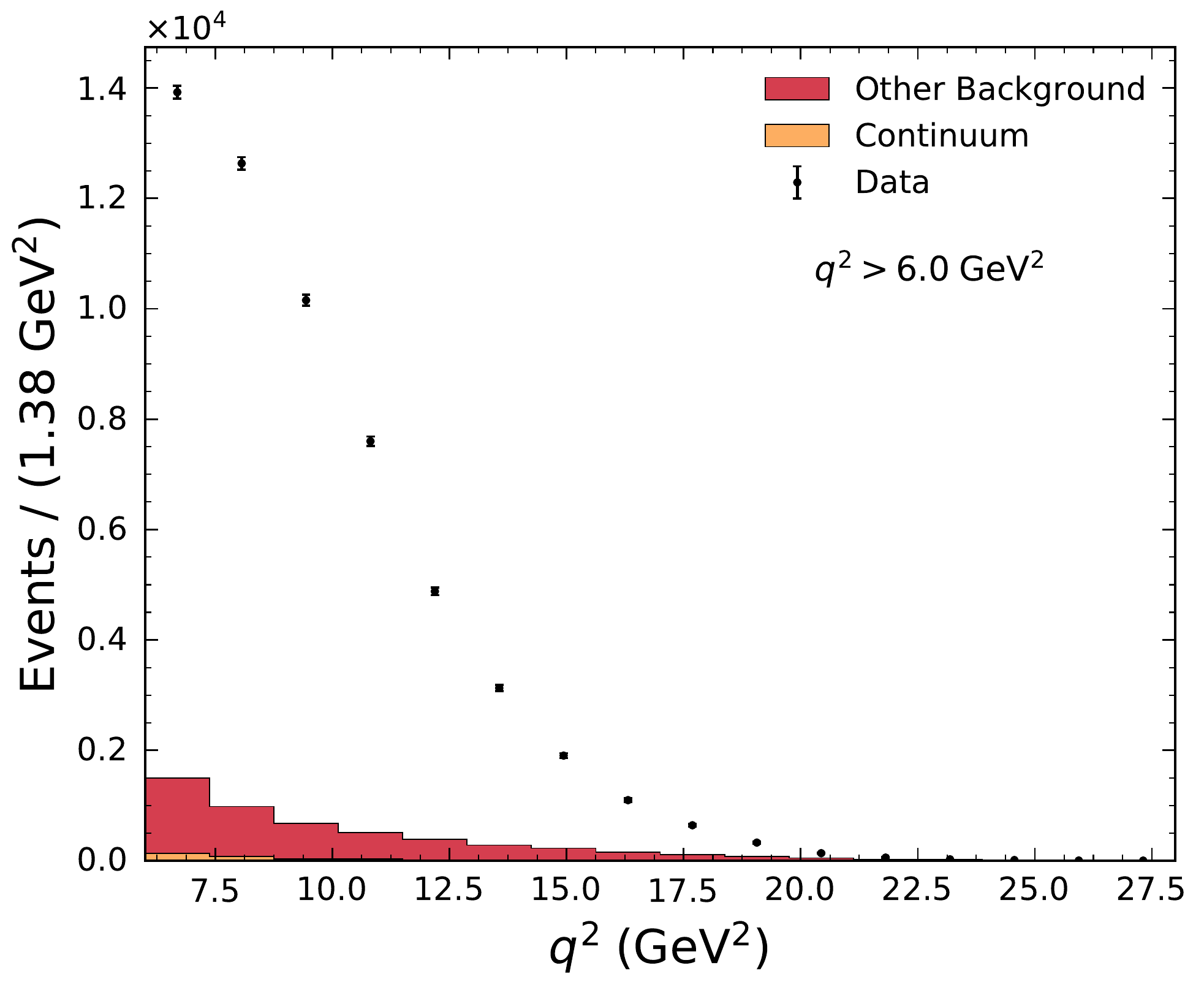}  
  \includegraphics[width=0.25\textwidth]{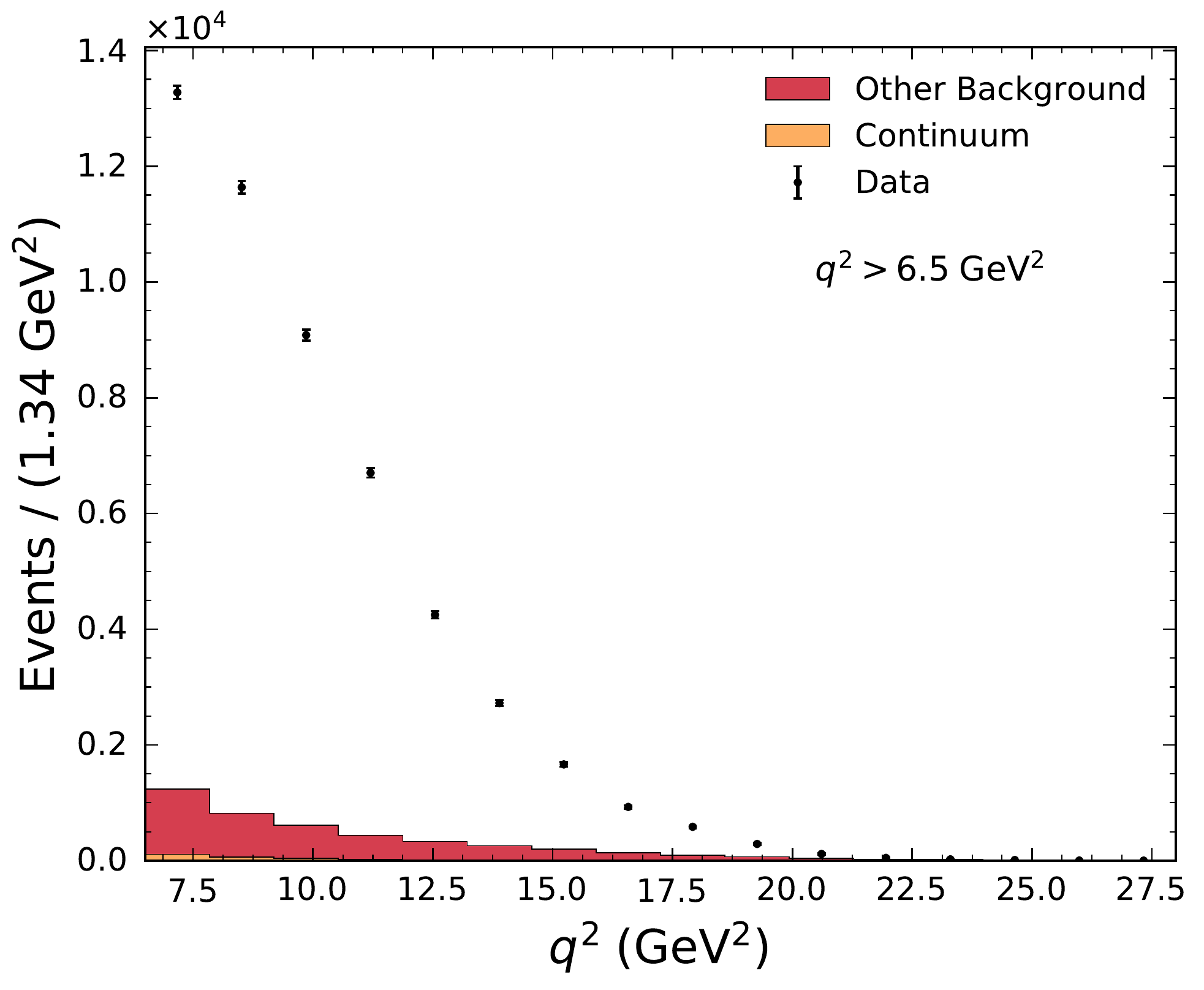}  
  \includegraphics[width=0.25\textwidth]{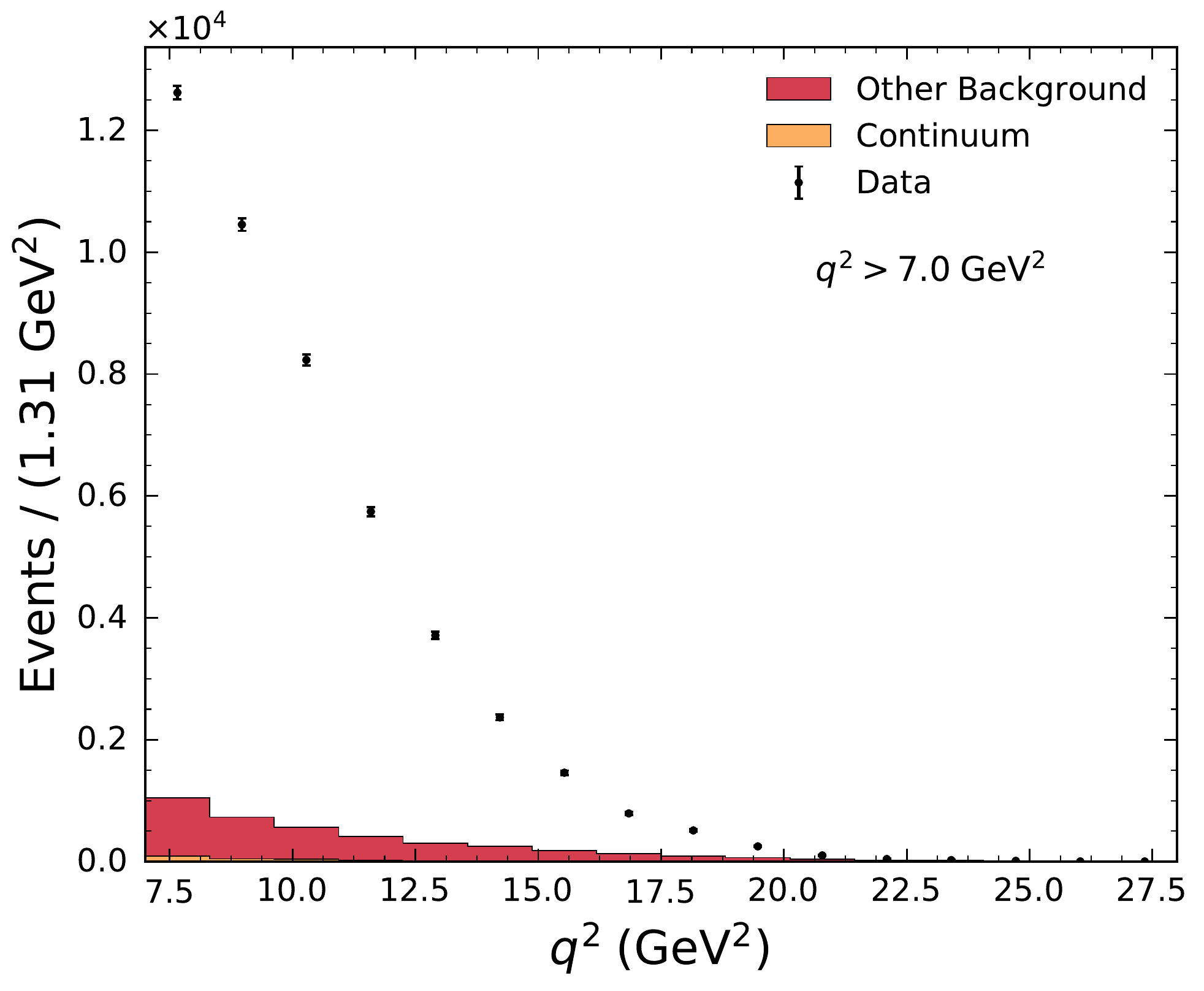}  
  \includegraphics[width=0.25\textwidth]{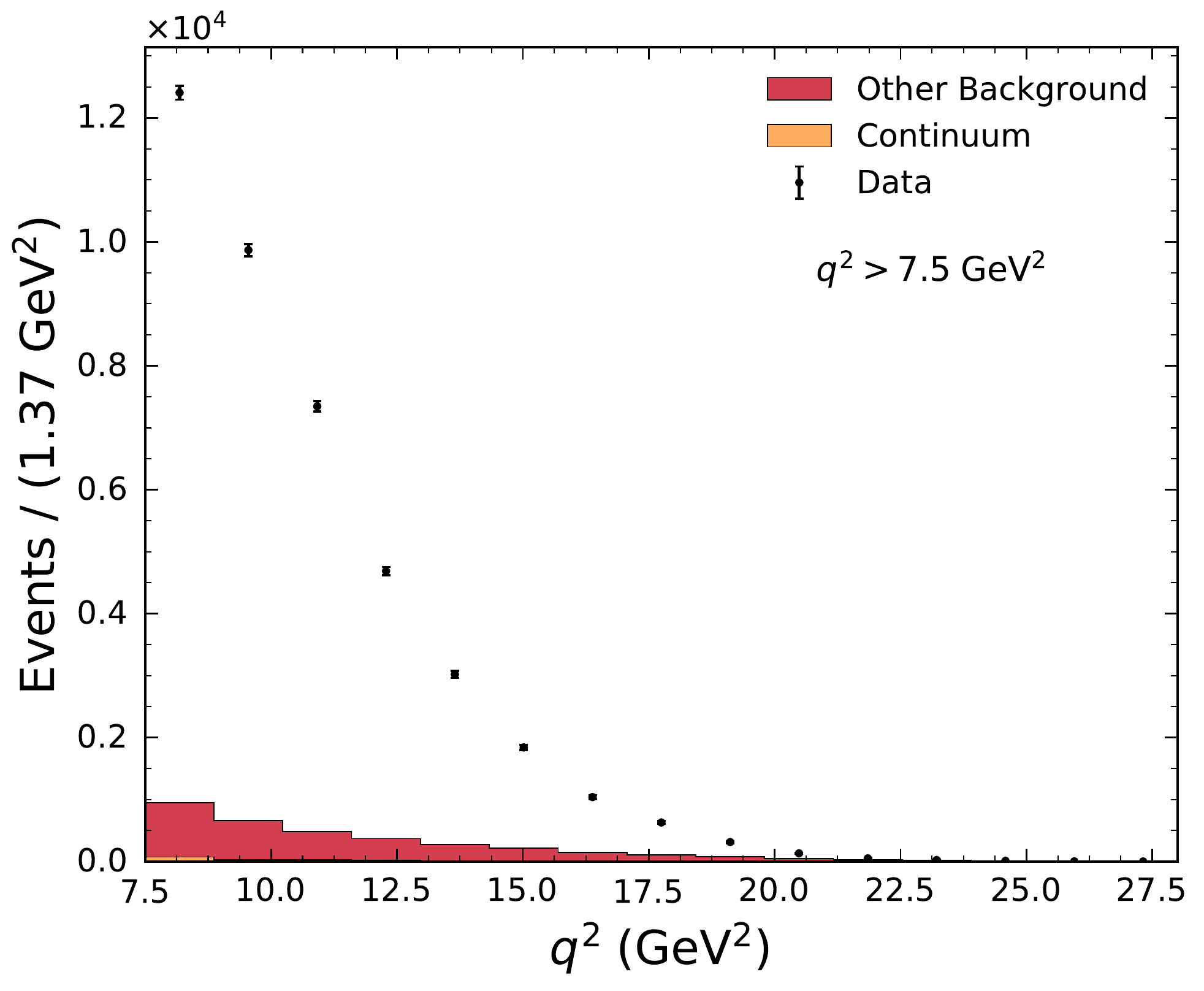}  
  \includegraphics[width=0.25\textwidth]{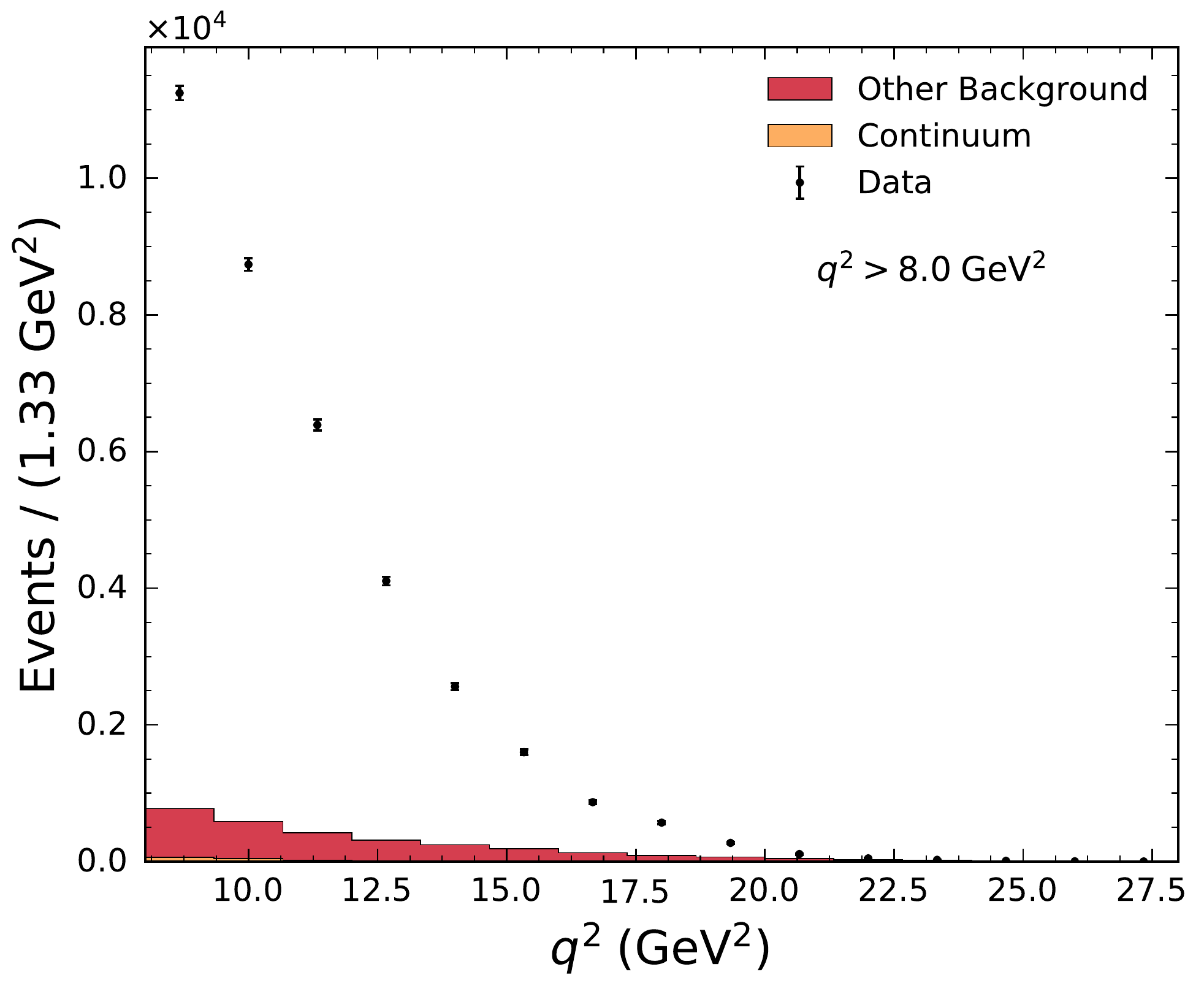}  
  \includegraphics[width=0.25\textwidth]{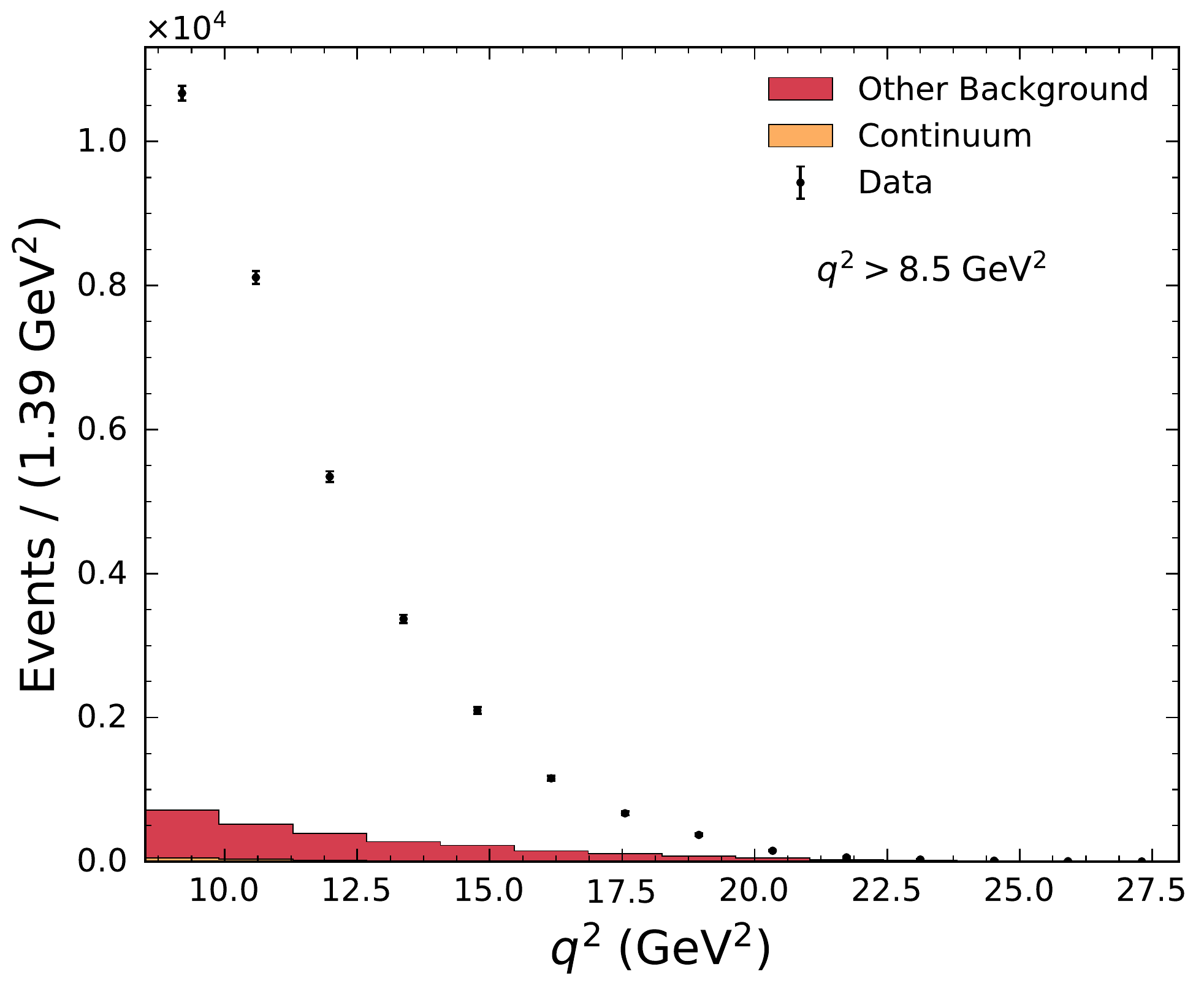}  
  \includegraphics[width=0.25\textwidth]{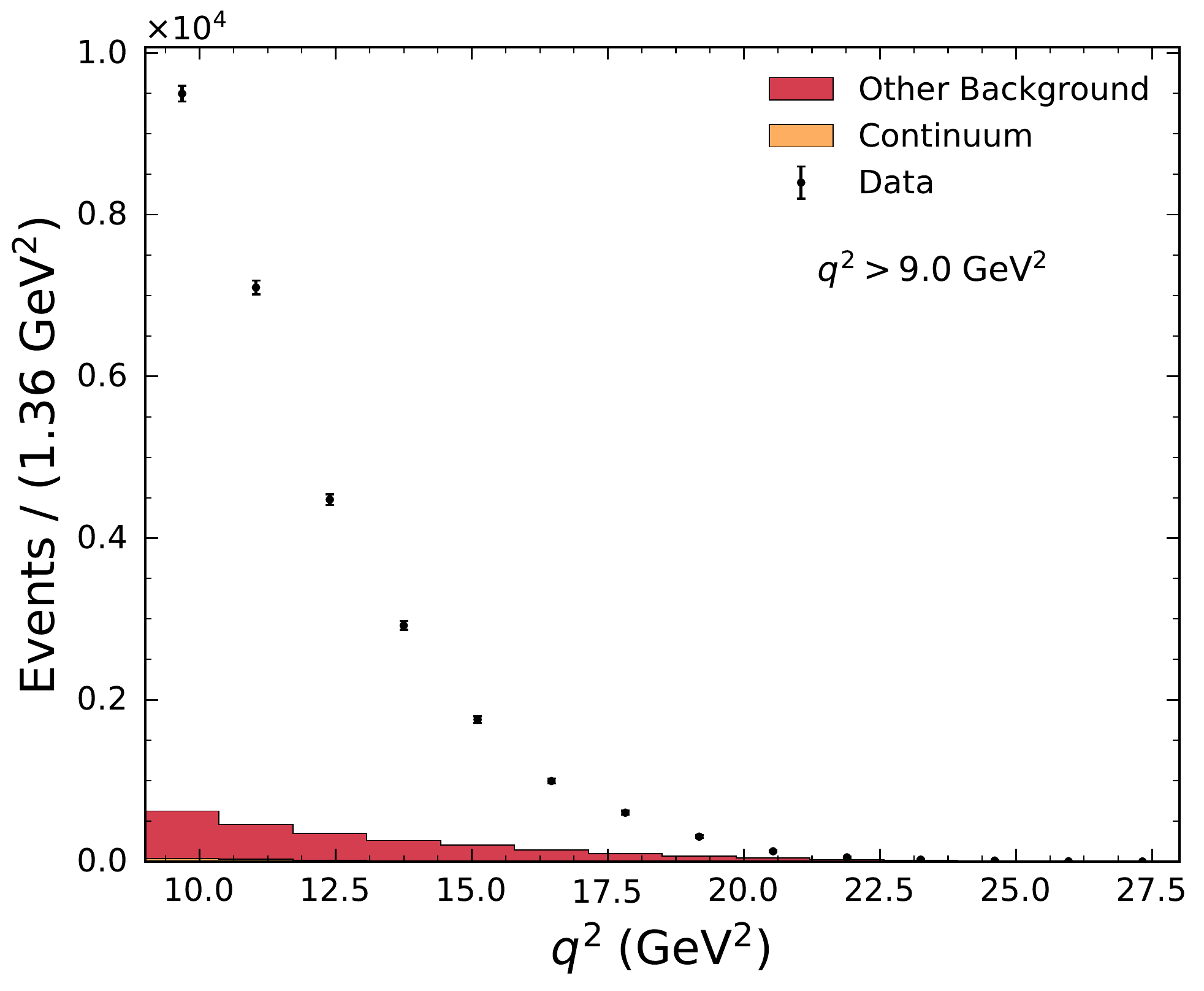}  
  \includegraphics[width=0.25\textwidth]{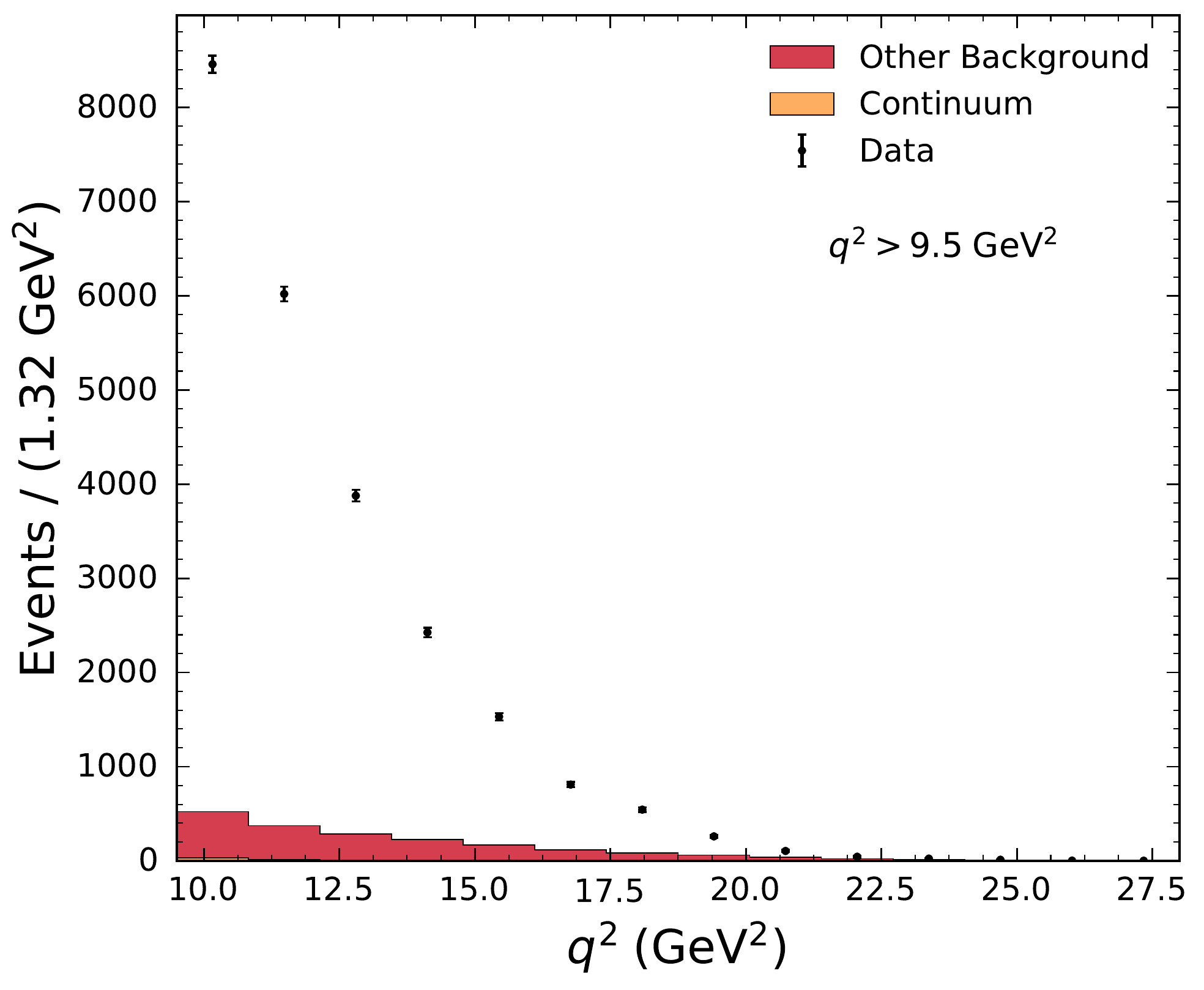}  
  \includegraphics[width=0.25\textwidth]{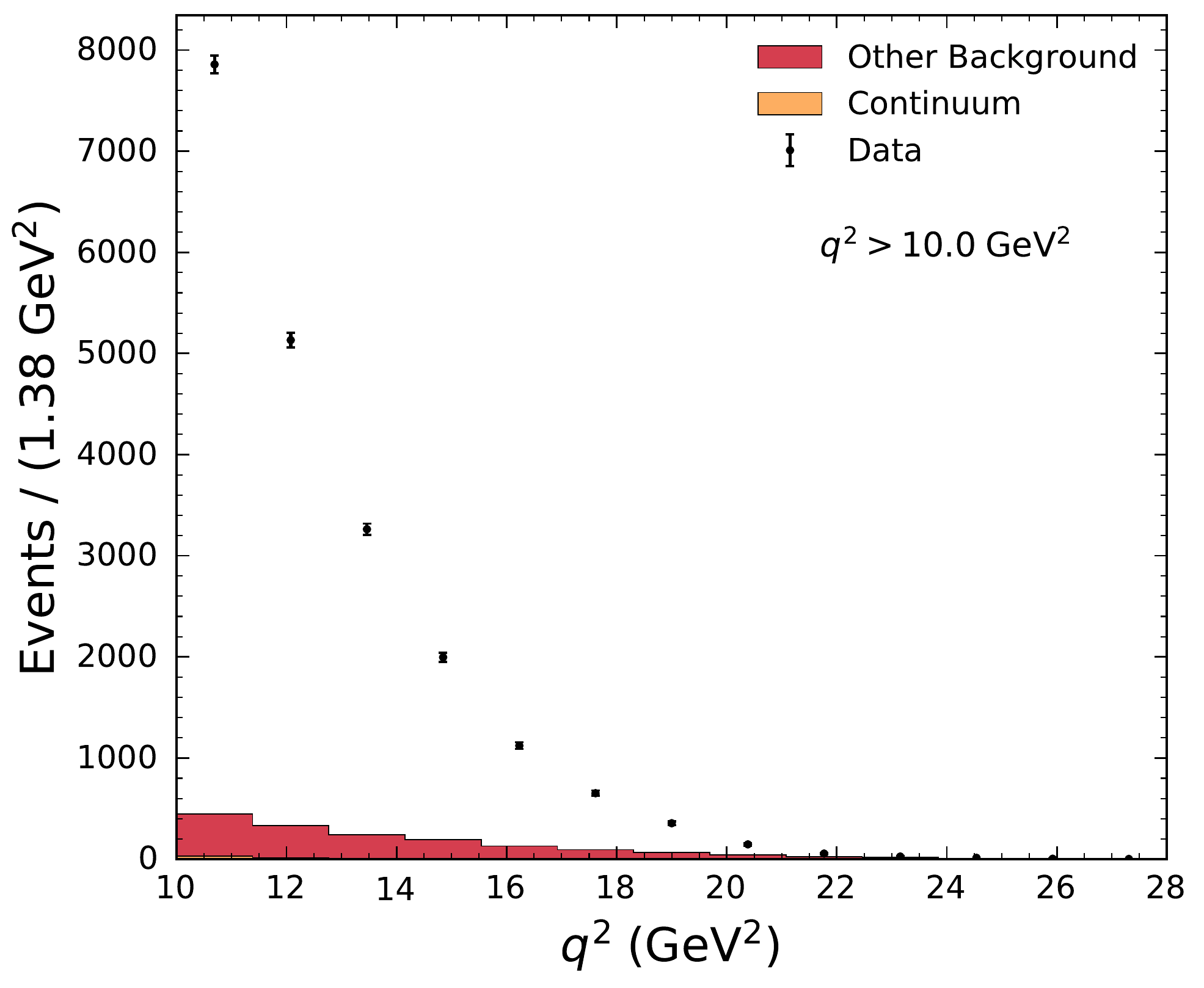}  
\caption{
 The determined background in the $q^{2}$ spectrum, after obtaining the expected background yields from the fit to the $M_X$ spectrum, for different $q^2$ threshold selections for muons are shown.
 }
\label{fig:bkg_sub_mu}
\end{figure}

\begin{figure}[b!]
  \includegraphics[width=0.25\textwidth]{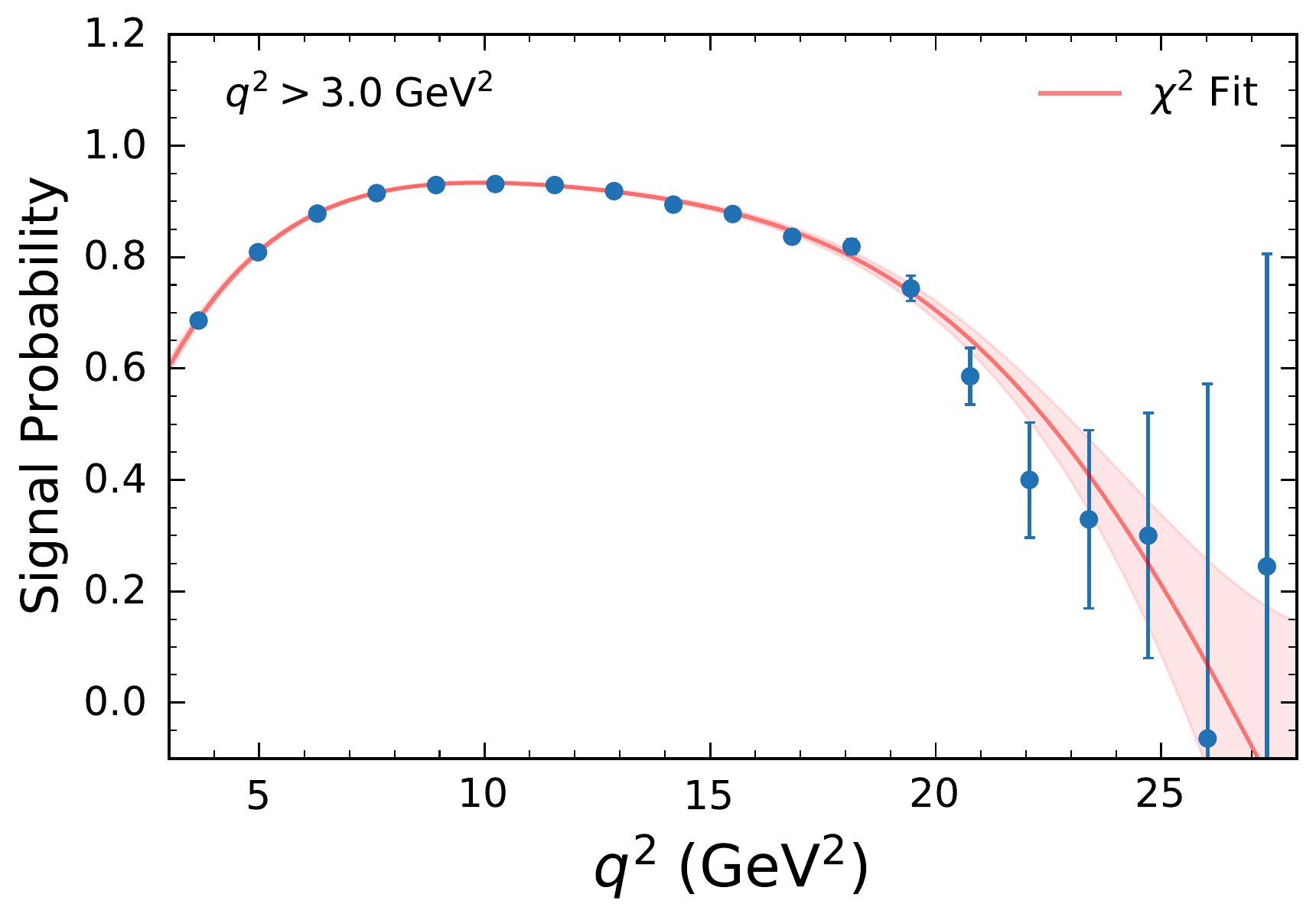}  
  \includegraphics[width=0.25\textwidth]{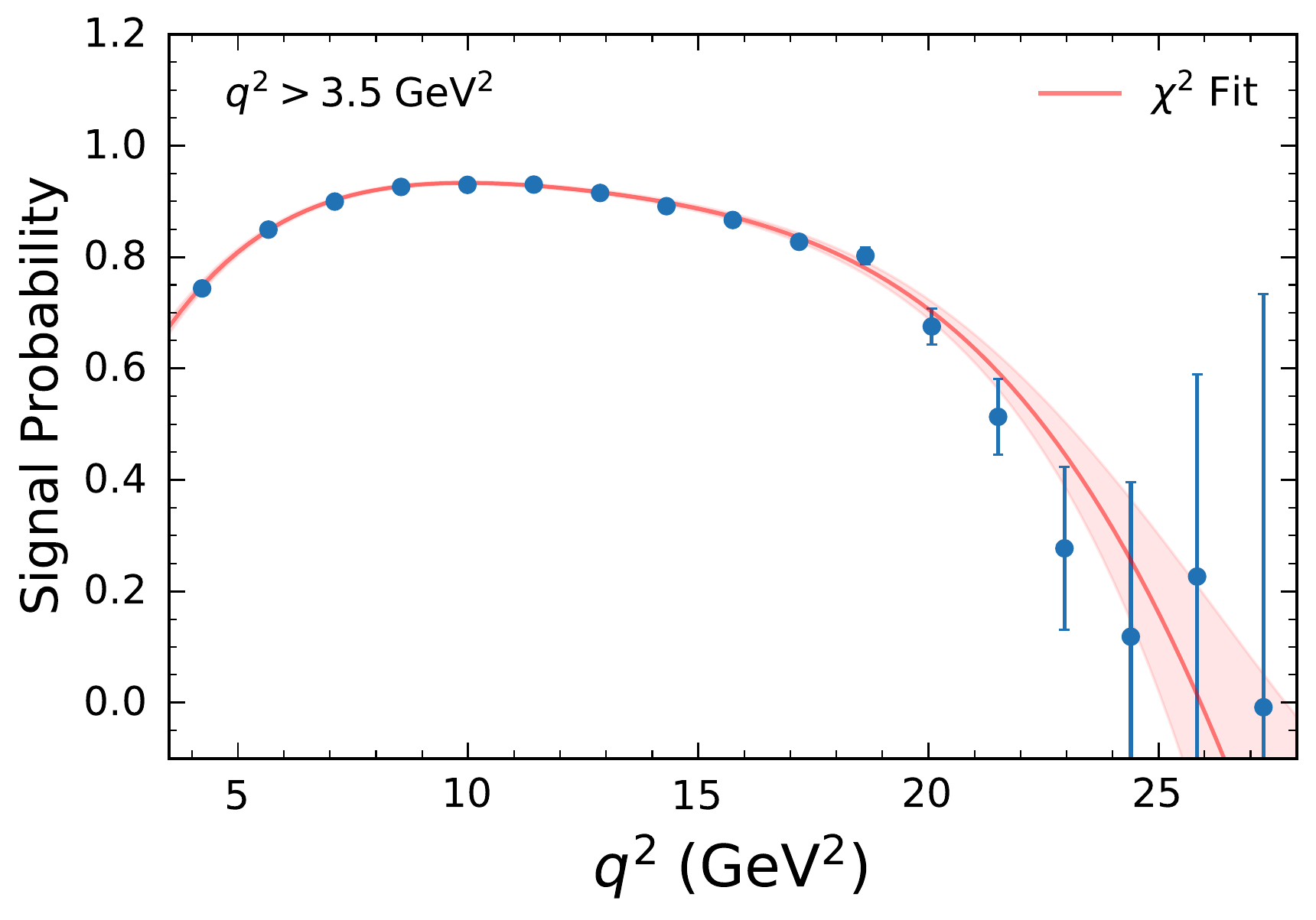}  
  \includegraphics[width=0.25\textwidth]{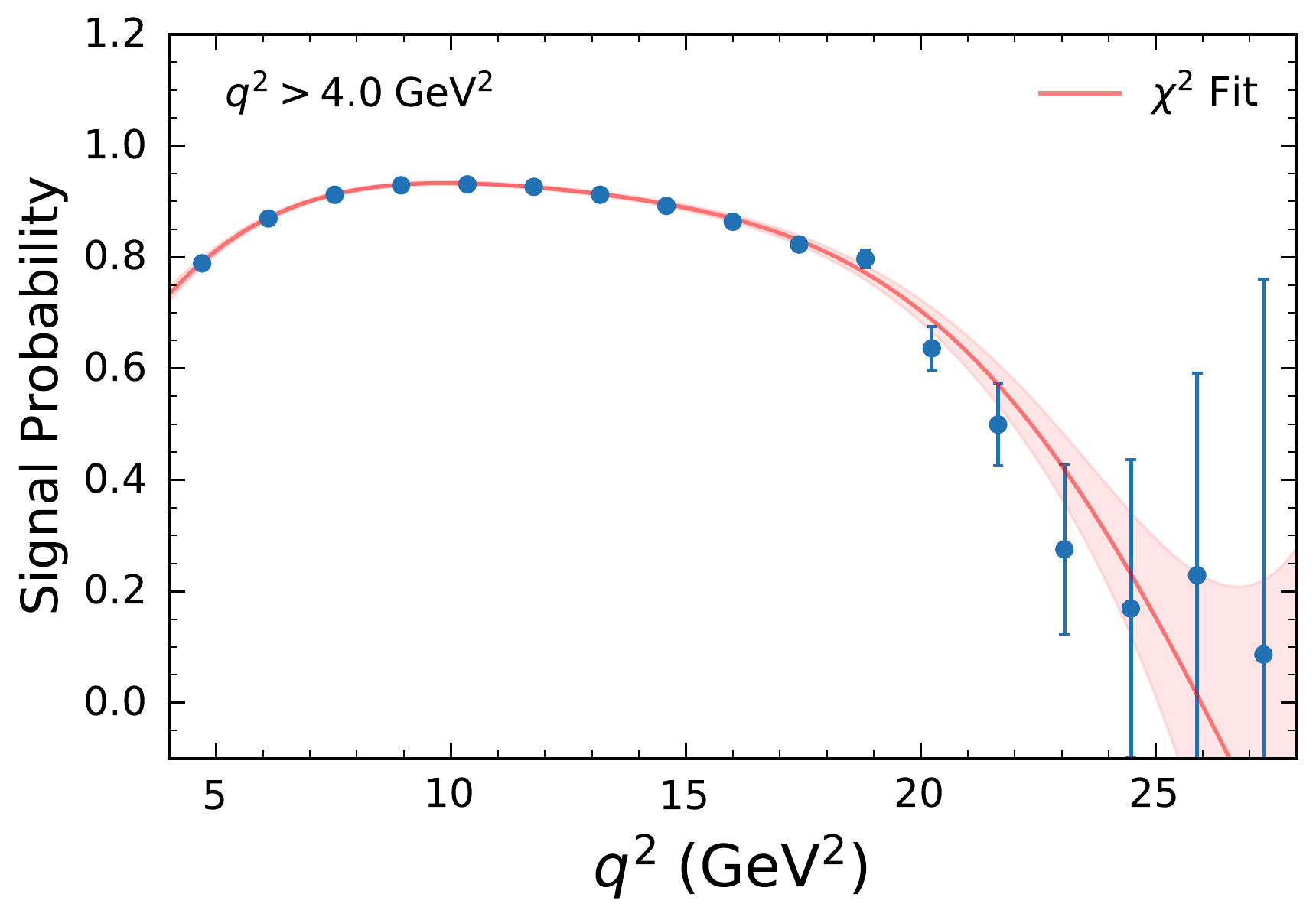}  
  \includegraphics[width=0.25\textwidth]{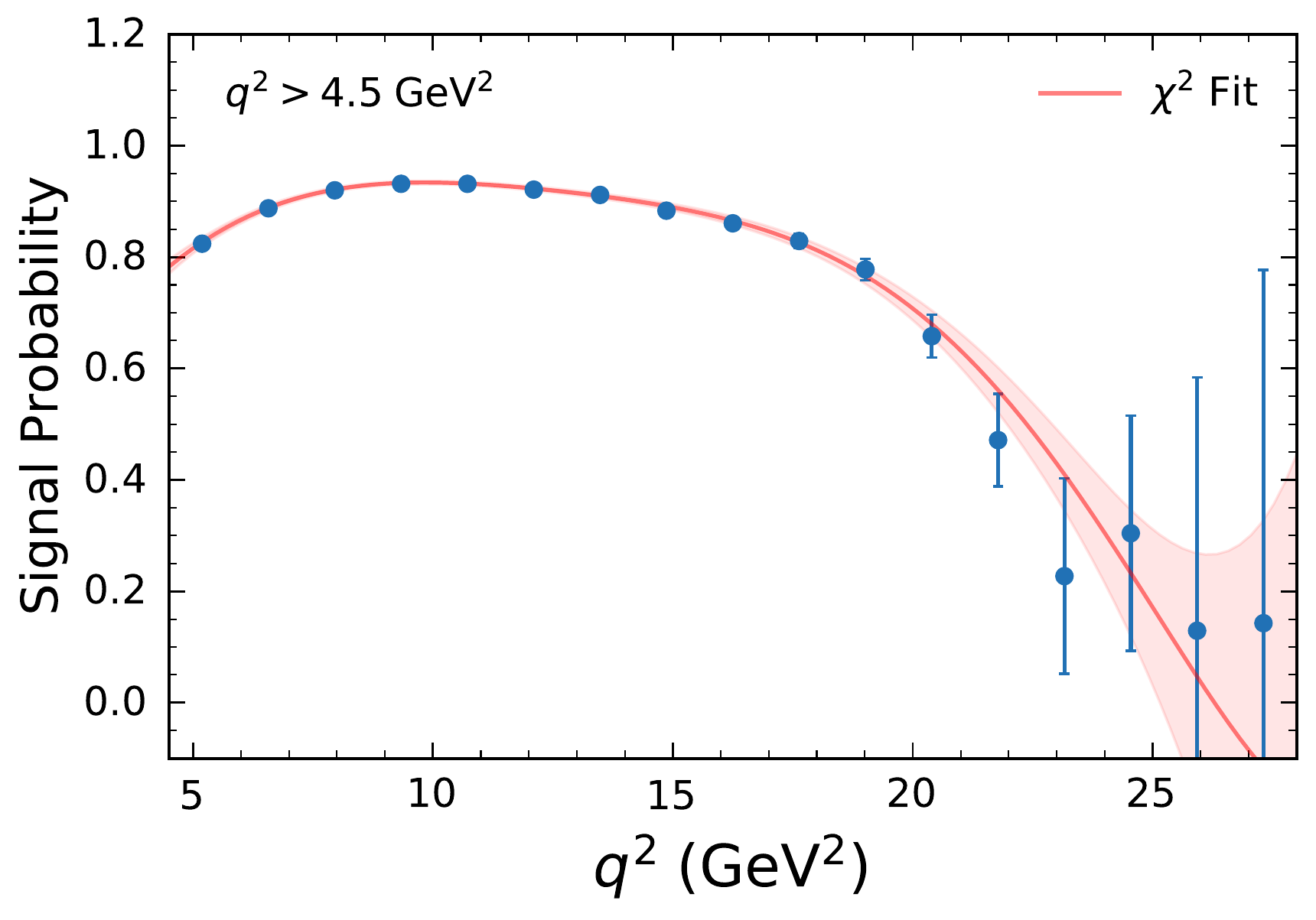}  
  \includegraphics[width=0.25\textwidth]{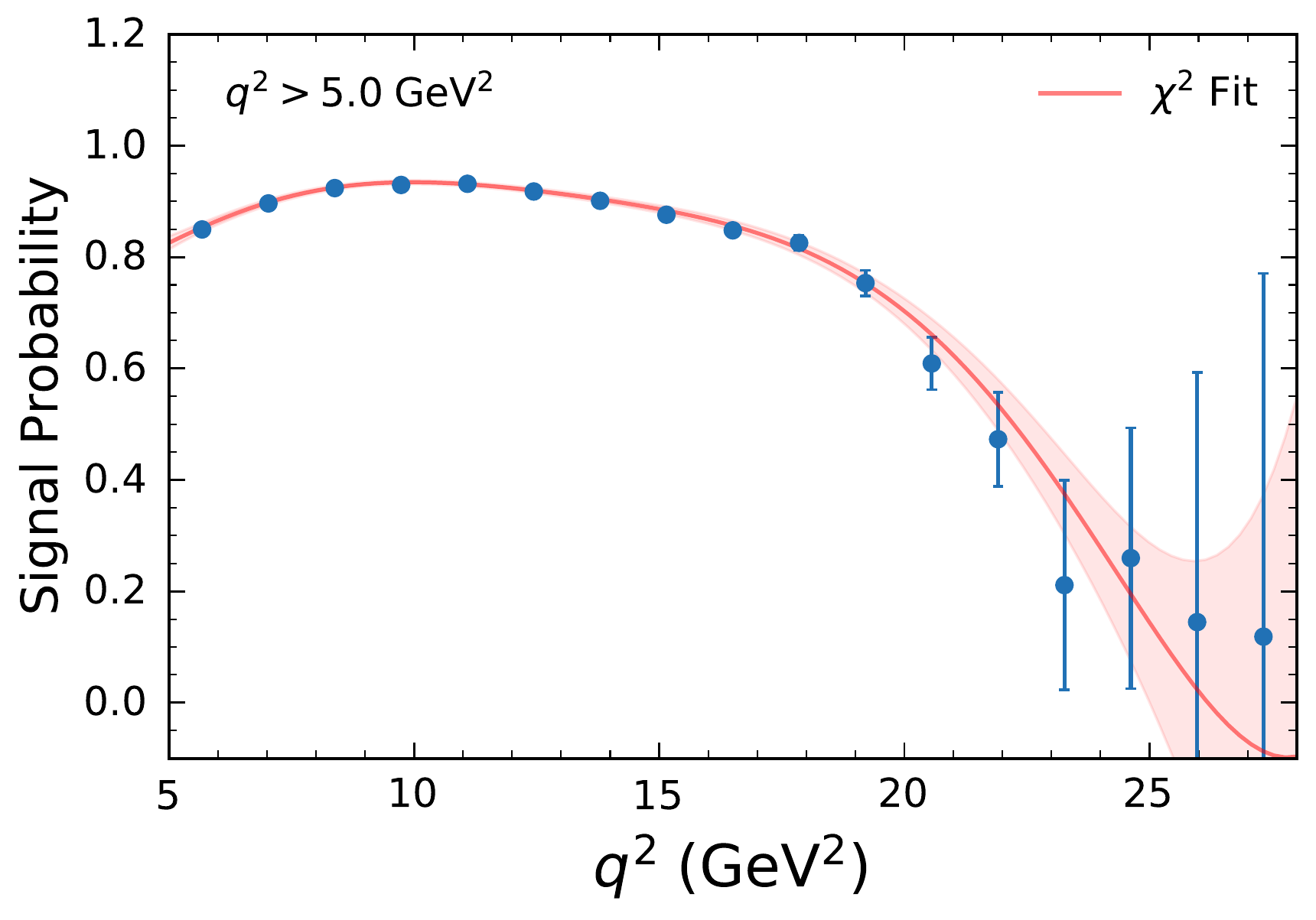}  
  \includegraphics[width=0.25\textwidth]{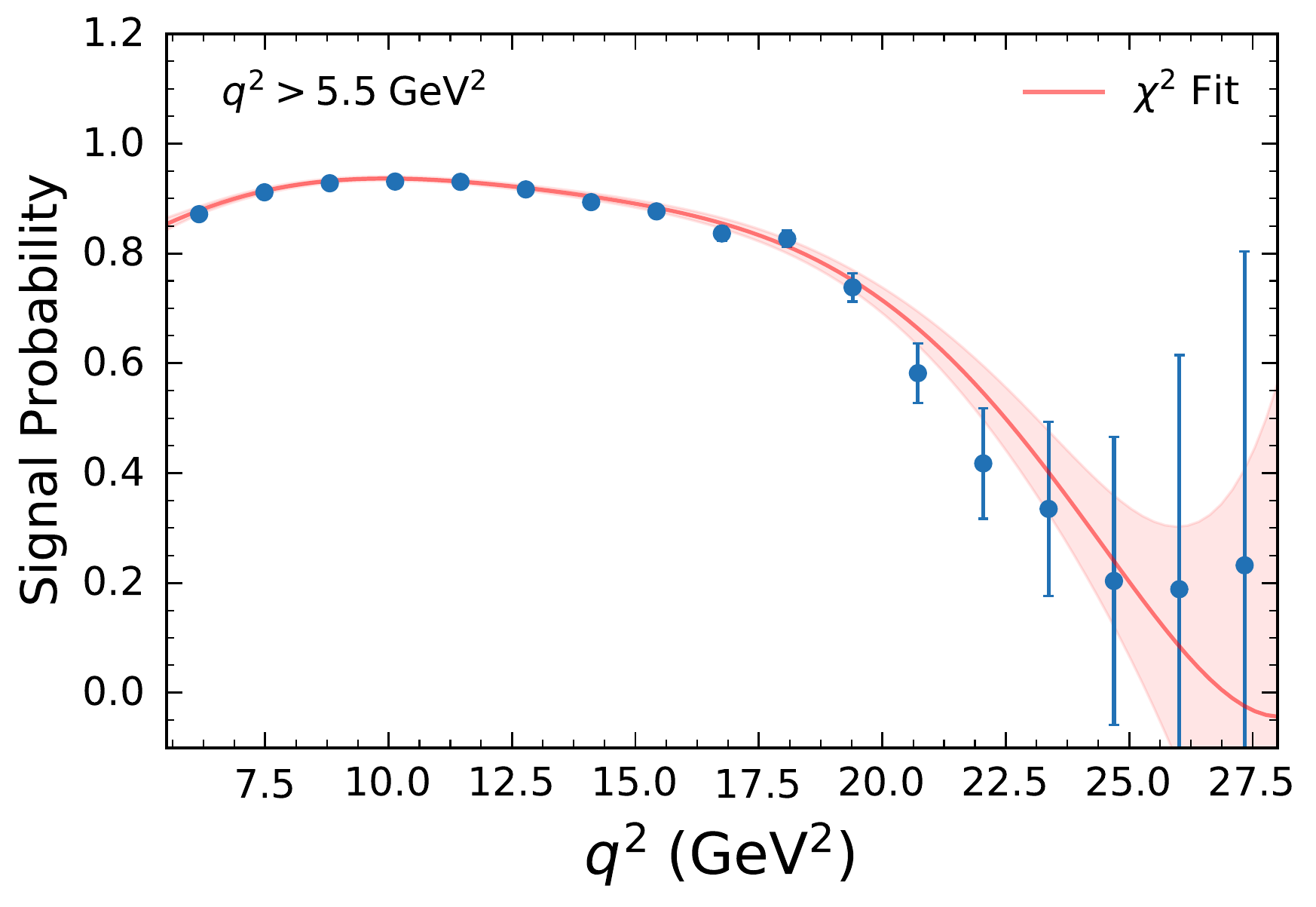}  
  \includegraphics[width=0.25\textwidth]{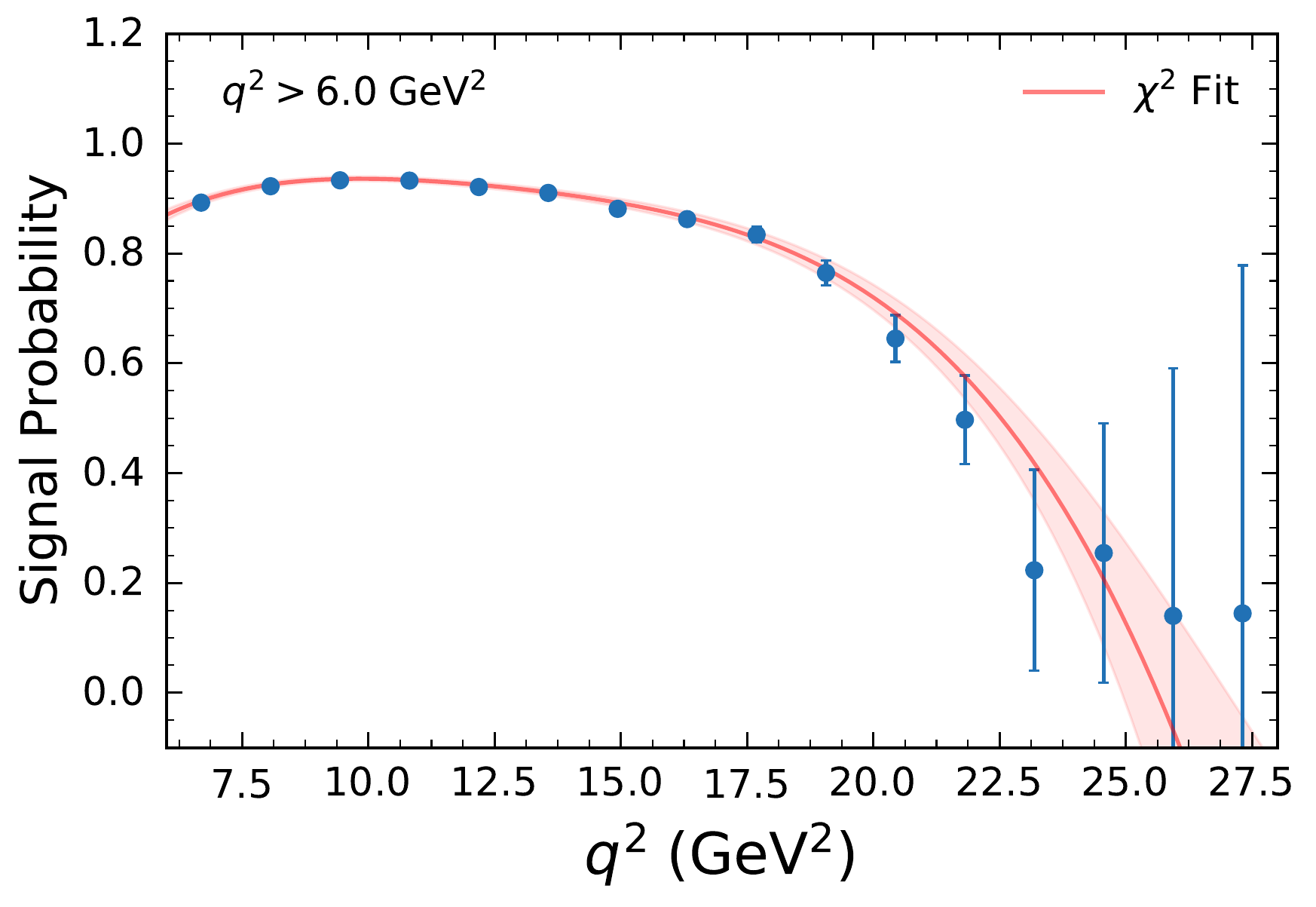}  
  \includegraphics[width=0.25\textwidth]{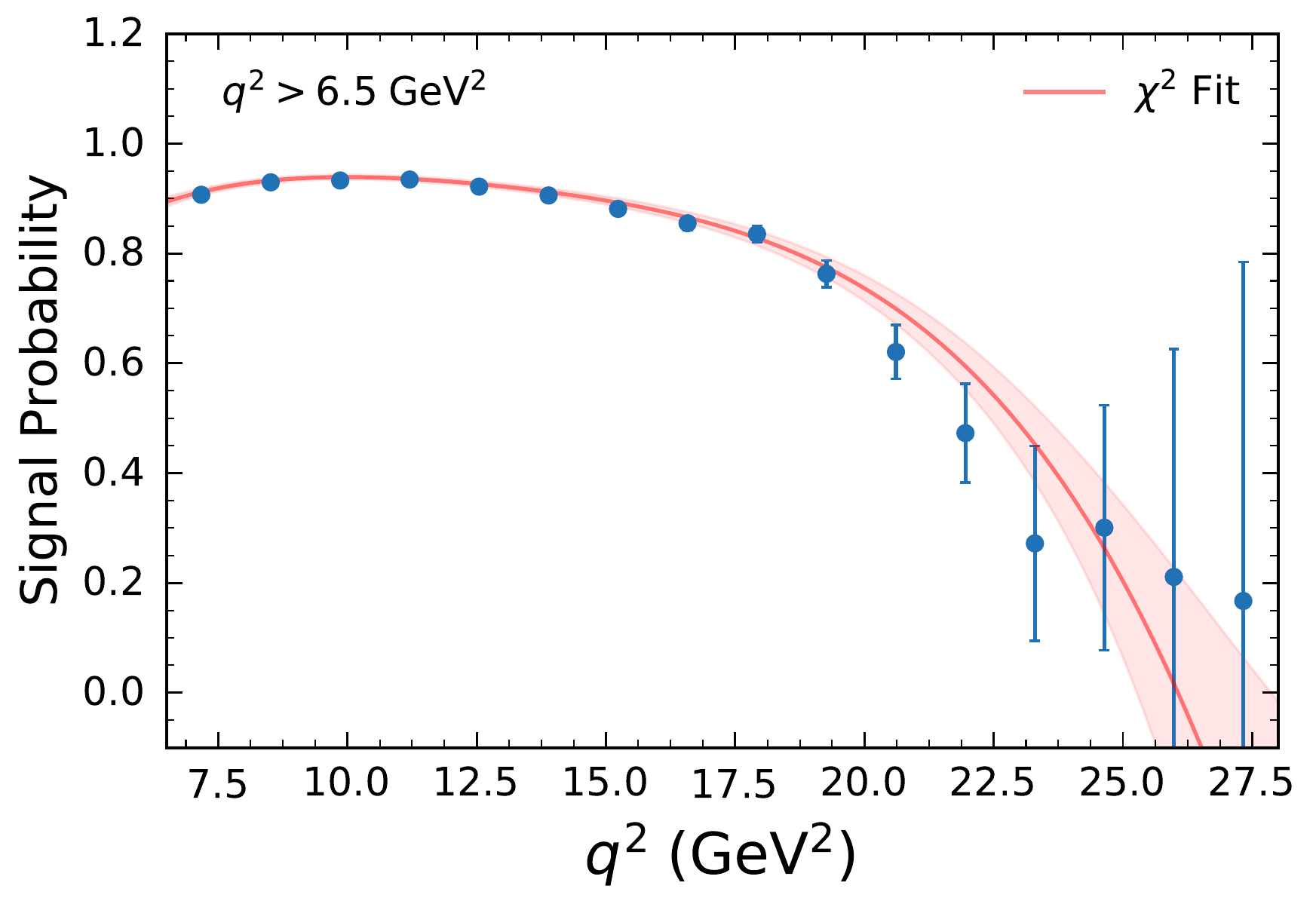}  
  \includegraphics[width=0.25\textwidth]{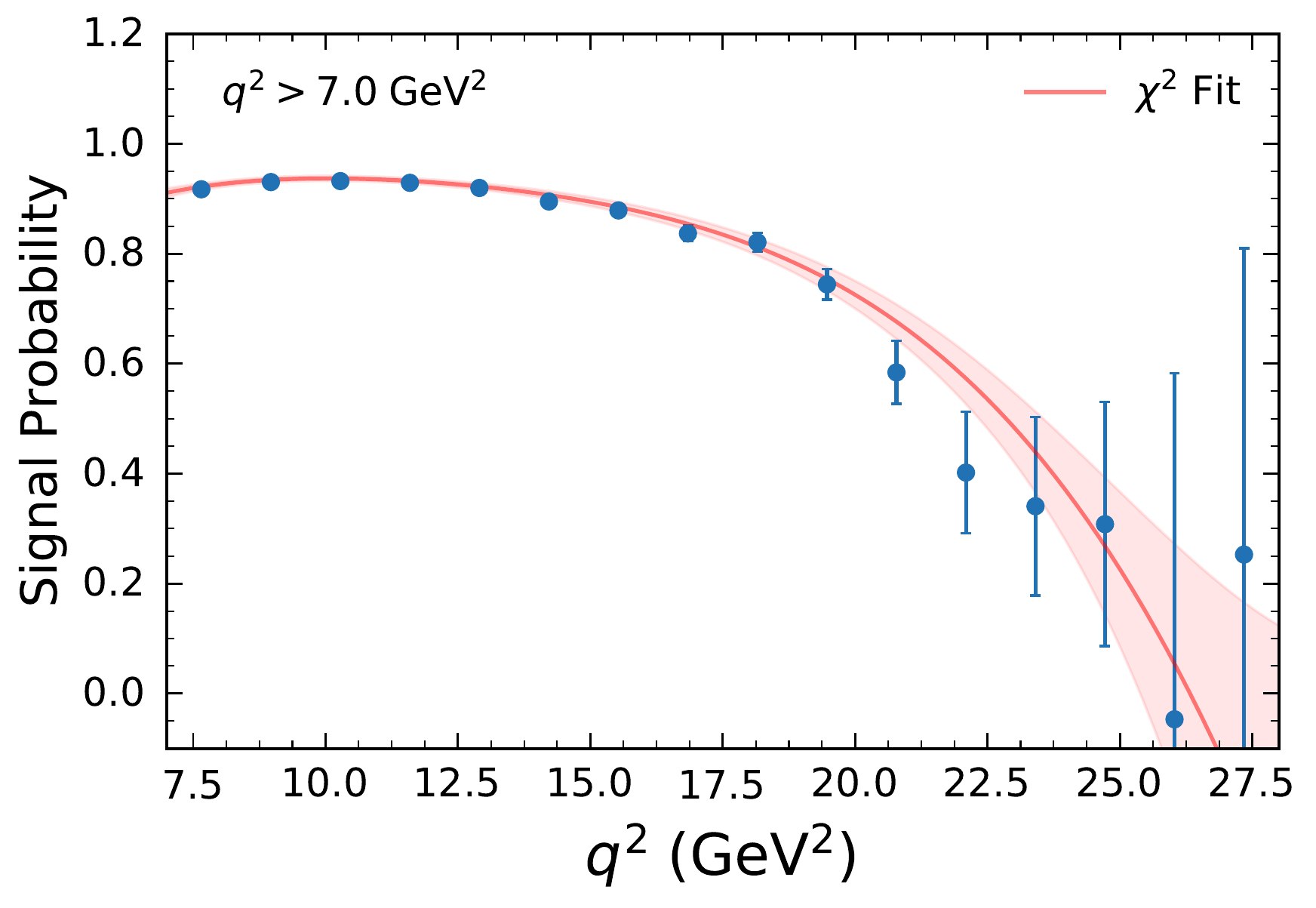}  
  \includegraphics[width=0.25\textwidth]{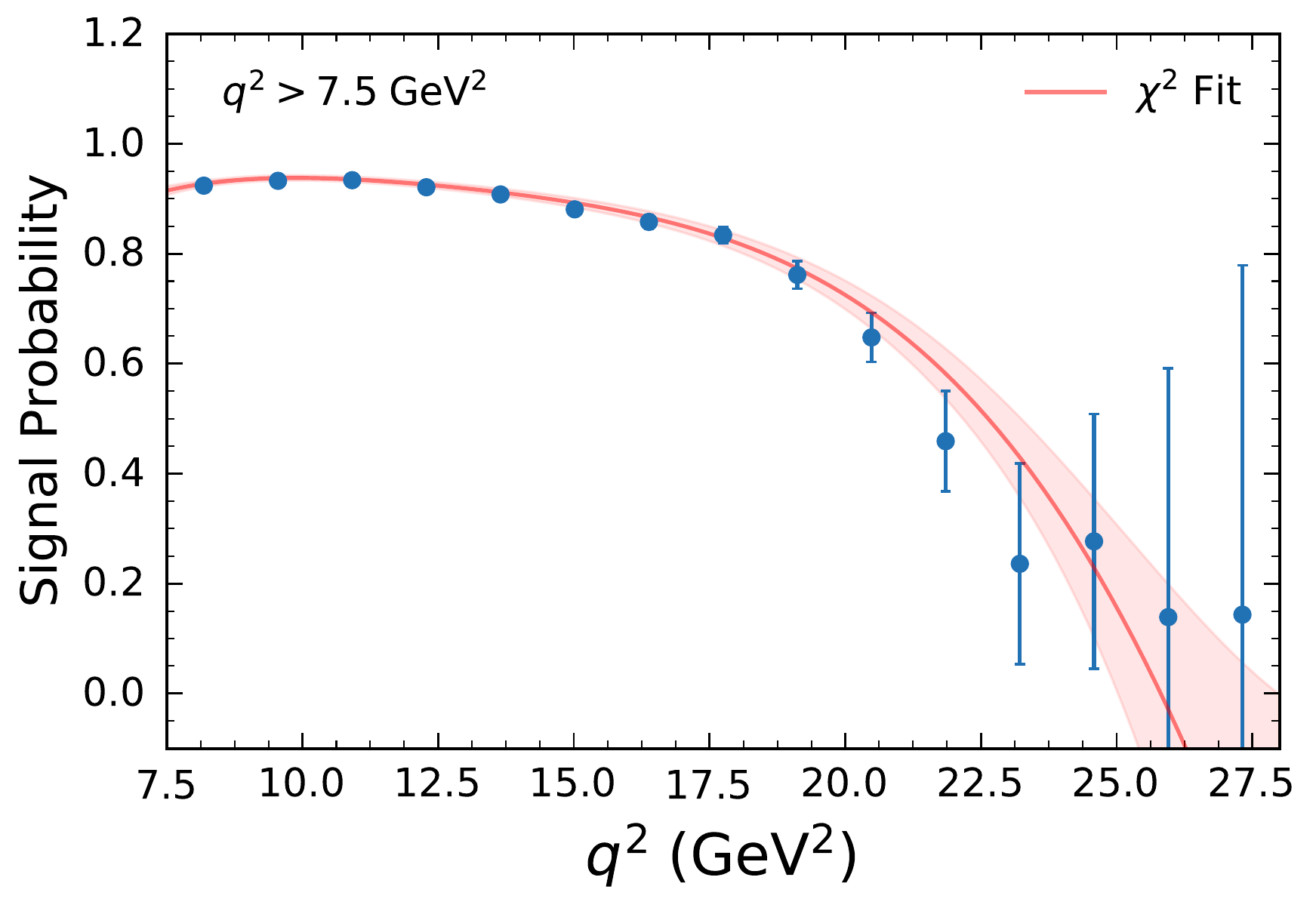}  
  \includegraphics[width=0.25\textwidth]{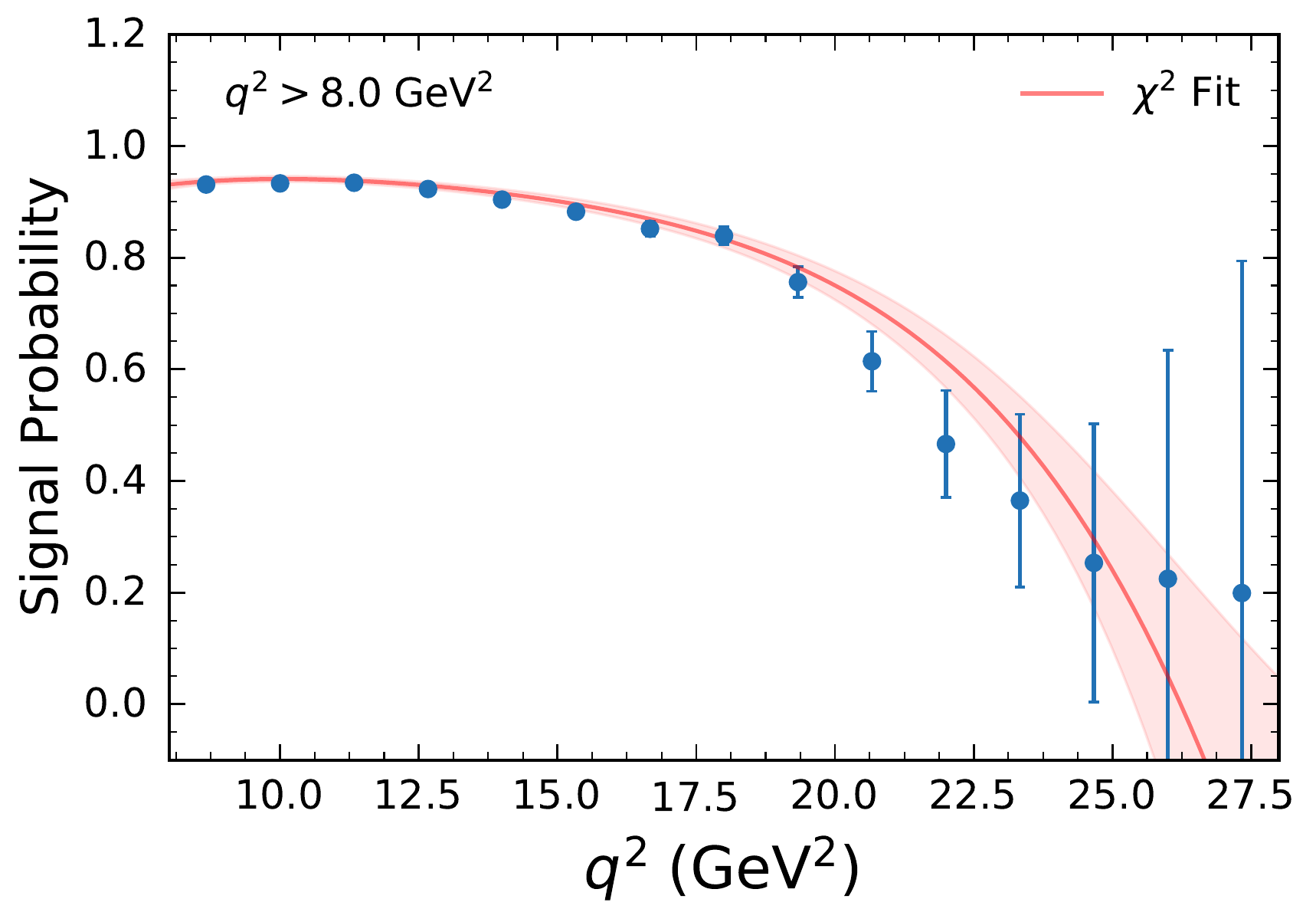}  
  \includegraphics[width=0.25\textwidth]{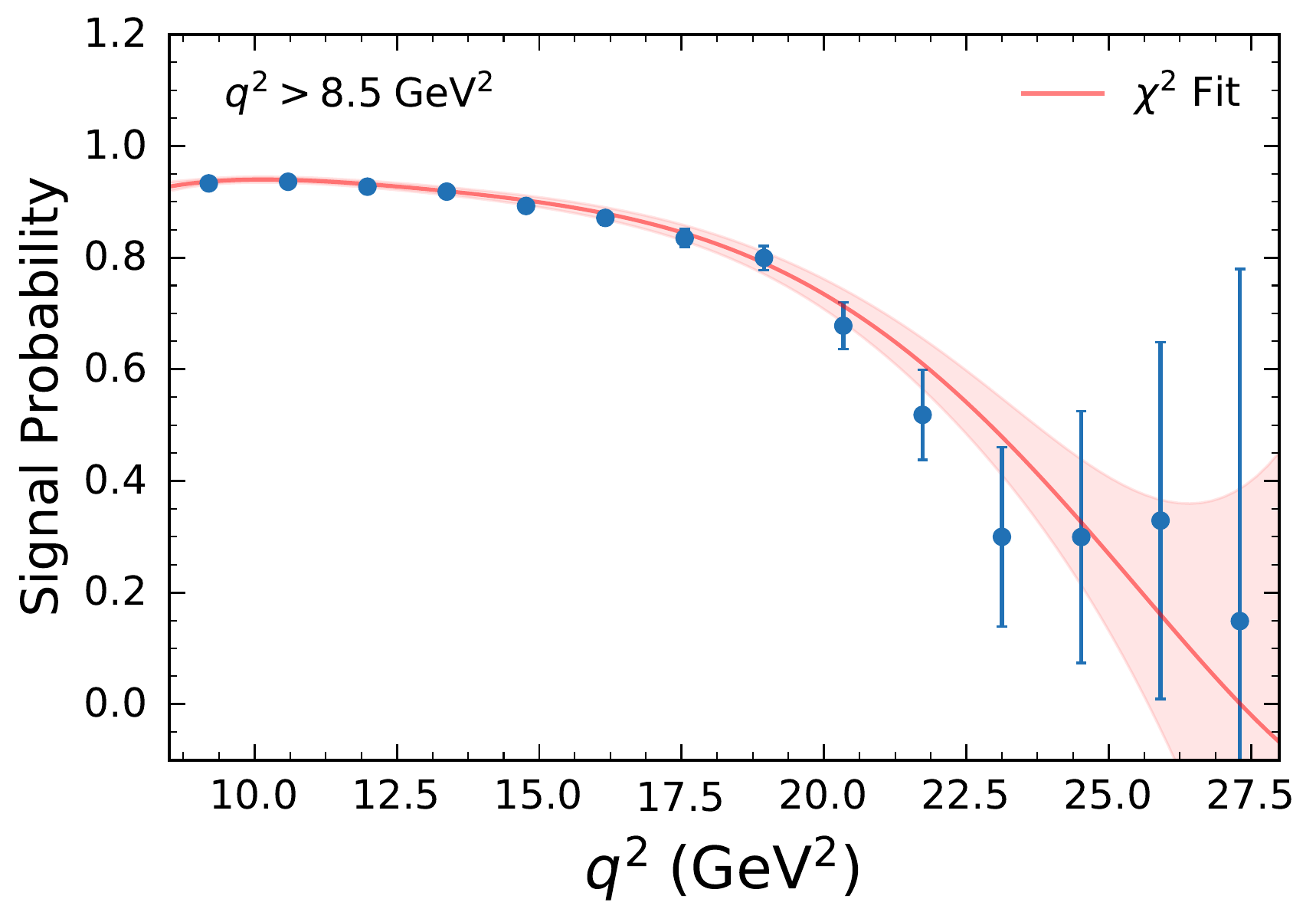}  
  \includegraphics[width=0.25\textwidth]{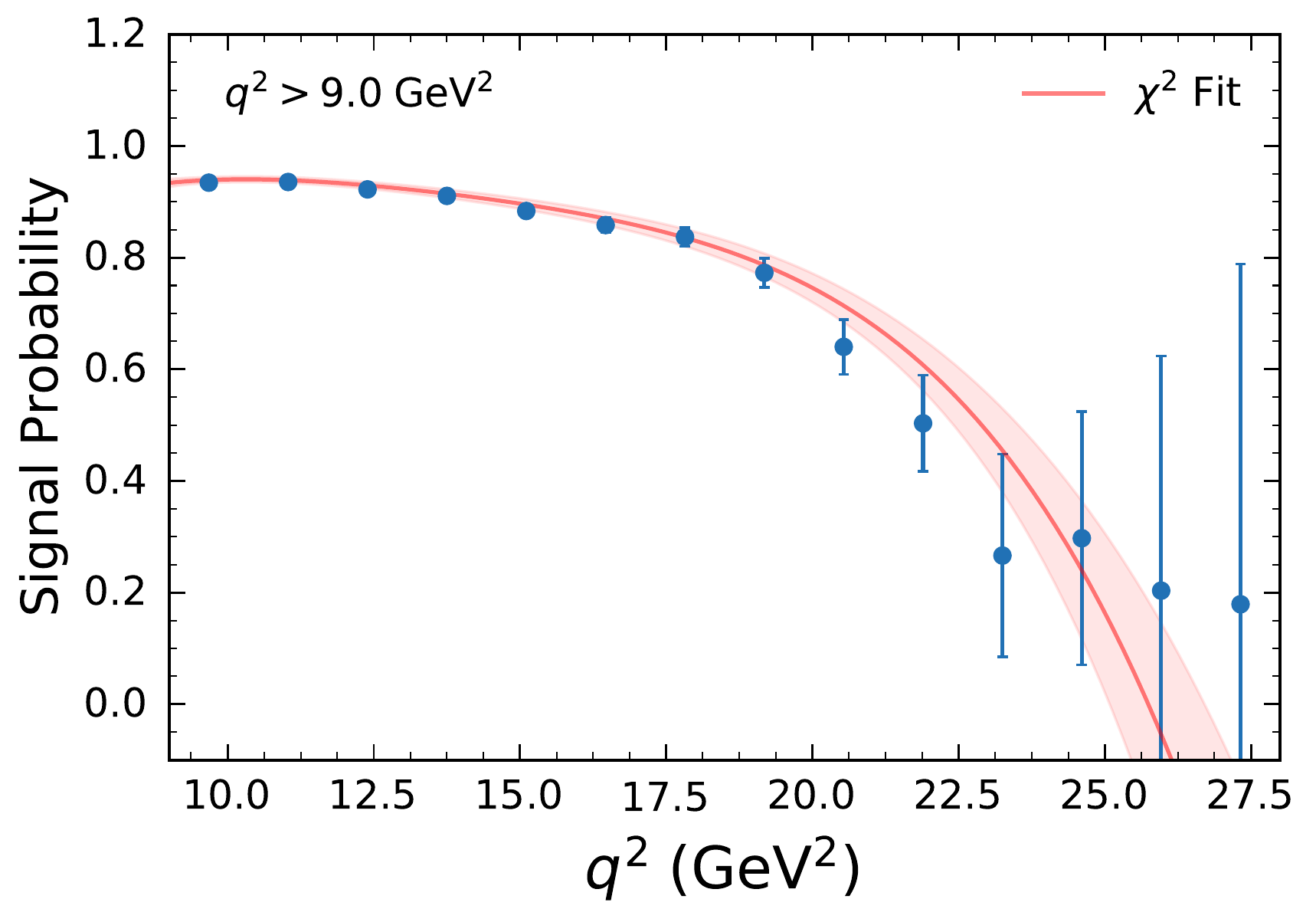}  
  \includegraphics[width=0.25\textwidth]{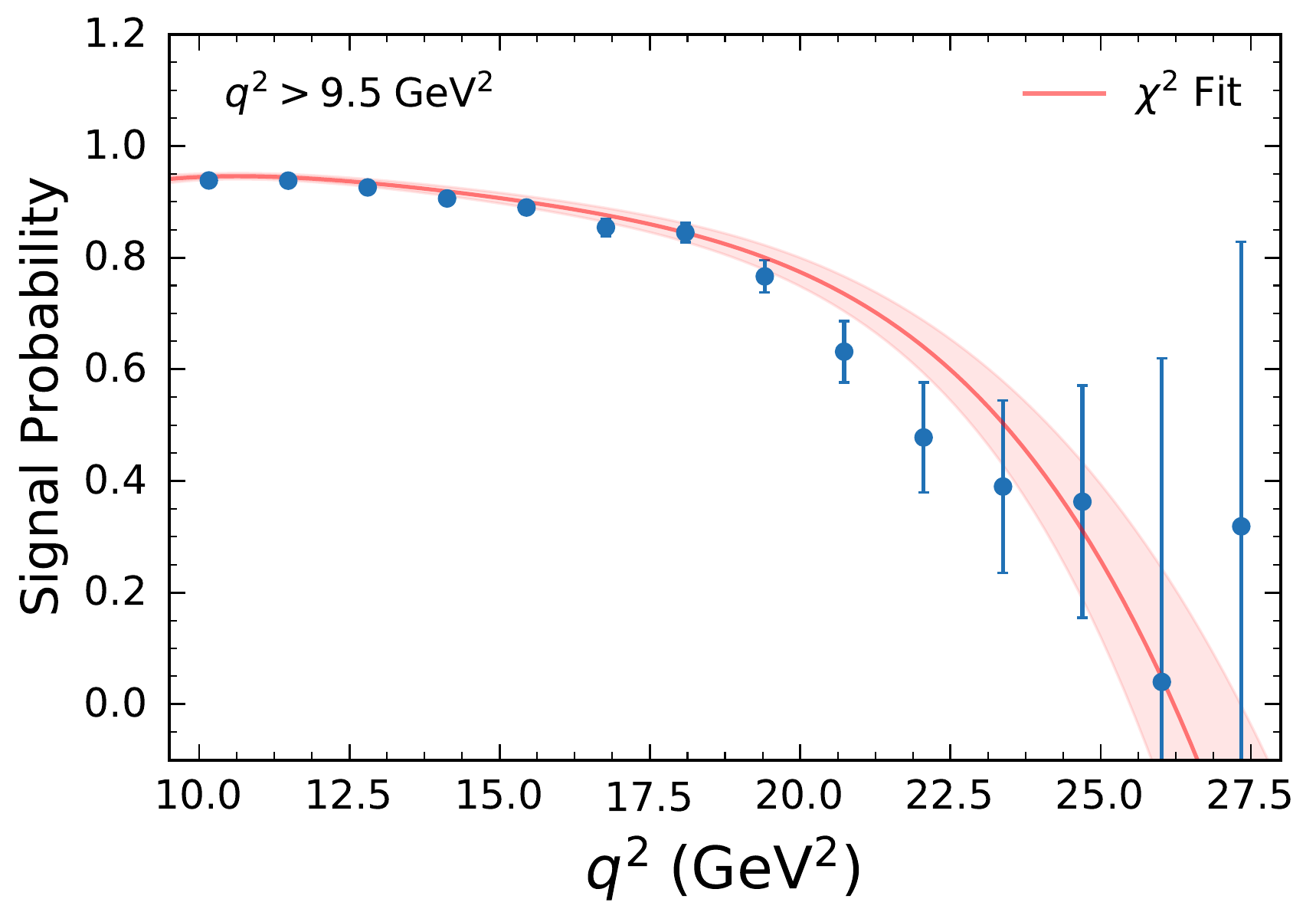} 
  \includegraphics[width=0.25\textwidth]{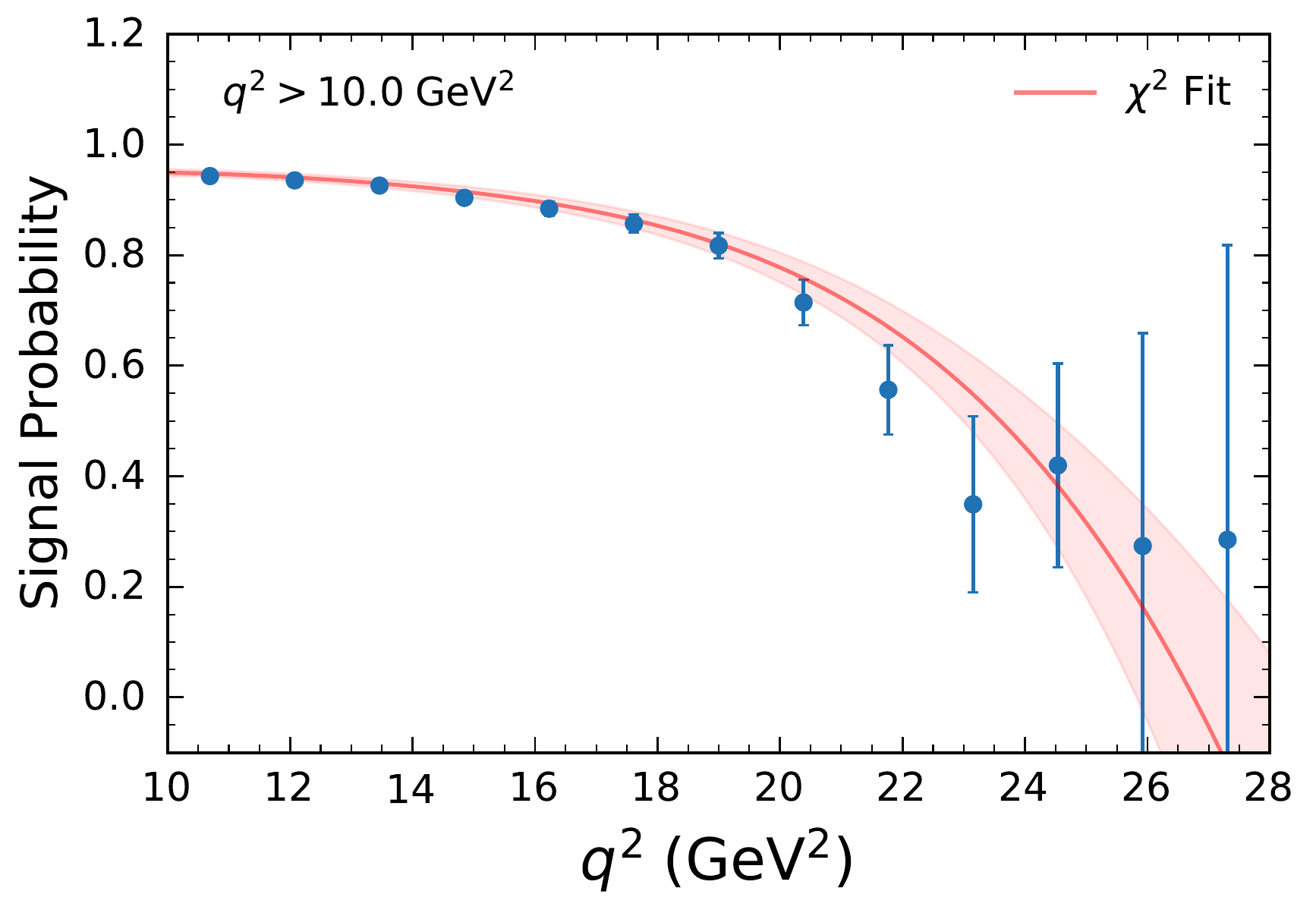}   
\caption{
 The determined signal probabilities for different $q^2$ threshold selections for muons are shown.
 }
\label{fig:w_mu}
\end{figure}

The determined background in the $q^{2}$ spectrum, after obtaining the expected background yields from the fit to the $M_X$ spectrum, and the signal probability weights are shown in Fig.~\ref{fig:bkg_sub_el}, \ref{fig:w_el}, \ref{fig:bkg_sub_mu}, and \ref{fig:w_mu}. 

\clearpage

\section*{Appendix C. Calibration Parameters, Calibration Curves and Calibration Factors}

\begin{figure}[b!]
  \includegraphics[width=0.39\textwidth]{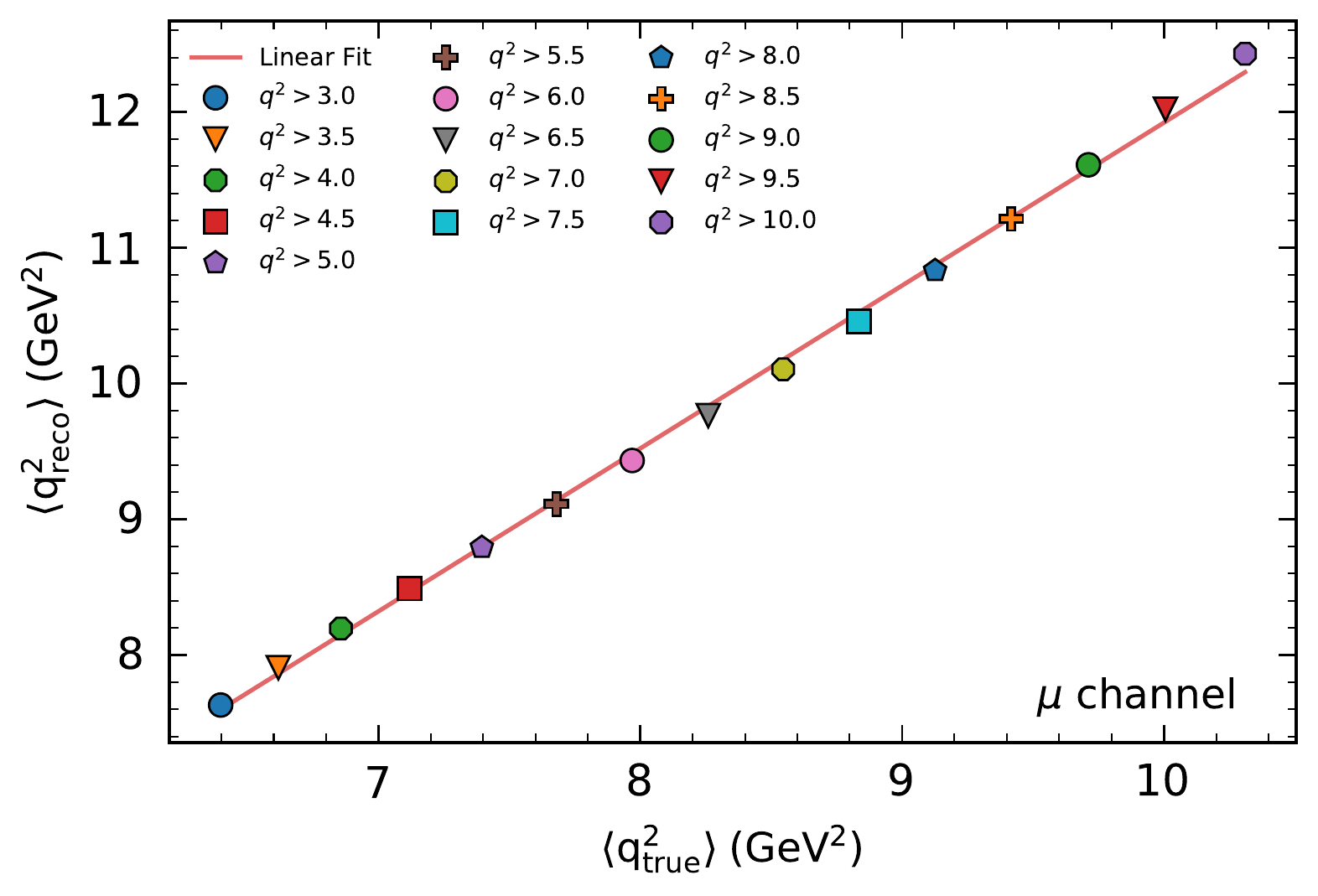}  
  \includegraphics[width=0.39\textwidth]{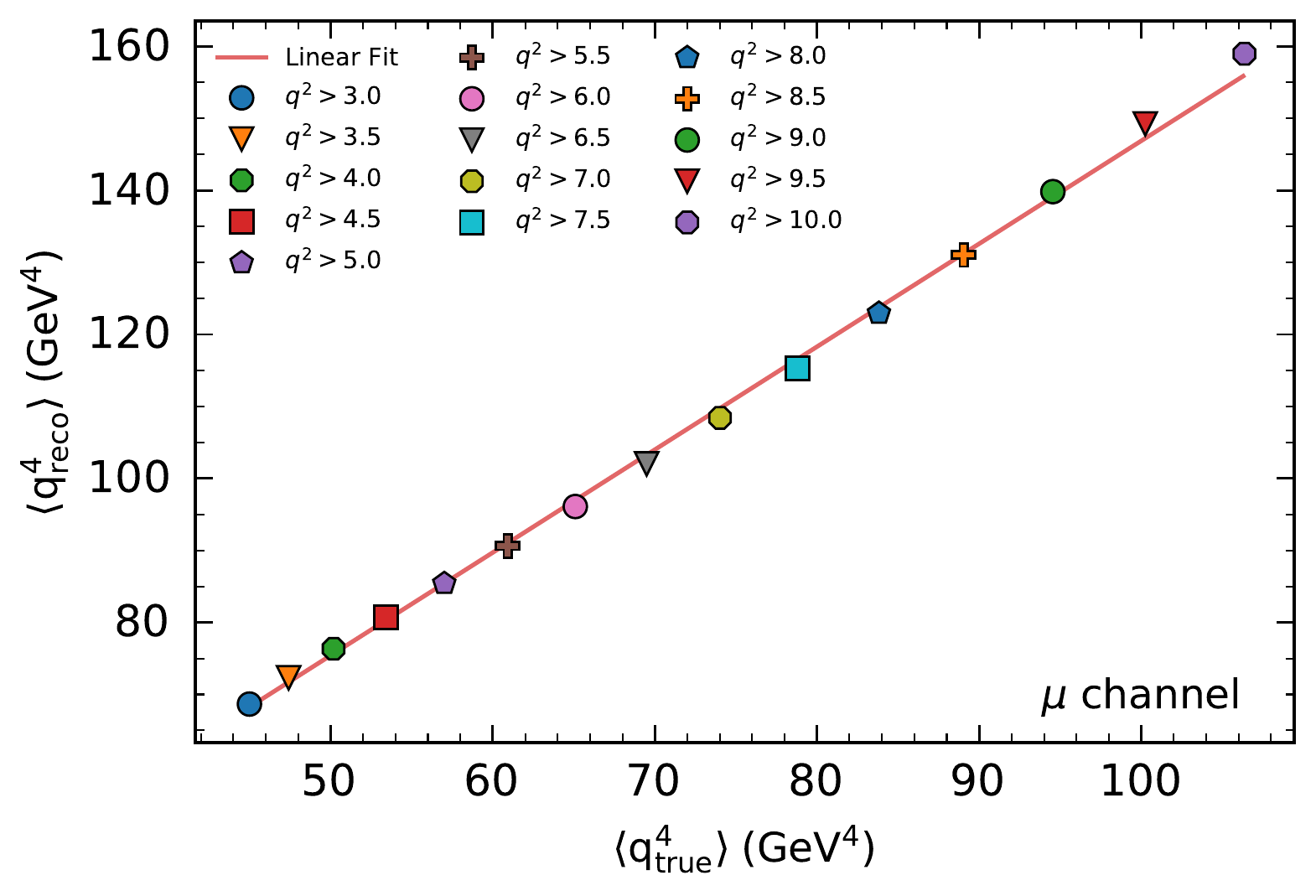}  
  \includegraphics[width=0.39\textwidth]{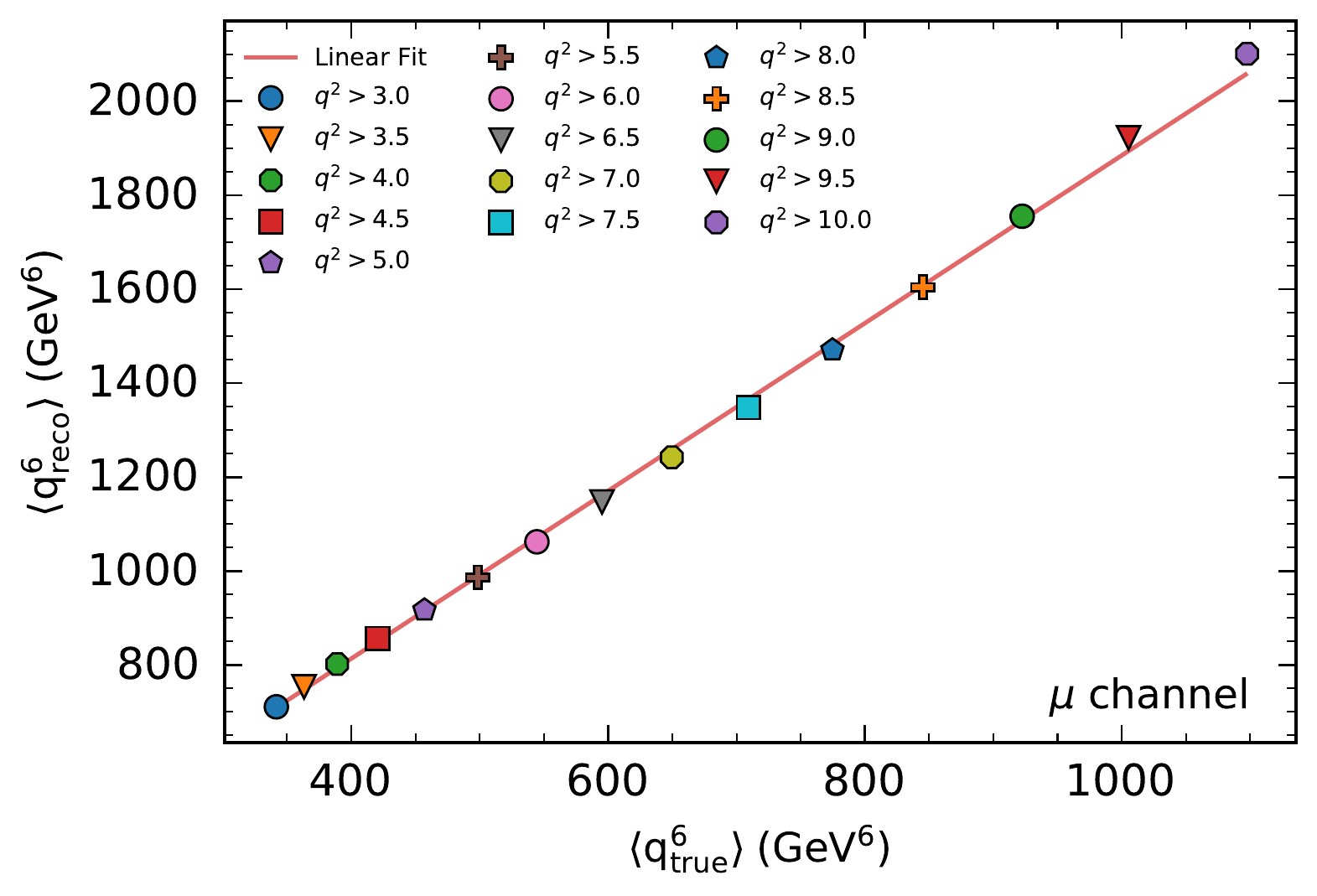}  
  \includegraphics[width=0.39\textwidth]{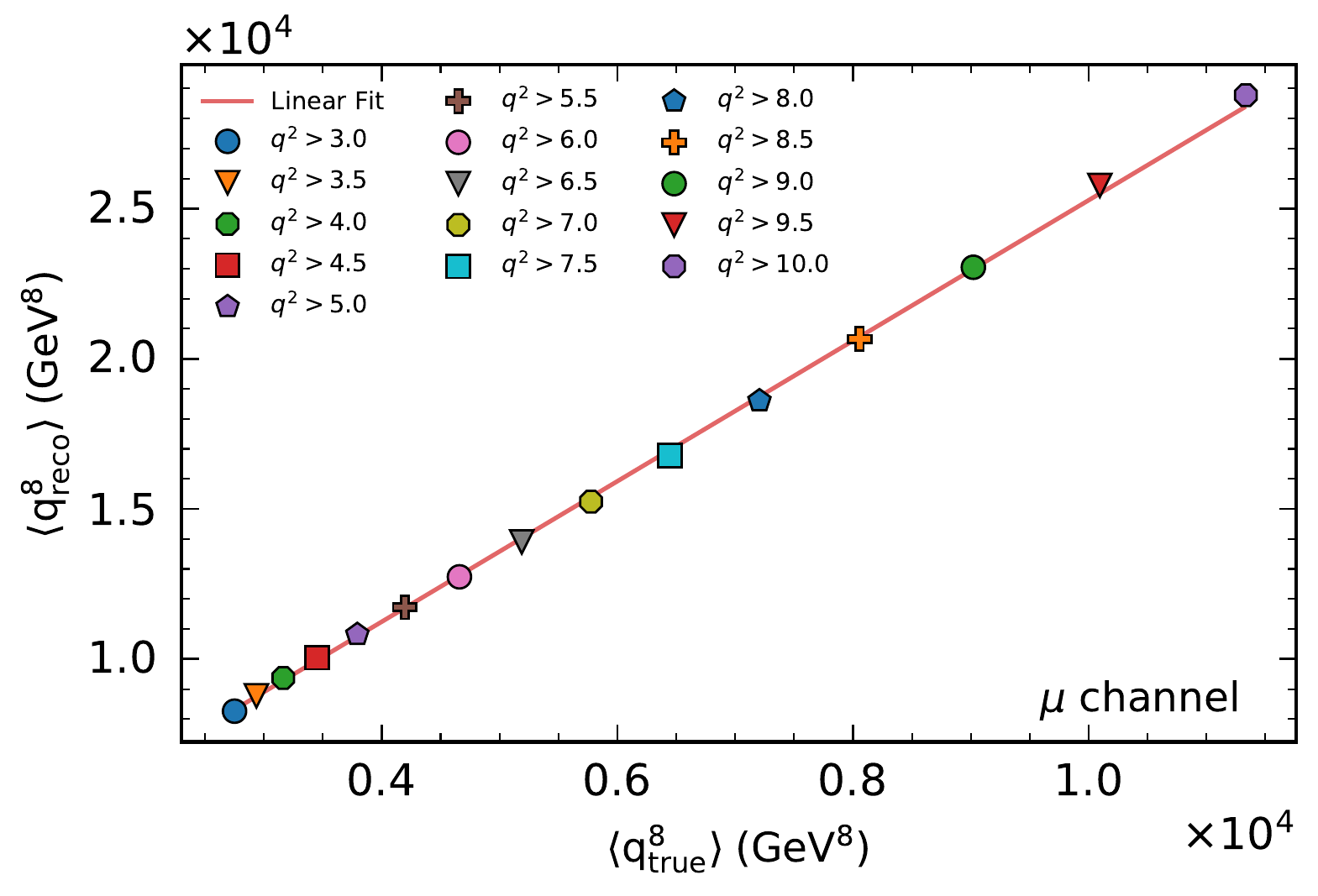} 
   \includegraphics[width=0.39\textwidth]{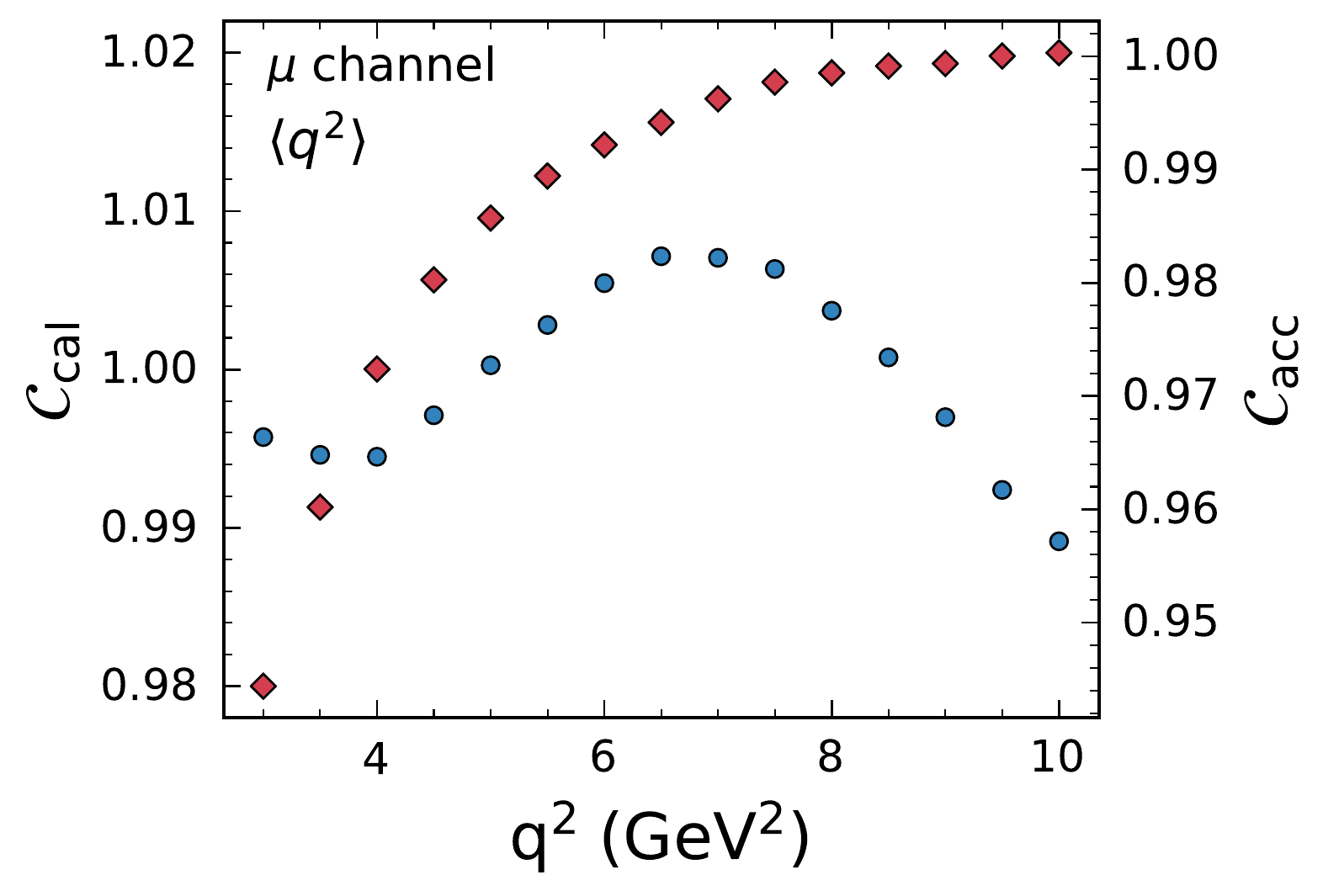}  
  \includegraphics[width=0.39\textwidth]{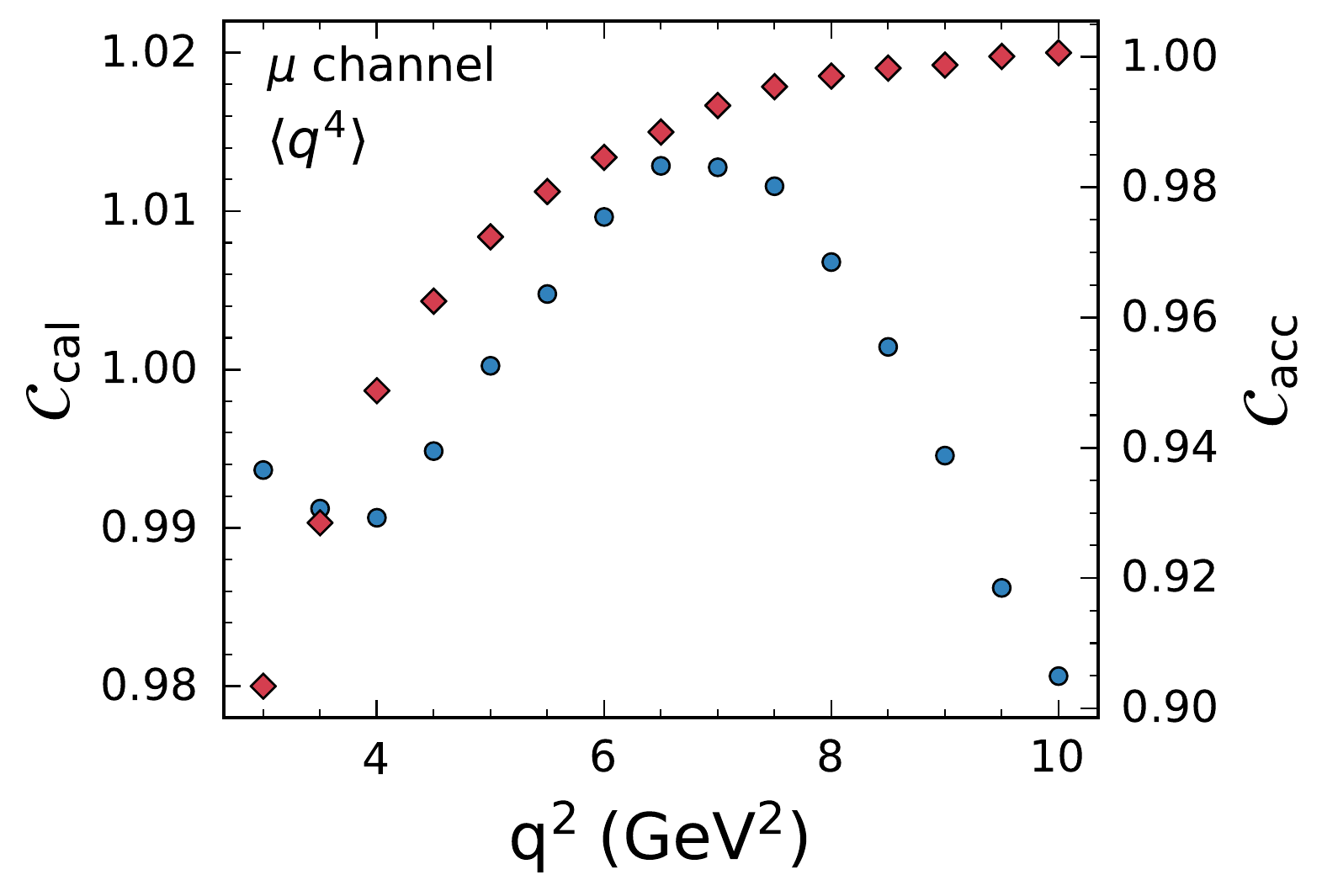}  
  \includegraphics[width=0.39\textwidth]{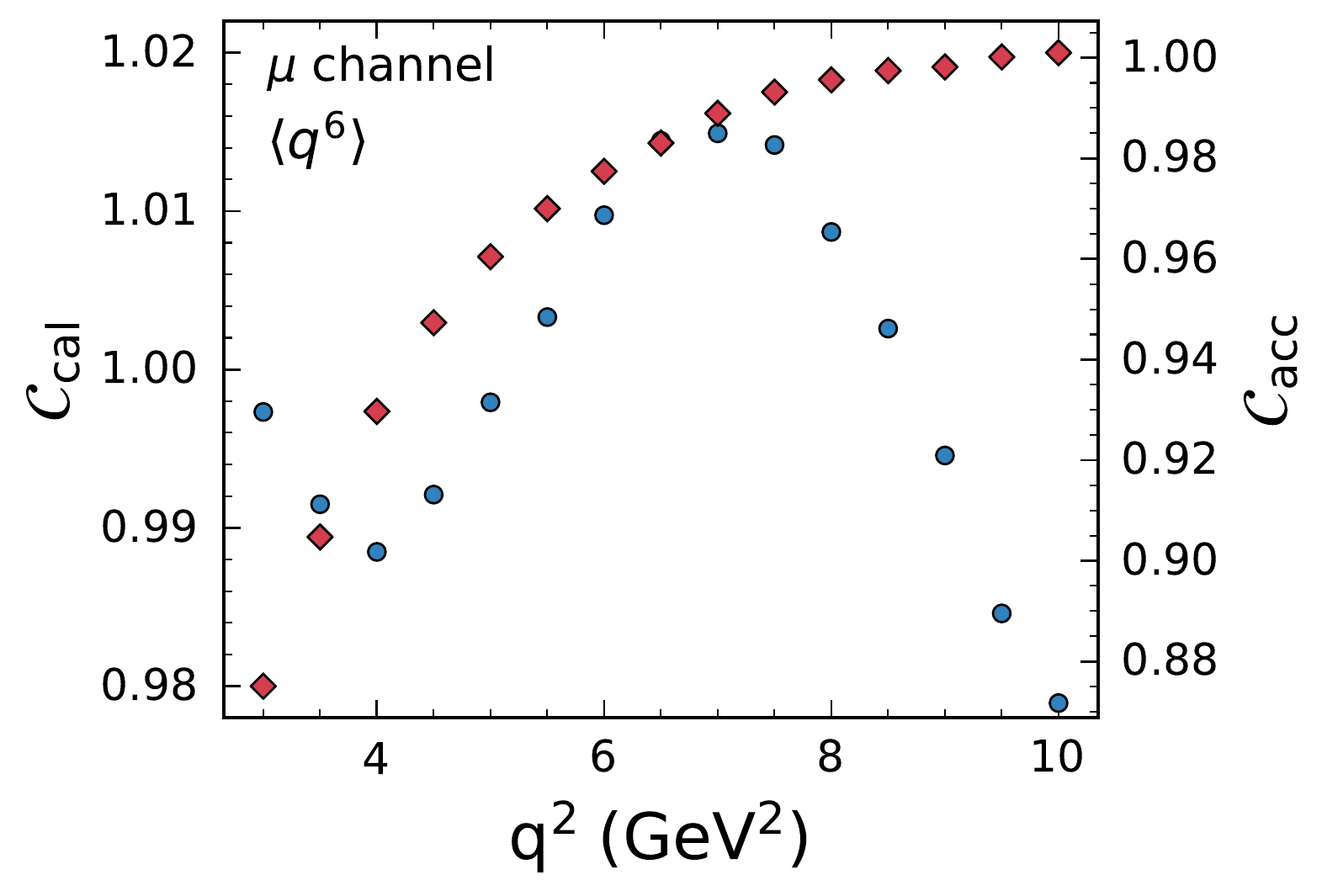}  
  \includegraphics[width=0.39\textwidth]{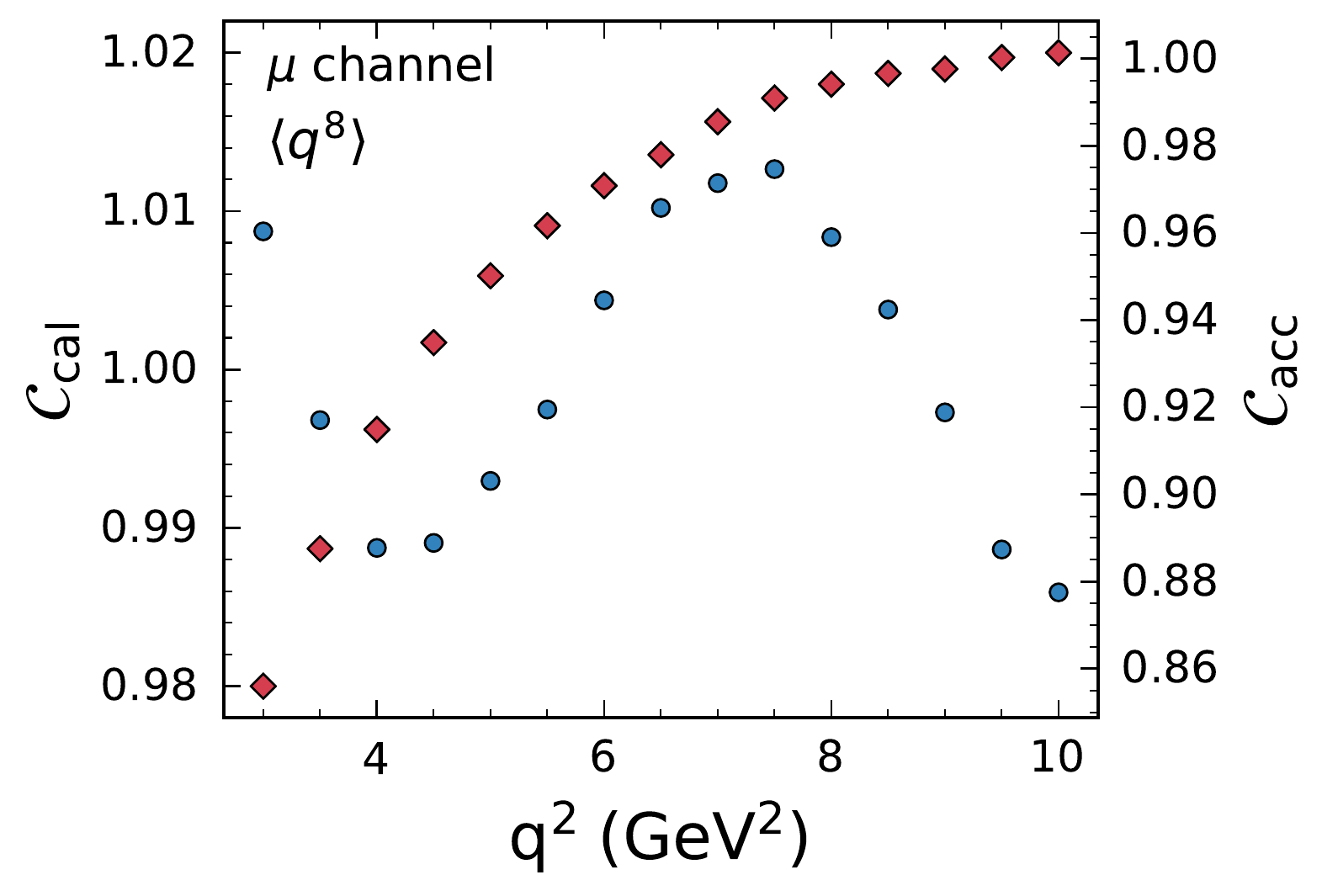}  
\caption{
 The calibration curves, bias and acceptance calibration factors for the first to the fourth moment of $q^2$ for muons.
 }
\label{fig:calib_1_2_3_4_mu}
\end{figure}

Fig.~\ref{fig:calib_1_2_3_4_mu} shows the calibration curves, the bias and acceptance calibration factors, $C_{\mathrm{cal}}$ and $C_{\mathrm{acc}}$, for muons for the various selections on $q^2$. Furthermore, Table~\ref{tab:calib_fit_results} summarizes the fitted parameters, $a_m$ and $b_m$, of the determined linear calibration functions for each order of moment $m$.

\begin{table}[]
\renewcommand\arraystretch{1.4}
\centering
    \caption{Summary of the fitted parameters of the determined linear calibration functions for electron and muon candidates with $a_m$ and $b_m$ denoting the offset and slope for moments of the order $m$. The values given in brackets denote the statistical uncertainty on the given parameters estimated from the post-fit covariance.}
    \label{tab:calib_fit_results}
    \begin{tabular}{lllll}
       \hline\hline
        & \multicolumn{2}{c}{Electrons} & \multicolumn{2}{c}{Muons} \\
        \hline
         $\langle q^{2m} \rangle$ & $a_{m}$ (GeV$^{2m}$)  & $b_{m}$ (GeV$^{2m}$)  & $a_{m}$ (GeV$^{2m}$)  & $b_{m}$ (GeV$^{2m}$)  \\
        \hline
        $\langle q^{2} \rangle$ & 1.245 (0.002) & -0.57 (0.02) & 1.200 (0.002) & -0.08 (0.02) \\
        $\langle q^{4} \rangle$ & 1.473 (0.004) & -1.6 (0.2) & 1.430 (0.004) & 3.9 (0.2) \\
        $\langle q^{6} \rangle$& 1.828 (0.006) & 33.0 (3.0) & 1.786 (0.006) & 99.0 (3.0) \\
        $\langle q^{8} \rangle$ & 2.386 (0.011) & 1027.0 (49.0) & 2.340 (0.010) & 1882.0 (50.0)  \\
        \hline\hline
    \end{tabular}
\end{table}

\section*{Appendix D. Selection efficiency of \bclnu\ components}
\begin{figure}[h!]
  \includegraphics[width=0.45\textwidth]{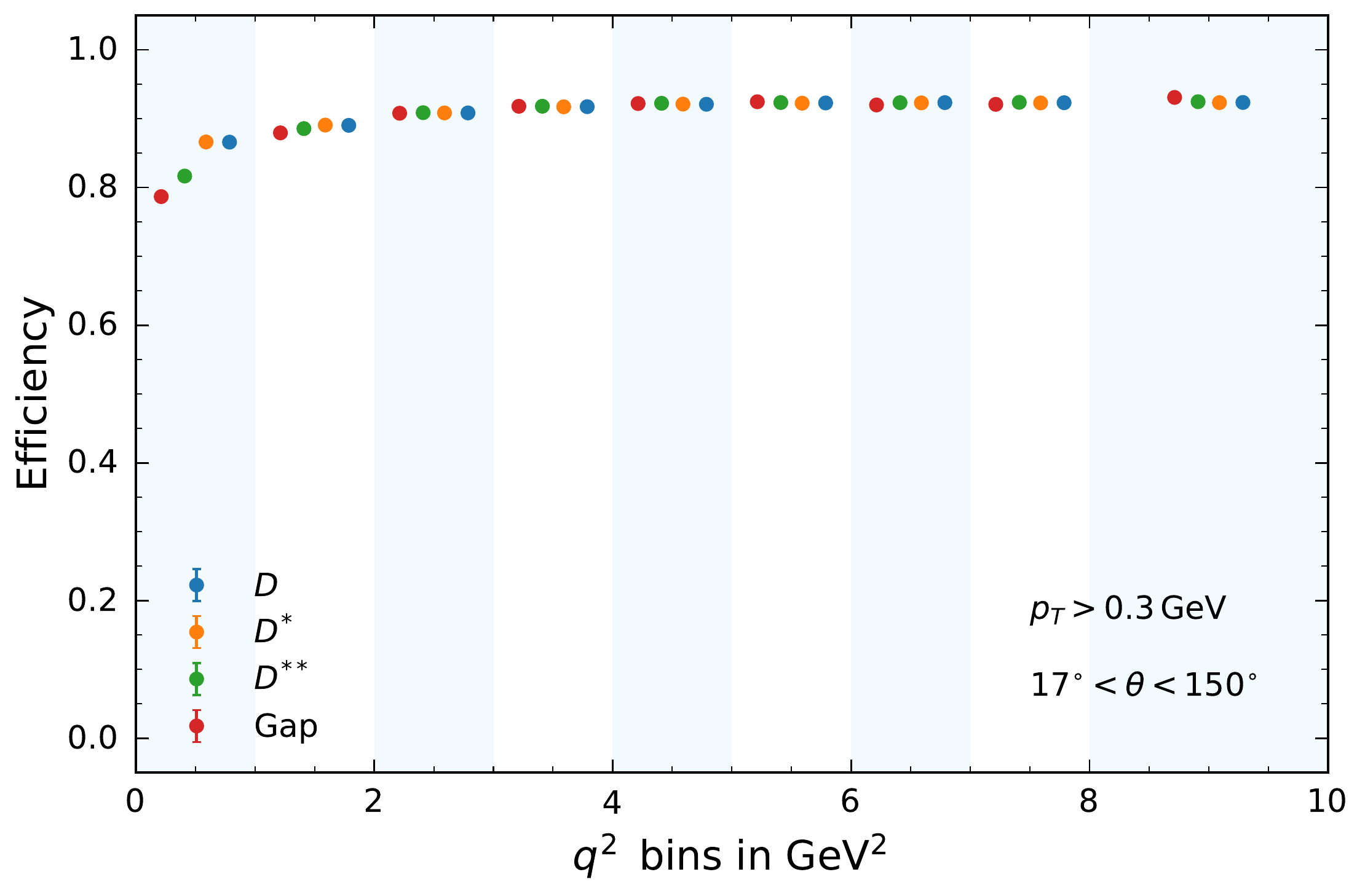}
 \includegraphics[width=0.45\textwidth]{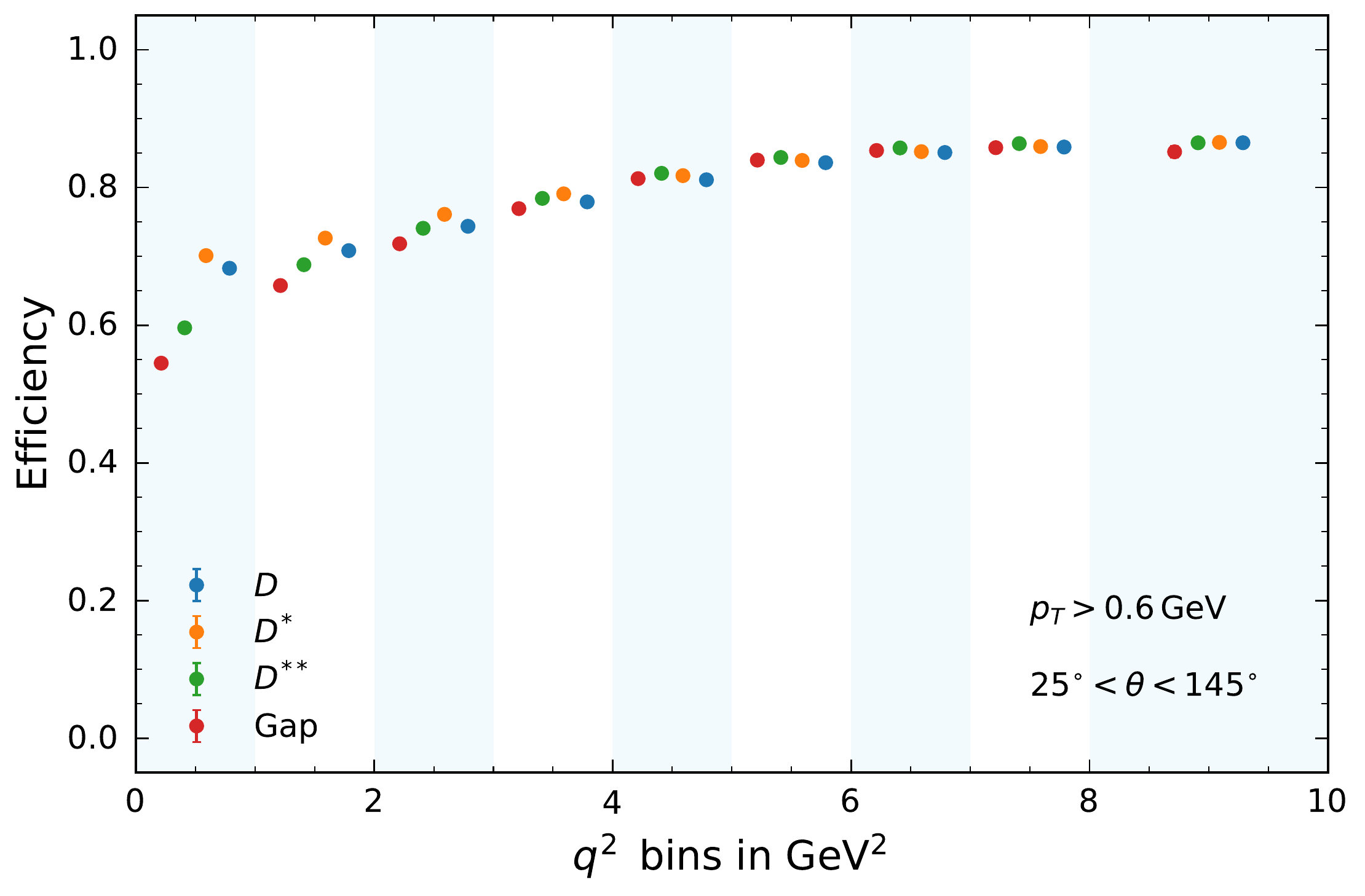}
\caption{Selection efficiencies of different \bclnu\  components for electron (left) and muon (right) final states with basic transverse momentum and angular acceptance requirements in bins of $q^{2}$.
}
\label{fig:efficiency}
\end{figure}

Selection efficiencies for different \bclnu\ components are shown in Fig.~\ref{fig:efficiency}  in bins of $q^{2}$. Events are required to pass basic transverse momentum and angular acceptance selection criteria. Different selection and acceptance efficiencies are observed in the low $q^{2}$ region for different \bclnu\ processes, especially for the non-resonant component.

\clearpage

\section*{Appendix E. Experimental Correlation Matrices}
\begin{figure}[b!]
  \includegraphics[width=1.\textwidth]{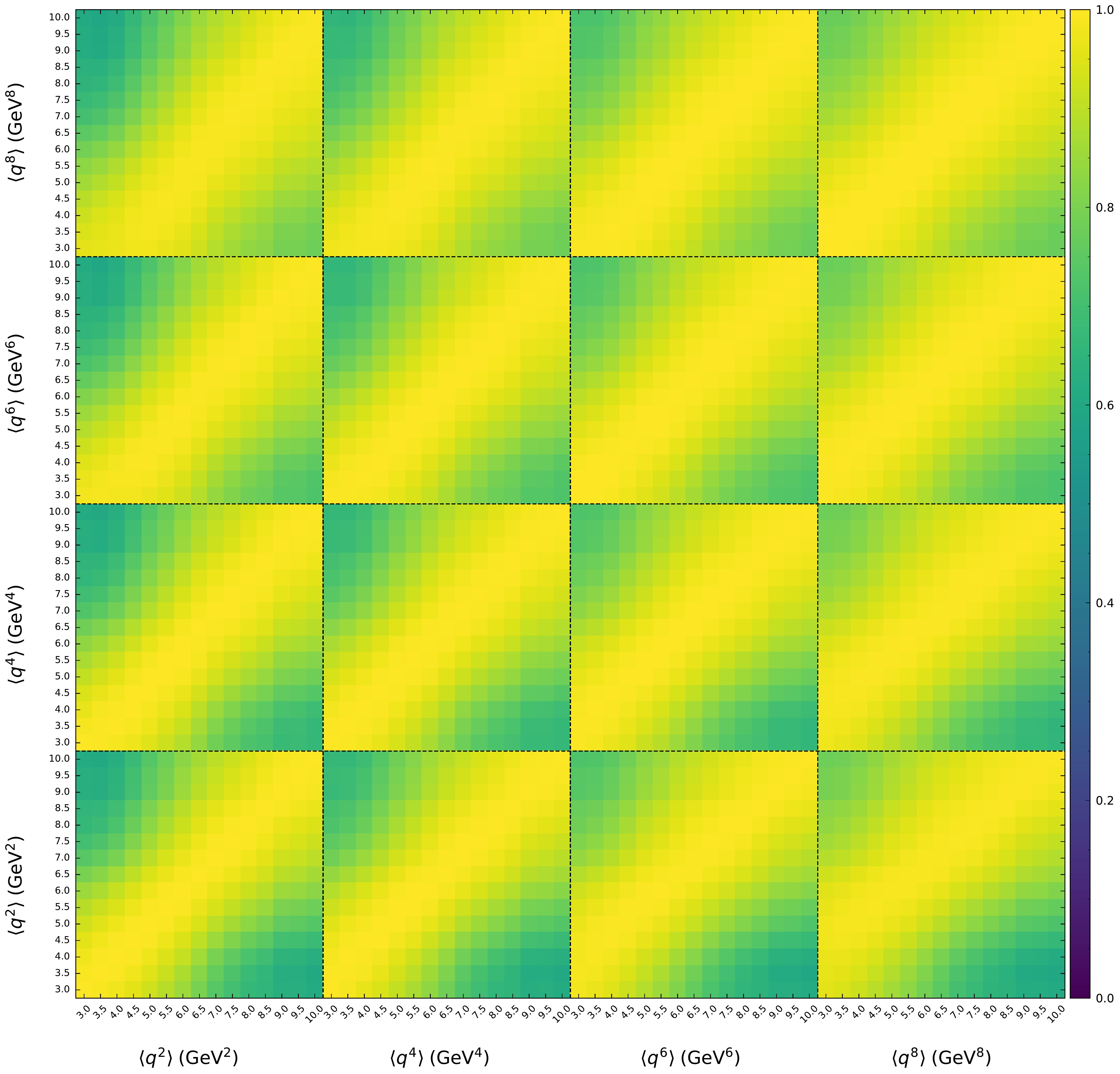}         
\caption{Color map of the full experimental correlation coefficients determined for the moments for the electron final state.
 }
\label{fig:sys_cov_el}
\end{figure}

\begin{figure}[b!]
  \includegraphics[width=1.\textwidth]{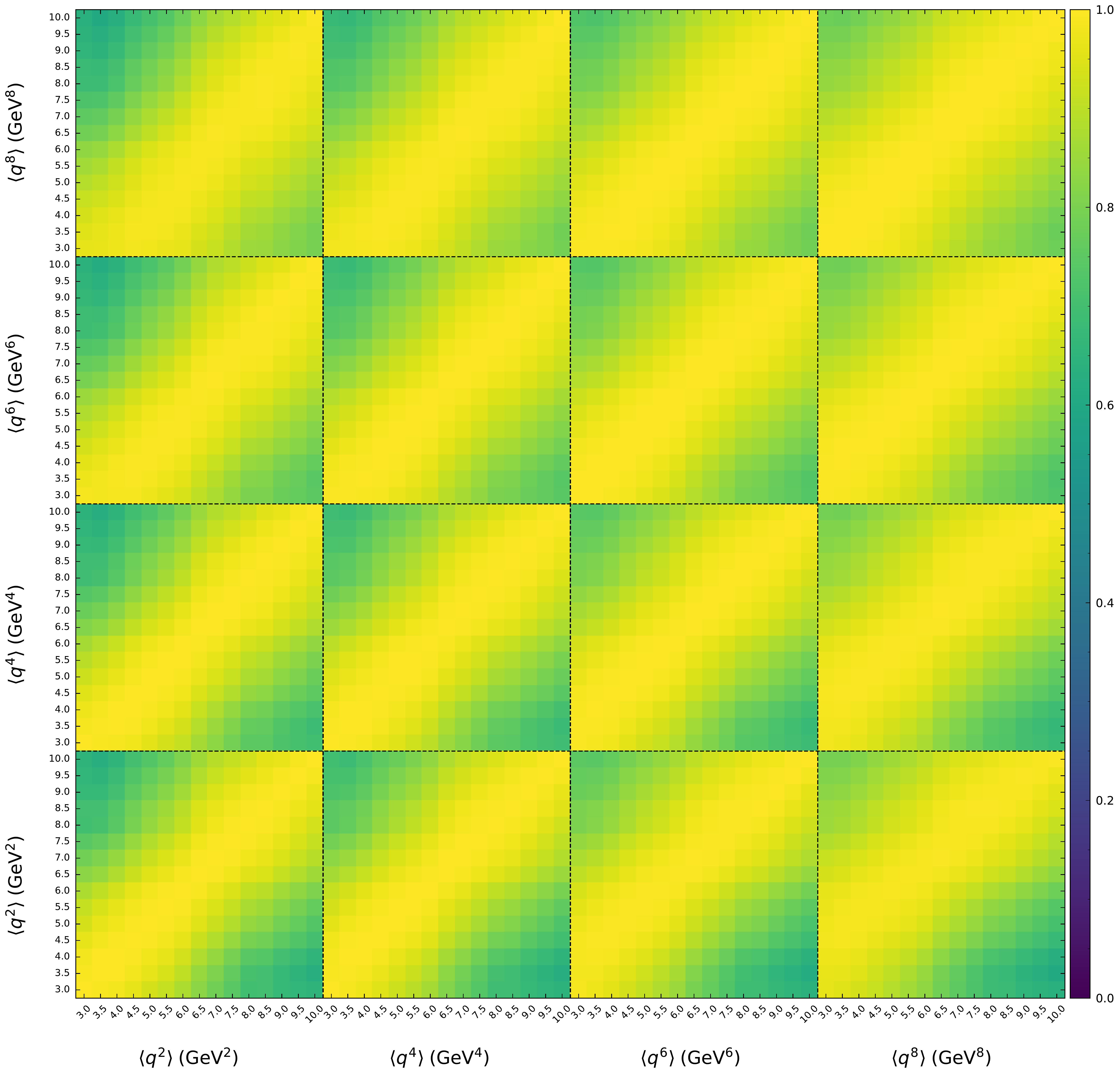}       
\caption{Color map of the systematic correlation coefficients determined for the moments for the muon final state.
 }
\label{fig:sys_cov_mu}
\end{figure}

The full systematic correlation matrices that are determined by combining the statistical and systematic covariance matrices are shown in Fig.~\ref{fig:sys_cov_el} and~\ref{fig:sys_cov_mu} for electrons and muons, respectively.

\clearpage

\section*{Appendix F. Central moments}
\begin{figure}[h!]
  \includegraphics[width=0.45\textwidth]{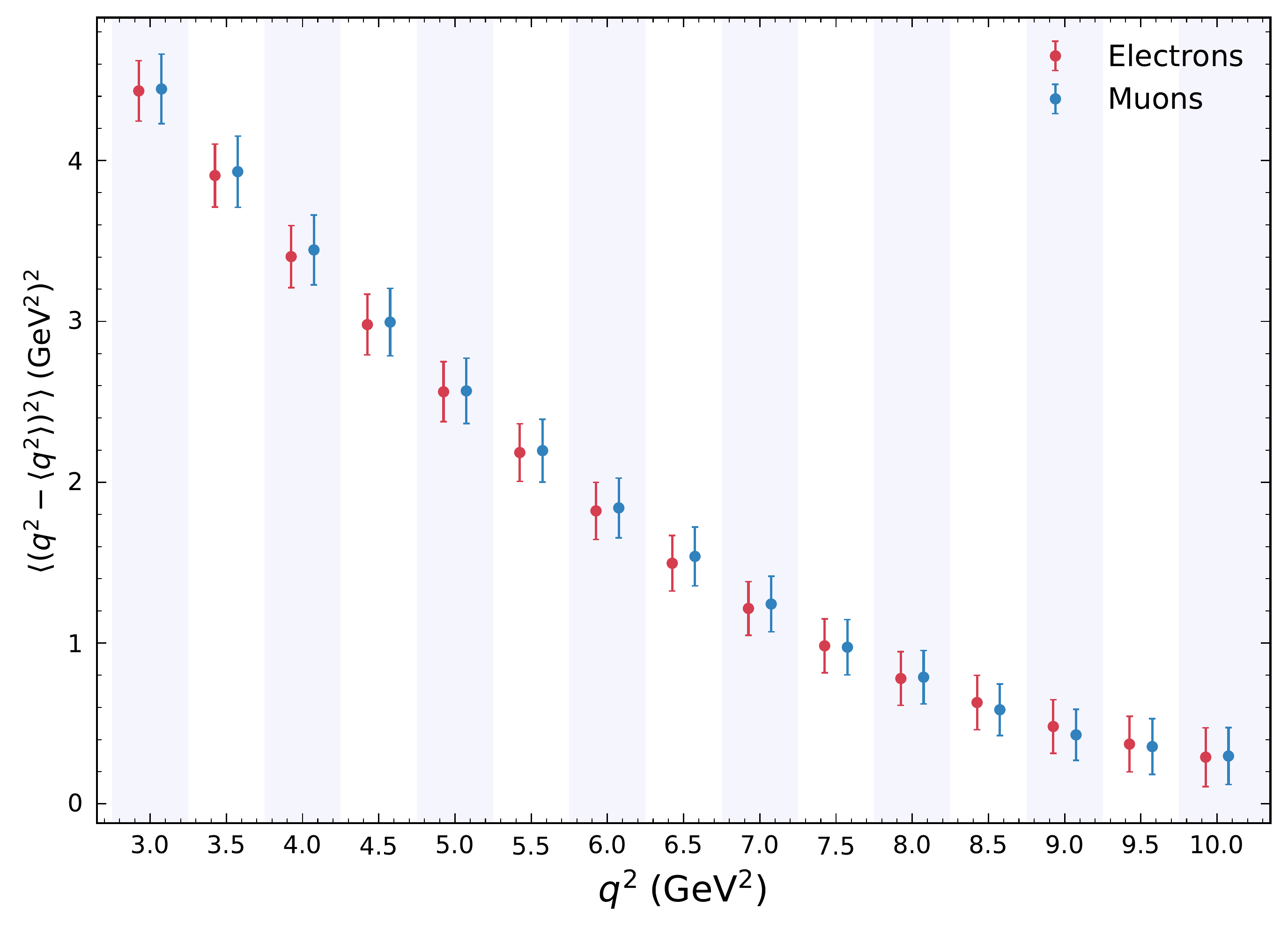}
 \includegraphics[width=0.45\textwidth]{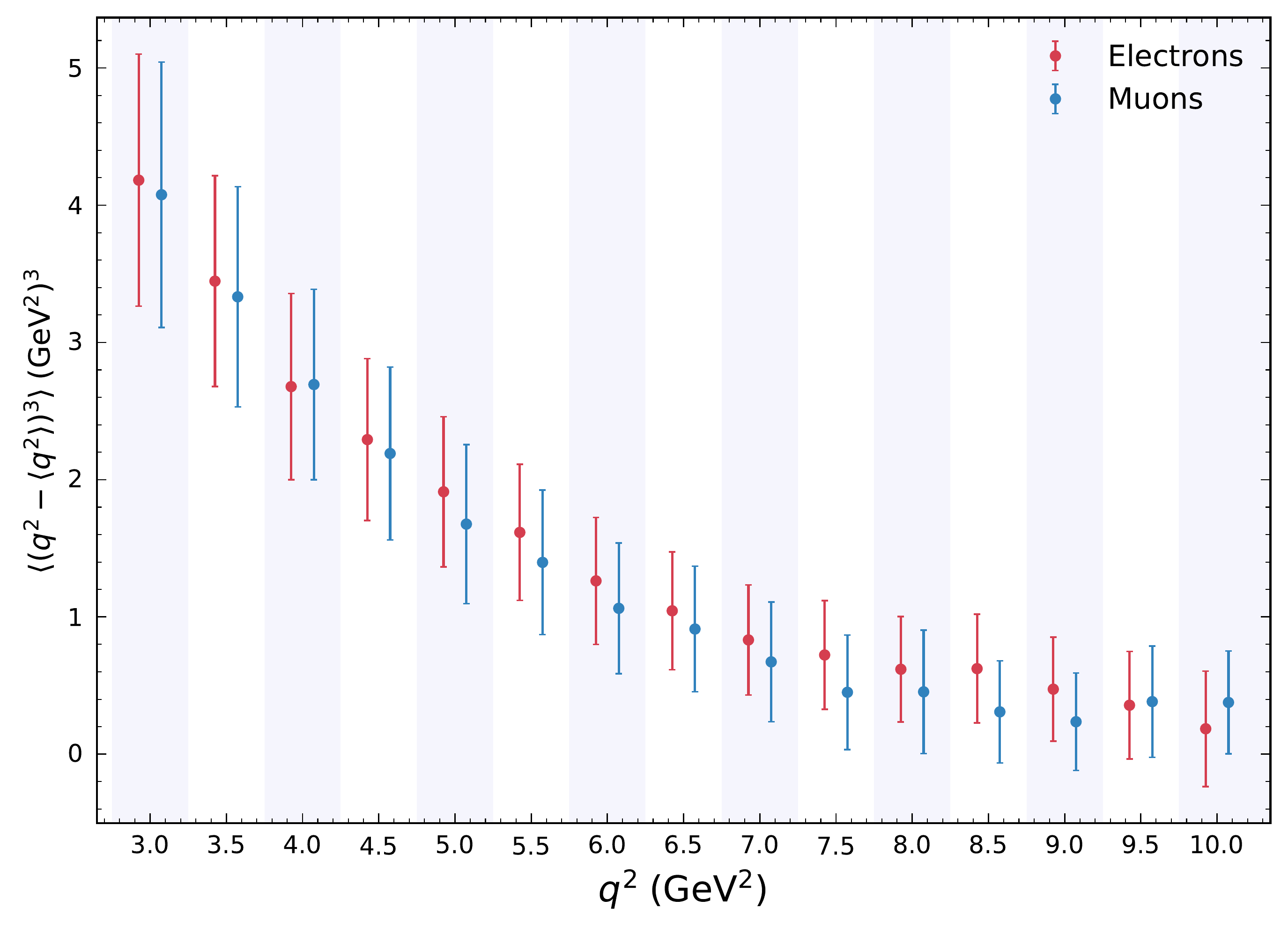}
 \includegraphics[width=0.45\textwidth]{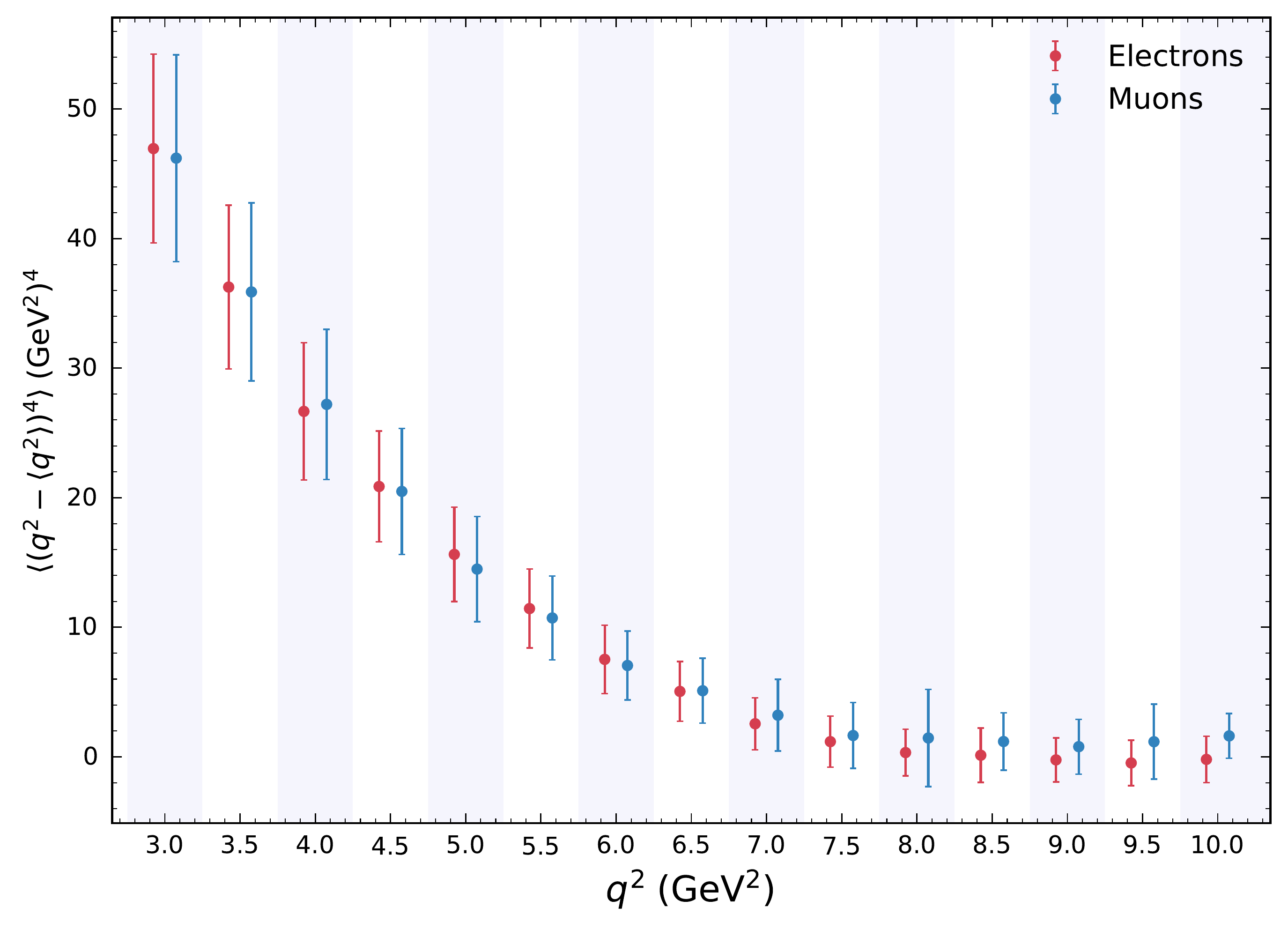}
\caption{The measured second (left), third (right), and fourth (bottom) central $q^{2}$ moments for both the electron and muon final states.
}
\label{fig:central_moments}
\end{figure}

The measured central $q^{2}$ moments, discussed in Section~\ref{sec:results}, are shown for both the electron and muon final states in Fig.~\ref{fig:central_moments}.

\end{appendix}


\end{document}